\documentclass[useAMS,usenatbib,usegraphicx]{mn2e}
\usepackage{amsmath}
\def \bea {\begin{eqnarray}}     
\def \ena {\end{eqnarray}}          
\def \bee {\begin{equation}}
\def \ene {\end{equation}}
\def    \bQ     {{\bf Q}}
\def    \be     {{\bf e}}

\def    \mc {\mbox{ cos }}
\def    \ms {\mbox{ sin }} 
\def    \mcs {\mbox{ cos}^{2}}
\def    \mss {\mbox{ sin}^{2}}

\def    \ma {\hat{\bf a}}               
\def    \me {\hat{\bf e}}
\def    \mr {{\bf r}}
\def    \mn {\hat{\bf N}}
\def    \mm {\hat{\bf n}}
\def    \mp {{\bf p}}
\title{Radiative Torques: Analytical Model and Basic Properties}

\author[A. Lazarian \& Thiem Hoang]{A. Lazarian
       $^1$\thanks{E-mail:lazarian@astro.wisc.edu}, and Thiem Hoang
       $^2$\thanks{E-mail: hoang@astro.wisc.edu} \\Department of Astronomy, University of Wisconsin, Madison,
       WI 53706, USA}

\begin{document}

\date{Draft version \today}            

\maketitle

\begin{abstract}
We attempt to get a physical insight into grain alignment processes by studying basic
properties of radiative torques (RATs). For
this purpose we consider a simple toy model of a helical              
grain that reproduces
 well the basic features of RATs.
The model grain consists of a spheroidal body with a mirror attached at an
angle to it. Being very simple, the model
allows analytical description of RATs that act upon it.
We show a good correspondence of RATs obtained for this model and those
of irregular grains calculated by DDSCAT. Our analysis of the role of 
different torque components for grain alignment reveals that one of the
three RAT components does not affect the alignment, but induces only
for grain precession. The other two components provide a generic alignment
with grain long axes perpendicular to the radiation direction, if the radiation
dominates the grain precession, and perpendicular to
magnetic field, otherwise.
The latter coincides with the famous predictions of the Davis-Greenstein
process, but our model does not invoke paramagnetic relaxation. In fact, we
identify a narrow range of angles between the radiation beam and
the magnetic field, for which the alignment is opposite to the
Davis-Greenstein
predictions. This range is likely to vanish, however,
in the presence of thermal wobbling of grains. In addition,
we find that a substantial part of
grains subjected to RATs gets aligned with low angular momentum,
which testifies, that most of the grains in diffuse interstellar medium
do not rotate fast, i.e. rotate with thermal or even sub-thermal velocities.
This tendency of RATs to decrease grain angular velocity
as a result of the RAT alignment  decreases the degree of polarization, by decreasing the degree of internal
alignment, i.e. the alignment of angular momentum with the grain axes.
For the radiation-dominated environments, we find
that the alignment can take place on the time
scale much shorter than the time of gaseous damping of grain rotation. This
effect makes grains a more reliable tracer of magnetic fields.
In addition, we study a self-similar
scaling of RATs as a function of $\lambda/a_{eff}$. We show that the self-similarity is useful for studying grain alignment by a broad spectrum of radiation, 
i.e. interstellar radiation field.
\end{abstract}


\section{Introduction}

Magnetic fields 
play a crucial role in many astrophysical processes, e.g. star formation,
accretion of matter, transport processes, including heat conduction and
propagation of cosmic rays. One of the easiest ways to study magnetic
field topology is via polarization of radiation arising from extinction
or/and emission by aligned dust grains. The new instruments like Scuba II (Bastien, Jenness \& Molnar 2005), Sharc II (Novak et
al. 2004) and an intended polarimeter for SOFIA open new horizons for tracing
of astrophysical magnetic fields with aligned grains.

In this situation it is unacceptable that the processes of grain alignment
are not completely understood (see review by Lazarian 2003). The enigma that
surrounds grain alignment since its discovery in 1949 (Hall 1949; Hilner 1949)
makes the interpretation of the polarization in terms of magnetic fields
somewhat unreliable. The failure of grains to
align at high optical depths was discussed, for instance, in Goodman (1995).

A recent progress in understanding of the grain alignment physics removed
many questions, but have not remedied the situation completely. Among the 
milestones let us mention the recent revival of interest to radiative torques (henceforth RATs).
Introduced by Dolginov \& Mytrophanov (1976), those torques, that arise from
the interaction of irregular grains with a flow of photons, 
were essentially forgotten till Draine \& Weingarter 
(1996, 1997, henceforth DW96, DW97, respectively) provided quantitative numerical studies. While Dolginov \& Mytrophanov (1976) were somewhat vague on what
makes RATs important for a grain, DW96 demonstrated that their arbitrary
chosen irregular grains exhibit dynamically important RATs when subjected to
a typical interstellar radiation field (ISRF). Very importantly, Bruce Draine 
incorporated RATs into the {\it publicly available} DDSCAT code (Draine \& Flatau
1994), which stimulated a further progress in the field. First laboratory
studies of RATs were reported in Abbas et al. (2004). 

The renewed interest to RATs coincided with a crisis of the paramagnetic
alignment as it is described in textbooks (Purcell 1979; Spitzer \& McGlynn 1979; Mathis 1986).
 Lazarian \& Draine (1999a) (hereafter LD99a) identified new  elements of grain dynamics,
which they termed ``thermal flipping'' and ``thermal trapping''. Due to
thermal wobbling arising from the dissipative 
coupling of grain vibrational and rotational degrees of freedom (Lazarian
1994; Lazarian \& Roberge 1997) grains smaller than a critical radius
$a_c$ flip frequently and thus average out uncompensated torques. These
torques, that were first
discussed by Purcell (1979), were considered essential to make otherwise 
inefficient paramagnetic alignment 
(Davis \& Greenstein 1951; Jones \& Spitzer 1967) to account for
the polarimetric observations. A new dissipative coupling mechanism related
to nuclear spins of constituent atoms, that was described in Lazarian \&
Draine (1999b) (henceforth LD99b), resulted in $a_c$ larger than the typical cut-off
scale for grains in diffuse interstellar medium (ISM). As the other
mechanisms, e.g. mechanical alignment (Gold 1951, review by Lazarian
2003 and references therein), have their limitations, this made the RAT alignment the only viable
mechanism to explain the ubiquity of interstellar polarization (DW97)
and possibly polarization arising from aligned dust in other astrophysical
environments (Lazarian 2007).  
       
The successes of RATs include a more recent work by
Cho \& Lazarian (2005),
where a substantial increase of the RAT efficiency with the grain
size was established. This work
 explained the sub-millimeter polarization data for quiescent starless cores
(Wart-Thompson et al. 2000) by
appealing to the differential RAT alignment of large grains. For such
cores the analysis of all other mechanisms in Lazarian, Goodman \& Myers (1997)
predicted only marginal degrees of alignment. The studies elaborating the
approach in Cho \& Lazarian (2005), e.g. Pelkonen, Juvela \& Padoan (2007),
Bethell et al. (2007), provided theory-motivated predictions of the degree of
alignment for numerically simulated molecular clouds and cores. 

However, the above explanation as well as other explanations (see Lazarian 2003) are based on the plausibility of arguments, rather than on
the rigorous RAT alignment theory. Indeed, DW96 considered
RATs as a means of spin-up. This induced a naive
explanation of RAT alignment action that could be perceived in some
of the papers that followed the DW96 study. 
There it was assumed that RATs were proxies of the
Purcell's torques (1979), that arise from the action of photons, rather
than from the action of H$_2$ formation over catalytic sites, as in the original mechanism. While the Purcell's torques
depend on the resurfacing and therefore short lived, RATs depend on
grain shape and can be long-lived. As a result, long-lived fast rotation of paramagnetic
grains should induce good paramagnetic alignment (Purcell 1979). This
understanding of RATs is not correct, as it is clear from a more careful
reading of DW96 and DW97.   

In fact, RATs can be subdivided into the parts that arise from isotropic
and anisotropic radiation fluxes. The part arising from anisotropic radiation,
for which we adopt a shorthand notation ``isotropic part'', is, indeed, similar
to the Purcell's torques. The ``anisotropic part'' is, however, both
usually stronger and has properties 
different from the Purcell's torques. The
major difference arises from the fact that RATs
are defined in the laboratory, rather
than in the grain coordinate system. Thus, the presence of even a small
anisotropic component of radiation, which is a natural
condition for any realistic astrophysical system, is bound to change
the dynamics of grain. Note, that the alignment by the anisotropic
radiation was first discussed by Doginov \& Mitraphanov (1976). They, however,
concluded that prolate and oblate grains can be aligned differently. Lazarian
(1995) took into account internal relaxation and claimed that both prolate
and oblate grains should be aligned with longer axes perpendicular to magnetic
field. Nevertheless, the theory lacked a proper description of RATs.

DW97 demonstrated numerically
that in the presence of anisotropic radiation
the grains can be aligned
by RATs in respect to magnetic field on the time scales much {\it shorter}
than the time scale for paramagnetic alignment\footnote{DW97 identifies this
time-scale with the gaseous damping time. In \S 5.5 we show that the
 alignment could happen much faster in the presence of strong radiation sources}. In general, the magnetic field for the RAT alignment
acts through inducing fast Larmor precession; the alignment 
potentially may happen both with long grain axes parallel and perpendicular
to magnetic field. Only the latter is consistent with polarimetric
observations, however (see Serkowski, Mathewson \& Ford 1975).

In the DW97 study, the alignment with longer grain axes perpendicular
to magnetic field (``right alignment'') happened more frequently
than the grain alignment with longer grain axes parallel to magnetic field 
(``wrong alignment'').
This experimental evidence, based on a limited sampling, raised
worrisome questions.
Is this a general property of radiative torques
or just a coincidence? Do we expect to see more of ``wrong alignment'' if the
grain environment is different from the interstellar one? What are the
chances that we are fooled by the ``wrong alignment'' while interpreting 
the polarimetry measurements in terms of the underlying magnetic fields?
It seems necessary to address these questions
if interpreting polarimetry data in terms
of underlying magnetic fields is sought.

Analytical calculations played an important role for formulating the models
of both paramagnetic and mechanical alignment (see Davis \& Greenstein 1951,
Jones \& Spitzer 1967, Purcell 1979, Spitzer \& McGlynn 1979).
 Although such calculations
 dealt with
intentionally idealized models of grains, they allowed deep insight into the relevant physics. 
Dolginov \& Mytrophanov (1976) attempted an analytical modeling
for RATs. They used a model
 grain containing two ellipsoids connected together at an angle. 
The radiative torques were calculated for such a model grain by assuming that
the wavelength is much larger than the grain size, i.e., in the Rayleigh-Hans approximation (Dolginov \& Silantev 1976). However, adopting their shape,
we could not reproduce numerically their analytical predictions for RATs.
This induce us to seek analytical models that would correspond to the 
DDSCAT calculations.

Our approach in the present paper is to provide a physical insight into basic
RAT properties. In \S 2 we explain why we consider only RATs due to anisotropic radiation.
Then, we describe a simple grain model that is subject to RATs and allows
analytical descriptions (\S \ref{amo1}). In \S \ref{sec33} we present calculations of RATs for 
a number irregular shapes and study the correspondence of their RATs with AMO.  We also briefly consider possible generalizations of our model (\S \ref{amo2}). In \S \ref{kalign}  we analyze the
alignment for both this model and irregular grains with respect to the radiation direction. In \S \ref{balign} we study the alignment of our model and irregular grains with respect to magnetic
field. Crossovers are studied in \S \ref{cross}, while we identify the conditions
for the magnetic field or the radiation direction
to act as the axis of alignment in \S \ref{sec9}.
As, even with modern computers, the calculations of RATs
for a variety of wavelengths is time consuming, 
we address the question of the accuracy of presenting radiative torques
as the function of the ratio of grain size to the wavelength $\lambda/a_{eff}$
in \S \ref{self}. 
The discussion of our results and the summary are provided in 
\S \ref{discuss} and \S 12, respectively.    

\section{Isotropic and Anisotropic RATs}

RATs can emerge even when the radiation field is isotropic. Devices similar to those used by Lebedev (1901) to measure radiation
pressure experience torques in the presence of the isotropic radiation (see also
the cartoon of a model with absorbing and reflecting strips in DW96). 

The dynamics of an irregular grain subjected to isotropic radiation is very similar
to a grain subjected to the Purcell's torques arising, for instance, from
H$_2$ formation. For instance, one would expect to have thermal trapping
of sufficiently small grains due to thermal fluctuations as described
in LD99a. Therefore RATs induced by isotropic radiation
(henceforth ``isotropic RATs'') only marginally alter
the problems that the paramagnetic alignment mechanism faces in explaining
observational data. In addition, as we mentioned above, the ``isotropic RATs'' are usually weaker
than those that arise when a grain is subjected to anisotropic radiation (see DW97).

Due to the situation described above, for the rest of the paper we shall 
associate RATs only with the part arising from anisotropic radiation, as, for instance, was done
in Cho \& Lazarian (2005). In other words, we treat the torques arising from isotropic radiation
as a particular realization of the Purcell's torques. 

\begin{figure}
\includegraphics[width=0.49\textwidth]{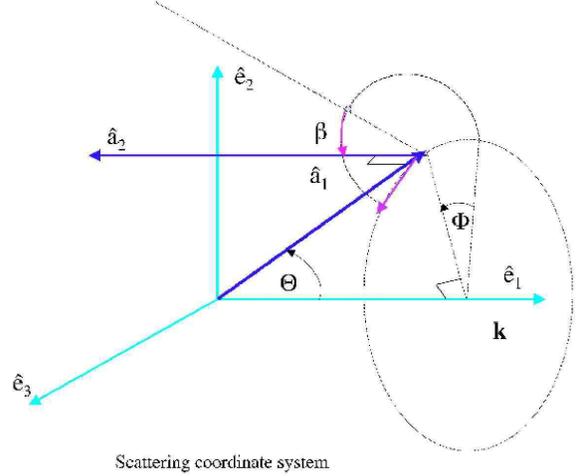}
\caption{The orientation of a grain, described by three principal axes
  $\hat{a}_{1},\hat{a}_{2}, \hat{a}_{3}$, in the laboratory coordinate system (scattering reference
system)
  $\hat{e}_{1},\hat{e}_{2}, \hat{e}_{3}$ is
  defined by three angles $\Theta, \beta, \Phi$. The
  direction of incident photon beam ${\bf k}$ is along $\hat{e}_{1}$.} 
\label{f1} 
\end{figure}

Let us now introduce briefly some basic definitions. RAT ${\bf \Gamma}_{rad}$ is defined by
\bea
{\bf \Gamma}_{rad}=\frac{\gamma u_{\lambda}\lambda a_{eff}^{2}}{2}{\bf Q}_{\Gamma},\label{eq1}
\ena
where ${\bf Q}_{\Gamma}$ is the RAT efficiency, $\gamma$ is the anisotropy degree, and  $u_{\lambda}$ is the energy density of
radiation field of the wavelength $\lambda$. Here $a_{eff}$ is the effective
size of the grain which is defined as the radius of a sphere of the same
volume with the irregular grain (similar to DW97). In general, ${\bf Q}_{\Gamma}$ is a
function of angles $\Theta, \beta, \Phi$ in which $\Theta$ is the angle
between the axis ${\bf a}_{1}$ corresponding to the maximal moment of inertia
(henceforth maximal inertia axis) with respect to the radiation
direction ${\bf k}$,
$\beta$ is the rotation angle of the grain around ${\bf a}_{1}$, and $\Phi$ is
the precession angle of ${\bf a}_{1}$ about ${\bf k}$ (see Fig. \ref{f1}). To help the
reader familiar with the earlier works on RATs, wherever possible, we use the
same the notations as in DW96 and DW97.

The RAT efficiency can be decomposed into components in the scattering system via
\begin{align}
{\bf Q}_{\Gamma}(\Theta, \beta,\Phi)&=Q_{e1}(\Theta,\beta,0)\hat{e}_{1}\nonumber\\
&+Q_{e2}(\Theta, \beta, 0)(\hat{e}_{2}\mc\Phi+\hat{e}_{3}\ms\Phi)\nonumber\\
&+Q_{e3}(\Theta, \beta, 0)(\hat{e}_{3}\mc\Phi-\hat{e}_{2}\ms\Phi), \label{eq2}
\end{align}
where $\hat{e}_{1}$, $\hat{e}_{2}$, $\hat{e}_{3}$ are shown in Fig.
\ref{f2}. In addition, for the sake of simplicity, we have denoted
$Q_{e1}(\Theta,\beta,0)\equiv{\bf Q}_{\Gamma}(\Theta,\beta,0).\hat{e}_{1}$,
$Q_{e2}(\Theta,\beta,0)\equiv{\bf  Q}_{\Gamma}(\Theta,\beta,0).\hat{e}_{2}$,
$Q_{e3}(\Theta,\beta,0)\equiv{\bf Q}_{\Gamma}(\Theta,\beta,0).\hat{e}_{3}$. In what
followings, we use $Q_{e1}, Q_{e2}, Q_{e3}$ for the RAT components and keep in mind that
they are functions of $\Theta, \beta$ at $\Phi=0$.  However, in some
particular cases, these angles will be explicitly written.

\section{Introducing Analytical model (AMO) of a helical grain}\label{amo1}

\begin{figure}
\includegraphics[width=0.49\textwidth]{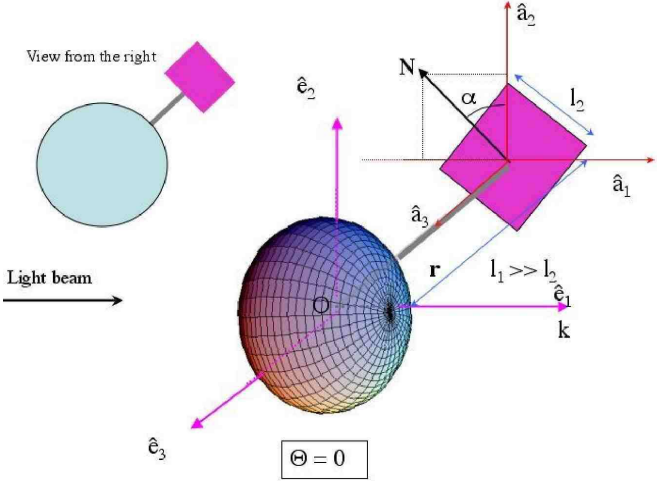}
\caption{Our grain model consists of a mirror connected to an oblate spheroid by a weightless 
rod. The distance between the mirror and the spheroid is assumed to be much
larger than the mirror's size. Both the mirror and the spheroid are perfectly reflecting.}
 \label{f2}
\end{figure}

Let us consider an asymmetric grain shape consisting of a reflecting
spheroid and a square mirror with the side $l_{2}$ attached on a pole of
the length  $l_{1}$. For the sake of simplicity, we assume that the mirror and
the pole are weightless. Also, both the mirror and the spheroid are assumed to
be perfectly reflecting. Moreover, we neglect
the shadowing of the mirror by the grain by assuming that $l_{1}\gg l_{2}$ (see
Fig. \ref{f2}).

\subsection{RATs from a reflecting spheroidal body} 
Consider first RATs acting on an oblate spheroidal body. As a consequence of
its symmetries, such a grain is not expected to
exhibit any spin-up arising from RATs, provided that the incident radiation is
not circularly polarized. The latter will be
our assumption for the rest of the paper.

Consider a photon beam of wavelength $\lambda$, propagating in the ${\bf k}$-direction parallel
to the axis $\hat{e}_{1}$ of the lab coordinate system (see Figs \ref{f1} and \ref{f2}). The
momentum of a photon is deposited to the grain as it reflects from the grain
surface. As a result, the grain experiences a net torque. 

For simplicity, we assume that the grain rotates fast around the maximal
inertia axis, so averaging over such a rotation is suitable. Also, from
equation ({\ref{eq2}) it follows that we only need to find RATs for
  $\Phi=0$. Therefore, the resulting RAT is
\begin{align}
{\bf \Gamma}_{rad}&=\frac{\gamma u_{rad}\lambda b^{2}}{2}(Q_{e1}\me_{1}+Q_{e2}\me_{2}+Q_{e3}\me_{3}),\label{eq3a}
\end{align}
where $b$ is the length of the major axis of the spheroid, RAT components are given by (see Appendix A for their derivation)
\begin{align}
Q_{e3}&=\frac{2ea}{\lambda}(s^{2}-1)K(\Theta, e)\ms2\Theta,\label{eq3}\\
Q_{e1}&=0,\label{eq4}\\
Q_{e2}&=0,\label{eq5}
\end{align} 
where $s=a/b<1$ is the ratio of the minor to the major axes, $K(\Theta, e)$ is
a fitting function depending on $\Theta$ and the eccentricity of the oblate spheroid (see the lower panel in Fig. \ref{ap1}).

Following equations (\ref{eq3})-(\ref{eq5}) we see that a reflecting spheroidal grain does not produce
any $Q_{e1}, Q_{e2}$, but $Q_{e3}$ only. 
Therefore, it is easy to see that 
the only effect of RATs on the spheroidal grain is to cause the precession
in the plane perpendicular to the radiation direction ${\bf k}\|\me_{1}$.

\subsection{Torques from a reflecting mirror}
Consider now torques that act upon the perfectly reflecting mirror attached at an angle to the oblate spheroid
(see Fig.~\ref{f2}). The pole is considered too thin to interact with the
radiation.\footnote{In our model, the only purpose of the existence of the
  pole is to minimize the effects of shadowing of the oblate grain core by the mirror.}
The normal unit vector $\mn$ which determines the orientation of the mirror in
the grain coordinate system is given by
\bea
\mn =n_{1}\ma_{1}+n_{2}\ma_{2},
\ena
where $\ma_{1}, \ma_{2},\ma_{3}$ are the principal axes of the grain
(i.e., the principal axes of the spheroid because the mirror and the pole are weightless). Here  $n_{1}=\ms
\alpha, n_{2}=\mc\alpha$ with $\alpha$ is the angle between $\mn$ and $\ma_{2}$ (see Fig. \ref{f2}).

Due to the rotation, the cross section of the mirror with the surface area
$A$, varies as (see Appendix B)
\begin{align}
A_{\perp}=A |\me_{1}. \mn|=A |n_{1} \mc\Theta-n_{2} \ms\Theta \mc\beta|,\label{eq7}
\end{align}

Following the same above procedure (see Appendix B for detail), we get RAT
\begin{align}
{\bf \Gamma}_{rad}&=\frac{\gamma u_{rad}\lambda l_{2}^{2}}{2}(Q_{e1}\me_{1}+Q_{e2}\me_{2}+Q_{e3}\me_{3}),\label{eq7b}
\end{align}
where  $l_{2}$ is the size
of the square mirror, $l_{1}$ is the length of the pole, and RAT components are given by
\begin{align}
Q_{e1}&=\frac{4l_{1}}{\lambda}|n_{1}\mc\Theta-n_{2}\ms\Theta\mc\beta|[n_{1}n_{2}\mcs\Theta\nonumber\\
&+\frac{n_{1}^{2}}{2}\mc\beta\ms2\Theta-\frac{n_{2}^{2}}{2}\mc\beta\ms2\Theta\nonumber\\
&-n_{1}n_{2}\mss\Theta\mcs\beta],\label{eq8}\\
Q_{e2}&=-\frac{4l_{1}}{\lambda}|n_{1}\mc\Theta-n_{2}\ms\Theta \mc\beta|[n_{1}^{2}\mc\beta\mcs\Theta\nonumber\\
&-\frac{n_{1}n_{2}}{2}\mcs\beta\ms2\Theta-\frac{n_{1}n_{2}}{2}\ms2\Theta+n_{2}^{2}\mc\beta\mss\Theta],\label{eq9}\\
Q_{e3}&=-\frac{4l_{1}}{\lambda}|n_{1} \mc\Theta-n_{2} \ms\Theta \mc\beta|
n_{1}\ms\beta[n_{1}\mc\Theta\nonumber\\
&-n_{2}\mc\beta\ms\Theta].\label{eq10}
\end{align}

 For $\Theta=0, \pi$, the RAT components for the mirror are 
\begin{align}
Q_{e1}&=\frac{4 l_{1} n_{1}^{2}n_{2}}{\lambda},\label{eq11}\\
Q_{e2}&=-\frac{4 l_{1} n_{1}^{3}}{\lambda}\mc\beta,\label{eq12} \\
Q_{e3}&=-\frac{4 l_{1} n_{1}^{3}}{\lambda}\ms\beta,\label{eq13}
\end{align}
Equations (\ref{eq11}), (\ref{eq12}) and (\ref{eq13}) reveal that $Q_{e1}$ does not depend on $\beta$, but
$Q_{e2}$, $Q_{e3}$ are periodic functions of $\beta$. As a result,
when averaging over the rotation angle $\beta$ from $0$ to $2\pi$, $Q_{e2},
Q_{e3}$ vanish for $\Theta=0, \pi$ (see a proof based on
 more general symmetry considerations
in \S \ref{sec42}).

For arbitrary $\Theta$, assuming that the rotation of the grain around the shortest axis is very fast, we can
average RATs over $\beta$ from $0$ to $2\pi$. The resulting RAT components are given by
\begin{align} 
Q_{e1}&=\frac{4\pi l_{1} n_{1}n_{2}}{\lambda} (3 \mbox{cos}^{2} \Theta - 1) f(\Theta, \alpha),\label{eq14}\\ 
Q_{e2}&=\frac{4\pi l_{1} n_{1}n_{2}}{\lambda} \ms 2\Theta g(\Theta, \alpha),\label{eq15}\\ 
Q_{e3}&=0,\label{eq16a}
\end{align} 
where $f(\Theta, \alpha), g(\Theta, \alpha) $ are fitting functions depending on
$\Theta, \alpha$ which characterize the influence of variation of cross section on RATs. The analytical approximations for them are given in Appendix B. Note, that the dependence on $l_{1}$ arises in equations (\ref{eq14}) and
(\ref{eq15}) due to the assumption $\lambda \ll l_{1}$. In the opposite limit
we expect $\lambda$ to act as an arm for torques and therefore no dependences of the RAT
efficiencies $Q_{e1}, Q_{e2}$ on $l_{1}/\lambda$ to exist.

\subsection{AMO: RATs' Properties}\label{amo3}

For the sake of simplicity, for the rest of
the paper, apart from the Appendices B2 and B3, we consider AMO for a single value 
of angle $\alpha=\pi/4$.  
Combining RATs produced by the reflecting oblate spheroid (see equations \ref{eq3}-
\ref{eq5}), and RATs induced by the reflecting mirror (see equations \ref{eq14}-
\ref{eq16a}), for $\alpha=\pi/4$ AMO has the following components 
\begin{align}
Q_{e1}&= \frac{4\pi l_{1} n_{1}n_{2}}{\lambda} (3 \mbox{cos}^{2} \Theta - 1)f(\Theta, \pi/4),\label{eq18}\\
Q_{e2}&= \frac{4\pi l_{1} n_{1}n_{2}}{\lambda} \ms 2\Theta g(\Theta, \pi/4),\label{eq19}\\ 
Q_{e3}&=\frac{2e a(s^{2}-1)}{\lambda}K(\Theta, e) \ms 2\Theta,\label{eq20}
\end{align}
where the analytical and numerical fitting functions $f(\Theta, \pi/4)$ and
 $g(\Theta, \pi/4)$ are shown in Figs B1 and B2, respectively.

However, as we see further, $Q_{e3}$ does not affect the alignment apart from inducing the precession. To roughly estimate the latter, one does not need to know
the exact form of $K(\Theta, e)$ (see \S \ref{kalign}).

Equation (\ref{eq18}) reveals clearly that $Q_{e1}$ is symmetric, while equation (\ref{eq19}) shows that 
$Q_{e2}$ is asymmetric. In addition, $Q_{e2}$ is zero for $\Theta=0, \pi$ and
$\pi/2$, i.e., when the maximal inertia
axis ${\bf a}_{1}$
is parallel or perpendicular to the
light direction (see also Fig. \ref{f6}). 

In particular,  two first components, $Q_{e1}, Q_{e2}$ depend on the product of projections of the normal vector,
i.e., $n_{1}n_{2}$, that suggests us to define the helicity of a grain. A grain of {\it right} helicity is defined so that, when 
the maximal inertia axis ${\bf a}_{1}$ is parallel to the radiation beam, 
it can rotate in clockwise sense around the radiation beam; a grain has {\it left} helicity if it rotates anti-clockwise around the radiation beam. With these definitions, we see from
equations (\ref{eq18}) and (\ref{eq19}) that
the grain with right helicity corresponds to the mirror being oriented so that
$n_{1}n_{2}>0$, and $n_{1}n_{2}<0$ for the grain with left helicity. 
It is straightforward to change the grain helicity from right to left by a
rotation of the mirror over an angle of $\pi/2$. Thus, grains with
 left and right helicity
are mirror symmetric, as expected. Naturally, the helicity of a spheroid is
zero.

Fig. \ref{f6} shows components of RAT normalized over the maximum of $|Q_{e1}|$, so that they exhibit the functional dependence of RATs
on $\Theta$. Since $Q_{e3}$ arises from the reflecting spheroid, not subject to the mirror size, in order for AMO to be self-consistent, we normalize $Q_{e3}$ so that $Q_{e3}^{max}=Q_{e1}^{max}$.
It is evident from examining the upper and lower panels of Fig.~\ref{f6}
that the transition from left-handed to right-handed grains induces a simultaneous change of the both $Q_{e1}$ and $Q_{e2}$ components. This can be easily understood based on the fact that both $Q_{e1}$ and $Q_{e2}$ depend on the product $n_{1}n_{2}$, which defines the helicity of grain. Therefore, as the helicity changes, i.e., $n_{1}n_{2}$ reverses the sign, $Q_{e1}, Q_{e2}$ change synchronically. We shall see that this property
is also present for arbitrary-chosen irregular grains that we study numerically
in \S \ref{sec33}. Fig. \ref{f6} also shows clearly that the third component $Q_{e3}$ that arises from
the spheroidal body of the grain does not change as the 
grain gets the opposite helicity. This exactly what we expect from a spheroid.

\begin{figure}
\includegraphics[width=0.49\textwidth]{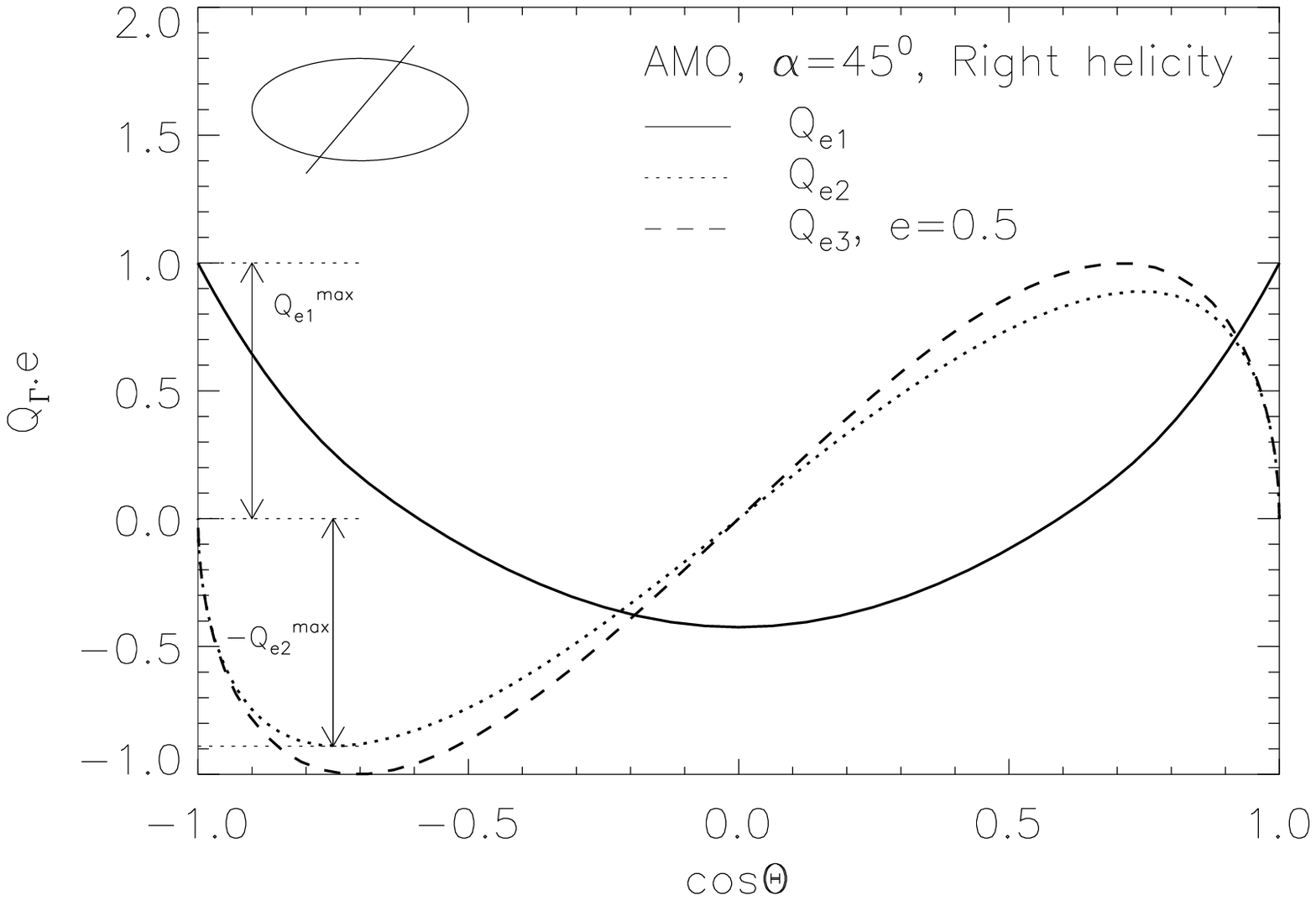} 
\includegraphics[width=0.49\textwidth]{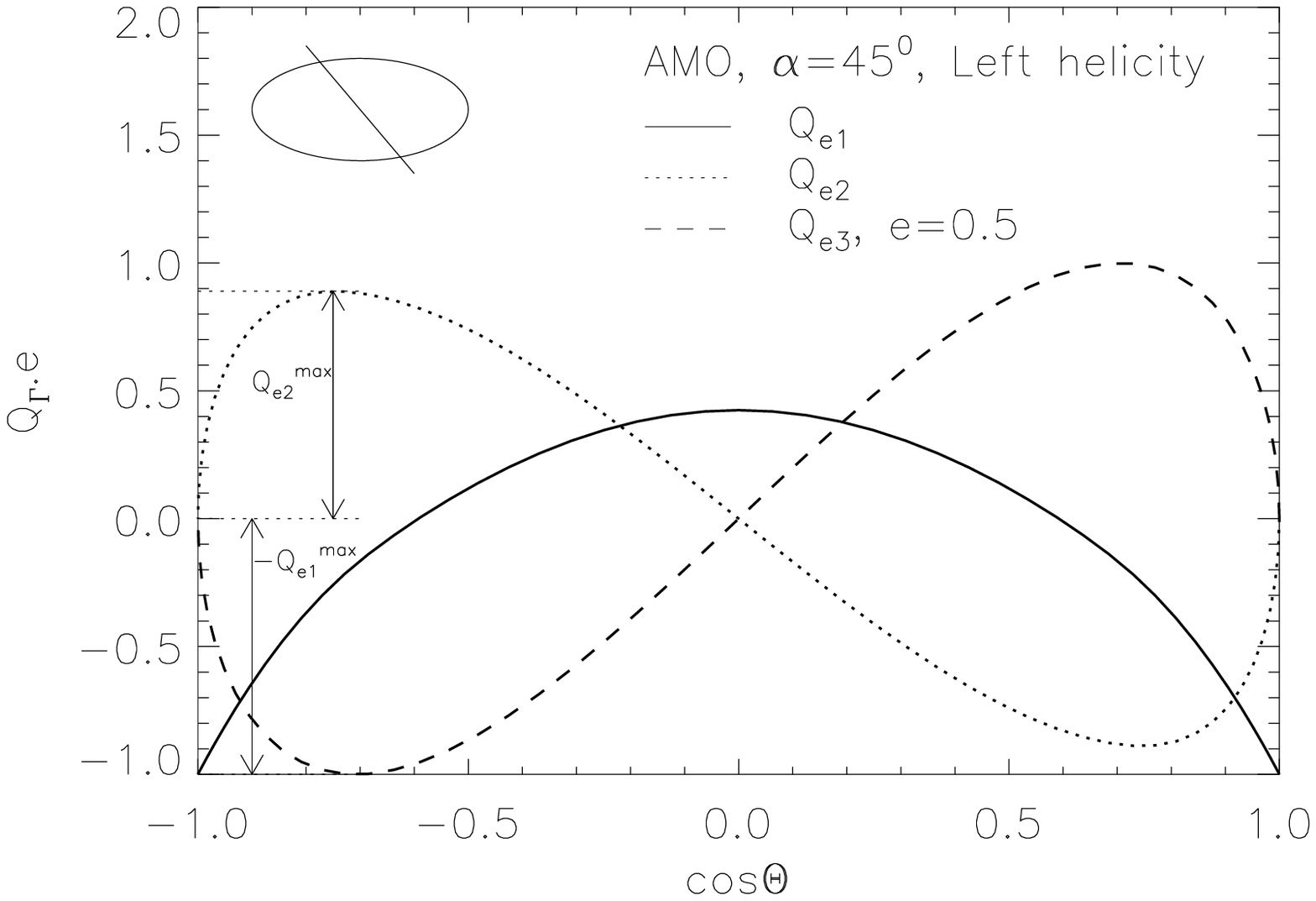}
\caption{The components of RAT normalized over the maximum of $|Q_{e1}|$ as a function of $\Theta$ for
right handed ({\it Upper Panel}) and left handed ({\it Lower Panel}) models of AMO. 
The helicity change
is obtained by twisting the mirror
by $\pi/2$. The maximum of $Q_{e3}$ is also normalized to be equal $Q_{e1}^{max}$. Figs exhibit 
  zeros of $Q_{e2}$, and symmetries of RATs that we study for irregular grains
  using DDSCAT in \S \ref{sec33}.}
\label{f6} 
\end{figure}

\section{RATs: AMO versus Irregular Grains}\label{sec33}

We justify AMO's utility
finding the correspondence of the functional form obtained
for the torques that our toy model experience with those exerted on
actual irregular grains. We start with finding generic properties
of RATs for irregular grains using general symmetry considerations and
follow further with numerical calculations.

\subsection{Symmetry considerations for RATs}
 
Here we show
that some properties of RATs follow from general considerations based on the
analysis of symmetries.
For instance,
we have observed that for AMO $Q_{e2}$, $Q_{e2}$ become 
 zero after
$\beta$-averaging at points $\Theta=0$ and $\pi$, while $Q_{e1}$ does not
depend on $\beta$. This property is valid for arbitrarily shaped grains.
Indeed, when $\Theta$ is either 0 or $\pi$ the radiation direction presents the
axis of symmetry. It is obvious, therefore that changes of $\beta$
cannot change the RAT component along the radiation direction (i.e. does
not change $Q_{e1}$), while any perpendicular component of RAT (i.e. both
$Q_{e2}$ and $Q_{e3}$), should vanish as the result of $\beta$-averaging\footnote{In fact, these considerations prove not only the zero values of $Q_{e2}$ and $Q_{e3}$ but also their periodicity as a function of $\beta$.}

Further on, we discuss the properties of $\beta$-averaged RATs.
We can observe, that, similar to the case of AMO, the component of RATs 
$Q_{e1}$ for irregular grains
 is symmetric with respect to 
$\Theta \to \pi-\Theta$ change (see Fig. \ref{f21}). This symmetry is not
exact, but it gets better for grains for which mutual shadowing of dipoles
gets less.
 The symmetry of $Q_{e1}$
ensures that the torque along ${\bf k}$ has the same sign and similar magnitude when the grain flips over. 
At the same time the RAT component   $Q_{e2}$ is anti-symmetric (see Fig. \ref{f21}), it changes the
sign for a transformation $\pi  
-\Theta$. This also corresponds to AMO. Similarly to $Q_{e3}$ the symmetry
of $Q_{e1}$ is only approximate.

\subsection{Zero points  of $Q_{e2}$ at $\Theta =\pi/2$}\label{sec42}

For AMO, according to equation (\ref{eq20}), $Q_{e2}$ is equal to zero for $\Theta=\pi/2$. A
similar property also exists for irregular grains, as obviously seen in Figs
\ref{f21} and \ref{f23}. There it is shown that
when the maximal inertia axis is perpendicular to the radiation direction, the
magnitude of $Q_{e2}$ is very small. This can be explained in terms of
the interaction
of the electric dipoles 
with the electric field vector of radiation as follows: the interaction between 
electric field and electric dipoles induces their rotation around
 ${\bf e}_{2}$ to emit circularly polarized photons. However since ${\bf a}_{1}$ is perpendicular to
${\bf k}$,  
the electric field is only able to induce the rotation of the electric dipoles in a plane
containing ${\bf e}_{2}$. As a result, the torque component vanishes.  
Quantitatively, according to equation (C5) (see Appendix C), it follows that, when $\Theta=90^{0}$, we have  
\bea
{\bf Q}_{abs}.{\bf e}_{2}\sim k \mbox{ cos } \beta  
\mbox{Re}\sum_{j}[p_{j}.E_{inc}]e^{ik x_{j}}.  
\ena
  
Since $x_{j}=r\mbox{sin }\alpha \mbox{sin }\beta$, it follows 
\bea  
{\bf Q}_{abs}.{\bf e}_{2} \sim \sum_{j}\mbox{ cos }\beta \mbox{ cos }(A\mbox{ 
  sin }\beta) [p_{j}.E_{inc}],  
\ena 
where the term $[p_{j}.E_{inc}]$ is a function that is independent of
$\beta$. 
It is obvious that ${\bf Q}_{abs}.{\bf e}_{2}$ is a function of $\beta$ which  
is zero when averaging is performed for $\beta$ over [$0$, $2 \pi$]. We calculated $Q_{abs}.e_{2}$ for
different $\lambda/a$ and grain shapes, and found that $Q_{e2}$ is indeed  
close to zero at $\mbox{cos}\Theta=0$, which is consistent with our analytical
expectation. 

\subsection{DDSCAT Calculations}\label{sec43}

\begin{figure*}
\includegraphics[width=0.32\textwidth]{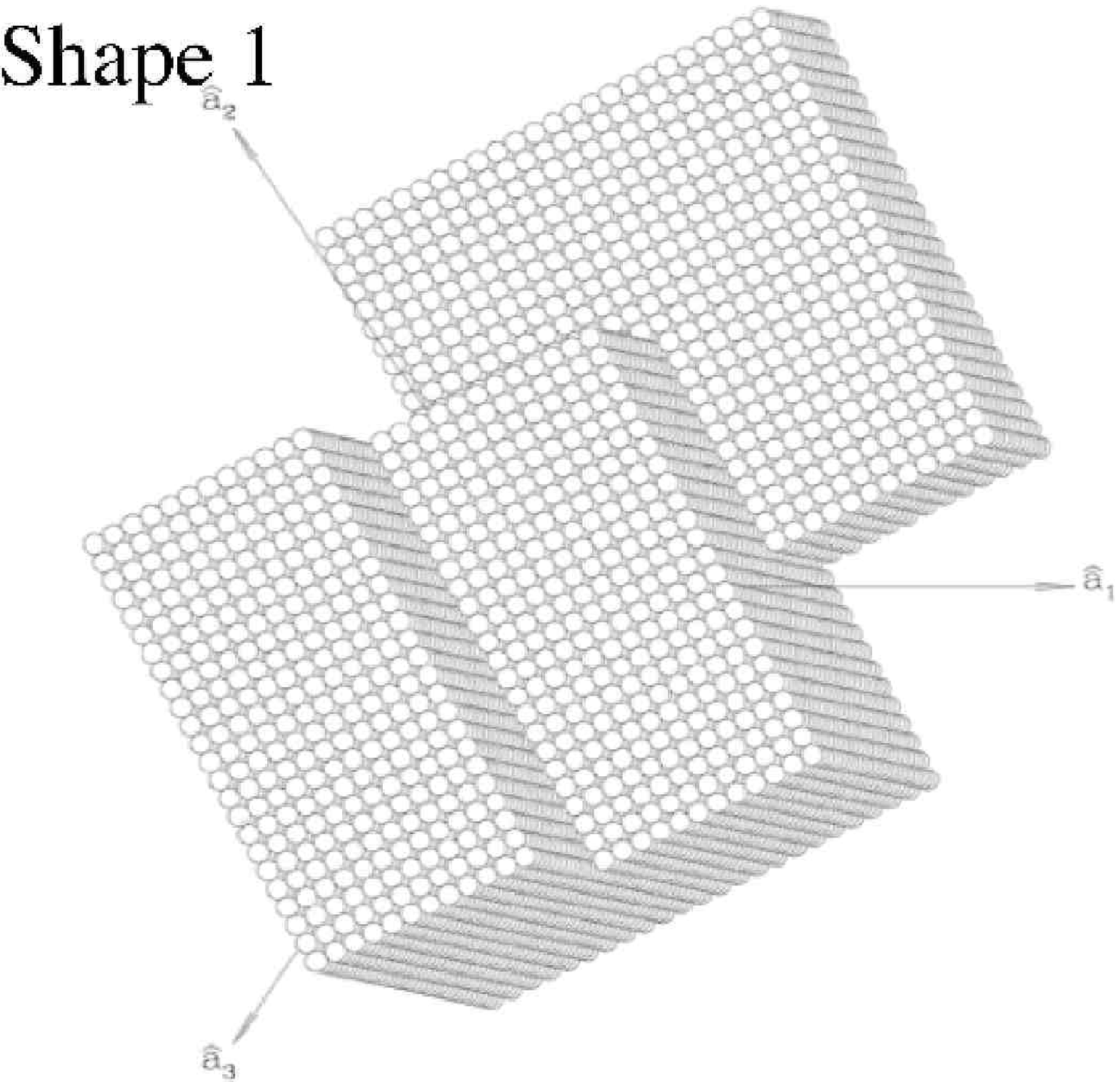} 
\includegraphics[width=0.32\textwidth]{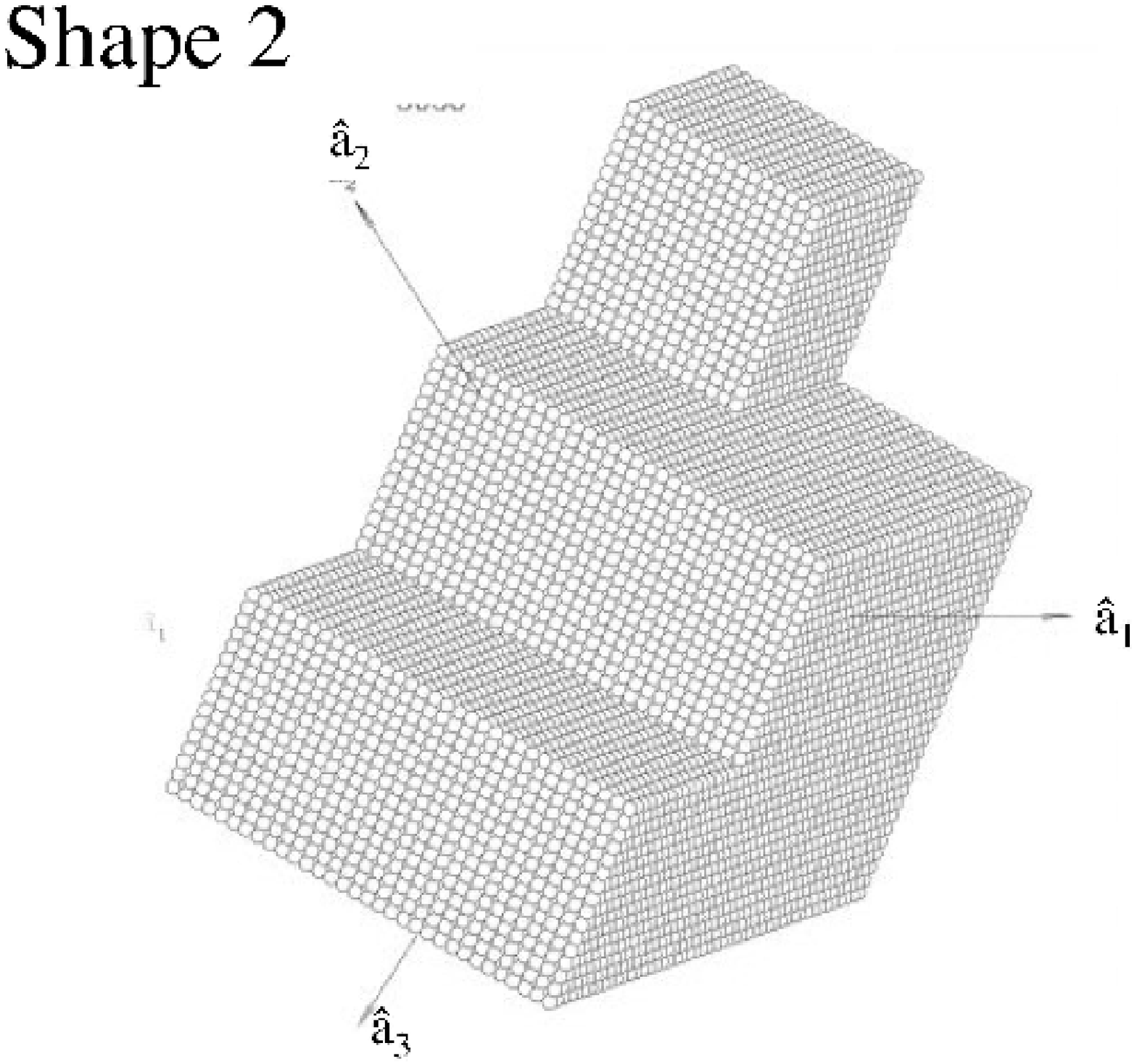}
\includegraphics[width=0.32\textwidth]{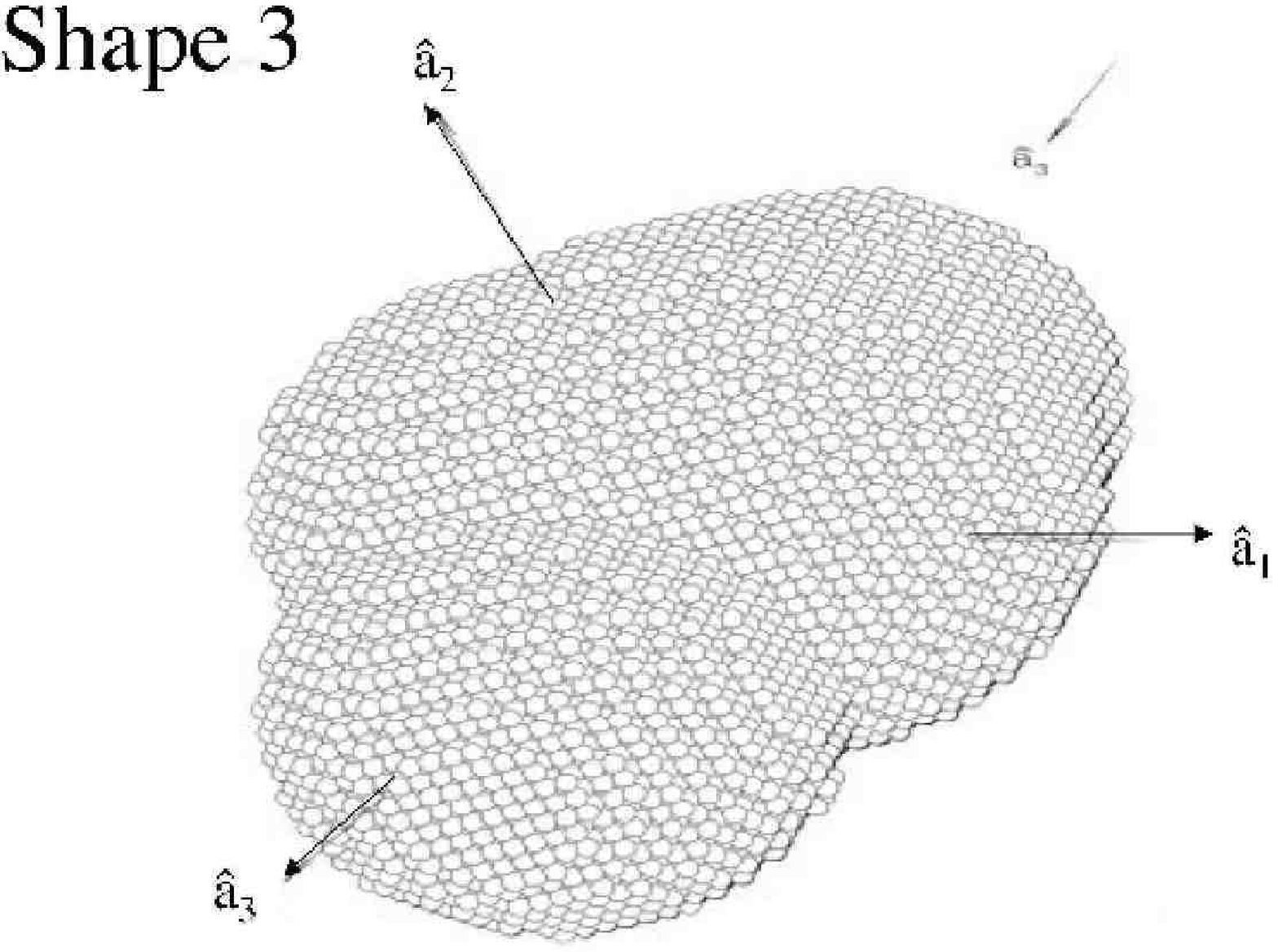} 
\includegraphics[width=0.32\textwidth]{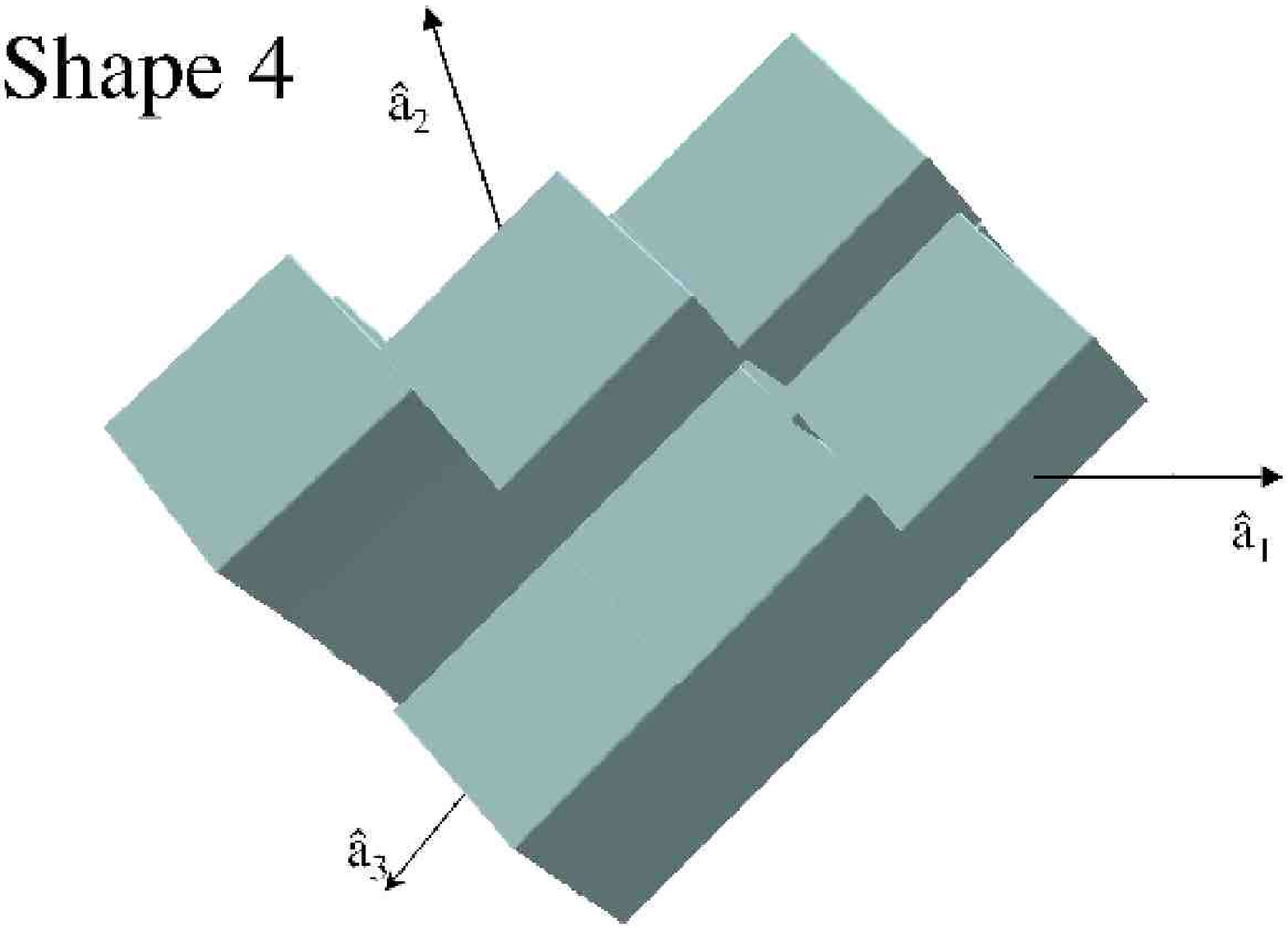} 
\includegraphics[width=0.32\textwidth]{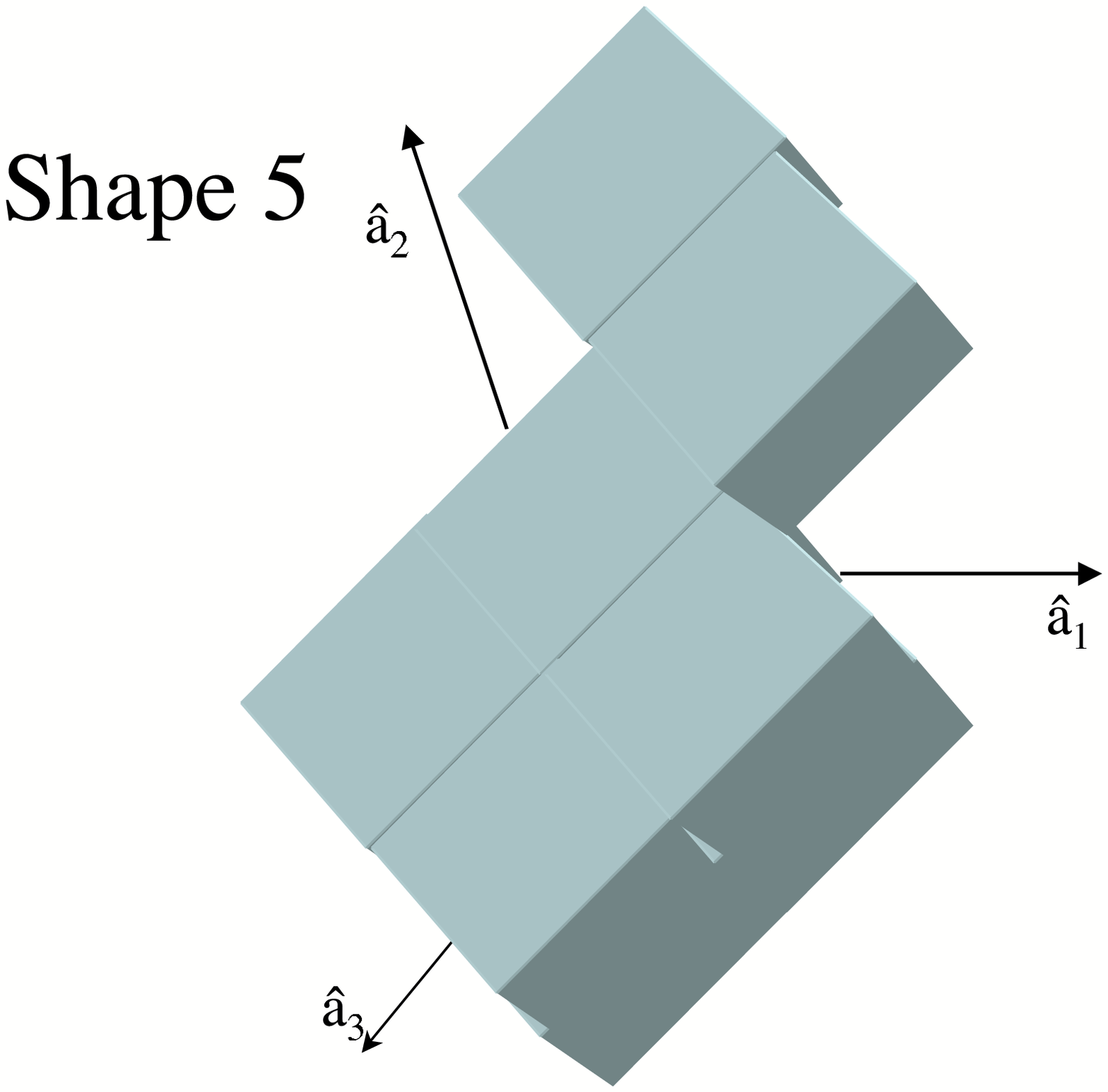}
\includegraphics[width=0.32\textwidth]{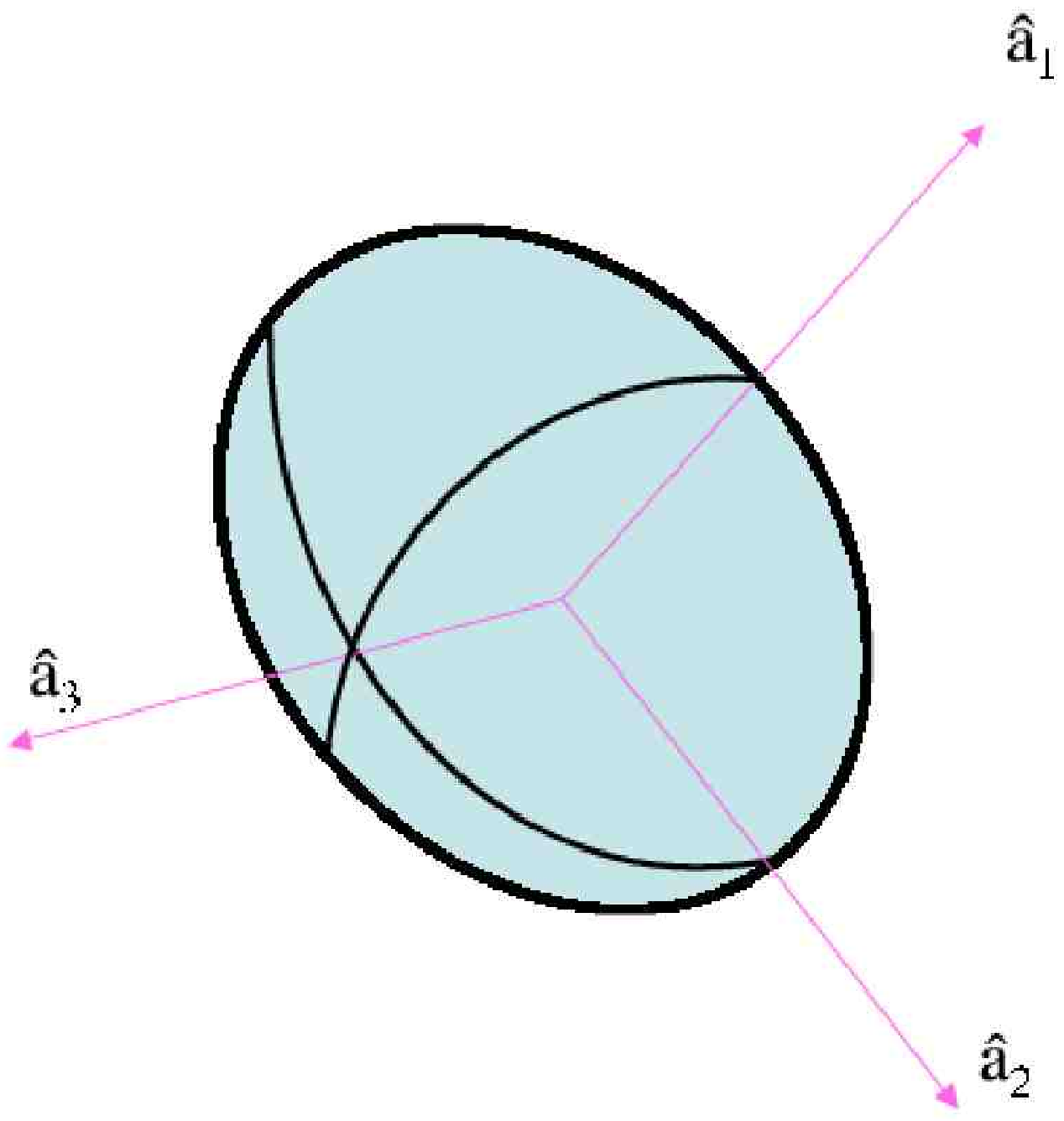}
 \caption{Geometry of grains under study: shape 1, 2, 3 are similar to those of
   DW97, shape 4, and 5 are created from 15 and 11 cubic blocks respectively,
   and an ellipsoidal shape.} 
 \label{f18} 
\end{figure*}

Fig.~\ref{f18} presents the test grain shapes that we have calculated RATs for using DDSCAT. Parameters for calculations are given in Table \ref{tab2}. Shapes 1, 2 and 3 have been
used in DW97. We added to them shapes 4 and 5. In addition, we created a mirror
symmetric shape of shape 1, namely, shape 1* and provided the DDSCAT 
calculations for a spheroidal grain (see more details in Table~\ref{tab2}). We
adopt dielectric functions for astronomical silicate in which a feature in
the ultraviolet is removed (see DW97; Weingartner \& Draine 2001; Cho \&
Lazarian 2005).

\begin{table*}
\caption{Grain shapes and parameters for calculation of RATs}
\begin{displaymath}
\begin{array}{rrrrr} \hline\hline\\
\multicolumn{1}{c}{\bf Grain~shapes} & \multicolumn{1}{c}{\bf Dipole~ \#} &{\bf Size ~(\mu m)} & \multicolumn{1}{c}{\bf Wavelength~ (\mu m)} &\multicolumn{1}{c}{\bf Helicity}\\[1mm]
\hline\\
{\rm Shape~ 1}&{\it 832000 }&{\it 0.05-0.2}&{\it ISRF}&{\it right}\\[1mm]
{\rm Shape~ 1^{*}}&{\it 53248}&{\it 0.2}&{\it 1.2}&{\it left}\\[1mm]
{\rm Shape~ 2}&{\it 45056}&{\it 0.2} &{\it ISRF}&{\it left}\\[1mm]
{\rm Shape~ 3}&{\it 102570} & {\it 0.2}&{\it ISRF} &{\it left} \\[1mm]
{\rm Shape~ 4}&{\it 15000}&{\it 0.2} &{\it 1.2}&{\it left}\\[1mm]
{\rm Shape~ 5}&{\it 11000}&{\it 0.2}&{\it 1.2} &{\it left} \\[1mm]
{\rm Hollow~1}&{\it 832000 }&{\it 1.0}&{\it 0.1}&{\it right}\\[1mm]
\hline\\[1mm]
 \\[1mm]\hline

\end{array}
\end{displaymath}
\label{tab2}
For all calculations here, we adopt the dielectric function for astronomical silicate.
 \end{table*}

\begin{figure}
\includegraphics[width=0.49\textwidth]{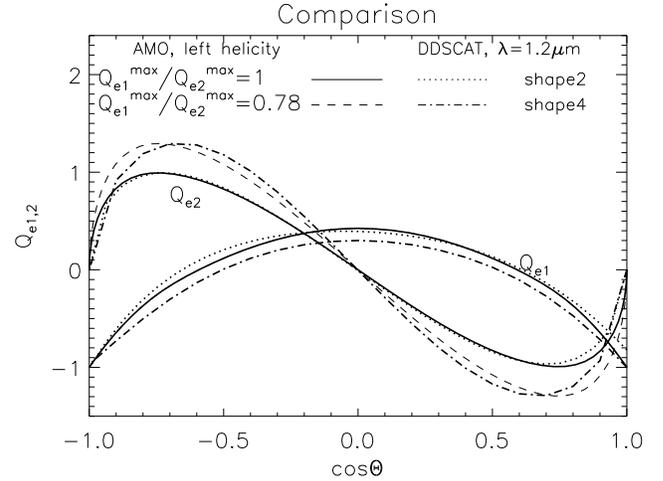}
\caption{Figure shows the comparison of RATs normalized over the maximum of $|Q_{e1}|$ between AMO (the left handed grain) and
 DDSCAT (for shapes 2 and 4 and monochromatic radiation of $\lambda=1.2\mu m$). Solid and dashed
 lines show normalized RATs corresponding to $Q_{e1}^{max}/Q_{e2}^{max}=0.78$ and $1$ in which the functional forms are obtained from the analytical approximation given by equations (\ref{eq18})
 and (\ref{eq19}) with tabulated functions $f, g$. Dot and dashed-dot
 lines show normalized RATs for shape 2 and shape 4, respectively.}
\label{f61} 
\end{figure}

We discussed for AMO, that the sign of helicity can be changed
by taking the mirror image of the grain.
 We performed a similar procedure
to the irregular grains and obtained results similar to the ones obtained
for AMO (see Figs. \ref{f23} and \ref{f23a}).

Note, that we observe that $Q_{e1}$ and $Q_{e2}$ change synchronously when we 
calculate torques for a mirror image of a grain. We see that the shape
1 has one type of helicity, while shapes 4, 5 and mirror symmetric image
of shape 1, i.e. shape 1*, have another type of helicity.
 
 \begin{figure}
\includegraphics[width=0.49\textwidth]{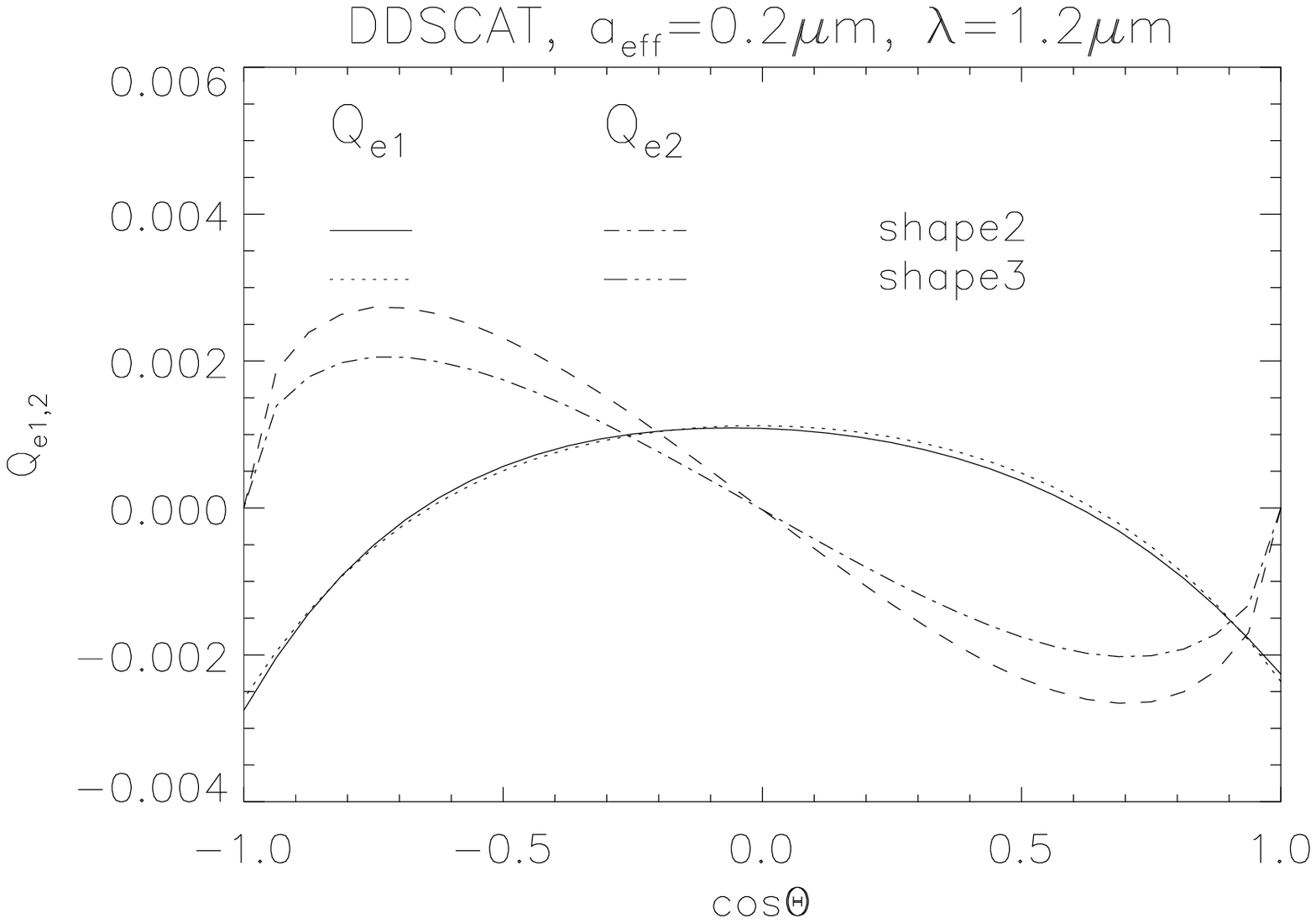}
\includegraphics[width=0.49\textwidth]{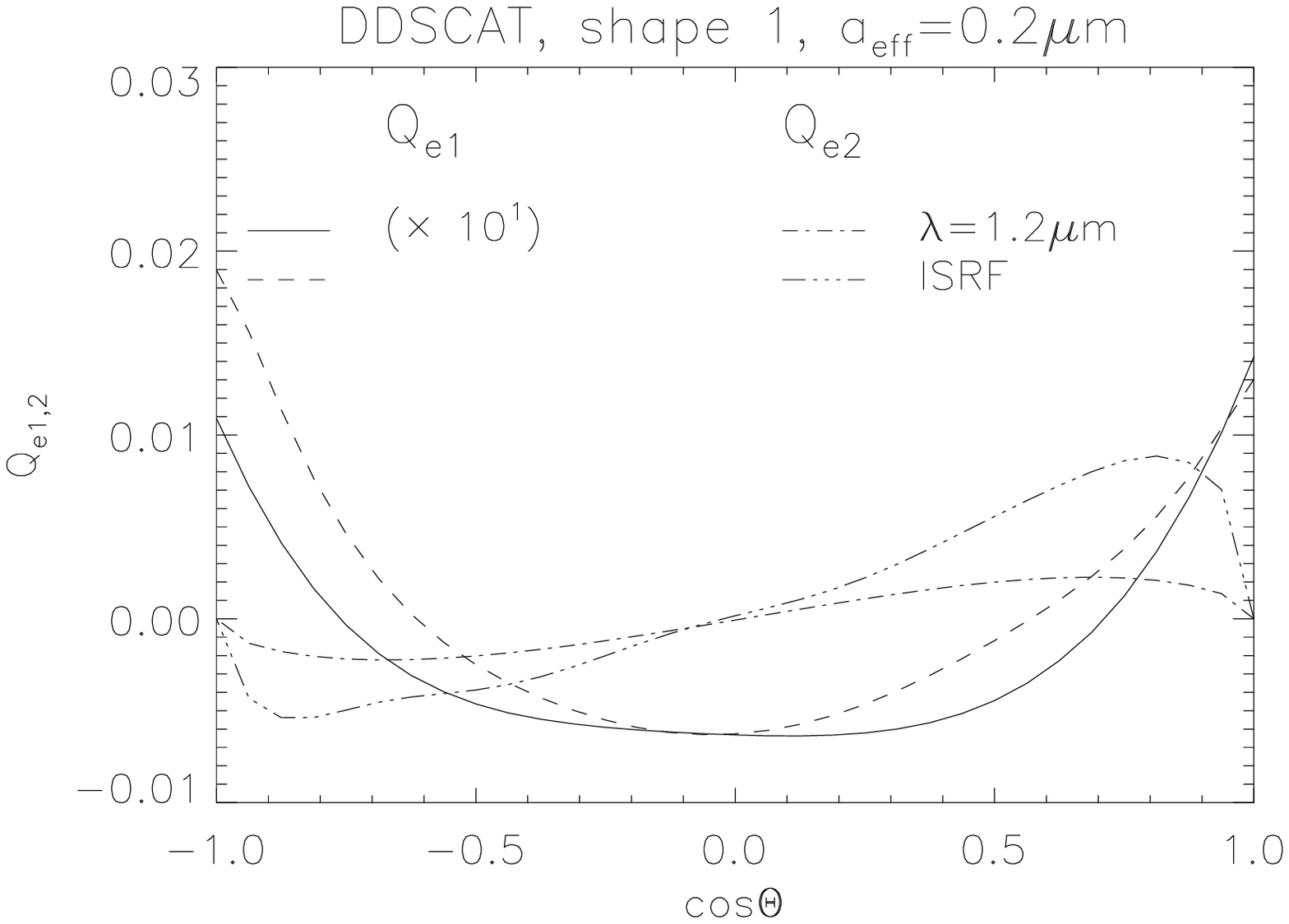}
 \caption{$Q_{e1}$ and $Q_{e2}$ for grains of left ({\it Upper Panel}) 
and right ({\it Lower Panel}) helicity. The symmetry of $Q_{e1}$ with the transformation $\pi -\Theta$ is clearly seen for all grain shapes. $Q_{e2}$ is antisymmetric
 with respect to the 
same transformation $\pi -\Theta$. A similarity with the torques
produced by AMO is evident (see also Fig \ref{f61}).} 
\label{f21}
\end{figure}

Fig.~\ref{f61} provides a comparison of normalized RATs between AMO and
DDSCAT calculations performed for two irregular grains induced by monochromatic radiation
field of $\lambda=0.2 \mu m$ (see more in \S \ref{para}). It can be seen that they possess the same symmetric properties as well as
zero points. Also, the functional form of normalized $Q_{e1}$ and $Q_{e2}$ calculated for
the irregular grains and AMO are remarkably similar, in particular for
the $Q_{e2}$ component. Typically, the RAT components for shape 2 are similar to those
of AMO
with $Q_{e1}^{max}/Q_{e2}^{max}=1$ ratio, while RATs of shape 4 are similar to those of AMO with $Q_{e1}^{max}/Q_{e2}^{max}=0.78$
. Hence, by changing the ratio of amplitudes of the RAT components for AMO, we
can obtain analytical expressions of RATs for a number of irregular
grains. To have RATs appropriate to irregular grains, it is necessary to use
DDSCAT to estimate the magnitude of RATs. Combining functional forms from AMO
and magnitude from DDSCAT, we can obtain analytical approximate expressions for
 RATs components of
irregular grains. Note, that in Figs \ref{f6} and \ref{f61}, we normalized RATs over
$|Q_{e1}^{max}|=|Q_{e1}(\Theta=0)|$, that gives rise to $Q_{e1}^{max}$
remained the same for all realizations of
AMO and the irregular grains. It is easy to see that with this choice, AMO reproduces very
well $Q_{e2}$ for irregular grains, but gets $Q_{e1}$, which is a bit larger at $\mc\Theta=0$ ($\Theta=\pi/2$) than that for irregular grains. Potentially, this may mean that 
more appropriate parametrization should include $Q_{e1}^{max}$, which is not
defined as $Q_{e1}$ at $|\mc\Theta|=1$ ($\Theta=0$ or $\pi$)  as we do in this paper, but, for instance, the amplitude value of $Q_{e1}$, which is $|Q_{e1}(\Theta=0)|+|Q_{e1}(\Theta=\pi/2)|$. We feel, however, that the our present parametrization has the advantage of 
simplicity and is sufficiently accurate.

 \subsection{Parameter study} \label{para}

Above we compared the properties of RATs in AMO with those
obtained numerically from DDSCAT for a few chosen grain shapes and radiation spectra.  
To see how general our numerical results are, 
we attempt a limited parameter study, namely, we study how the properties of
RATs vary with  the spectrum of the incident radiation for different grain
shapes. One can view the AMO formulae as a {\it physically motivated} fit to RATs
acting  on astrophysical grain with $a_{eff}/\lambda<1$. The parameter study
is intended to find out how good is this fit.

Fig. \ref{f23} shows $Q_{e1}$ and $Q_{e2}$ for the shape 1 produced by different
radiation fields. There the upper panel show that when monochromatic radiation
fields of $\lambda/a_{eff}$ increases, the symmetry of $Q_{e1}$ and zeros of
$Q_{e2}$ do not change. However, their amplitude decreases. In addition, the
symmetric property of $Q_{e1}$ and zeros of $Q_{e2}$ also remain unchanged
when being
averaged over different radiation spectrum (see the lower panel in
Fig. \ref{f23}).

Now let consider RAT properties for different irregular grains. The upper panel in Fig. \ref{f23a} shows RATs for different
shapes: shape 1* which is a mirror symmetric copy of shape 1, and shapes 4 and 5 are
built from 15 and 11 cubic blocks, respectively. 
We also see clearly that $Q_{e1}$ exhibits the symmetry, and $Q_{e2}$ exhibits
the asymmetry that we have already seen with AMO and other grain shapes.

From Figs \ref{f23} and \ref{f23a}, it follows that the form of $Q_{e1}, Q_{e2}$
for shape 1 is mirror-symmetric to the corresponding
RAT components applied to shape 1*. This mirror
symmetry is also evident when we compare $Q_{e1}, Q_{e2}$ with those 
of shapes 4, 5 (see Fig.~\ref{f18}).
 This implies, similar to AMO, irregular grains may be of right and left
helicities. A comparison between Figs~ \ref{f21}, 
\ref{f23}{\it upper}, and \ref{f23a}{\it upper} shows that shapes 1*, 2, 3, 4, 5 are of left helicity, while the shape 1 is
of right helicity. Note, that Fig. \ref{f23}{\it lower} clearly shows that the
helicity is independent of wavelength and is intrinsic attribute of
grains which is associated to their shape. Especially, we can obtain a
grain with the opposite helicity by performing a mirror symmetric 
transformation,
which is illustrated by AMO in Fig.~\ref{f6}. We remind the reader, that the
correspondence between $Q_{e1}$ and $Q_{e2}$ for AMO and shape 2 and 4 is
illustrated by Fig.~\ref{f61}.
  \begin{figure}
\includegraphics[width=0.49\textwidth]{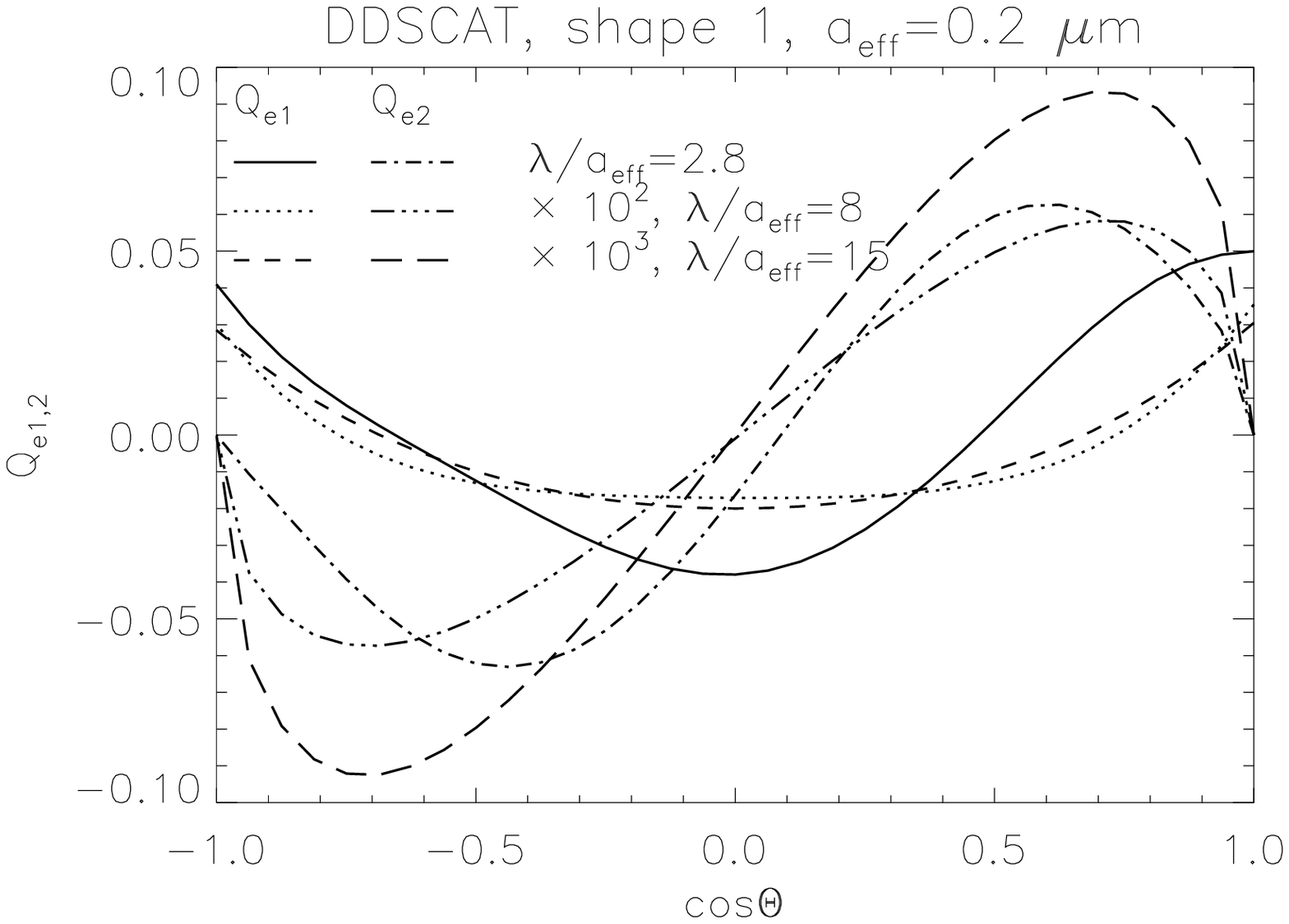}  
\includegraphics[width=0.49\textwidth]{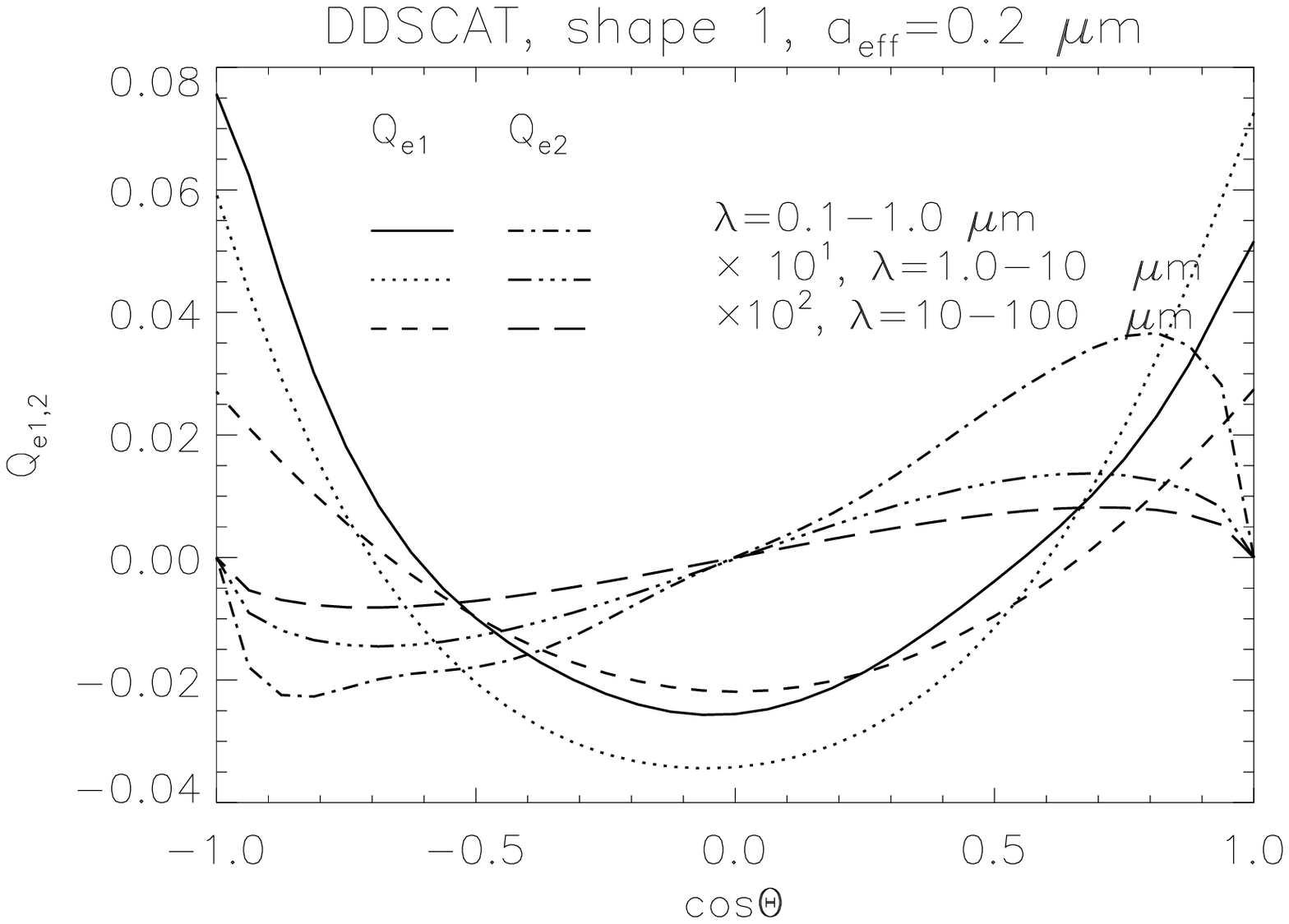}   
 \caption{RATs for the shape 1 corresponding to various $\lambda/a_{eff}$ ({\it Upper
Panel}) and
   RATs averaged over a range of wavelengths ({\it Lower Panel}).}  
 \label{f23} 
\end{figure}

\begin{figure}  
\includegraphics[width=0.49\textwidth]{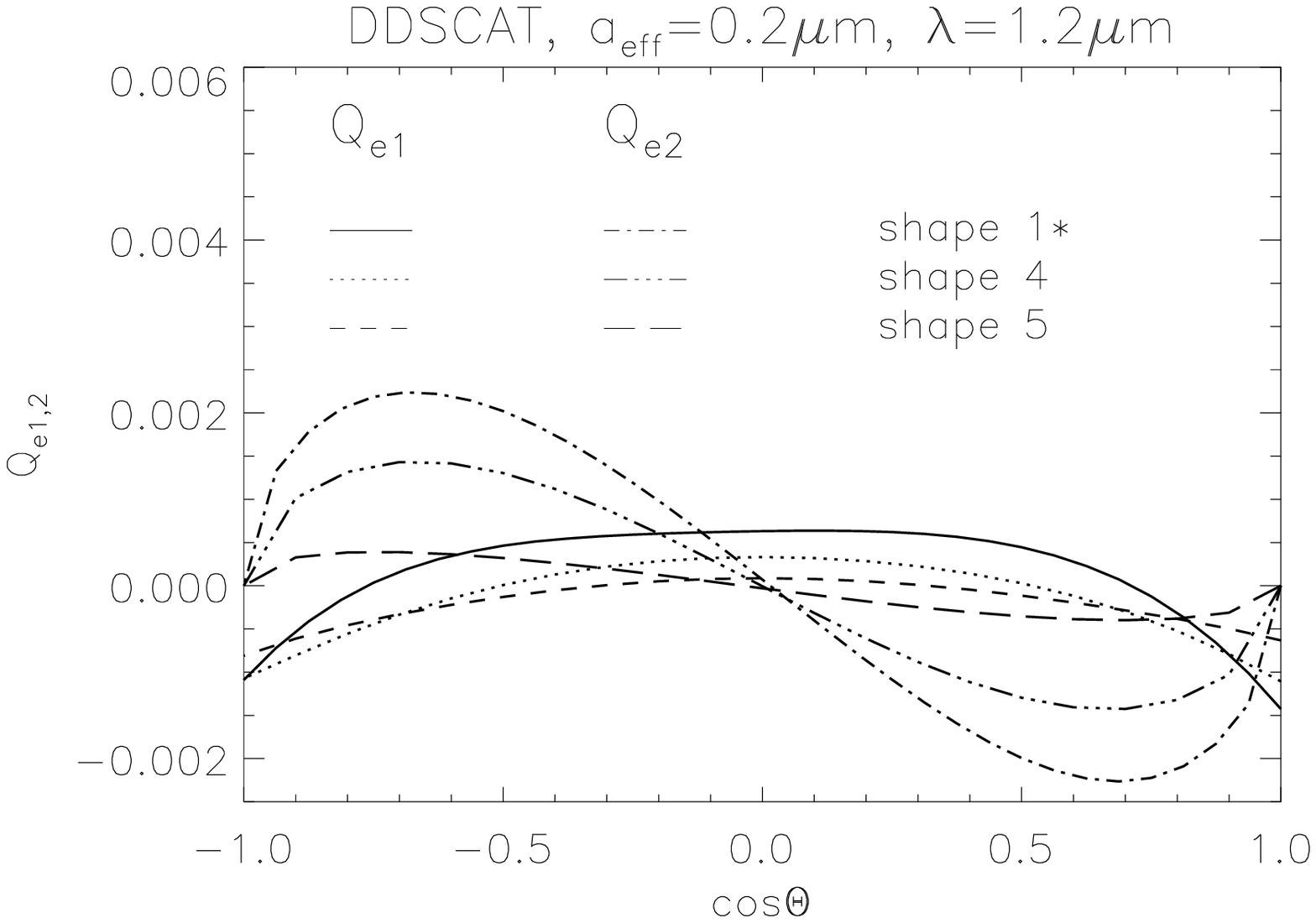} 
\includegraphics[width=0.49\textwidth]{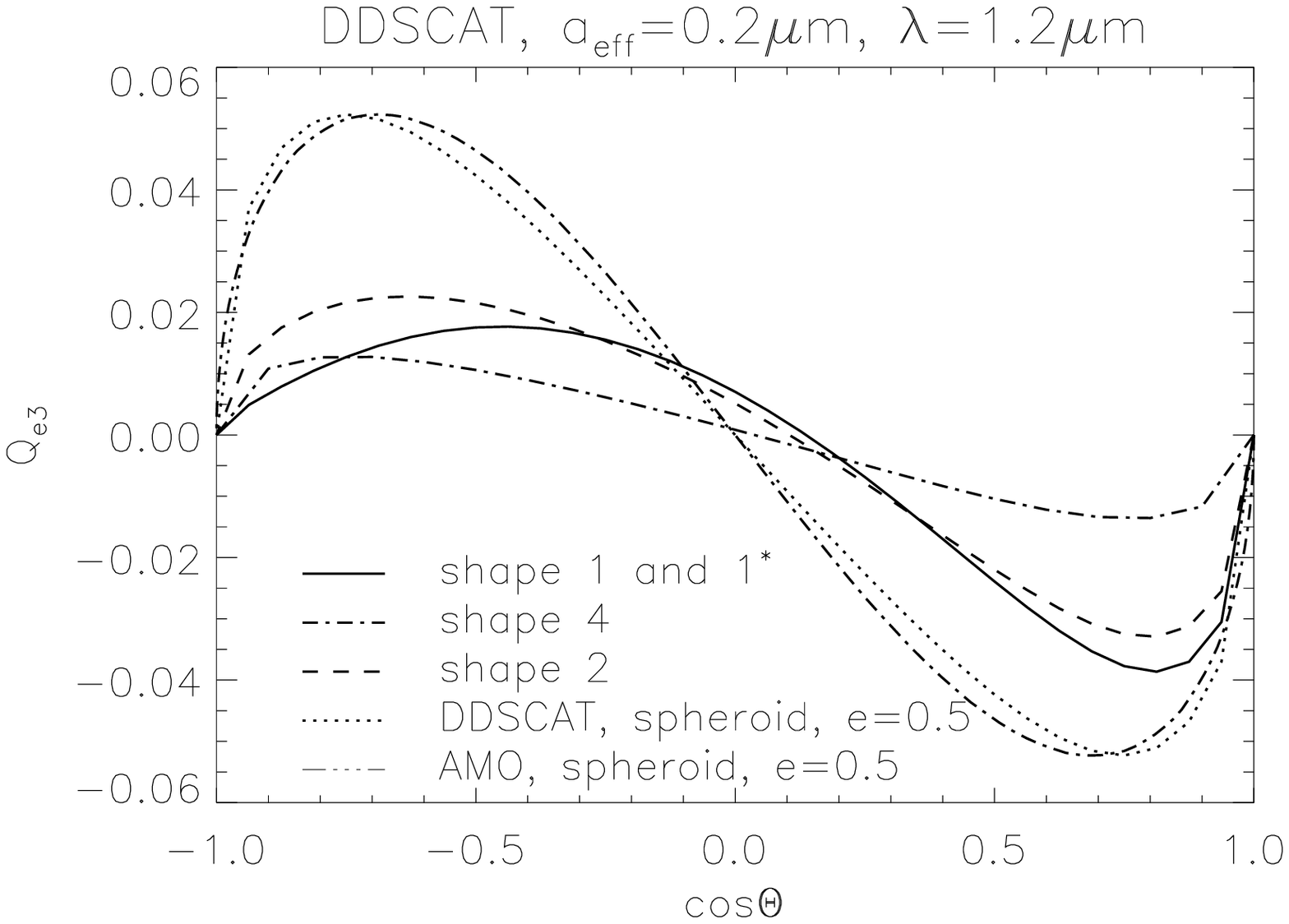} 
\caption{{\it Upper panel:} RATs for different grain shapes. Shape 1* corresponds to mirror-symmetric image of
shape 1. Symmetric features of $Q_{e1}$ and zeros of $Q_{e2}$ are clearly
found. {\it Lower panel:} The third component of RATs $Q_{e3}$ for different shapes are
shown together with that of a spheroid with $e=0.5$ predicted by AMO. An analogy exists between the zeros of $Q_{e2}$ and $Q_{e3}$. Also, the shape of $Q_{e3}$ is not affected by the change of grain helicity.}
 \label{f23a} 
\end{figure}
For the third RAT component $Q_{e3}$, it exhibits analogous properties with
$Q_{e2}$ obviously seen in the lower panel in Fig. \ref{f23a}. Also, for
an axisymmetric shape, i.e., spheroid, $Q_{e3}$ is still significant. Note, that its functional form obtained by DDSCAT is very consistent with that predicted by AMO in see \S \ref{amo1} (see dot and dot-dashed line in Fig. \ref{f23a}). 
Furthermore, we see
that it has similar forms for left (shape 2 and 3) and right (shape 1) helical
grains (see Fig.~\ref{f23a}{\it lower}). While the dependences of
$Q_{e1}$ and $Q_{e2}$ undergo a transformation when shape 1 is substituted
by the shape 1*, having the opposite helicity, a comparison of lower and
upper panels of Fig.~\ref{f23a} shows that the shape of $Q_{e3}$ component
stays the same.
This is expected, as
$Q_{e3}$ does not depend on the helicity of grains.

Also, the $Q_{e3}$ component has zeros
at $\Theta=0, \pi$, similar to $Q_{e2}$ (see the lower panel in Fig. \ref{f23a}). However, the anti-symmetry of $Q_{e3}$ is less prominent than for $Q_{e2}$.  

As we will see that the ratio $Q_{e1}^{max}/Q_{e2}^{max}$ is an important
parameter that determines the existence of high-$J$ attractor points. For
irregular grains, this ratio is a function of the ratio of wavelength to grain size, as  shown in
  Fig. \ref{f23*} for three irregular grains. The
peak of $Q_{e1}^{max}, Q_{e2}^{max}$ is different for different
shapes. The form of the curve for sufficiently large ratios of
$\lambda/a_{eff}$ can be approximated as $10 \frac{a_{eff}}{\lambda}$. This dependence can be used to
reduce the number of DDSCAT calculations necessary for determining the alignment
for arbitrary radiation fields.

\begin{figure}  
\includegraphics[width=0.49\textwidth]{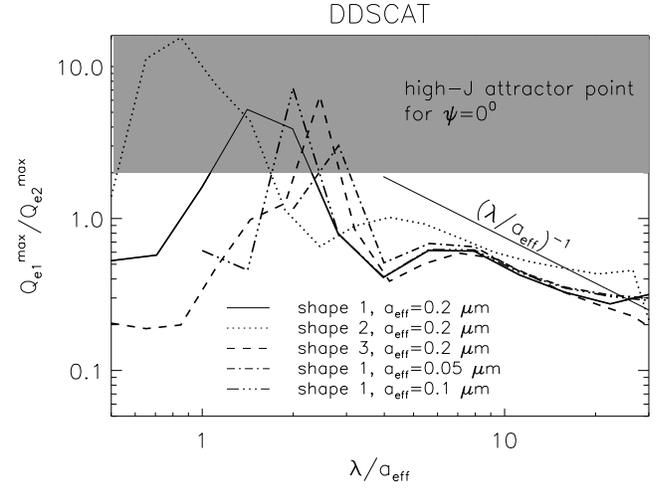}
\caption{Ratio of $Q_{e1}^{max}, Q_{e2}^{max}$ as function of
wavelength to grain size $\lambda/a_{eff}$ for different grain shapes and sizes. The shaded region corresponds to the parameter space in
which the high-$J$ attractor point is present for the alignment in respect
to the beam direction (see \S \ref{kalign}).}
 \label{f23*} 
\end{figure}

In general, we observe strong similarities between the plots of RAT components
obtained for very different grains. Thus we can expect that the RAT alignment
should be similar for such grains. We will discuss the alignment for AMO and irregular grains in \S~ \ref{kalign} and \S~\ref{balign}.

 \subsection{RATs: Comparison with AMO}\label{sec94}

The actual grains are not perfectly reflecting particles and the scattering that
they induce cannot be described by
geometric optics that we employ for AMO. Therefore the justification of the AMO utility
can be obtained via a comparison of the functional form of the torque components obtained for irregular grains with the AMO predictions. Naturally, one should not expect to see
the amplitudes of the torques to be the same. Therefore the comparison should be done
for the normalized torque components. However, we preserve the ratio of the components.

Naturally, our sample of RATs acting on grains studied with DDSCAT is
limited. It includes several grain sizes. For instance, for shape 1, we
studied for grain sizes of $0.05, 0.08, 0.1$ and $0.2 \mu m$. For other
shapes, the size $a_{eff}=0.2 \mu m$ is studied, except the hollow shape 1
with $a_{eff}=1 \mu m$. We calculated RATs for the entire spectrum of ISRF
corresponding to 21 wavelengths in the range  $\lambda=0.1 \mu m$ to $100\mu m$, for
shape 1, 2 and 3, and the monochromatic radiation with $\lambda=1.2\mu m$ for
other shapes. This provides us with RATs calculated for 130 realizations of
grains and radiation fields \footnote{Each realization corresponds to a given grain
  size and a given wavelength}. This makes it the most extensive sample of RATs
studied numerically. It is obvious, that in our paper we cannot present plots
 of the RATs for all the realizations that we calculated (see e.g. Fig. \ref{f61}). A quantitative comparison based on the deviation testing for normalized RATs of all realizations and AMO
will be presented below. 

We show in Fig. \ref{f23*} that, for irregular grains, the relative
amplitude of $Q_{e1}$ versus $Q_{e2}$ changes both with the grain shape and
wavelengths. However, our studies in this paper shows that the
functional form of the RAT components for all the cases we studied is still
well represented by AMO (with different ratio of $Q_{e1}^{max}/Q_{e2}^{max}$, e.g., Fig. \ref{f61}). In other words, while DDSCAT studies of alignment for grains of a
few chosen shapes cannot reveal the generic properties of the RAT alignment,
revealing the correspondence of the functional dependences of the torques
between irregular grains and AMO provides a deep insight into the
alignment. 

Since we are only interested in the functional forms of RATs, let us introduce the mean deviation over $\Theta$ for the components $Q_{e1}$ and $Q_{e2}$ as followings 
\begin{align}
\langle\Delta^{2}\rangle(Q_{ei})&=\frac{1}{\pi( Q_{ei}^{max})^{2}}\int_{0}^{\pi}(Q_{ei}^{DDSCAT}(\Theta)-Q_{ei}^{AMO}(\Theta))^{2} d\Theta,\label{eq82a}
\end{align}
where $Q_{ei}^{DDSCAT}(\Theta)$ denote $Q_{e1}(\Theta), Q_{e2}(\Theta)$ for irregular grains, $Q_{ei}^{AMO}(\Theta)$ denotes the torque components for AMO in which the relative magnitude are rescaled to have the same ratio  $Q_{e1}^{max}/Q_{e2}^{max}$ with each realization of irregular grains\footnote{We note again that, throughout this paper, apart from Appendices B2 and B3, the functional forms of RAT components for AMO corresponds AMO with $\alpha=45^{0}$, and the ratio of their maximum is adjustable.}. In equation (\ref{eq82a}), $Q_{ei}^{max}$ is the maximum of $Q_{e1}$ and $Q_{e2}$, that is chosen the same for both AMO and irregular grains.

We perform $\langle\Delta^{2}\rangle$ testing for our sample consisting of $130$ realizations of irregular grain shape, size and wavelength.  To see the correspondence of AMO with different grain shape, size and wavelength, in Fig. \ref{chi_test0} we show $\langle\Delta^{2}\rangle$ as a function of $\lambda/a_{eff}$.

\begin{figure}
\includegraphics[width=0.49\textwidth]{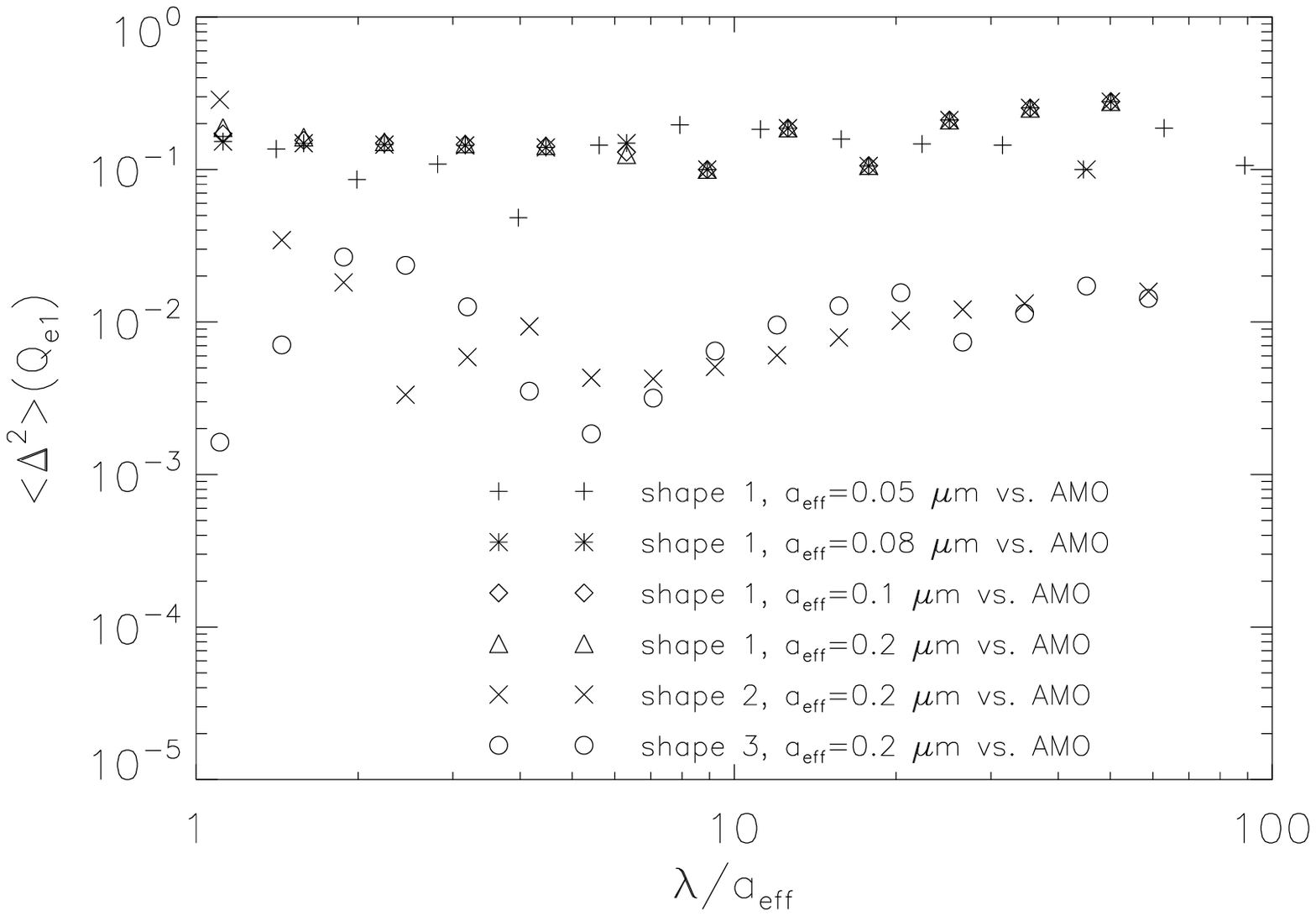}
\includegraphics[width=0.49\textwidth]{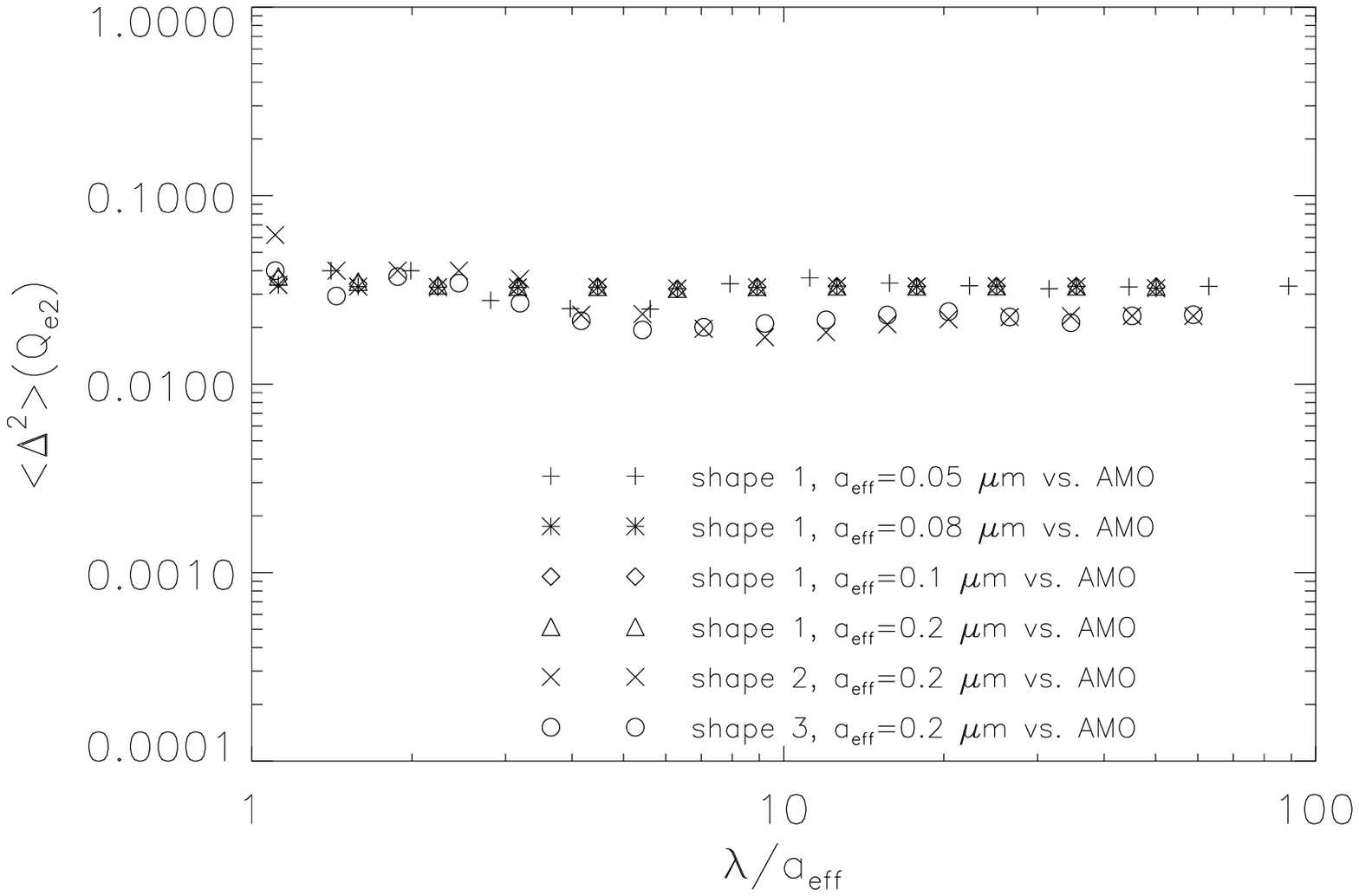}
\caption{Figs show $\langle \Delta^{2}\rangle$ testing as function of $\lambda/a_{eff}$ for normalized $Q_{e1}$ (upper panel) and $Q_{e2}$
  (lower panel) between irregular grain shapes and sizes with AMO.}
\label{chi_test0}
\end{figure}
Fig. \ref{chi_test0}{\it upper} shows a good correspondence for  the component $Q_{e1}$ between irregular grains and AMO. The value of $\langle\Delta^{2}\rangle$ ranges from as small as $10^{-3}$ to $2\times 10^{-1}$. 
In addition, Fig. \ref{chi_test0}{\it lower} shows an extremely  good correspondence for the component $Q_{e2}$ between all cases of irregular grains and AMO. The value of for different sizes of shape 1 are nearly the same, but it changes with grain shapes. For instance, shape 2 and 3 have the better correspondence with AMO than shape 1. The corresponding value of $\langle\Delta^{2}\rangle$ is about $2\times10^{-2}$ and $3\times 10^{-2}$ for shape 2, 3 and 1, respectively (see Fig \ref{chi_test0}{\it lower}). For both components and all grain shapes and sizes, $\langle\Delta^{2}\rangle$ do not change so much with respect to wavelength of radiation field for $\lambda/a_{eff}>3$. However, it increases when $\lambda/a_{eff}$ decreases. It is clear that the smaller $\langle\Delta^{2}\rangle$ is, the better correspondence is. Therefore, the observed worse correspondence for the range $\lambda/a_{eff}<3$ can been easily explained in terms of the RAT properties for irregular grains. In fact, following Fig. \ref{f23}{\it upper}, it can be seen that the components $Q_{e1}, Q_{e2}$ for $\lambda/a_{eff}=2.8$ are indeed less symmetric than for $\lambda/a_{eff}=8$ and $15$. In other words, the symmetry of RATs decreases toward small $\lambda/a_{eff}$. 

Note, that lower values of $\langle\Delta^{2}\rangle$ can be achieved for small $\lambda/a_{eff}$
ratios, if $Q_{e1}$ is considered separately for $\Theta$ within ranges
$[\pi, \pi/2]$ and $[\pi/2,0]$. This procedure can compensate for the 
differences in the surface characteristics on the side of the grain towards the
light direction and opposite to it. As we discuss in \S \ref{amo2} such
differences are natural for less idealized models of helical grains. We do not persue this approach within this paper, prefering a simple model
with a reasonable fit to a more complex model that can provide a better fit. 
Nevertheless, in future, dealing with particular cases, e.g.
 graphite grains, the approach of considering 
different ranges of $\Theta$ may prove advantageous.

Therefore, our quantitative comparison of the RATs based on $\langle\Delta^{2}\rangle$ for irregular grains with AMO implies  that the functional form of RATs obtained
in the limit $\lambda\ll a_{eff}$ (AMO) is also valid for RATs in the limit
$\lambda\sim a_{eff}$ and $\lambda>a_{eff}$ (for irregular grains). This allows us to use AMO as a
representative model for describing RATs of realistic astrophysical grains,
e.g. study grain alignment using AMO.


\section{AMO: Definition and Generalizations}\label{amo2}

Our studies above show that the model of AMO corresponding to $\alpha=\pi/4$
corresponds well to the results of numerical calculations of RATs, provided that we treat $Q_{e1}^{max}/Q_{e2}^{max}$ as an adjustable parameter. Indeed, while we show in Appendix B that this ratio changes with $\alpha$, the range of the ratio variations is lower than for irregular grains (compare Fig. \ref{f5} and Fig \ref{f23*}). 

This adjustment is natural, as we do not really expect the scattering by
irregular dielectric grains to be entirely equivalent to scattering by mirrors
of our toy model of a grain. 


Our analytical studies are based on AMO, which is provides a simple model
of RATs. If necessary,
AMO can be trivially generalized by adding additional mirrors to grain
surface and accounting for a partial shadowing of the mirrors by the
spheroidal grain body. In addition, one can consider not perfectly 
reflective models, but models with refractive grain body and also refractive
plate instead of a mirror. This allows us to vary AMO's properties.

An arbitrary attachment of mirrors may make $Q_{e1}$
less symmetric.  For instance, a grain
shown in the upper panel of Fig. \ref{f17} has a mirror attached to one of its surfaces. Naturally,
turning this surface towards the beam of radiation produces torques that
are different from the case when the mirror is hidden by the elliptical body
of the grain. 

\begin{figure}
\includegraphics[width=0.49\textwidth]{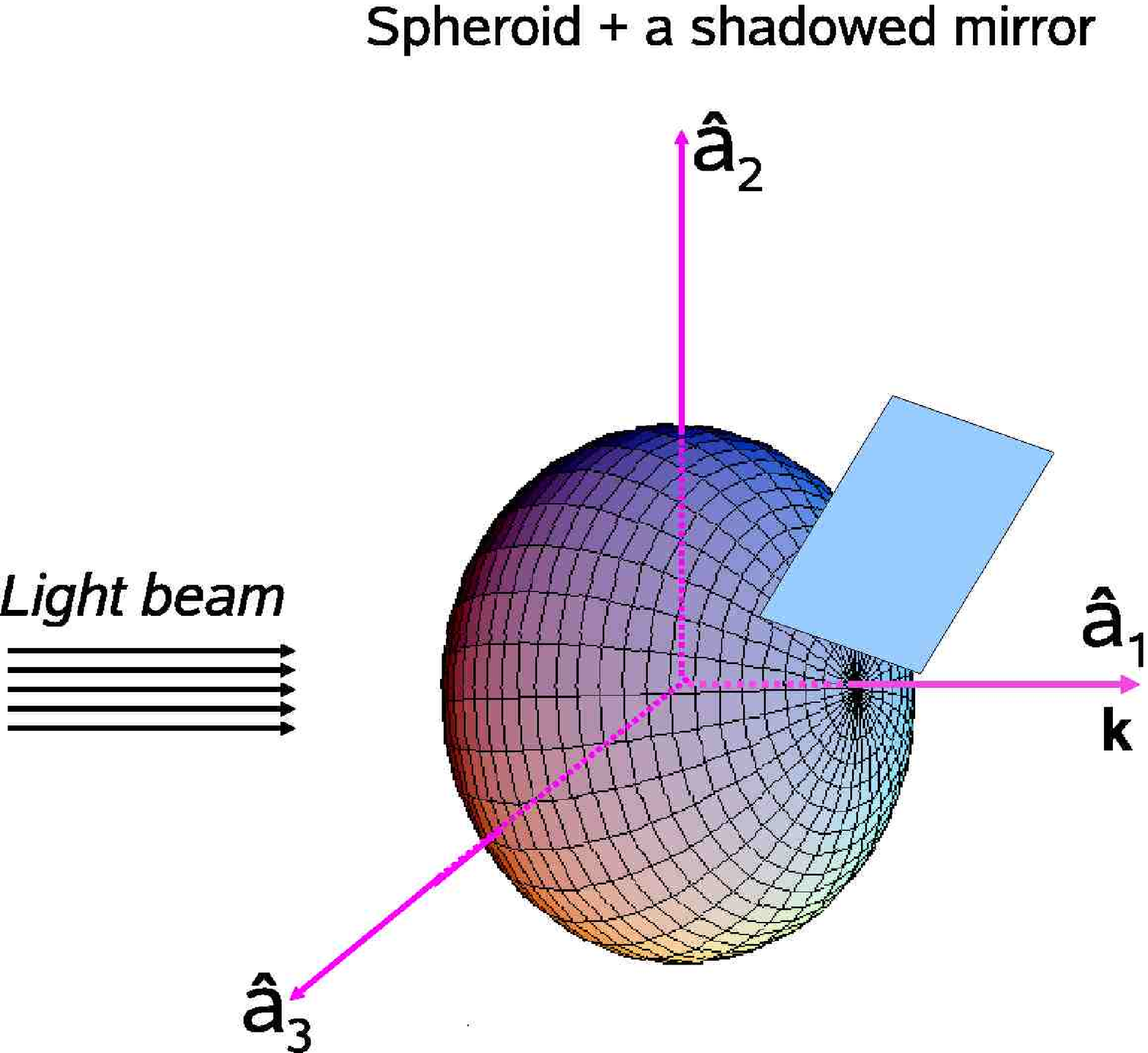} 
\includegraphics[width=0.49\textwidth]{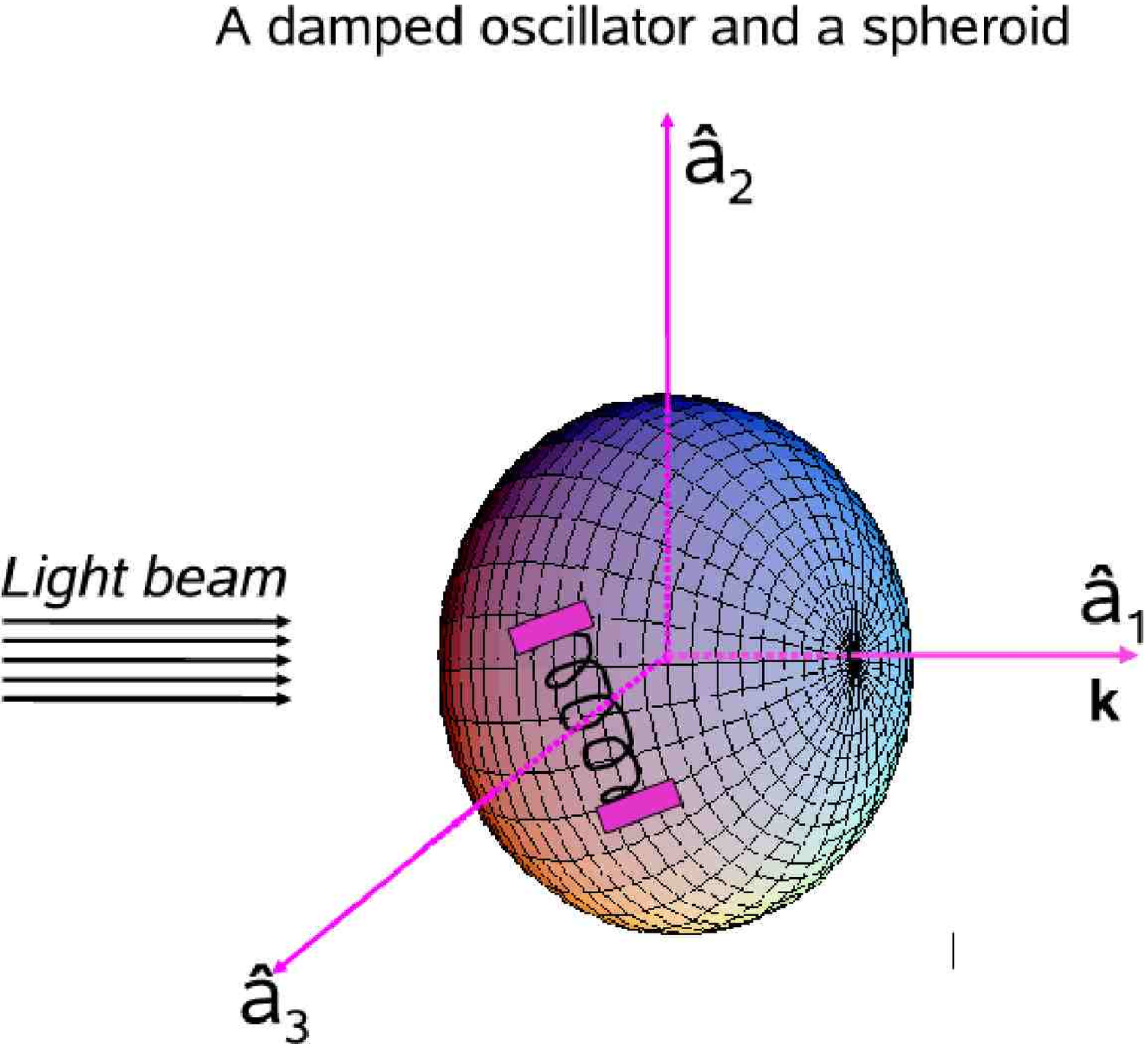} 
\caption{Models of a grain shape consisting of a mirror shadowed partially by
  the oblate spheroid (upper panel) and a partially damped oscillator
attached to the oblate spheroid (lower panel).} 
 \label{f17}                                                              
\end{figure} 

In addition, the adopted AMO does not demonstrate account for
torques arising from absorption. If, however, one adds to the spheroidal
grain a one-dimensional damped oscillator with the axis at an angle to the 
spheroid symmetry axis (see the lower panel in Fig. \ref{f17}), such a grain 
would have a non-zero torque component $Q_{e2}$ arising from light absorption.
Note, that just adding absorbing component to one of the surfaces of the 
reflecting mirror would distort the symmetry of torques $Q_{e1}$, which would
correspond to RATs calculated for irregular grains for small $\lambda/a_{eff}$
ratio.

While adding such effects could provide us with better fits of RATs obtained numerically for particular shapes, this would increase the
complexity of our model. We do not pursue this path here, therefore. However,
we can see that 
in some cases, e.g. dealing with highly absorbing grains, e.g. graphite
grains, one may have to make the corresponding analytical model more 
sophisticated.

Note, that some modifications do not require making the model
 more complex.
Historically, spheroidal grains were used to demonstrate alignment. However,
for demonstrating the effects of H$_2$ torques Purcell (1979) considered a
``brick-shaped'' grain. If we want to study the effects of H$_2$ torques
for AMO, it is natural to consider not a ``spheroid with a wing'', as
we have done in this paper, but a ``brick with a wing''. In terms of AMO this would not change the properties, apart from the value of 
$Q_{e3}$ torque, which is not important for the alignment, anyhow.

Intentionally, our AMO is based on an intuitive macroscopic model.
 Microscopic models based on
the analytical treatment of RATs are possible also. For instance,
light scattering by optically active (chiral)
sphere was analytically studied by Bohren (1974). The author derived analytical solutions for scattering matrix. Due to the optically activeness, such spherical grain can produce RATs. This model
would correspond to the Dolginov (1972) suggestion of quartz grains being spin up 
and aligned by RATs. 

An important generalization of AMO is to consider that its inertial
 properties are given not by a spheroid but by a triaxial ellipsoid. We
expect changes to result in difference in dynamics during periods of crossovers. The corresponding effects together with the modification of inertial properties of AMO are considered in Hoang \& Lazarian (2007).

All in all, while it is rather easy to make AMO more sophisticated and provide
a better fit for RATs within different ranges of $\lambda$ and $a_{eff}$,
for the rest of the paper, we adopt a very simple model of AMO. This model
provides a reasonable fit to generic properties of 
RATs acting on actual irregular grains. Thus studying grain alignment with AMO 
should provide insight into the alignment processes acting upon
 astrophysical grains.

\section{Alignment with respect to light direction}\label{kalign}

While the earlier studies dealt with the alignment in respect to magnetic
field,
in this section we show that RATs can align grains on their own, without any
influence of magnetic fields. In this case, the direction of radiation ${\bf k}$ is 
the axis of alignment. As we further discuss in \S \ref{sec91} such alignment
happens in the presence of magnetic field when the rate of precession induced
by radiation is faster than the Larmor precession rate.
 Dealing with this simple case also prepares
us for dealing with a more complicated cases of alignment in respect to magnetic field in \S \ref{balign}. 

\subsection{RATs: spin-up, alignment and precession}

To understand the role of the RAT components, we calculate torques 
 that spin up, align, and induce grain precession. A RAT component that acts to spin up grains, $H$, is directed
along $\hat{{\bf a}}_{1}$, the component that aligns
grains, $F$, is directed along $\hat{\Theta}$, and RAT that causes precession, $G$,
is along $\hat{\Phi}$.  
They are respectively given by (see DW97) 
 \begin{align}
 H(\Theta, \Phi)&=\mbox{\bQ}_{\Gamma}.\mbox{\be}_{1}(\Theta,\Phi)\mbox{cos}\Theta+
 \mbox{\bQ}_{\Gamma}.\mbox{\be}_{2}(\Theta,\Phi)\mbox{sin}\Theta\mbox{cos}\Phi\nonumber\\
&+\mbox{\bQ}_{\Gamma}.\mbox{\be}_{3}(\Theta,\Phi)\mbox{sin}\Theta\mbox{sin}\Phi\label{eq21},\\ 
F(\Theta, \Phi)&=-\mbox{\bQ}_{\Gamma}.\mbox{\be}_{1}(\Theta,\Phi)\mbox{sin}\Theta+
\mbox{\bQ}_{\Gamma}.\mbox{\be}_{2}(\Theta,\Phi)\mbox{cos}\Theta\mbox{cos}\Phi\nonumber\\
&+\mbox{\bQ}_{\Gamma}.\mbox{\be}_{3}(\Theta,\Phi)\mbox{cos}\Theta\mbox{sin}\Phi\label{eq22},\\
G(\Theta, \Phi)&=-\mbox{\bQ}_{\Gamma}.\mbox{\be}_{2}(\Theta,\Phi)\mbox{sin}\Phi+\mbox{\bQ}_{\Gamma}.\mbox{\be}_{3}(\Theta,\Phi)\mbox{cos}\Phi\label{eq23}. 
\end{align}
  
 On the other hand, following equation (\ref{eq2}), RATs at a precession angle  $\Phi$ are related to
which at  $\Phi=0$ via  
 \begin{align}
\mbox{\bQ}_{\Gamma}.\mbox{\be}_{1}(\Theta,\Phi)&=Q_{e1}(\Theta,0),\label{eq24}\\
\mbox{\bQ}_{\Gamma}.\mbox{\be}_{2}(\Theta,\Phi)&=Q_{e2}(\Theta,0)\mbox{cos}\Phi-Q_{e3}(\Theta,0)\mbox{sin}\Phi,\label{eq25}
\\ 
 \mbox{\bQ}_{\Gamma}.\mbox{\be}_{3}(\Theta,\Phi)&=Q_{e2}(\Theta,0)\mbox{sin}\Phi+Q_{e3}(\Theta,0)\mbox{cos}\Phi.\label{eq26} 
\end{align}
 
 Plugging the above equations into (\ref{eq21}), (\ref{eq22}) and
 (\ref{eq23}) we get
 \begin{align}
G(\Theta, \Phi)&=Q_{e3}(\Theta,0)\label{eq27},\\
 H(\Theta, \Phi)&=Q_{e1}(\Theta,0)
 \mbox{ cos}\Theta + Q_{e2}(\Theta,0)\mbox{ 
  sin}\Theta,\label{eq28} \\ 
F(\Theta, \Phi)&=-Q_{e1}(\Theta,0)\mbox{ 
  sin}\Theta + Q_{e2}(\Theta,0)\mbox{ cos}\Theta.\label{eq29}
 \end{align}    

If for $G(\Theta, \Phi)$ we are mostly interested in its amplitude, the
functional form of $F(\Theta, \Phi)$ and $H(\Theta, \Phi)$ is essential for
grain alignment. The problem is that $F(\Theta, \Phi)$ and $H(\Theta, \Phi)$
as well as their counterparts obtained in the presence of magnetic field (see
equations \ref{eq61} and \ref{eq62}), vary substantially from one grain to another. For AMO
different grains correspond to different ratio $Q_{e1}^{max}/Q_{e2}^{max}$. As we mentioned in \S \ref{sec43} and \S \ref{para}, irregular grains are different in terms of RATs for the radiation of different wavelengths and
different grain sizes. However, for both AMO and irregular grains, the generic properties of the RAT components (i.e., symmetry of $Q_{e1}$, as well as the asymmetry and zeroes of $Q_{e2}, Q_{e3}$) is always unchanged. Therefore, unlike $Q_{e1}$ and $Q_{e2}$, the components
$F(\Theta, \Phi)$ and  $H(\Theta, \Phi)$ do not demonstrate a universal
behavior and play an auxiliary role in our study.

We see that the precessing torque depends only on the third component
 $Q_{e3}(\Theta,0)$, while the aligning  and 
 spinning torques are related to two first components, namely
 $Q_{e1}(\Theta,0), Q_{e2}(\Theta,0)$. We note that the functions $F, G, H$
 are the functions of only variable $\Theta$, in this case.

For AMO with $\alpha=45^{0}$, substituting analytical expressions $f=f_{\pi/2}$ and $g$ given by equations (\ref{eq16})
and (\ref{eq17b}) into equations (\ref{eq14}) and (\ref{eq15}), we get

\begin{align}
Q_{e1}(\Theta, 0)&=\frac{16 l_{1} n_{1}n_{2}|n_{2}|}{3\lambda}(5\mc^{2}\Theta-2),\label{eq30a}\\
 Q_{e2}(\Theta, 0)&=\frac{40 l_{1} n_{1}n_{2}|n_{2}|}{3\lambda}\ms2\Theta(1.191+0.1382\mcs\Theta).\label{eq31a}
\end{align} 
Therefore, equations (\ref{eq27})-(\ref{eq29}) becomes
\begin{align}
G(\Theta)&=-\frac{2e a (1-s)}{\lambda} K(\Theta, e)\ms 2\Theta \label{eq30},\\
H(\Theta)&=\frac{8 l_{1}n_{1}n_{2}|n_{2}|}{3\lambda}\mc\Theta[1+6.91\mss\Theta\nonumber\\
&+\mcs\Theta(5+1.382\mss\Theta)],\label{eq31} \\ 
F(\Theta)&=\frac{8 l_{1}
  n_{1}n_{2}|n_{2}|}{3\lambda}\ms\Theta[-1+6.91\mcs\Theta\nonumber\\
&+1.382\mbox{cos}^4\Theta+5\mss\Theta].\label{eq32} 
\end{align}

\subsection{Simplified Treatment of Crossovers} \label{cross}
In general, the maximal inertia axis ${\bf a}_{1}$ of our
 model grain can precess about the vector of the angular  momentum
${\bf J}$. In the present paper, however, for the sake of simplicity, 
we assume a perfect internal alignment, i.e. ${\bf J} \| 
{\bf a}_{1}$. This assumption coincides with that in DW97 and
can be justified by the high efficiency of the internal relaxation within a
 wobbling grain.
This relaxation stems from the Barnett relaxation discovered by Purcell (1979)
and/or nuclear relaxation introduced in LD99b.
However, these relaxation processes provide a good
coupling only when $J \gg J_{th}\approx (kT_{d} I_{1})^{1/2}$ where $T_{d}$
is the dust temperature, $I_{1}$ is the maximal moment of inertia of grain (Lazarian 1994), 
i.e. when a grain rotates with suprathermal velocities. This condition 
is not satisfied as a grain approaches crossover points, 
i.e. as $J\rightarrow J_{th}$.

We adopt below a simplified treatment of crossovers, which is different, however, from the treatment of crossovers in DW97. There it was assumed that
${\bf J} \| {\bf a}_{1}$ up to the moment of the angular velocity getting zero.
After that DW97 assumed that ${\bf J}$ changes its direction to the opposite,
while the direction of ${\bf a}_{1}$ is preserved.
Such a model of crossovers differs from the earlier work on crossovers, which
suggests that, as the angular velocity goes to zero, the grain undergoes a flip,
i.e., ${\bf J}$ preserves its direction, while the direction of ${\bf a}_{1}$
changes to the opposite (Spitzer \& McGlynn 1979; Lazarian \& Draine 1997). Therefore, in what follows, we adopt a model in which ${\bf J} \| {\bf a}_{1}$
up to a crossover; at the crossover ${\bf a}_{1}$ flips, while the direction of
${\bf J}$ is preserved. This makes the dynamics of grains very different from
that in DW97\footnote{We may mention parenthetically, that the dynamics that
  we get has similarities to the dynamics of grains with thermal fluctuations
  accounted for, as in WD03. There, however, it was interpreted as a new
  effect related to thermal fluctuations.}. Although we accept that our
model is not precise when $J \sim J_{th}$, we claim the our simplified model
is ``roughly true''. The latter point is justified for regular crossovers
(i.e., crossovers without thermal fluctuations) in \S \ref{cross}.
 
\subsection{Stationary and singular points}

\subsubsection{Equations of motion}
To find out whether grains can be aligned with respect to ${\bf k}$, we need to
find stationary states for grain motion exerted by RATs.
We start with equations of motion
 
 \begin{align} 
 \frac{dJ}{dt}&=M H(\Theta)-J,\label{eq33}\\
 \frac{d\Theta}{dt}&=M \frac{ F(\Theta)}{J},\label{eq34}
 \end{align}
where time $t$ and angular momentum $J$ are scaled in the units of the gas
 damping time $t_{gas}$ (see Table \ref{tab1}) and the thermal angular
 momentum $I_{1}\omega_{T}$. Here $I_{1}$ is the maximal
 moment  of inertia and $\omega_{T}=(\frac{kT_{gas}}{I_{1}})^{1/2}$ is
the thermal angular velocity with $T_{gas}$ is the gas temperature. In
 equations (\ref{eq33}) and (\ref{eq34})
\bea
 M=\frac{\gamma u_{rad}\lambda a_{eff}^{2}t_{gas}}{2I_{1}\omega_{T}}
\ena
contains important parameters of the physical problem. There $u_{rad}$ is the energy density of
radiation field, $\lambda$ is the wavelength.  In the present paper, for our estimates, we use the values
 provided in Table \ref{tab1} for
interstellar medium. It will be explicitly stated in case other values of these parameters are used.
 \begin{table}
\caption{Physical parameters for diffuse ISM }
\begin{displaymath}
\begin{array}{rrrr} \hline\hline\\
\multicolumn{1}{c}{\it Definitions} & \multicolumn{1}{c}{\it Typical~ Values} \\[1mm]
\hline\\
{\rm Gas~ density}& {\rm n=30~ cm^{-3}}\\[1mm]
{\rm Gas~temperature}& {\rm T_{gas}=100~K}\\[1mm]
{\rm Gas~damping~time}& {\rm t_{gas}=4.6\times
  10^{12}(\frac{\hat{\rho}}{\hat{n}\hat{T}_{gas}^{1/2}}) a_{-5}~ s}\\[1mm]
{\rm Dust~temperature}& {\rm T_{d}=20~K}\\[1mm]
{\rm Anisotropy~degree}&{\rm \gamma=0.1}\\[1mm]
{\rm Grain~ size}&{\rm a_{eff}=0.2 \mu m}\\[1mm]
{\rm Mean~wavelength}&{\rm \overline{\lambda}=1.2~ \mu m}\\[1mm]
{\rm Mean~density~of~ISRF}&{\rm {u}_{rad}=8.64 \times 10^{-13}~ erg~ cm^{-3}}\\[1mm]
\\[1mm]
\hline\hline\\
\end{array}
\end{displaymath}
The normalized values that we also use
are  $\hat{T}_{gas}=T_{gas}/100~ K,~\hat{n}=n/30~ g~ cm^{-3}$, and
$a_{-5}=a_{eff}/10^{-5}~ cm $, and normalized grain density
$\hat{\rho}=\rho/3~ g~ cm^{-3}$.
\label{tab1}
\end{table}

\subsubsection{Stationary points}

The stationary point is determined by setting the equations of
motion (\ref{eq33}) and (\ref{eq34}) equal to zero. This point may be an attractor or a repellor point. We remind the
reader that for a stationary point to be an attractor point, it requires (see Appendix C)
\begin{align}
\left. \frac{d F(\Theta)}{H(\Theta) d\Theta}\right|_{\Theta=\Theta_{s}} &<1,\label{eq44}\\
\left. H(\Theta)
  \frac{dF(\Theta)}{d\Theta}\right|_{\Theta=\Theta_{s}}&<\left. F(\Theta)\frac{d H(\Theta)}{d\Theta}\right|_{\Theta=\Theta_{s}}.\label{eq45}
\end{align}

At the stationary points, $ F(\Theta_{s})=0$, so equations (\ref{eq44}) and (\ref{eq45}) reduce to 
\bea
\left. \frac{d F(\Theta)}{d\Theta}\right|_{\Theta=\Theta_{s}}<0,\mbox{ for } H(\Theta_{s})>0,\label{eq46} 
\ena
and
\bea
\left. \frac{d F(\Theta)}{d\Theta}\right|_{\Theta=\Theta_{s}}>0,\mbox{ for } H(\Theta_{s})<0.\label{eq47} 
\ena

Consider that initially a grain has the maximal inertia axis ${\bf a}_{1}$ parallel to
 ${\bf k}$, i.e., $\Theta_{0}=0$, then $ F(\Theta_{0})=0, H(\Theta_{0})=Q_{e1}(\Theta_{0})$. These assumptions are
 usually used to estimate the grain angular velocity induced by RATs (see
 DW97; Cho \& Lazarian 2005).  Therefore, the equations of motion become
\begin{align}
\frac{d\Theta}{dt}&=0,\\
\frac{dJ}{dt}&=M Q_{e1}-J.
\end{align}
Solutions for the above equations are easily found,
\begin{align}
\Theta&=0,\label{eq38}\\
J&=M Q_{e1}(\Theta)+(J_{0}-M Q_{e1}(\Theta_{0}))e^{-t},\label{eq39}
\end{align}
where $J_{0}$ is the initial angular momentum of grains.
Equation (\ref{eq39}) shows that as $t \gg 1, J =M Q_{e1}(\Theta=0)$. In other words, grains
 initially parallel to ${\bf k}$ are spun up by RATs and aligned with ${\bf
 k}$ at angular momentum $J_{s}=M Q_{e1}(\Theta_{s}=0)$, regardless of its initial angular momentum $J_{0}$. However, the particular point of phase
 trajectory map may not provide stable orientation of the grain, i.e., may not be
attractor points. \footnote{The shaded area with diagonal lines in Fig \ref{f13} corresponds to the existence of such points. We see that many wavelengths the condition is not satisfied.}   
 
For $\ms\Theta_{0} \ne 0$ , substituting $ F(\Theta), H(\Theta)$ from
 equations (\ref{eq31}) and (\ref{eq32}) into equations (\ref{eq33}) and (\ref{eq34}), we obtain
 \begin{align}
 \frac{dJ}{dt}&=\frac{8 l_{1} n_{1}n_{2}|n_{2}|}{3\lambda}M\mc\Theta[1+6.91\mss\Theta\nonumber\\
&+\mcs\Theta(5+1.382\mss\Theta)]-J,\label{eq40}\\
 \frac{d\Theta}{dt}&=\frac{8 l_{1} n_{1}n_{2}|n_{2}|M}{3\lambda J}\ms\Theta[-1+6.91\mcs\Theta\nonumber\\
&+1.382\mbox{cos}^4\Theta+5\mss\Theta].\label{eq41}
 \end{align}
Hence, we can easily find a stationary point\footnote{Another stationary point $\Theta=\pi, J<0$ is forbidden since $J$ can not be negative. Note, that the analysis in this section is applied for the right helical grain, i.e., $n_{1}n_{2}>0$.}
\begin{align}
\Theta_{s}&= 0,\label{eq42}\\
J_{s}&=\frac{48 l_{1} n_{1}n_{2}|n_{2}|M}{3\lambda}.\label{eq43}
\end{align}
From equations (\ref{eq31}) and (\ref{eq32}) we get
\bea
\left. \frac{d  F(\Theta)}{d\Theta}\right|_{\Theta_{s}=0}=\frac{90.232 l_{1}
  n_{1}n_{2}|n_{2}|}{3\lambda},\label{eq48}
\ena
and $H(\Theta_{s}=0)>0$. As a result, the stationary point $\Theta_{s}=0$ is a repellor point.

In a general case, $ F(\Theta), H(\Theta) $ are given by equations (\ref{eq28}) and (\ref{eq29}). Thus, substituting
 them into equations (\ref{eq46}) and (\ref{eq47}), we get a criteria for a stationary point
 $\Theta_{s}$ being an attractor point, namely,
 \bea
 \frac{-Q'_{e1}\ms\Theta_{s}+Q'_{e2}\mc\Theta_{s}}{Q_{e1}\mc\Theta_{s}+Q_{e2}\ms\Theta_{s}} <1,\label{eq50} 
 \ena
where $Q'_{e1}=\frac{dQ_{e1}}{d\Theta},Q'_{e2}=\frac{dQ_{e2}}{d\Theta}$. However, our
approximate formulae above catches the phenomenon correctly and we still get
repellor points for $\Theta_s=0$ with the exact expressions for AMO.

As the range for ratios of $Q_{e1}/Q_{e_2}$ for an arbitrarily chosen irregular grain
may be different from AMO, for some irregular grains we still may have attractor
points for $\Theta_s=0$, as it is shown in Fig~\ref{f13} for shape~1
subjected to interstellar radiation field. There the relevant data corresponds
to $\psi=0$, as in this situation the direction of magnetic field and the light coincide
and thus our considerations about the alignment in respect to ${\bf k}$ are applicable.

\subsubsection{Singular points: crossovers}

When $J=0$, we have a singular point (see equation \ref{eq34}).
In terms of our simplified
equations it corresponds to a crossover.
As we discussed earlier, a crossover is a period in which a
  grain spins down to the point that the component of angular momentum
  parallel to ${\bf a}_{1}$ gets zero. In the presence of strong internal
  relaxation that tends to align ${\bf J}$ and ${\bf a}_{1}$, this means that
  $J$ should get small (Spitzer \& McGlynn 1979). Our equations above are
  derived in the assumption of ${\bf J}\| {\bf a}_{1}$ and therefore can not
  treat grain crossovers (cf. \S6~\ref{cross}). However, they can still trace the grain
  dynamics as the grain phase trajectory approaches the crossover and $J \to
  J_{th}$. Assuming that initial angular momentum of the grain $J_{0} \gg J_{th}$, we disregard the difference
  between $J=J_{th}$ and $J=0$. Thus, from equation~(\ref{eq34}) it follows that, in order to have a physical crossover,
 grains must have $J =0$, and the aligning
  torque $F(\Theta)$ must be zero. The latter condition is
  naturally satisfied since we found that $ F(\Theta)=0$ at stationary
  points. The former one is satisfied as RATs act to decelerate the
  grain rotation. Indeed, equation~(\ref{eq32}) shows that $  F(\Theta)>0$, i.e., it acts to increase $\Theta$, for every angles
 $\Theta<\pi$. Yet, equation (\ref{eq31}) shows that $ H(\Theta)<0$ for $\mbox{cos }\Theta <0$ and $ H(\Theta) >0$ for
 $\mbox{cos }\Theta >0$. Therefore, if initially the angular momentum of grains makes an
 angle $\mc \Theta <0$, then their angular momentum is 
decreased due to $ H(\Theta)< 0$; so grains approach directly to the state of $J =0$ (see
 the upper panel in Fig. \ref{f8} for right helical grain). 
On the other hand, for grains which initially have $\mc \Theta >0$,
the aligning torque acts
  to increase $\Theta$, while the spinning torque increases their angular
  momentum. Eventually, grains attain the angle $\mc\Theta<0$ for which the
  spinning torque changes the sign, so they are decelerated to the state of $J=0$ (see
 Fig. \ref{f8} for phase maps of left and right helical grains).

One can observe that
 when grains get to the crossover with very low $J$, their maximal inertia axis
 $\ma_{1}$ flips with respect to ${\bf J}$ to enter the opposite flipping
 state, i.e., grains flip from the upper to lower panel in Fig. \ref{f8}{\it upper}, for
 instance. Right after that, grains flip back to the initial state (upper
 panel). This back and forth flipping process takes place frequently. As a
 result, the crossover point, in our approach, can be treated as the attractor
 point at zero angular momentum, hereafter, called low-$J$ attractor
 point\footnote{In Hoang \& Lazarian (2007), we will show that  zero angular
 momentum attractor points become attractor points at thermal angular momentum
 due to  thermal wobbling}, to be distinguished from high-$J$ attractor points.

 \subsection{RATs: alignment by one component}  \label{sec4p4}

 The grain alignment for the case at hand is uniquely related to two 
components $Q_{e1}$ and $Q_{e2}$. To understand which component causes the
alignment, let us study the role of $Q_{e1}$ and $Q_{e2}$ separately. 

If $Q_{e2}=0$, then $\frac{dQ_{e2}}{d\Theta}(\Theta_{s}=0, \pi)=0$, equation
(\ref{eq50}) for attractor points becomes
\bea
Q_{e1}(\Theta_{s}=0)>0,\label{eq51} \\
Q_{e1}(\Theta_{s}=\pi)<0.\label{eq51*}
\ena
According to AMO, $Q_{e1}(\Theta_{s}=0, \pi)=\frac{16 l_{1} n_{1}n_{2}|n_{2}|}{3\lambda}(5\mbox{cos}^2\Theta_{s}-2)=1>0$. As a result, we
expect that the stationary point $\Theta_{s}=0$ is an attractor point,
i.e., grains are perfectly aligned with respect to ${\bf k}$. While, the stationary point $\Theta_{s}=\pi$ is a repellor point.\\

For the case $Q_{e1}=0$, besides stationary points $\Theta_{s}=0, \pi$ as
discussed above, there is another stationary point corresponding to
$\Theta_{s}=\pi/2$ (see equation \ref{eq29}). For $\Theta_{s}=0, \pi$, equation (\ref{eq50}) reduces to
\bea
\frac{dQ_{e2}}{d\Theta}(\Theta_{s}=0)<0, \label{eq52} \\
\frac{dQ_{e2}}{d\Theta}(\Theta_{s}=\pi)>0. \label{eq52*}
\ena
For AMO with $\alpha=45^{0}$, thus, $Q_{e2}=\frac{40 l_{1} n_{1}n_{2}|n_{2}|}{3\lambda}\ms 2\Theta (1.191+0.1382\mcs\Theta)$, we have $
\frac{dQ_{e2}}{d\Theta}(\Theta_{s}=0, \pi)=\frac{106.2 l_{1} n_{1} n_{2}|n_{2}|}{3\lambda}\mc2\Theta_{s} >0$. It means
that the condition (\ref{eq52}) is not satisfied, but equation (\ref{eq52*}) is fulfilled. In other words, the stationary point $\Theta_{s}=0$ is a repellor points, while the stationary point $\Theta_{s}=\pi$ is an attractor point.

For a particular stationary point $\Theta_{s}=\pi/2$, equation (\ref{eq52}) in this case becomes
\bea
\frac{Q'_{e2}\mc\Theta_{s}}{Q_{e2}\ms\Theta_{s}}=0<1.\label{eq53}
\ena
Therefore, the stationary point $\Theta_{s}=\pi/2$ is, indeed, an
attractor point. Note, equation (\ref{eq53}) which indicates that
$\Theta_{s}=\pi/2$ is the attractor point is valid for an arbitrary form of $Q_{e2}$. These results show that the component $Q_{e2}$ acts to align
grains in the direction perpendicular to the radiation beam for arbitrary grain shapes.

\subsection{Phase trajectories}

To show our predictions from the above analysis, we construct phase trajectory maps in which each RAT component acts separately. The difference of our
phase maps from those in DW97 is that here we present the alignment in
respect to ${\bf k}$ rather than to magnetic field. In addition, we
treat crossovers differently from how they are treated in DW97. With the 
exception of Figs \ref{f12*} and \ref{f15}, to avoid over-crowding of our phase maps we do
not draw arrows with $t_{gas}$ intervals.

Throughout the present paper, the phase trajectory map represents the evolution of $J$ and angle between ${\bf J}$ and ${\bf k}$ or ${\bf B}$ 
 \footnote{We note that the angle between ${\bf J}$ and ${\bf k}$ or ${\bf B}$ is shown rather
 than the angle between ${\bf a}_{1}$ and ${\bf k}$ or ${\bf B}$ as in
 DW97.}. Each phase map 
has upper and lower panels, which correspond to ${\bf J}$ parallel and anti-parallel to
 ${\bf a}_{1}$, respectively. Also, a circle denotes an attractor point and a
 cross denotes a repellor point. 

Moreover, the title of the trajectory map presents information about the model (AMO) or grain shape, size for irregular grains. For trajectory maps labeled with AMO on their titles, it is a default that the RAT components are taken from the exact calculations of equations (\ref{eq8}-\ref{eq10}) for $\alpha=45^{0}$ which have $Q_{e1}^{max}/Q_{e2}^{max}=1.2$. For the maps of AMO in which the ratio $Q_{e1}^{max}/Q_{e2}^{max}$ is explicitly shown, it means that the functional forms of RATs are similar for the case $\alpha=45^{0}$, but the relative amplitudes are rescaled to the shown value. 

 \begin{figure}
\includegraphics[width=0.49\textwidth]{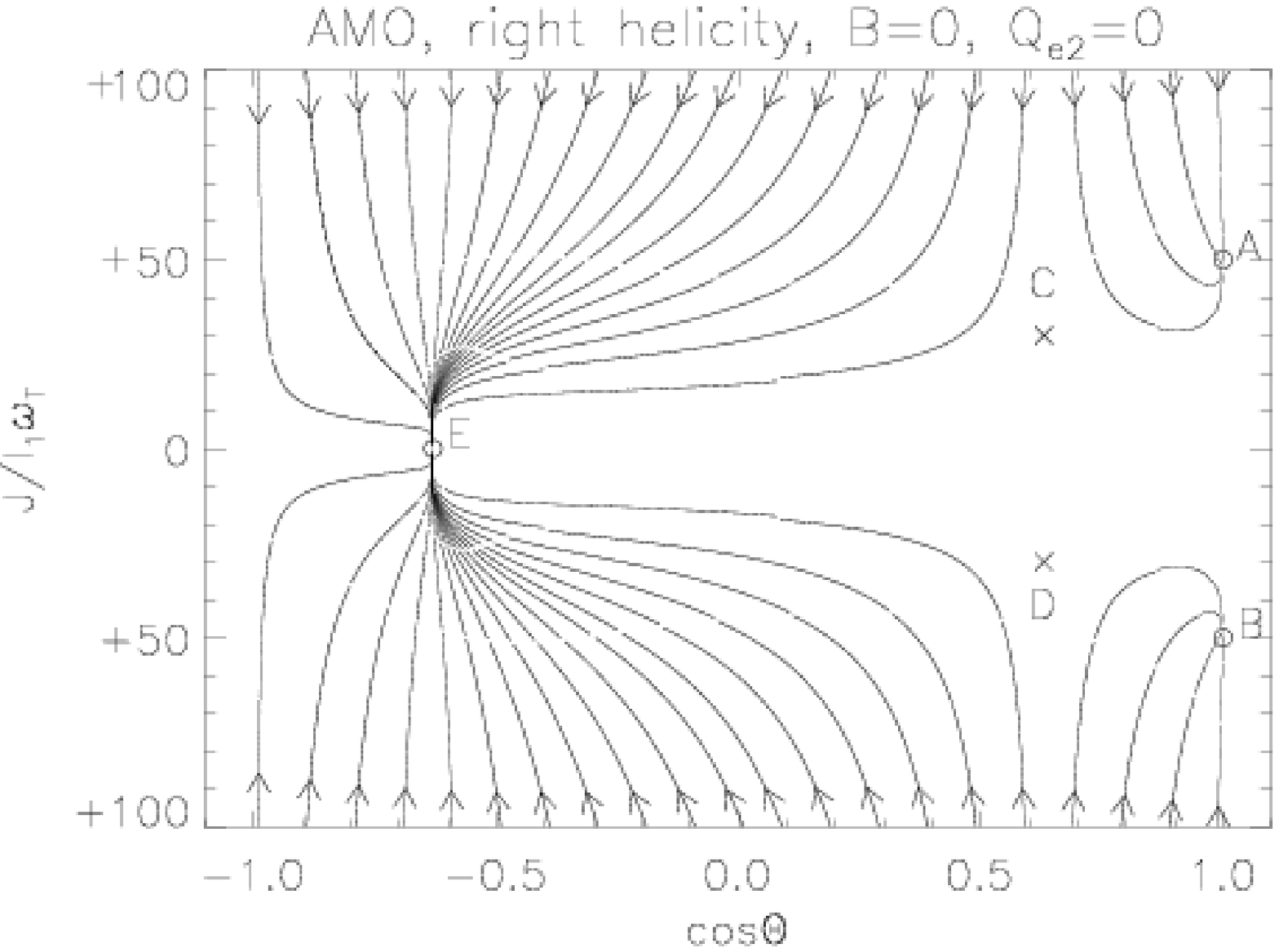} 
\hfill
 \includegraphics[width=0.49\textwidth]{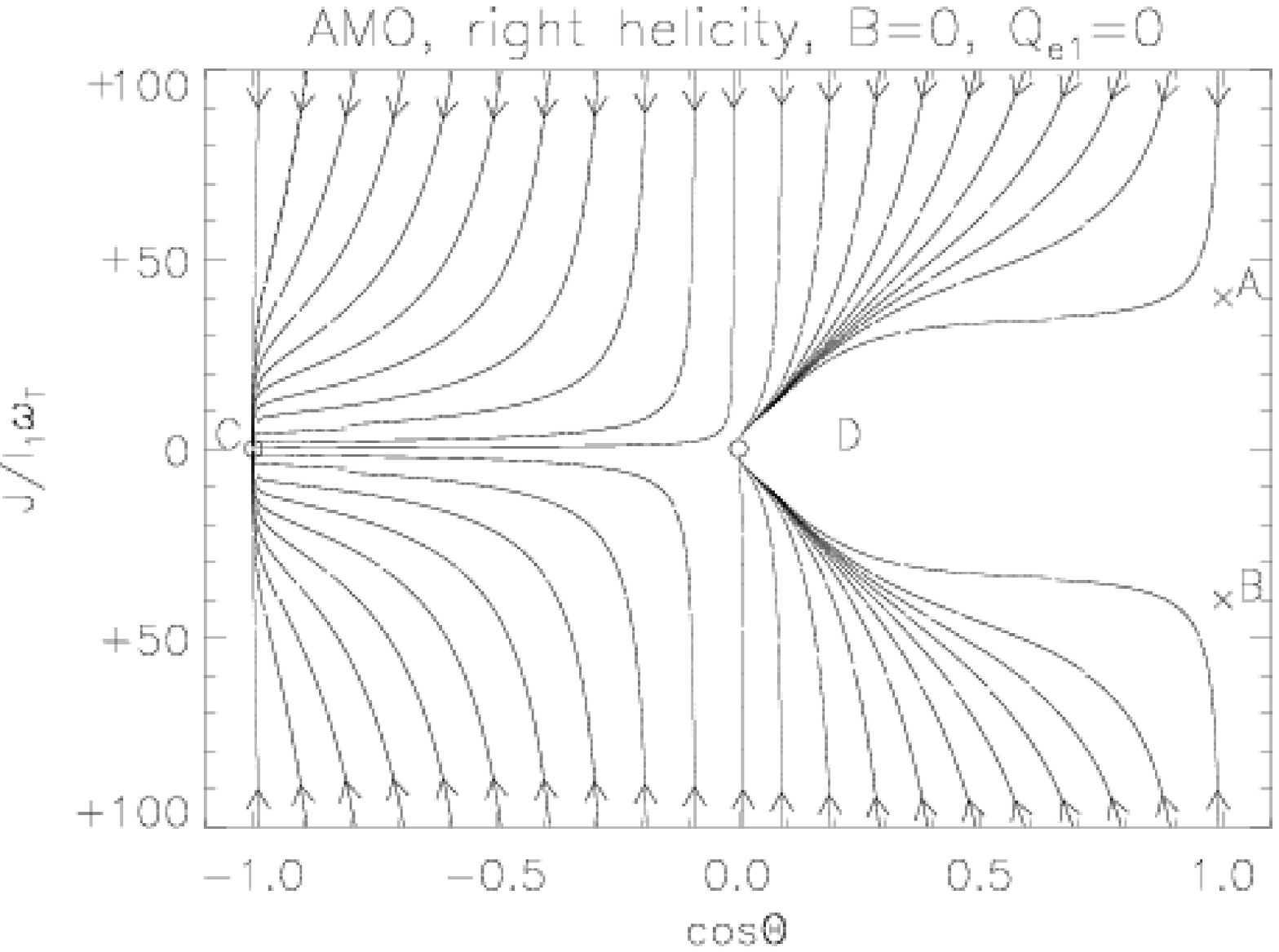}
\caption{Phase trajectory maps of grains in the absence of magnetic
field corresponding to
 $Q_{e2}=0$ ({\it Upper Panel}) and 
  $Q_{e1}=0$ ({\it Lower Panel}). The map in upper panel exhibits two attractor
  points A, B corresponding to a perfect alignment as pointed out by the
 analysis; E is a low-$J$ attractor point and two repellor points C, D. 
  The lower panel shows two repellor
  points A, B and two attractor points C, D in which D corresponds to ``wrong''
 alignment.} 
 \label{f7}
 
 \end{figure}
 
The map for the case when  only $Q_{e1}$ is present has three attractor points:
  A, B  correspond to $\Theta_{s}=0$, and E corresponds to zero of $Q_{e1}$ at $\mbox{cos}\Theta=-0.6$ (see the upper panel in
  Fig.  \ref{f7}). Therefore, $Q_{e1}$  acts to align grains with
  ${\bf J}$ parallel to ${\bf k}$, i.e, ${\bf a}_{1}$ parallel or anti
  parallel to ${\bf k}$ . In addition, there are two repellor points C,
  D. These features show clearly our predictions obtained with AMO.

  When $Q_{e1}=0$, Fig. \ref{f7} (lower panel) shows that the phase trajectory map has one attractor point D at
  $\Theta_{s}=\pi/2$, which corresponds to a perpendicular alignment, one
  low-$J$ attractor point C at $\Theta_{s}=\pi$, and two repellor points A, B as predicted by our above analysis. So
$Q_{e2}$ can align some grains with the maximal inertia axis ${\bf a}_{1}$ perpendicular to the light direction. However, we
 discussed in \S 2 that $Q_{e2}$ is equal zero at the attractor point
$\Theta_{s}=\pi/2$, while $Q_{e1}$ is different to zero at this point; thus the  alignment with the long axes parallel to the direction of light does not
 occurs\footnote{We checked that even the addition of $Q_{e1}$ with an
amplitude of $10^{-9}$ of the amplitude of $Q_{e2}$ can destroy such
an alignment.} for the case of alignment with respect to ${\bf k}$ (see \S~\ref{sec5}). 

\subsection{RATs: alignment by joint action of torques, role of
$Q_{e1}^{max}/Q_{e2}^{max}$ ratio}\label{qratio}

 So far we have dealt with the alignment by one component of RATs, and pointed out
 that $Q_{e1}$ acts to align grains with long axes $\perp$ to ${\bf k}$, while $Q_{e2}$
can produce the alignment for some grains with long axes $\|$ to ${\bf k}$. In reality, grains are
 driven simultaneously by both components, so the alignment of grains depends
 on which component of RATs is predominant.

As discussed in previous sections, for AMO, we have the stationary point
which is determined by equations (\ref{eq42}) and (\ref{eq43}). Also, we have analytically shown that the
stationary point $\Theta_{s}= 0$ ($\pi$ for left helical AMO) is the repellor point. To test our expectation, we produce phase trajectory maps of grains driven
by RATs with the precessing, spinning-up and aligning components given by equations (\ref{eq30})-(\ref{eq32}). Fig. \ref{f8} shows the maps for right and left helical AMOs. It can be 
seen that the phase maps have two stationary points which positions are exactly given by analytical expressions
(\ref{eq42}) and (\ref{eq43}). Among these stationary points, there are two repellor points
A, B at $\Theta_{s}=0$ (or $\pi$ for lower panel), and one low-$J$ attractor point C at $\Theta_{s}=\pi$ (or $\Theta=0$ for lower panel). 
In addition, we see that for the right and left helical AMOs, the trajectory maps
are related through a transformation: $\Theta \to \pi-\Theta$.    
 
\begin{figure}
\includegraphics[width=0.49\textwidth]{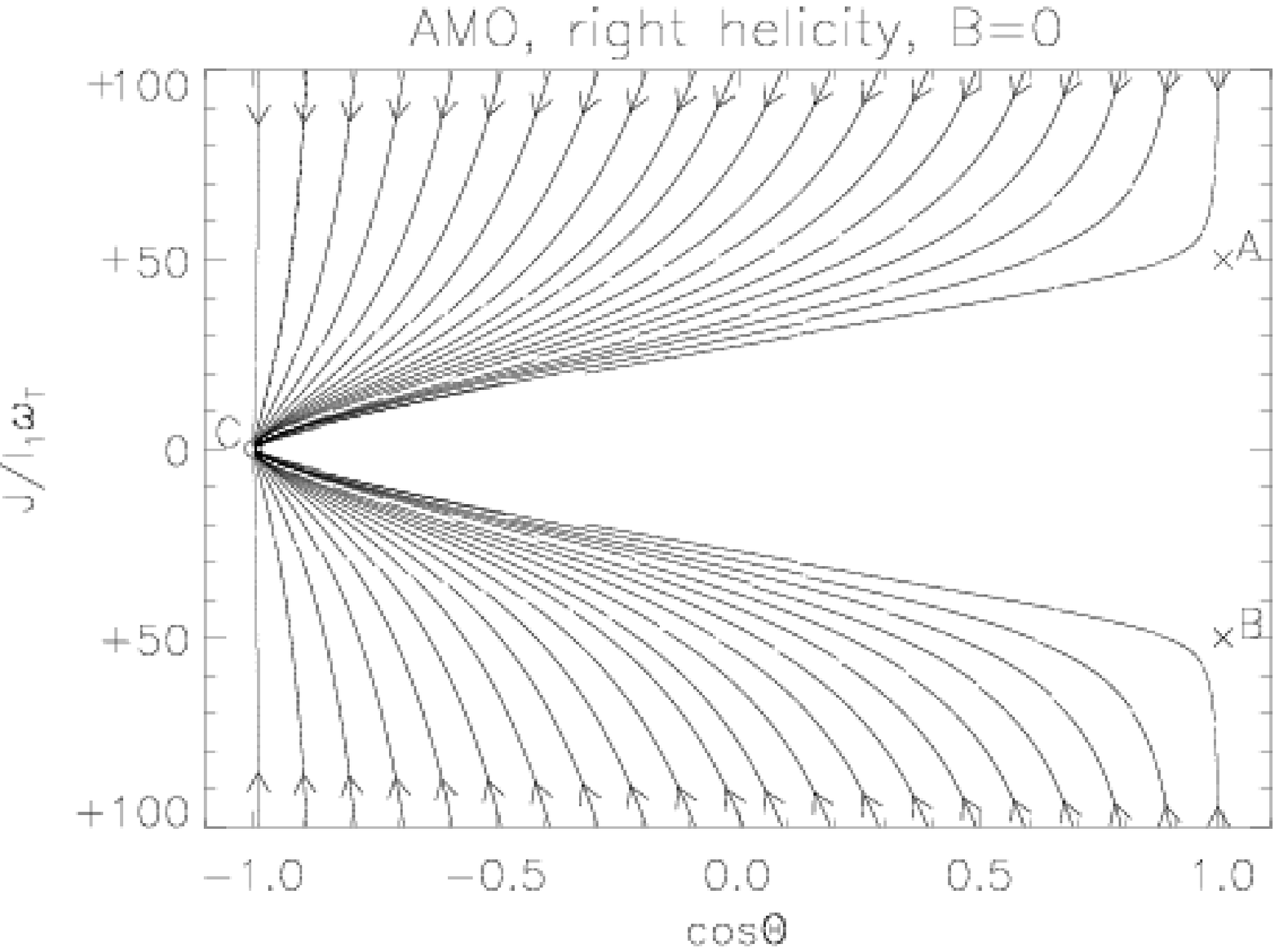}
\hfill
\includegraphics[width=0.49\textwidth]{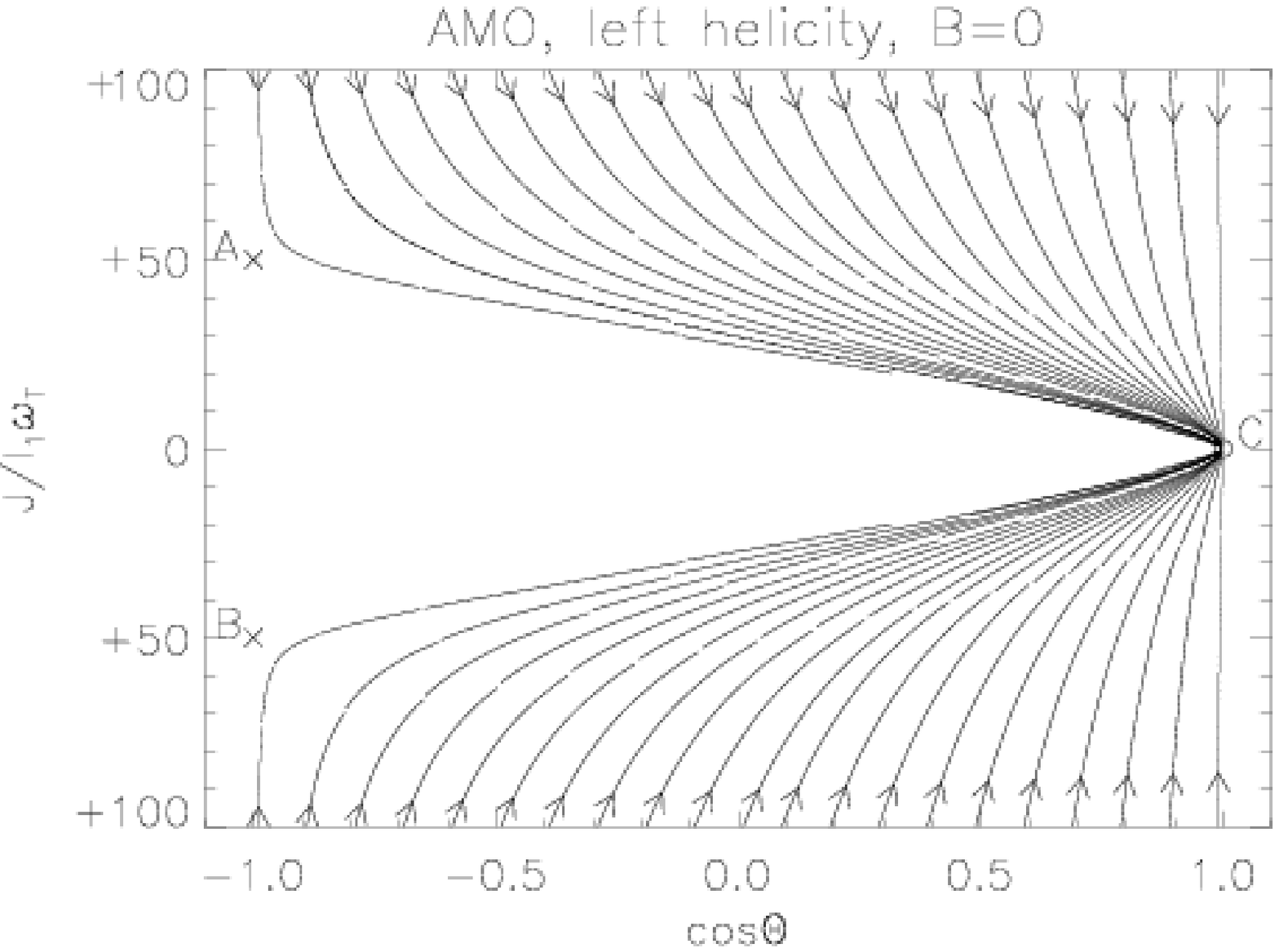}
\caption{Upper and lower panels show the phase trajectory maps for grains with right and
 left helicity, respectively. Figs show that grains have a perfect alignment 
with respect to ${\bf
  k}$ with one low-$J$ attractor point C at $\Theta=\pi$ (upper panel) and
$\Theta=0$ (lower panel). Also, two repellor points A,
  B are shown as predicted in our analysis. The maps for upper and lower panels are
  mirror-symmetric.}
\label{f8}
\end{figure}

We have seen that , in both cases of grain alignment by one component and joint actions of
components, the important features (e.g., attractor and repellor points) present in the trajectory maps  that are constructed with exact RATs are consistent
with the predictions based on the approximate formulae using the fitting
function $f_{\pi/2}$ and $g$. It indicates that, though the fitting functions
do not give the best fit in some particular angles $\Theta$, they can
provide us, intuitively, the alignment property of grains by RATs.

We found in the case of alignment by one component that if only $Q_{e1}$ is at
work, it aligns grains with two attractor points with $J \gg J_{th}$, $\Theta=0$, i.e., high-$J$ attractor points. In contrast, when only
$Q_{e2}$ acts, those points are repellor points; Instead $Q_{e2}$ produces an attractor point
at $\Theta=\pi/2$. When both components act simultaneously, are there high-$J$
attractor points? Obviously, we may conjecture that if $Q_{e1}$ is predominant
over $Q_{e2}$, then high-$J$ attractor points should still appear. Otherwise,
high-$J$ stationary points are repellor points.

For our default AMO with $\alpha=\pi/4$, we have predicted that the stationary points $\Theta=0$ are
always repellor points. This can also be understood  in terms of the ratio of $Q_{e1}^{max}/Q_{e2}^{max} \sim 1$, for this case, i.e, the
dominant criteria is not satisfied (see equations \ref{eq18} and \ref{eq19}).

To study when we have high-$J$ attractor points, let us write RATs for AMO in
a simplified form
\begin{align}
Q_{e1}&=\frac{Q_{e1}^{max}}{3}(5\mcs\Theta-2),\label{eq54}\\
Q_{e2}&=Q_{e2}^{max}\ms2\Theta,\label{eq55}
\end{align}
where $Q_{e1}^{max}, Q_{e2}^{max}$ are maximal values of $Q_{e1}, Q_{e2}$.

Clearly, stationary points for this model are $\Theta_{s}=0$ because aligning
torque $F(\Theta_{s})=-Q_{e1}\ms\Theta_{s}+Q_{e2}\mc\Theta_{s}=0$ at
$\Theta_{s}=0$.\\
Using the criteria of an attractor point (i.e., equation \ref{eq50}), it follows that
$\Theta_{s}=0$ is an attractor point if
\bea
\frac{1}{Q_{e1}}\frac{dQ_{e2}}{d\Theta}(\Theta_{s}=0)<1,\label{eq56}
\ena
Plugging equations (\ref{eq54}) and (\ref{eq55}) into equation (\ref{eq56}), we get 
\bea
Q_{e1}^{max}> 2 Q_{e2}^{max}.\label{eq57}
\ena 
From Fig. \ref{f5} and equation (\ref{eq57}), it follows
that for the original AMO, i.e., AMO in which the relative magnitude of RAT component is not rescaled yet, the maximum of the ratio $max[Q_{e1}^{max}/Q_{e2}^{max}]$ is $1.3$, and
therefore the
stationary points $\Theta_{s}=0$ (corresponding to high $J$)
 are always repellor points. However, as we have discussed earlier, their relative magnitude is adjustable. Thus, AMO can produce the phase map with high attractor points for $\psi=0^{0}$, provided that it satisfies equation (\ref{eq57}). Note, that the criteria (\ref{eq57}) is only applicable for the case of alignment with respect to ${\bf k}$ or $\psi=0^{0}$ (i.e., ${\bf k}\| {\bf B}$). For an arbitrary angle $\psi$, the criteria is shown in the lower panel of Fig.~\ref{f13}. There it can be seen that for some irregular grains (i.e. shape~1, ISRF)
$\Theta_s=0$ can correspond
to high attractor points.

\subsection{Alignment for irregular grains}

Similar to the AMO case, we consider first the alignment of dust in 
respect to the radiation direction ${\bf k}$. In particular, to compare the
action of RATs for AMO and an irregular grain, we consider the alignment that
is induced by individual torque components. 

To calculate the phase map, we use again the parameters from 
Table~{\ref{tab1}. We exemplify the alignment for irregular grains using
shape 1, which shows the maximum deviations from AMO predictions.

Fig. \ref{f24} shows the phase trajectory maps of an irregular 
grain (shape 1) driven by RATs calculated by DDSCAT  with either $Q_{e2}$ or $Q_{e1}$ taken to be
zero.

\begin{figure}
\includegraphics[width=0.49\textwidth]{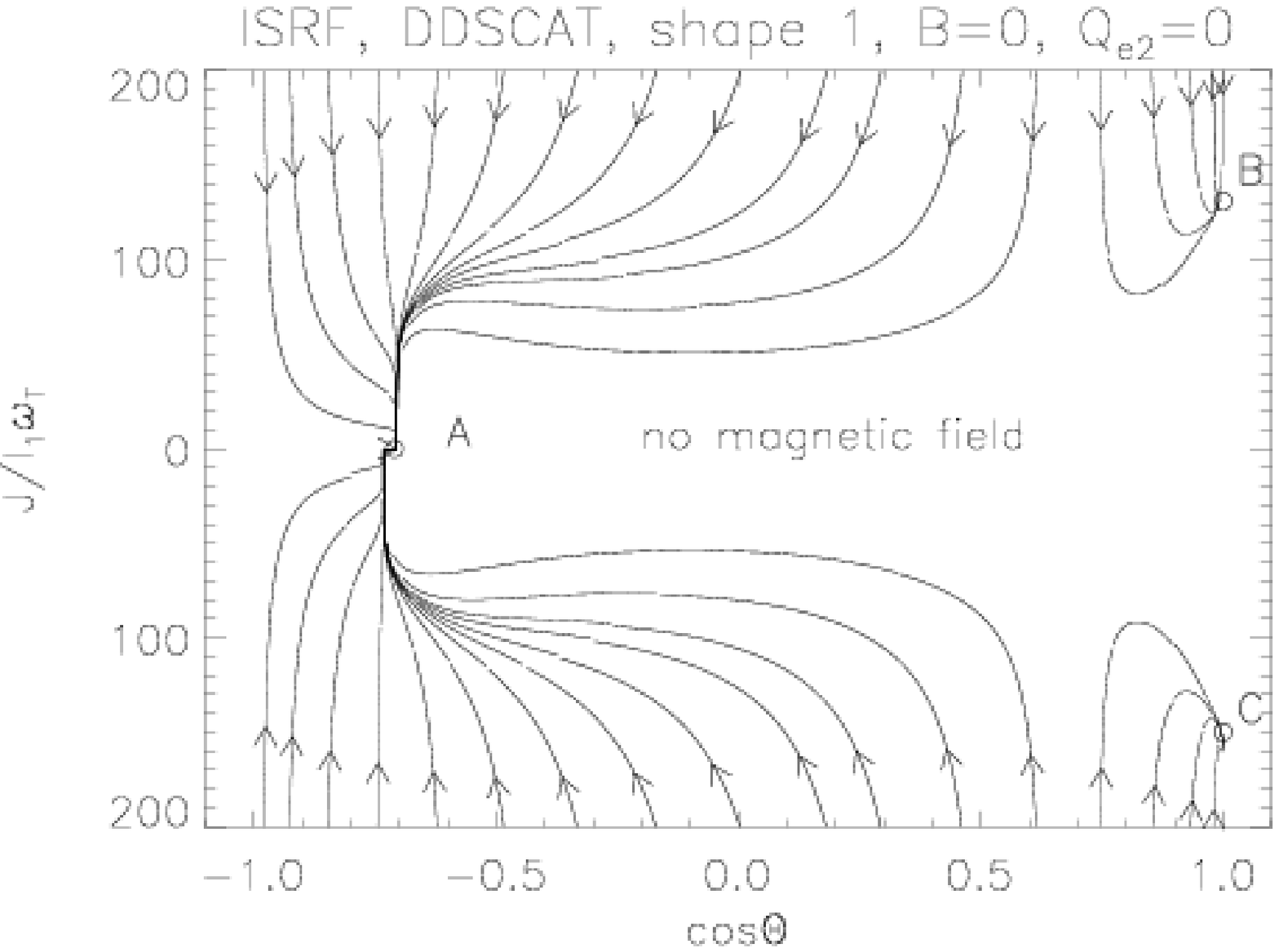}
\hfill
\includegraphics[width=0.49\textwidth]{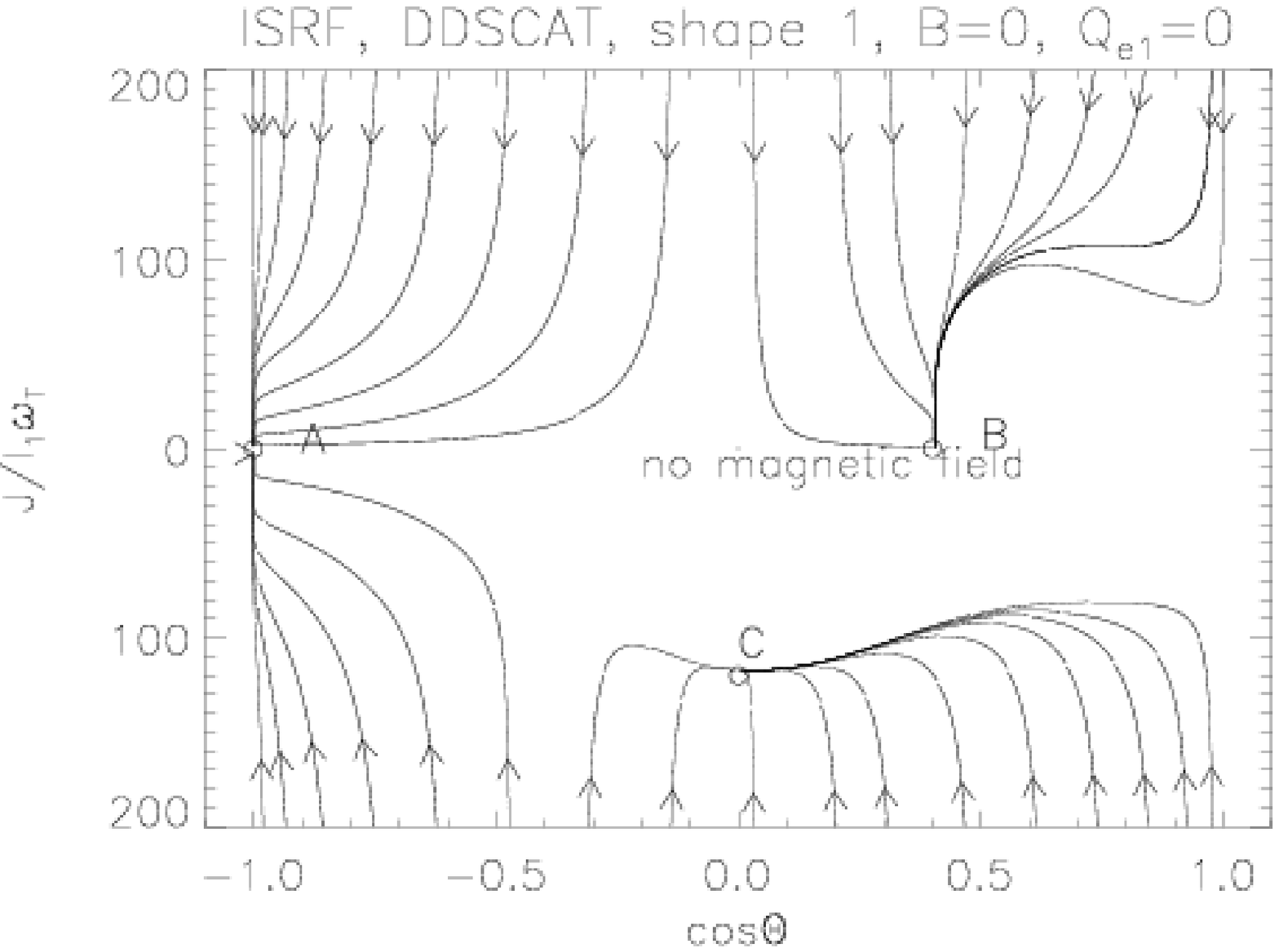}
\caption{Phase trajectory maps in the absence of magnetic field corresponding to
  $Q_{e2}=0$ ({\it Upper Panel}) and  $Q_{e1}=0$ ({\it Lower Panel}). 
  The map in the {\it Upper Panel} has
    three attractor points A, B, C in which B, C correspond to
high angular momentum. The map in {\it Lower Panel} has one attractor
point C corresponding to ``wrong'' alignment and two attractor points with $J=0$.}
  \label{f24}
 \end{figure}

We see that these trajectory maps are similar to those constructed by RATs
from AMO (see Fig. \ref{f7}). Indeed, for $Q_{e1}=0$ case, the map for AMO in Fig. \ref{f7} shows two attractor points at $\mc \Theta= 1$, which are
found in Fig. \ref{f24}. 
However, in the situation when only $Q_{e2}$ acts, the upper panel in Fig.
\ref{f24} shows an attractor point at $\Theta=\pi/2$ and $J\gg J_{th}$, which is somewhat
different from
Fig. \ref{f7}. This difference stems from the fact that $Q_{e2}$ for shape 1 is though small, but not equal to zero at $\Theta=\pi/2$. Therefore, RATs are still able to
spin up grains and align them there. 
 \begin{figure}
 \includegraphics[width=0.49\textwidth]{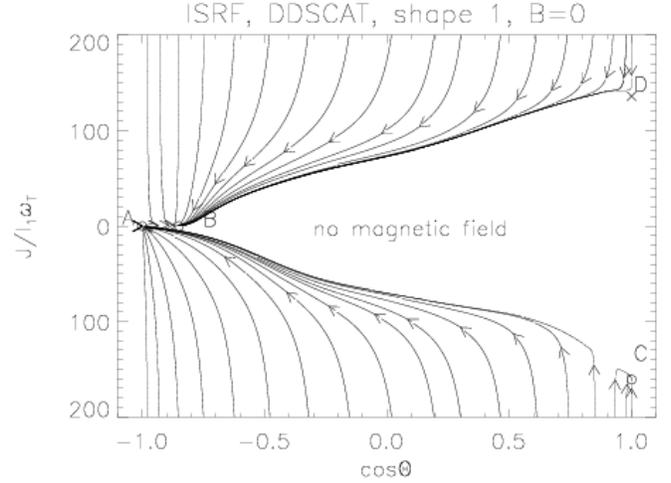} 
\caption{Phase trajectory map in the absence of magnetic field when all torque components act together shows the alignment with one high-$J$ attractor point C , two low-$J$ attractor points A and B corresponding to perfect alignment with ${\bf k}$, and one repellor point D.}
  \label{f25}
 \end{figure}

Fig. \ref{f25} shows the phase map when all torque components are at work corresponding to shape 1 and ISRF.
It is shown that the trajectory map does not have the attractor point at $\Theta=\pi/2$. Thus, similarly to AMO, the attractor point corresponding to aligning grains
with long axes parallel to the direction of radiation 
disappears when non-zero $Q_{e1}$ is accounted for.

Moreover, from Fig.~\ref{f25} (shape 1) and the upper panel in Fig.
\ref{f8} for AMO, it is seen that both maps have the same repellor point at
$\mc \Theta=1$ in the upper panel. However, the point C in Fig. \ref{f25} is an attractor point, rather
than a repellor point B as seen for AMO. This difference stems
from the fact that for AMO, $Q_{e1}=\frac{4\pi l_{1}n_{1} n_{2}|n_{2}|}{\lambda}\frac{4}{3\pi}(5\mbox{cos}^{2}\Theta-2)$
is completely symmetric, i.e., $Q_{e1}(\Theta=0)=Q_{e1}(\Theta=\pi)$, while
for the Shape 1, Fig.~\ref{f21} (the solid-dot line) shows that
$Q_{e1}(\Theta=0)<Q_{e1}(\Theta=\pi)$. Therefore, the stationary point
$\Theta=0$ in the upper panel, does not satisfy the criteria for attractor points, i.e., it is a
repellor point, while it is an attractor point in the lower panel.

\section{Alignment with respect to ${\bf B}$}\label{balign}

We showed in \S \ref{kalign} that grains can be aligned with respect to ${\bf k}$. 
Below we consider the case when magnetic field is essential in terms of grain
precession (see Fig.~\ref{f3}). As earlier, we disregard the paramagnetic relaxation. We make an extensive use of the physical insight obtained with
a more simple case of alignment in \S \ref{kalign}. Indeed, because of the precession about magnetic field the analytical treatment of the corresponding processes gets
less transparent here compared to that in \S \ref{kalign}.
  
\subsection{Equations of motion in presence of ${\bf B}$}

 \begin{figure}
\includegraphics[width=0.49\textwidth]{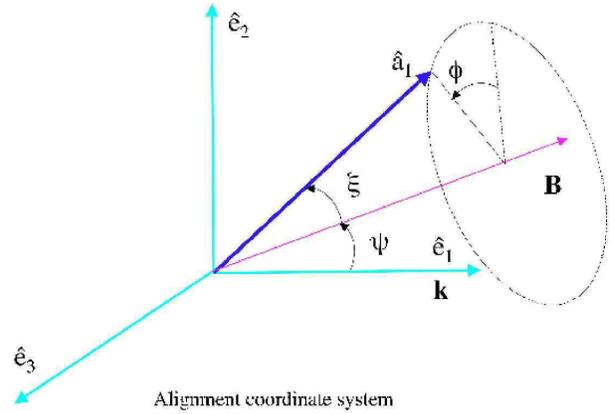} 
\caption{Coordinate systems used to study grain alignment in which the
  external magnetic field {\bf B} defines the alignment axis.} 
 \label{f3} 
\end{figure} 
In the presence of magnetic field, equations of motion in dimensionless units
become
\begin{align}
\frac{d\phi}{dt}&= \frac{M}{\ms\xi}G(\xi, \psi,\phi)-\Omega_{B},\label{eq58}\\
\frac{d\xi}{dt}&=M \frac{F(\xi, \psi,\phi)}{J},\label{eq59}\\
\frac{dJ}{dt}&=M H(\xi, \psi,\phi)-J,\label{eq60}
\end{align} 
where $\Omega_{B}$ is the Larmor precession rate of ${\bf J}$ around the magnetic
field ${\bf B}$. Here $F, H, G$ are RAT components projected to three axes
$\hat{\xi}, \hat{J}, \hat{\phi}$, which are given by (see DW97)
\begin{align}
F(\xi,\psi,\phi)&=Q_{e1}(\Theta,0)[-\ms\psi \mc\xi \mc\phi-\mc \psi \ms\xi]\nonumber\\
&+Q_{e2}(\Theta, 0)[\mc \Phi(\mc \psi \mc\xi\mc\phi\nonumber\\                 
&-\ms\psi \ms\xi)+\ms\Phi \mc\xi \ms\phi]\nonumber\\
&+Q_{e3}(\Theta, 0)[\mc\Phi \mc\xi \ms\phi +\ms\Phi(\ms\psi \ms\xi\nonumber\\
&-\mc\psi \mc\xi \mc\phi)],\label{eq61}\\
H(\xi,\psi,\phi)&=Q_{e1}(\Theta, 0)[-\ms \psi \ms\xi\mc\phi+\mc\psi \mc\xi]\nonumber\\
&+Q_{e2}(\Theta, 0)[\mc\Phi(\ms\psi \mc\xi\nonumber\\
&+\mc\psi \ms\xi \mc\phi)+\ms\Phi \ms\xi \ms\phi]\label{eq62},\\
G(\xi,\psi,\phi)&=Q_{e1}(\Theta, 0)[\ms \psi \ms\phi]\nonumber\\
&+Q_{e2}[\ms\Phi\mc\phi-\mc\Phi\mc\psi \ms\phi]\nonumber\\
&+Q_{e3}[\mc\Phi \mc\phi+\ms\Phi \mc\psi \ms\phi]\label{eq62b}.
\end{align}
 Here $\Theta$ and $\Phi$ are related to $\xi$, $\psi$, $\phi$ via
\begin{align}
\mc \Theta&=\mc\xi \mc\psi -\ms\xi \ms\psi \mc\phi,\label{eq63}\\
\Phi&=2\mbox{tan}^{-1}\frac{\ms\Theta-\ms\xi \ms\psi}{\ms\xi \ms\phi}.\label{eq64}
\end{align}
Equation (\ref{eq62}) reveals explicitely that the component $Q_{e3}(\Theta, 0)$ does not
contribute to spinning up grains. On the other hand, we found numerically that
the last term containing $Q_{e3}(\Theta, 0)$ in equation (\ref{eq61}) goes to
zero after averaging over the precession angle $\phi$. Therefore, similar to
the case of alignment with respect to ${\bf k}$, the only effect of
$Q_{e3}(\Theta, 0)$ is to induce the grain precession.

After averaging over the precession angle $\phi$, the equations of
motion (\ref{eq58})-(\ref{eq60})
are reduced to two equations for $\xi$ and $J$, whereas $F(\xi, \psi, \phi), H(\xi, \psi, \phi)$ are replaced by
$\langle F(\xi, \psi)\rangle_{\phi}, \langle H(\xi, \psi)\rangle_{\phi}$.

\subsection{Stationary points for arbitrary shaped grains} 
While in this section we deal with AMO, some results can be obtained in a
general case of arbitrary shaped grains. More results of this nature are
presented in \S 8.

In the presence of magnetic field, aligning and spinning torques are complicated
functions of RATs, involving $\psi, \xi, \phi$ variables. Therefore, it's not
easy to derive general analytical expressions for stationary points. However, we can find some particular physically
interesting situations that
correspond to stationary points.

 For instance, the perfect alignment 
corresponds to the maximal inertia axis ${\bf a}_{1}$ parallel to the magnetic
field, i.e., $\ms\xi_{s} =0$. For this angle,  from equations (\ref{eq63}) and (\ref{eq64}), we have $\Theta=\psi$, and
 $\Phi=0$ or $\pi$. Hence, $Q_{e1}(\Theta,0)=Q_{e1}(\psi,0)$, $Q_{e2}(\Theta,0)=Q_{e2}(\psi,0)$. Equation~(\ref{eq61}) becomes
 \begin{align}
 F(\xi_{s},\psi,\phi)&=Q_{e1}(\psi,0)\ms\psi \mc\phi+Q_{e2}(\psi,0)\mc\Phi \mc\phi \nonumber\\
&+Q_{e3}(\xi, 0)\mc\Phi\ms\phi. \label{eq65}
 \end{align}

 Obviously, $F(\xi_{s},\psi,\phi)$ is a function of the precession
 angle $\phi$ about the magnetic field ${\bf B}$. Thus, if the grain precesses rapidly around ${\bf B}$, then we can average $F(\xi_{s},\psi,\phi)$ over $\phi$ from $0$ to $2\pi$. As a result,
  \begin{align}
\langle F(\xi_{s},\psi)\rangle_{\phi}&=Q_{e1}(\psi,0)\ms\psi \int_{0}^{2\pi}\mc\phi
 d\phi \nonumber\\
&+Q_{e2}(\psi,0)\mc\Phi \int_{0}^{2\pi}\mc\phi
 d\phi\nonumber\\
&+Q_{e3}\mc\Phi\int_{0}^{2\pi}\ms\phi d\phi=0,
\label{eq66}
 \end{align}
which implies that for {\it grains of an arbitrary shape}, and
for {\it arbitrary direction of light with respect to the magnetic field},
there are always two stationary  points at $\xi_{s}=0$ and $\pi$. This very
fact makes the alignment of the grains with long axes perpendicular to ${\bf B}$ in some sense 
the expected one, although it does not present a sufficient condition for
such an alignment.

If attractor points exist for $\xi_{s}$ different from $0$ or $\pi$, the 
alignment may get ``wrong'', i.e. with the maximal inertia axis of the grain
tending to be parallel to the magnetic field. Here and below we adopt the convention
that the alignment is ``right'' if it corresponds to the Davis-Greenstein predictions,
which made the Davis-Greenstein mechanism so popular even in spite of its inefficiency.
Needless to say, that
 RATs we seek ``right'' alignment, i.e. with long grain axes perpendicular
to magnetic fields without appealing for paramagnetic relation.

The ``wrong'' alignment may be expected, for instance,
when the radiation beam is perpendicular to the
magnetic field, i.e. $\psi=\pi/2$. For this $\psi$, consider
the direction of ``wrong'' alignment corresponding to $\xi_{s}=\pi/2$.
For  $\psi=\pi/2$, and $\xi=\pi/2$, we have $\mc \Theta=-\mc \phi$, and
 $\Phi=\pi/2$ for $\phi <\pi$ and $-\pi/2$ for $\phi>\pi$. Therefore, equation
 (\ref{eq61}) becomes 
 \begin{align}
F(\xi_{s},\psi,\phi)&=Q_{e3}(\Theta,0)\ms \Phi.\label{eq67}
 \end{align}
 Thus
 \bea 
\langle F(\xi_{s},\psi)\rangle_{\phi}=\int_{0}^{2\pi}Q_{e3}(\phi,0)d\phi=0.\label{eq68}
\ena
Here we use the property $Q_{e3}(\phi, 0)\sim \ms2\phi$ in calculating the integral.

\subsection{"Right" and "wrong" alignment for AMO}

The introduction of fast precession arising from ${\bf B}$ makes the dynamics
of grains more interesting. For instance, it allows for a parameter space
for ``wrong'' alignment, high attractor points for AMO, shifts of the crossover points.

\subsubsection{Torques considerations}

Above, we found that in the presence of ${\bf B}$, two permanent
stationary points are $\xi_{s}=0, \pi$. If being attractor points, they correspond to the ``right''
alignment. 

Consider a case of suspected ``wrong'' alignment for AMO at $\xi_{s}\sim \pi/2$.  
Since as $\psi=\pi/2$, $\mc\Theta=-\mc\phi$, so RATs
(see equations \ref{eq30a} and \ref{eq31a}) become 
\begin{align}
Q_{e1}&=\frac{16 l_{1} n_{1}n_{2}|n_{2}|}{3\lambda}(5\mc^{2}
\phi-2),\label{eq69}\\
 Q_{e2}&=\frac{40 l_{1} n_{1}n_{2}|n_{2}|}{3\lambda}\ms2\phi(1.191+0.1382\mcs\phi),\label{eq70}
\end{align}
Substituting equations (\ref{eq69}) and (\ref{eq70}) into equation
 (\ref{eq62}) and averaging over the precession angle $\phi$, we get
\begin{align}
\langle H(\xi, \psi)\rangle_{\phi}&=\frac{16 l_{1} n_{1}n_{2}|n_{2}|}{3\lambda}\int_{0}^{2\pi}(5\mc^{2}\phi-2)\mc\phi
 d\phi\nonumber\\
&+\frac{40 l_{1} n_{1}n_{2}|n_{2}|}{3\lambda}[\int_{0}^{2\pi} \ms2\phi \ms
 \phi(1.191)d\phi\nonumber\\
&+\int_{0}^{2\pi}\ms2\phi(0.1382\mcs\phi)  d\phi]=0.\label{eq71}
\end{align}
The fact that the integral (\ref{eq71}) is equal to zero means 
that RATs do not spin up grains when they are perpendicular to the magnetic field. 

Now, let us study whether $\xi_{s}=\pi/2$ satisfies the condition of an attractor
points. Fig.~\ref{f10}{\it upper} shows the spinning and aligning torques for
$\psi=89.9^{0} \sim 90^{0}$ \footnote{Here we take $\psi=89.9^{0}$ to avoid
  the singularity that may appear due to zero of $\langle H\rangle$ as shown in
  equation (\ref{eq71}) when considering condition of attractor points} for
the state ${\bf a}_{1}\| {\bf J}$. It shows that, there are  stationary points at
$\mc\xi=\pm 1$ and $\mc\xi_{s}=0.1, -0.1$ for the ${\bf J}$ parallel and anti-parallel to ${\bf a}_{1}$  corresponding to zeroes of $\langle F(\xi) \rangle_{\phi}$. For the stationary point C with $\mc\xi_{s}=0.1$, Fig. \ref{f10} shows $\left. \frac{d\langle
    F(\xi)\rangle_{\phi}}{d\xi}\right|_{\xi_{s}}>0$, and $\langle H(\xi_{s})\rangle_{\phi}>0$. Hence, this stationary point does not satisfy
equation (\ref{eq46}), i.e, it is a repellor point. Meanwhile, the stationary point C' at $\mc\xi_{s}=-0.1$, we have $\left. \frac{d\langle
    F(\xi)\rangle_{\phi}}{d\xi}\right|_{\xi_{s}}<0$ and $\langle
H(\xi_{s})\rangle_{\phi}>0$. As a result, the stationary point C' is an
attractor point. Similarly, the point A at $\mc\xi_{s}=1$ is an attractor
point since $\left. \frac{d\langle F(\xi)\rangle_{\phi}}{d\xi}\right|_{\xi_{s}}<0$ and $\langle
H(\xi_{s})\rangle_{\phi}>0$; also, the point B is a low-$J$ attractor
point. However, the points A' and B' are repellor points in the case ${\bf
  J}$ anti-parallel to ${\bf a}_{1}$.

Therefore, the possibility of existence of ``wrong'' alignment is feasible, but it
happens at a low angular momentum. So, RATs from anisotropic radiation field
themselves can not maintain the ``wrong'' alignment with respect to magnetic field
in the presence of thermal wobbling and the bombardment by the ambient gas (see Hoang \& Lazarian 2007).

\subsubsection{Effect of isotropic torques on ``wrong'' alignment}

If grains are sufficiently large not to experience frequent thermal flips (see LD99a),
they
can be subjected to regular isotropic torques, which include both Purcell's torques
(Purcell 1979) and those from isotropic radiation flux (DW96). According to
LD99b, the this corresponds to grains larger than $10^{-4}$~cm, which is much larger
than the typical size of the grains in our Table~2. However, such large
grains are relevant
to many astrophysical environments, e.g. comets, dark clouds, accretion disks.

The ``wrong'' alignment gets modified when we take into
account RATs induced by isotropic radiation, or equivalently, Purcell's $H_{2}$ torques. Indeed, ``isotropic'' RATs and Purcell's spin-wheel are both parallel to the
maximal inertia axis, i.e, ${\bf a}_{1}$. Therefore, the total spinning and
aligning torques are $
\langle H(\xi, \psi)\rangle_{\phi}+Q_{iso}, \langle F(\xi,
\psi)\rangle_{\phi}$, respectively.

Since the aligning torque $\langle F(\xi, \psi)\rangle_{\phi}$ depends uniquely on
RATs induced by anisotropic radiation, the positions of ``wrong''
attractor points do not change. Meanwhile, their angular momentum are added by
a term resulted from ``isotropic'' RATs or Purcell pinwheel.

Now, let introduce $H_{2}$ torques. $H_{2}$ torques, after averaging over the
grain rotation around the maximal inertia axis, are given by
\bea
{\bf \Gamma}_{H_{2}}={\bf a}_{1}\frac{I_{1}\omega_{H_{2}}}{t_{gas}}p,
\ena
where
\bea
\omega_{H_{2}}\sim 5\times 10^{7} \left(\frac{f}{a_{-5}^{2}}\right)\left(\frac{l}{10
  A^{0}}\right) \left(\frac{E}{0.2 eV}\right)^{1/2}\left[\frac{n(H)}{n}\right] s^{-1}.
\ena
Here $l^{2}$ is the surface area of a catalytic site of the grain surface,
$f$ is the efficiency of $H_{2}$ formation, $n(H)$ is the density of atomic hydrogen, $E$ is the kinetic energy of the
escaping $H_{2}$ molecule, and $p$ is a random variable (see DW97).
In addition to parameters given in Table \ref{tab1}, here we assume that $n(H)/n=1, f=\frac{1}{3}, l=10 A^{0}$, and $E=0.2 eV$; we consider then the role of $H_{2}$ torques for
$p=1, -1$ corresponding to ${\bf \Gamma}_{H_{2}}$ parallel and anti-parallel
to ${\bf a}_{1}$. 

\subsubsection{Trajectory maps}

To illustrate the alignment with respect to the magnetic field for AMO, 
we construct phase trajectory maps using RATs obtained by averaging equations
(\ref{eq8})-(\ref{eq10}), for different radiation directions
$\psi$. To do this we adopt the parameters from Table \ref{tab1}. 
For $\psi=0^{0}$, i.e., ${\bf B}\| {\bf k}$, the phase map shown in the upper
panel of Fig. \ref{f8} exhibits a perfect alignment of ${\bf J}$ with respect to ${\bf B}$.

\begin{figure}
\includegraphics[width=0.49\textwidth]{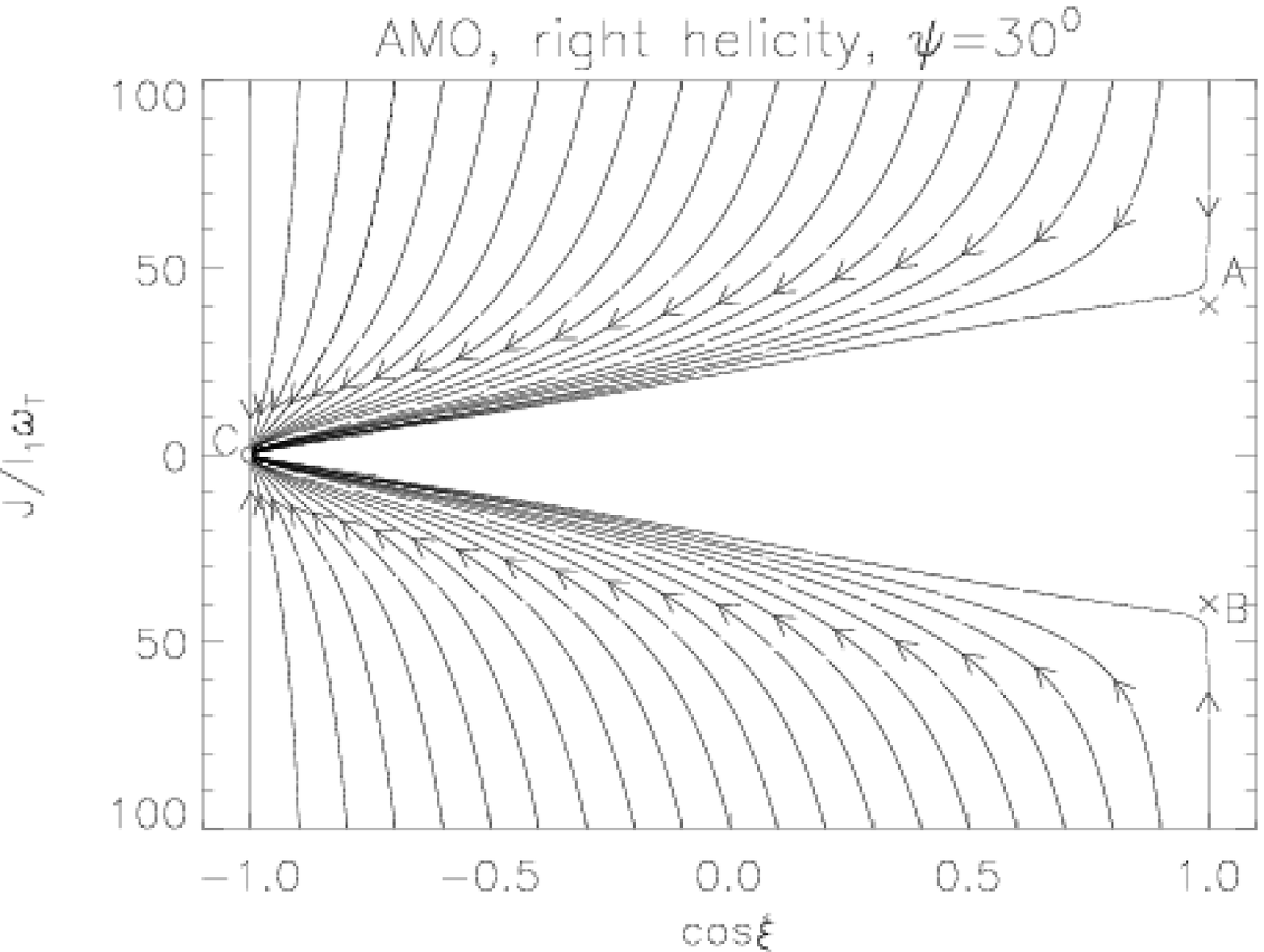}
\includegraphics[width=0.49\textwidth]{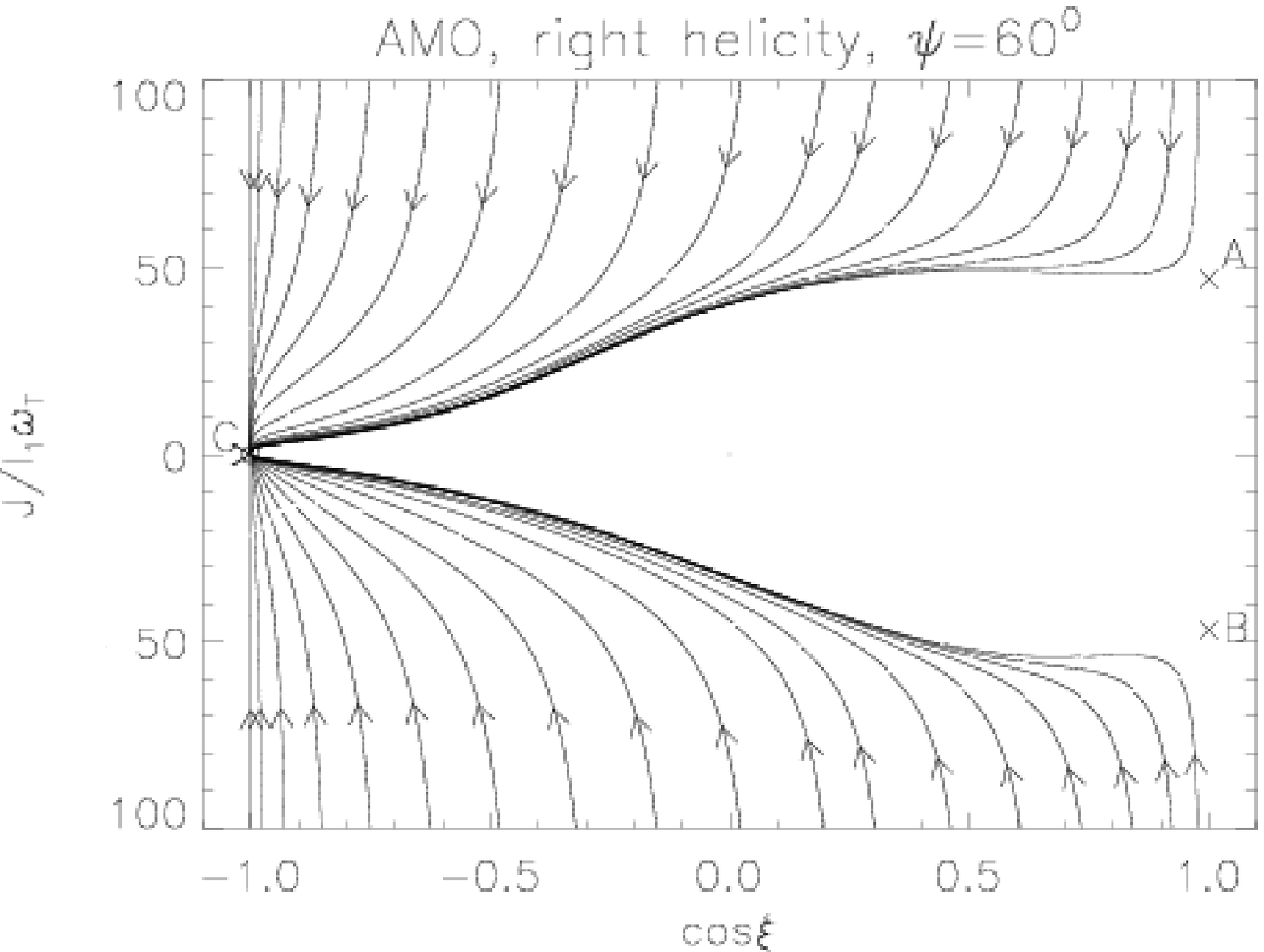}
\caption{Phase trajectory maps for AMO for $\psi=30^{0}$ ({\it
    Upper Panel})
  and $\psi=60^{0}$ ({\it Lower Panel})  both have zero attractor points C,
  and two repellor points A, B. }
\label{f9}
\end{figure} 

Fig.~\ref{f9} shows the phase trajectory maps for $\psi=30, 60^{0}$ with
two repellors A and B, and one low-$J$ attractor point C.

\begin{figure}
\includegraphics[width=0.49\textwidth]{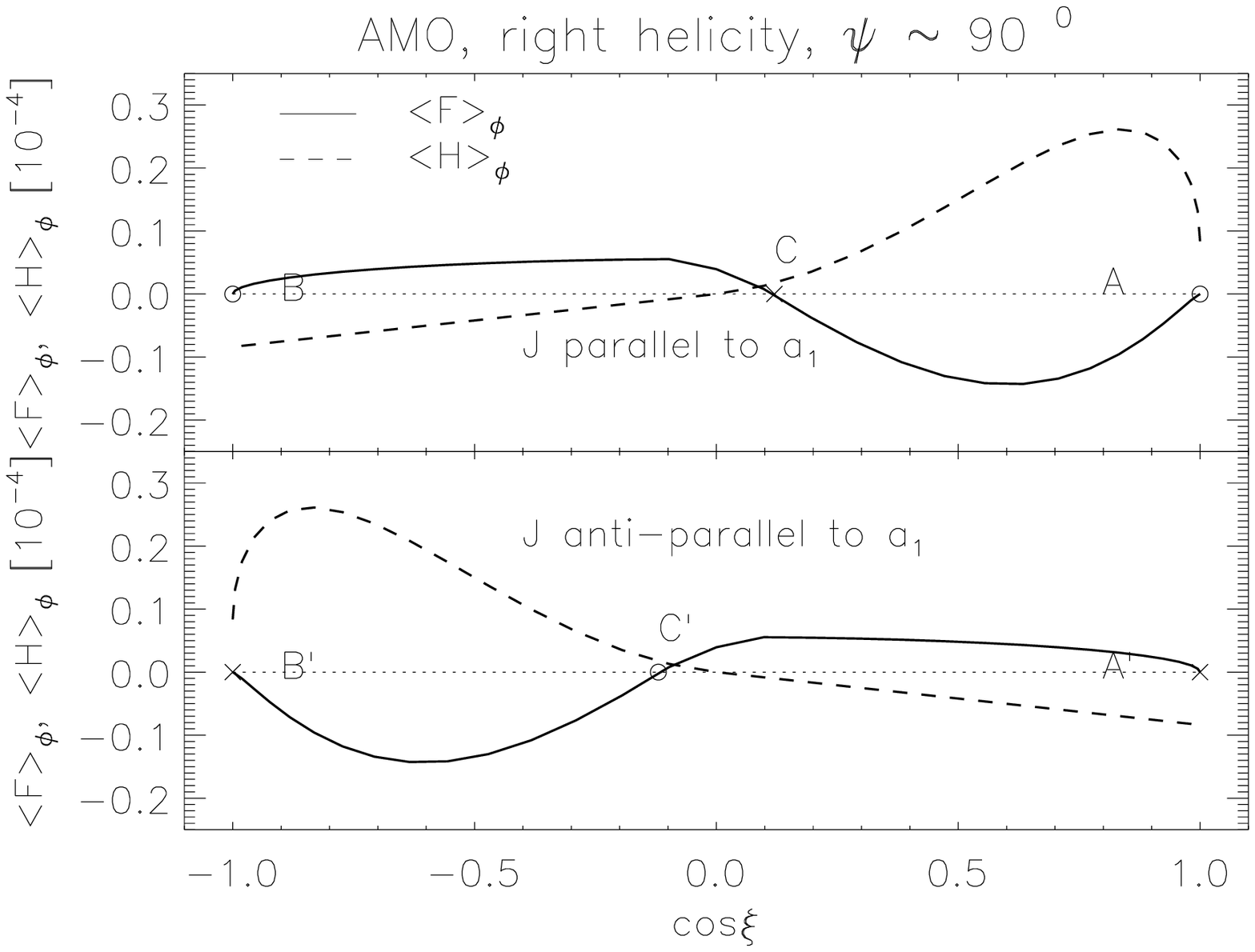}
\hfill
\includegraphics[width=0.49\textwidth]{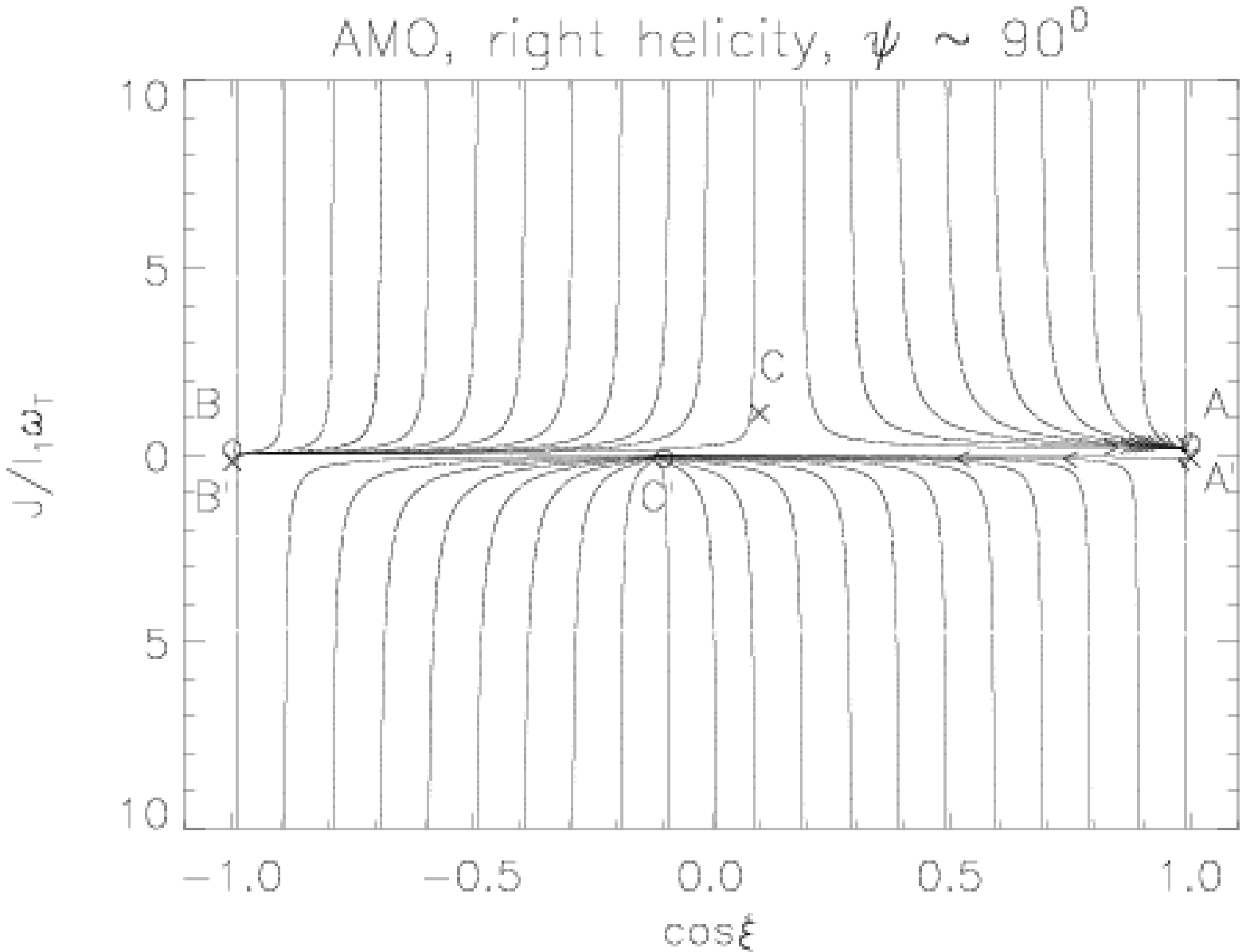}
\caption{Aligning and spinning torques for the particular direction of light
  $\psi\sim 90^{0}$ and the corresponding phase map for AMO. {\it Upper Panel}: 
Solid line shows the aligning torque $\langle F(\xi, \psi)\rangle_{\phi}$ with
two zeroes corresponding to two stationary
  points at $\mc\xi=0.1, -0.1$ corresponding to the case ${\bf J}$ parallel and anti-parallel to ${\bf a}_{1}$, besides two zeroes $\mc\xi=\pm1$, while the
dashed  line
  shows the spinning up torque $\langle H(\xi, \psi)\rangle_{\phi}$. {\it Lower
    panel}: Phase map corresponding to RATs in the upper panel shows one
  ``wrong'' attractor point C' at very low $J/I_{1}\omega_{T}\sim 0$ and $\mc\xi=-0.1$.}
\label{f10}
\end{figure} 

 \begin{figure}
\includegraphics[width=0.49\textwidth]{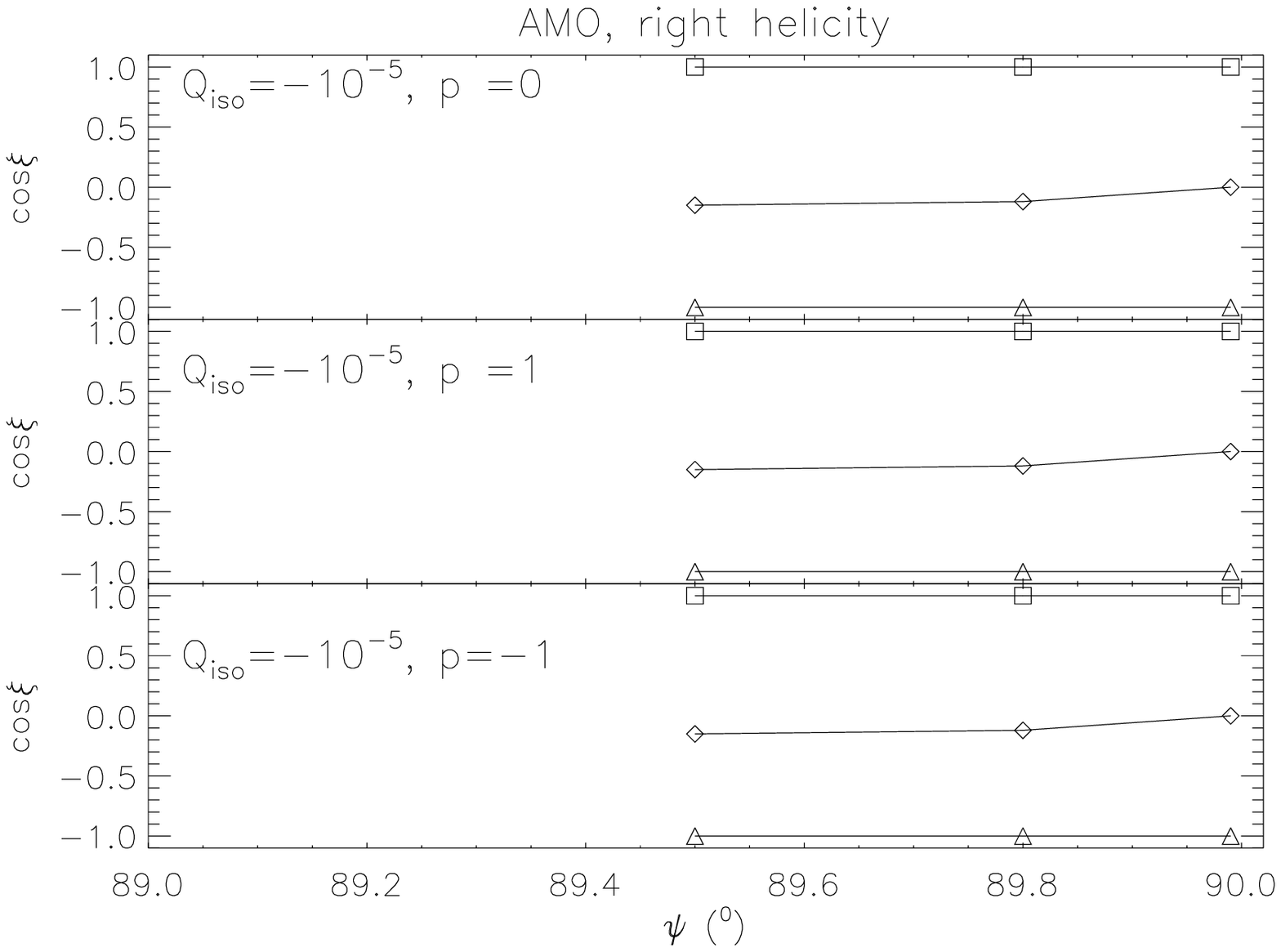}
\space
\includegraphics[width=0.49\textwidth]{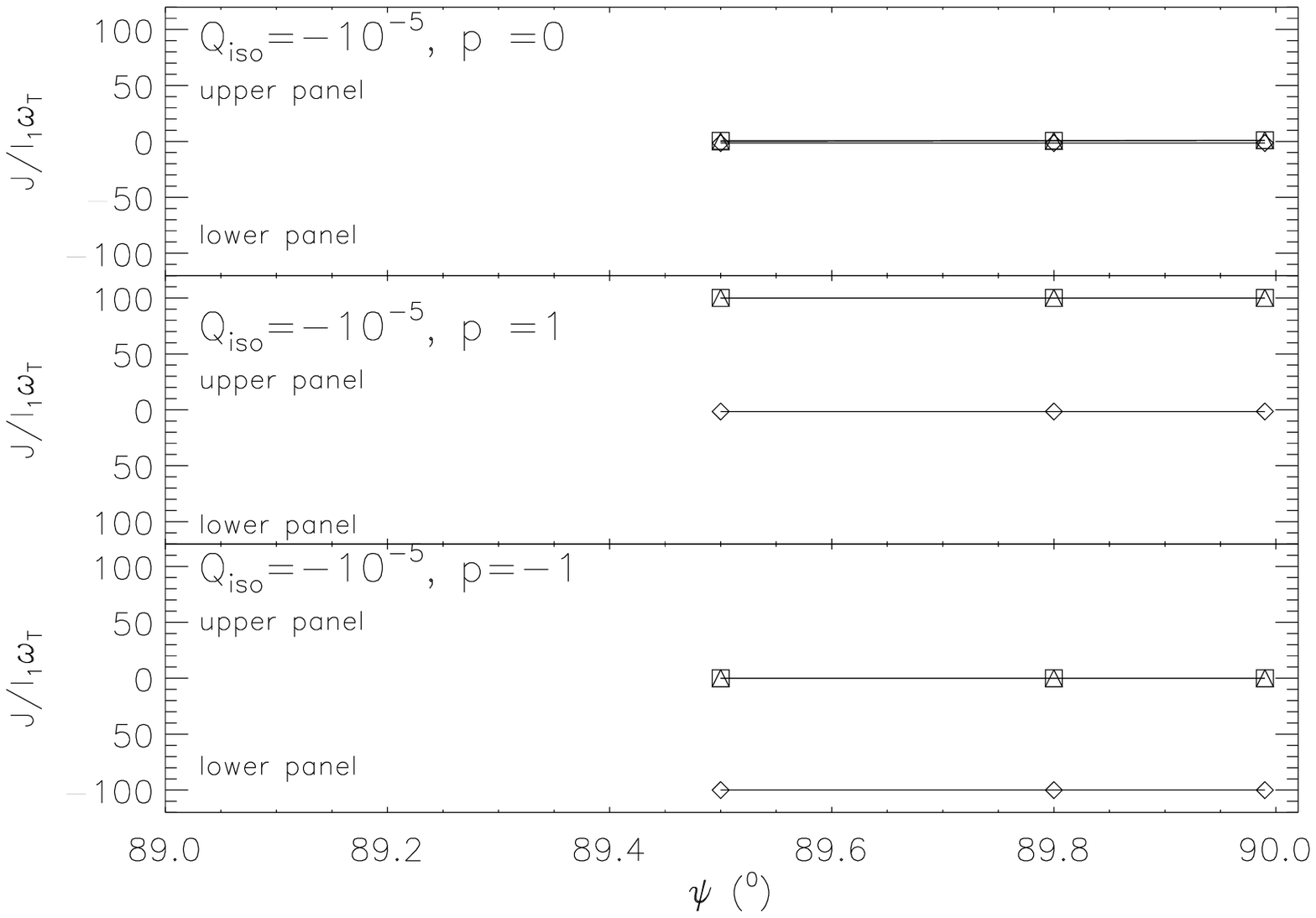}
\caption{Diagram for the position ({\it Upper panel}) and angular momentum ({\it
    Lower panel}) of attractor points as function of $\psi$ for
  three cases,
  $Q_{iso}=-10^{-5}, p=0$ (no $H_{2}$ torques), $Q_{iso}=-10^{-5}, p=1$ ($\Gamma_{H2}$ is parallel to
  ${\bf a}_{1}$), and $Q_{iso}=-10^{-5}, p=-1$
  ($\Gamma_{H2}$ is anti-parallel to ${\bf a}_{1}$). Each symbol (diamond,
  triangle, and square)  denotes an
  attractor point, and a symbol-line  shows the extend of the ``wrong'' 
alignment as a function of $\psi$. }
 \label{f11}
 \end{figure}

For $\psi \sim 90^{0}$, the lower panel in  Fig. \ref{f10} shows an attractor
point C' at $\mc\xi=-0.1$ corresponding to
$J \sim 0$ and two repellor points at $\mc\xi=1$,
respectively. It is shown that the phase map is not symmetric between upper
and lower panels compared to the maps for $\psi<90^{0}$. This stems from the
fact that spinning and aligning torques are no longer symmetric. The existence of the attractor point C' indicates
that there is, indeed, a ``wrong'' alignment
situation as predicted by our analysis above. However, to what extend of $\psi$ the "wrong" alignment occurs?

Fig. \ref{f11} shows the angle of attractor points (upper) and their corresponding angular momentum (lower) for a range $\psi=[89.5^{0},90^{0}]$. In each frame of Fig. \ref{f11}{\it lower}, we show angular momentum of attractor points present in both upper and lower panels of a trajectory map (upper and lower panel are labeled).

For the case $Q_{iso}=10^{-5}, p=0$, the upper panel in Fig. \ref{f11} shows that there are always three attractor
 points in which one attractor point happens at $\mc\xi=-0.1, -0.05$ and $0$ for $\psi=89.5^{0}, 89.8^{0}$ and $90^{0}$, corresponding to "wrong" alignment, two other with $\mc\xi=\pm 1$. However, their angular momentum is very low, about $1.5 I_{1}\omega_{T}$ (top frame in Fig. \ref{f11}{\it lower}). When $H_{2}$ torques are taken into account, if $H_{2}$ torques are parallel to ${\bf a}_{1}$, i.e., $p=1$, then the attractor points $\mc\xi=\pm 1$ are lifted to $J/I_{1}\omega_{T}\sim 100$, while the angular momentum of the "wrong" attractor point is unchanged (see the middle frame in Fig. \ref{f11}{\it lower}). In contrast, if $p=-1$, the angular momentum of the
"wrong" attractor point can be increased by $H_{2}$ torques to $J/I_{1}\omega_{T}=100$, but the attractor points  $\mc=\pm 1$ are unchanged (see lower frame in Fig. \ref{f11}{\it lower}). Note, that the shift of "wrong" attractor point position toward $\mc\xi=0$ is also seen on the diagram when $\psi \to 90^0$ (Fig. \ref{f11}{\it upper}).

In summary, the presence of "wrong" alignment with respect to magnetic fields can be expected, but it happens in a very narrow range of $\psi$ in the vicinity of $\psi=\pi/2$. In addition, their angular momentum is very low in the absence of isotropic torques such as $H_{2}$ torques and "isotropic" RATs. 
Naturally, the effects of isotropic torques are negligible if grains are flipping,
and therefore are thermally trapped
as it is discussed in LD99a.

\subsection{Alignment for Irregular Grains}

Similarly as in \S \ref{kalign} we present the phase trajectory maps for shape 1,
which is the ``most irregular'' in terms of RATs.
Consider first phase maps for shape 1 obtained for and $\psi=90^{0}$ first.
(see Fig. \ref{f25a}). For this case, RATs for shape 1 create an attractor point at
$\xi=90^{0}$, $J/I_{1}\omega_{T}=2$ in the phase map, which corresponds to
{\it wrong alignment} (see Fig. \ref{f25a}). 
AMO also does produce such a "wrong" alignment but at much smaller angular
velocity, $J/I_{1}\omega_{T} \sim 0$ (see Fig. \ref{f10}). This difference
is resulted from the property of $Q_{e2}$ component. Following AMO, $Q_{e2}=0$ at
$\Theta=\pi/2$, while for shape 1, this component is not zero at this angle.   

All in all, for various grains shapes studied, we found that there is a narrow range $\psi=85^{0}\to 90^{0}$
in which there exist "wrong" alignment.
 
\begin{figure}
\includegraphics[width=0.49\textwidth]{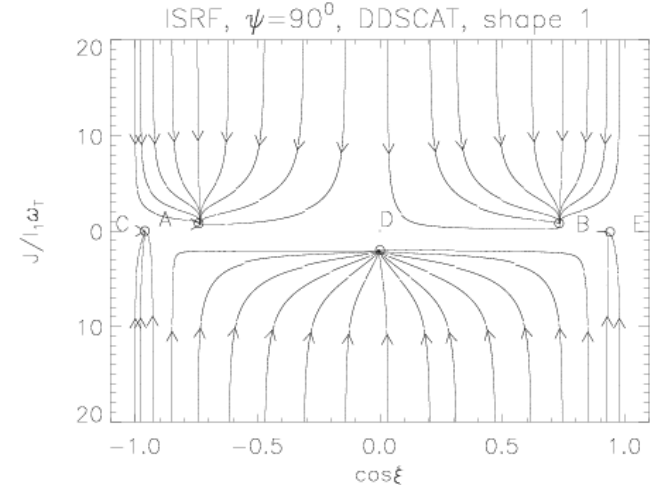}
\caption{For $\psi=90^{0}$, the map shows three high-$J$ attractor point A, B and D and two low-$J$ attractor point C and E. Most grains in the lower frame of the map align on D with long axes parallel to ${\bf B}$, i.e., "wrong" alignment. }
 \label{f25a} 
\end{figure}

In addition, we illustrate the alignment in respect to magnetic field with
phase trajectories obtained for $\psi=30^{0}$. For shape 1, similar studies were
performed in DW97. However, their treatment of crossovers was different from 
ours. Thus the phase trajectories that we observe are different. We do not see cyclic maps with
grains emerging from a crossover to get accelerated by RATs. Instead, we see,
similar to the case of AMO (see Fig. \ref{f9}{\it upper}), the low-$J$ attractor points (see Fig \ref{f25a}). 
\begin{figure}
\includegraphics[width=0.49\textwidth]{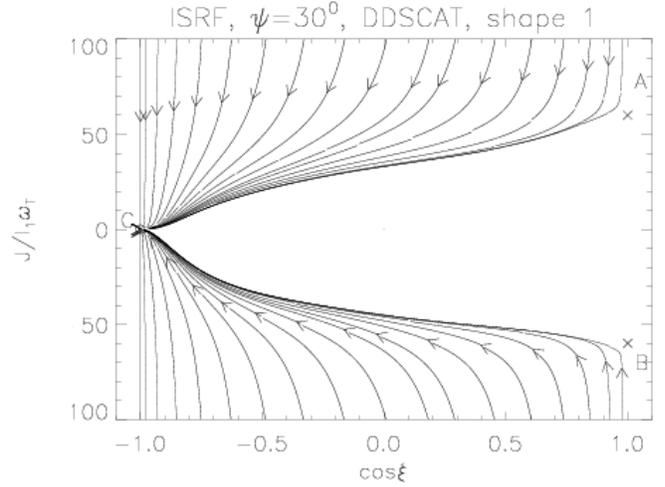} 
\caption{Phase trajectory map for grain shape 1 and $\psi=30^{0}$ shows the alignment with two repellor points A and B an one low attractor point C. }
\label{f25b}
\end{figure}

Our study provides important insight into the role of $Q_{e1}$ and $Q_{e2}$ components.
For instance,
in the presence of magnetic field $Q_{e1}$ tends to provide the
alignment with long axes perpendicular to magnetic field, i.e. the
``right'' alignment. The possibility of ``wrong'' alignment within a limited
range of $\xi$ is related to $Q_{e2}$. Similarly, the ratio of $Q_{e1}^{max}/Q_{e2}^{max}$
determines the existence of the high-$J$ attractor points.

\subsection{Ratio of $Q_{e1}^{max}/Q_{e2}^{max}$: existence of high-$J$ and shift of low-$J$ attractor points }\label{sec5}

\begin{figure}
\includegraphics[width=0.49\textwidth]{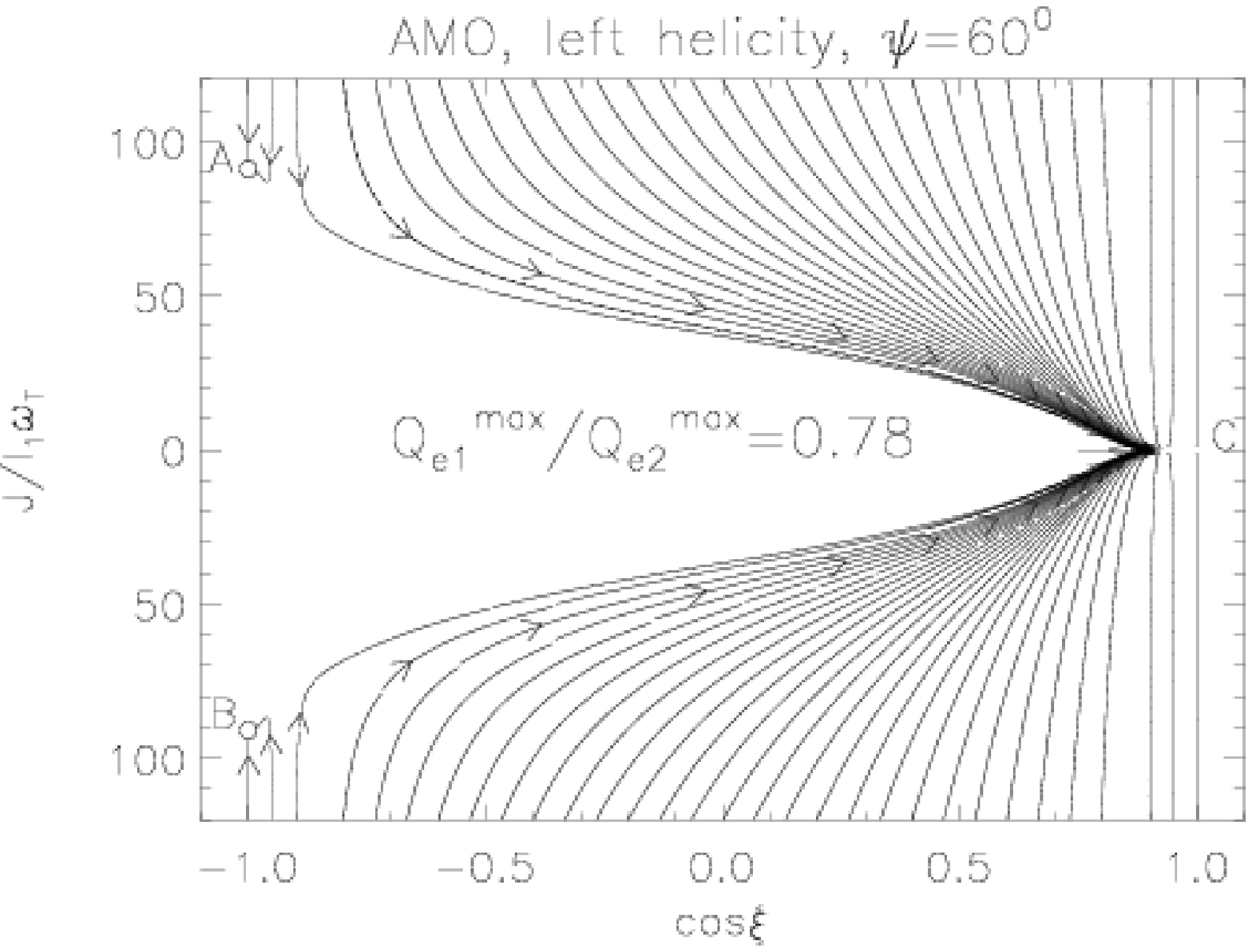}
\includegraphics[width=0.49\textwidth]{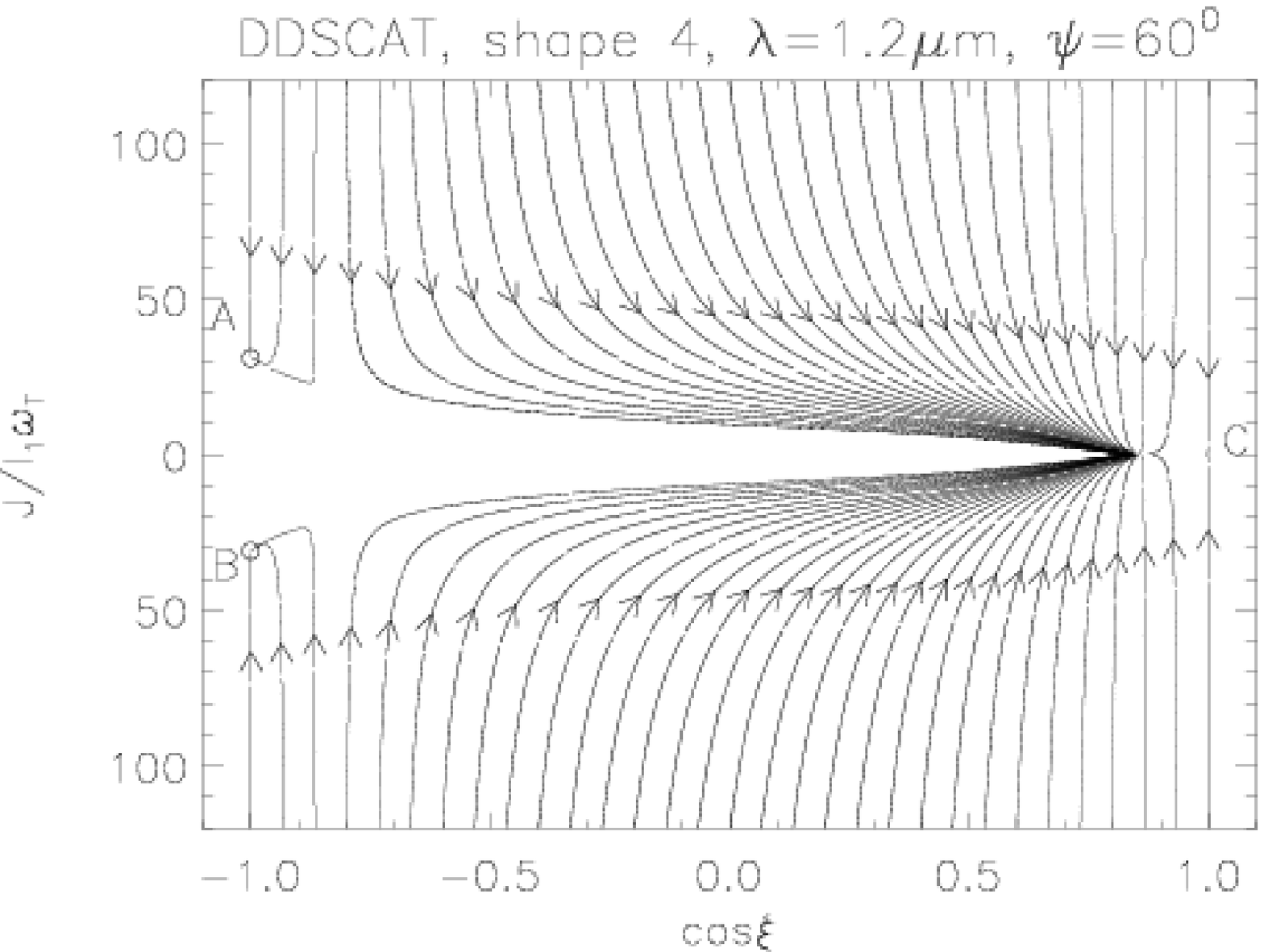}
\caption{Phase trajectory maps with $\psi=60^{0}$ for the case of AMO ({\it Upper Panel})
  and for the shape 4 (see Fig. \ref{f18}) corresponding to the same ratio
  $Q_{e1}^{max}/Q_{e2}^{max}=0.78$  ({\it Lower Panel}). The maps show that both AMO
  and DDSCAT give rise two high-$J$ attractor points A, B at $\mc \xi=-1$ and one low J
  attractor point C.}
\label{f12}
\end{figure}
\begin{figure}
\includegraphics[width=0.48\textwidth]{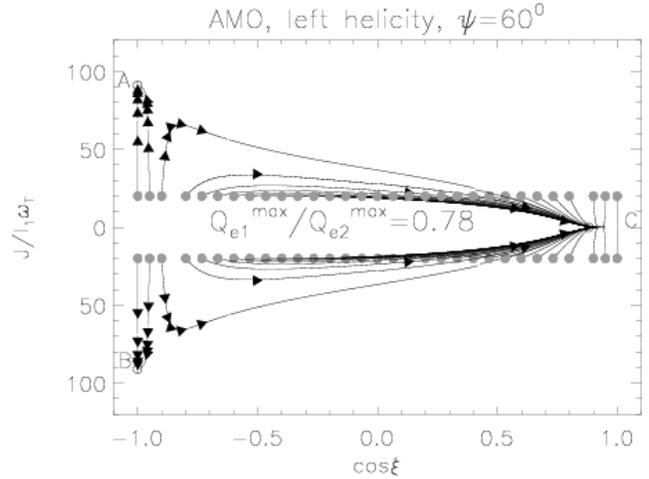}
\caption{Similar to the upper panel in Fig. \ref{f12}, but grains start from
angular momentum smaller than the value of high-$J$ attractor point. Filed circles show initial position of grains. Filled arrows show the time interval of $0.5 t_{gas}$. Some grains get aligned at low attractor point
  over a short time $t<0.5 t_{gas}$, some other get to the high J attractor points over $t \sim 3 t_{gas}$.}
\label{f12*}
\end{figure}

The role of the ratio $Q_{e1}^{max}/Q_{e2}^{max}$ on the grain alignment has been studied in \S \ref{qratio}
for the case of no magnetic field (see \S \ref{kalign} when this is applicable). Below
we consider the role of this ratio when the magnetic field is present.
We shall show that this ratio is important in determining whether
grains have high-$J$ attractor points.

In previous subsections we have derived the expressions for the stationary points,
and presented the trajectory maps for the RAT alignment for a particular $\alpha$,
namely,
$\alpha=45^{0}$. It is clearly shown that stationary points $\mc\xi=\pm 1$ do
not depend on the magnitude of $Q_{e1}, Q_{e2}$. However, their properties,
i.e., whether they are attractor or repellor points do so.

Let us first consider a realization of AMO in which the functional forms are
established for $\alpha=45^{0}$ and their magnitudes are rescaled to have $Q_{e1}^{max}/Q_{e2}^{max}= 0.78$ (see Fig. \ref{f5}). Note, that this ratio is similar to the
ratio of RATs obtained by DDSCAT for the irregular shape 4 with radiation
field of wavelength  $\lambda=1.2 \mu m$ (see Fig. \ref{f61}).

Fig. \ref{f12} shows the obtained trajectory maps for $\psi=60^{0}$ for both
AMO and shape 4. We see that both AMO and shape 4 produce the maps with two
attractor points  A, B at $\mc\xi=-1$ and one low-$J$ attractor point C, but
one difference is that the percentage of grains on A and B for shape 4 is
higher than for AMO. For AMO and $\psi=60^{0}$, the existence of high-$J$
attractor points A and B with $Q_{e1}^{max}/Q_{e2}^{max}= 0.78$ is not found in the case $\alpha=45^{0}$ that has $Q_{e1}^{max}/Q_{e2}^{max}= 1.2>1$ in which these points are repellor points (see Fig. \ref{f9}). In other words, the existence of high-$J$ attractor points depends on
the value of $Q_{e1}^{max}/Q_{e2}^{max}$ as predicted in \S~\ref{qratio}. Furthermore, the similarity in the
phase maps between the AMO and shape 4, which have the similar ratio
$Q_{e1}^{max}/Q_{e2}^{max}=0.78$, indicates a good
correspondence of AMO with irregular grains in terms of grain dynamics.

While in most cases, we show the phase trajectory maps that start from high
values of $J$, in Fig. \ref{f12*} we also show the phase trajectories starting
at $J=J_{th}$. The dynamics with this initial conditions are similar. However,
the alignment can be achieved faster (see further discussion in \S \ref{fast}).
 The filled arrows in Fig. \ref{f12*} mark time scales of $0.5 t_{gas}$. We see that it takes approximately less than $0.5 t_{gas}$ to get
to the lower-$J$ attractor point C for grains with initial orientation close to C, for the typical 
interstellar diffuse gas conditions. Also, it takes about $3 t_{gas}$ to get to the high-$J$ attractor points A and B (see Fig. \ref{f12*}). These time scales  are usually much smaller than that for 
paramagnetic damping invoked in the Davis-Greenstein mechanism

Now let us use the approximate functional form of RATs
given by equations (\ref{eq54}) and
(\ref{eq55}), and seek the range of $Q_{e1}^{max}/Q_{e2}^{max}$,
in which the phase map has high J attractor points.
Fig. \ref{f13} shows the ratio of
$Q_{e1}^{max}/Q_{e2}^{max}$ for which there exist high-$J$ attractor
points in the phase trajectory map of grains. It is shown that for $\psi <45^{0}$, $Q_{e1}$ is required to be
dominant over $Q_{e2}$, at least $Q_{e1}^{max}=2Q_{e2}^{max}$,  to have high-$J$ attractor points; their ratio is an
increasing function of $\psi$. At $\psi=45^{0}$, one does not have high-$J$
attractor points because $Q_{e1}, Q_{e2}$ are equally projected to ${\bf B}$
and to the direction ${\bf \xi}$ perpendicular to ${\bf B}$.

For $\psi>45^{0}$, high-$J$ attractor points occur when $Q_{e2}$
becomes predominant, i.e., $Q_{e1}^{max}/Q_{e2}^{max}<1$ (see the lower panel
in Fig. \ref{f13}).  
Also in Fig. \ref{f13}, the intermediate region with parallel lines correspond
to the range in which the phase
trajectory map
has only low-$J$ attractor points, while high-$J$ stationary points are repellors.

Fig. \ref{f13} shows also the existence of high J attractor points for
irregular shapes 1 (both monochromatic radiation field and ISRF), 2 and 4 (only
monochromatic radiation). For shape 1 and radiation of $\lambda=1.2\mu m$, its phase map has low attractor points
for $\psi<60^{0}$  and high J attractor point for $\psi\ge 60^{0}$. But, the
map for ISRF has J attractor point only in a small range
$\psi<10^{0}$. Interestingly enough, for this wavelength, the ratio of
$Q_{e1}^{max}/Q_{e2}^{max}$ exceeds those possible for the "original" AMO. As a result, as we mentioned
earlier, an attractor point is possible for $\psi=0$, which also describe the
situation of grain alignment in the absence of magnetic field. For such situations
AMO has only repellor points (see \S\ref{kalign}).

 \begin{figure}
\includegraphics[width=0.5\textwidth]{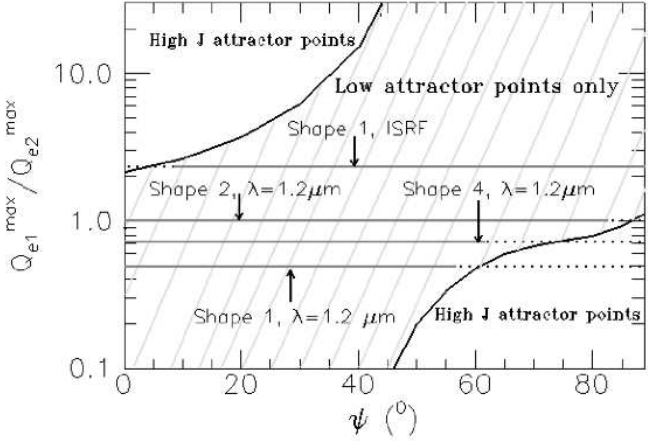}
\caption{Ratio of $Q_{e1}^{max}/Q_{e2}^{max}$ for which the phase map has
  high-J attractor points or low-$J$ attractor points as a function of $\psi$. Curved solid lines show
  predictions by AMO, solid lines (maps with low-$J$ attractor points only), dot lines
  (maps with high J attractor points) show the result for irregular grains.}
\label{f13}
\end{figure}

\begin{figure}
\includegraphics[width=0.49\textwidth]{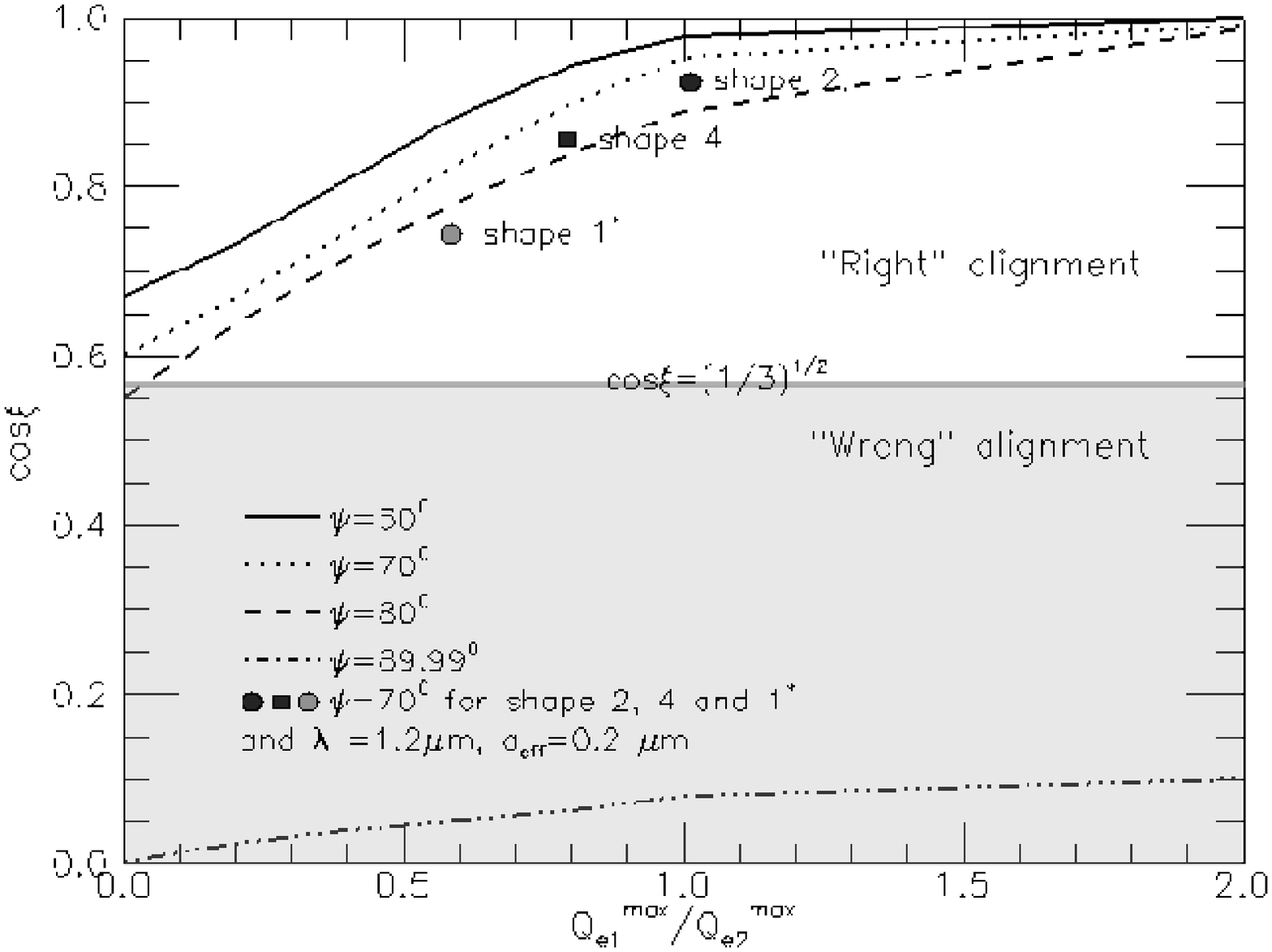}
\includegraphics[width=0.49\textwidth]{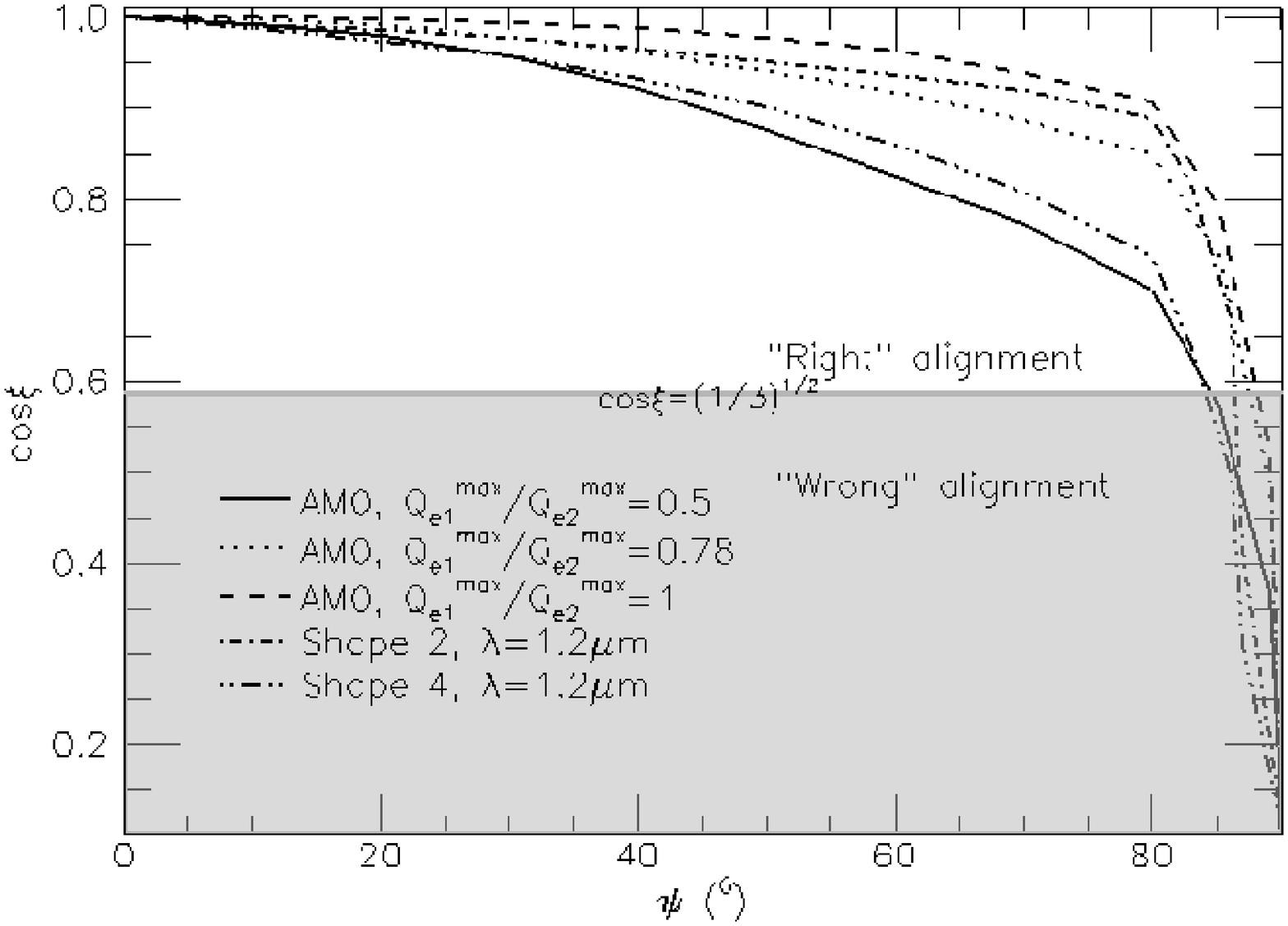}
\caption{{\it Upper panel:} The shift of position of low-$J$
attractor point as function of $Q_{e1}^{max}/Q_{e2}^{max}$ for different $\psi$.  {\it Lower panel:} The shift of position of low-$J$
attractor point with respect to $\psi$ for AMO with three different ratio  $Q_{e1}^{max}/Q_{e2}^{max}$  and
  for irregular grains. The shaded area corresponding to the degree of
  alignment 
  $R=1.5\mcs\xi-0.5\le0$ shows the range of ``wrong'' alignment.}
\label{f13*}
\end{figure}

The differences in the range of $\psi$ in which
the maps have high J attractor points predicted by AMO and irregular
grains (i.e., dot lines extend outside the region surrounded by thick
solid lines in Fig. \ref{f13}) exist, but it is moderate.
In fact, the correspondence between the ranges of $\psi$ for the
existence of the high-$J$
attractor point revealed between our predictions and
the actual irregular grains allows one to find out whether
the high attractor point is expected if only $Q_{e1}^{max}/Q_{e2}^{max}$ ratio is
known. Note, that the existence of high attractor points is important for evaluating
the degree of grain alignment.

Another new effect related to $Q_{e1}^{max}/Q_{e2}^{max}$ ratio is a regular
shift of
the position of the crossover point, which,
as we discussed below, is also low-$J$ attractor point.
Fig.~\ref{f13*}{\it upper} shows this shift as a function of the ratio $Q_{e1}^{max}/Q_{e2}^{max}$ for different $\psi$ predicted by AMO. It is shown that the low-$J$ attractor point tends to shift to $\mc\xi=0$ as $Q_{e2}^{max}$ increases. Particularly, for $\psi=89.9^{0}\sim 90^{0}$, the low-$J$ attractor point coincides to $\mc\xi=0$, i.e., grains are aligned with long axes perpendicular to magnetic field.
The above observed tendency is consistent with our earlier discussion in \S~\ref{sec4p4} that $Q_{e2}$ act to align grains perpendicular to  {\bf B}.

Fig.~\ref{f13*}{\it lower} presents the shift as a function of $\psi$ for particular angles $\alpha$ of AMO and
for irregular grains. It is clearly seen that, the low-$J$ attractor points for AMO and irregular grains are always at $\mc\xi=1$ for $\psi=0^{0}$, which means that no any "wrong" alignment possibility exists in this case. As $\psi$ increases, they shift to $\mc\xi=0$, and finally, fall on the perpendicular alignment angle $\mc\xi=0$ at $\psi\sim 90^{0}$. For shape~4, however, the low-$J$ attractor point can produce ``wrong''
alignment earlier than for shape 2, i.e., when $\psi\sim 85^{0}$. This is because the shape 4 has the smaller ratio $Q_{e1}^{max}/Q_{e2}^{max}$ than the shape 2.
 
We feel that $Q_{e1}^{max}/Q_{e2}^{max}$ provides a sufficiently good
parametrization for torques of irregular grains. Therefore, in terms of
practical calculations, establishing this ratio may be sufficient for
describing the alignment of realistic irregular shapes. This would require
much less computational efforts compared with obtaining the shapes for
the entire range of $\Theta$.

\section{Regular crossover}\label{cross}

The dynamics of AMO that we observed above
was very different from that of grains
in DW97. As we can see in \S 8, the properties of the RAT components are similar
for AMO and irregular grains, including those studied by DW97. Therefore
the difference stems from the different treatment of crossovers in our and
the DW97 models.

The most striking difference between our and DW97 trajectory maps is that 
our crossover points correspond to $J=0$ and grains cannot get out of these
points. Therefore our crossover points are also the attractor points. Note,
that, unlike DW97, we do not observe cyclic maps. The latter are the artifact
of their model of crossovers adopted there.   

In what follows, we first discuss crossovers in the most general terms and later consider
a particular case of a crossover which is not affected by thermal wobbling
(cf. WD03). Then we briefly discuss the possible effects of thermal wobbling.

The crossover dynamics has two distinct regimes. The first one, described
in Lazarian \& Draine (1997) takes place when the time of the crossover
$t_{cros}\sim J_{\bot}/Q$ where $J_{\bot}$ is the angular momentum component
perpendicular to ${\bf a}_{1}$ (Spitzer \& McGlynn 1979), i.e.
the time during which the grain undergoes a regular flipping subjected
to the torque $Q$, is shorter than the time of internal relaxation 
$t_{relax}$. For $H_2$ torques Lazarian \& Draine (1997) obtained
that this regime is fulfilled for grains larger that $a_c\sim 10^{-5}$~cm.
When LD99b introduced nuclear relaxation it became clear that
the typical critical size for interstellar grains  $\sim 10^{-4}$~cm. This size
depends on the value of the torques rather weakly, thus we may also
accept it as the critical size of the grains in diffuse ISM
 in the presence of radiative torques. However, in the vicinity of really strong
radiation sources the critical size gets smaller. Note, that $a_c$ is larger
than the ``typical'' $a_{eff}$ in Table~\ref{tab1}, but as we mentioned earlier, grains
typical for diffuse interstellar gas are different from grains typical for other
astrophysical environments.

The randomization of grains during a crossover happens due to random
processes associated with atomic bombardment and H$_2$ formation. This is
a random walk process in which the squared deviation in angular momentum $(\delta J)^2$
scales with the crossover time $t_{cros}$. Thus the deviation of the angular momentum
in the process of a crossover scales as 
$\delta J/J_{th}\sim 1/J_{th}Q^{1/2}$.
As $J_{th}\sim (kT_{d} I_1)^{1/2} \sim a_{eff}^{5/2}$ and assuming that $Q$ is proportional to the cross-section $\sim a_{eff}^2$ (see \S 10), we get the deviation $\delta J/J\sim a_{eff}^{-7/2}$, that decreases quickly with the grain size. Therefore the assumption of
{\it no randomization during crossovers} should initially be accurate for grains much
larger than $a_c$. Such grains according to Cho \& Lazarian (2005) are 
responsible for far infrared polarization emanating from dark starless cores.

For grains less than the critical size $a_c$ 
LD99a showed that 
the physics is different. As the angular momentum
of a grain get comparable with $J_{th}$, such grain wobbles fast due to
coupling of rotational and vibrational degrees of freedom by the internal
relaxation (see Lazarian \& Roberge 1997). For H$_2$ torques this
results in flipping that reverses the direction of torques and gets the
grains ``thermally trapped'' (see LD99a).

All in all, if 
grain dynamics is presented in the axes of $J/J_{th}$ and $\cos \gamma$ with $\gamma$ being the angle between ${\bf a}_{1}$ and ${\bf J}$, 
the assumption of ${\bf J}\|{\bf a}_{1}$ is accurate for $J/J_{th}\gg 1$. In
the range of $J/J_{th}= [0,1]$ the actual crossover physics should be
accounted for.  

In view of the discussion above, consider a regular crossover  first, i.e. assume that
$a_{eff}\gg a_c$. The assumption that ${\bf a}_{1}$ is always parallel to 
${\bf J}$ breaks inevitably as  $J_{\|}\rightarrow 0$. Indeed,
whatever is the efficiency of the internal relaxation mechanism, there
will be some residual $J_{\perp}$. Lazarian \& Draine (1997) showed
that the mean value of $J_{\perp}$ during a regular crossover cannot be smaller than $J_{th}$.
In addition, as $J$ gets small, the efficiency
of relaxation drops and the gaseous bombardment increases $J_{\bot}$ further
in accordance with
the original Spitzer \& McGlynn (1979) theory of crossovers. 

To simplify our treatment we disregard all internal
relaxation processes during the crossover.
This is justifiable as for $a_{eff}\gg a_c$ the crossover happens on the time
scale shorter than the internal (e.g. nuclear) relaxation time.
Since the mirror is assumed weightless, 
for AMO the dynamics of free rotation coincides with that of a spheroid. As a result, for a given $J$, ${\bf a}_{1}$ precesses around  ${\bf J}$ with a constant angle $\gamma$. The state of the grain is completely determined
by describing ${\bf J}$ in the lab and body systems. Equations of motion for
 this case are
\begin{align}
\frac{d{\bf J}}{dt}&={\bf \Gamma}-\frac{{\bf J}}{t_{gas}},\label{cr1}\\
\frac{J d\mc\gamma}{dt}&=-\frac{dJ}{dt}\mc\gamma+\frac{dJ_{\|}}{dt},\label{cr2}
\end{align}
where ${\bf \Gamma}$ is  RAT, and $J_\|=J \mc\gamma$ is the
component of angular momentum along the maximal inertia axis.
In the coordinate system $J, \xi, \phi$, equation (\ref{cr1}) returns to the set of equations of motion (\ref{eq58})-(\ref{eq60}). Averaging over the angles, $\Phi$, and $\phi$ corresponding to the precession
of ${\bf a}_{1}$ around ${\bf J}$ and ${\bf J}$ around ${\bf B}$, 
respectively, and solving three resulting equations for $J, \xi,\gamma$, we can obtain trajectory maps during the regime of low J and regular crossover. 

\begin{figure}
\includegraphics[width=0.49\textwidth]{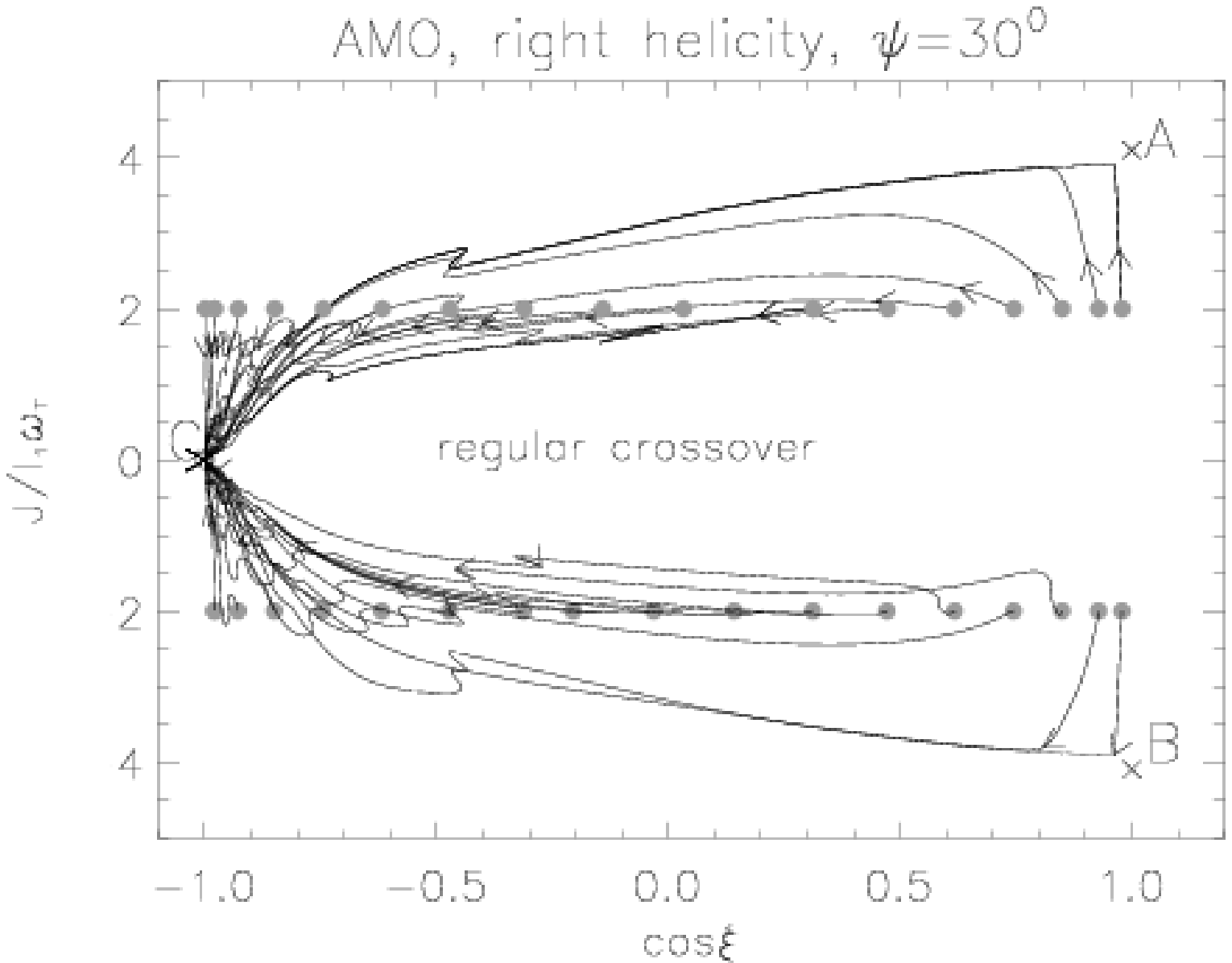}
\includegraphics[width=0.49\textwidth]{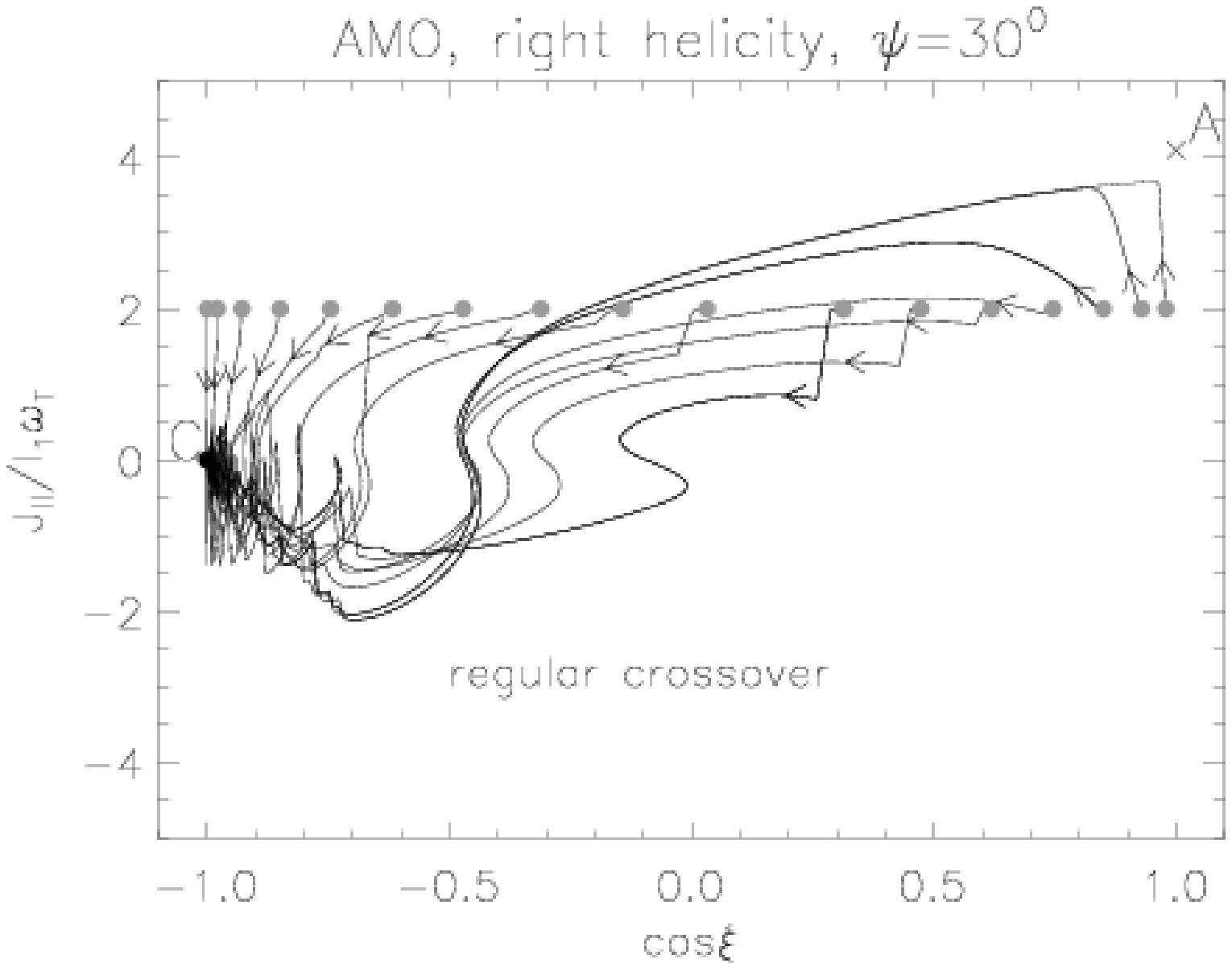}
\caption{Phase maps for AMO with $\psi=30^{0}$. {\it Upper panel} shows that all
 grains approach an attractor point C of $J=0$ corresponding to the same angle $\mc\xi=-1$. However regular
 crossovers corresponding to $J_{\|}=0$ occur at different angles as
illustrated by {\it lower panel}.} 
\label{f15}
 
 \end{figure}
In Fig. \ref{f15} we show the phase map for the case $\psi=30^{0}$.
The most striking feature observed there is that the grain experiences not
a single, but multiple crossovers. Unlike Purcell's torques,
as the grain flips, RATs do not change their direction. As a result,
they drive $J$ further down. In other words, the dynamics of grains 
roughly corresponds to what we observed in the earlier sections of the
 present paper (cf. DW97 where grains are spun up after passing $J=0$). 

Depending on magnetic field value, the crossover may happen on the
time scale smaller than the Larmor precession period. In this case, different
grains will undergo crossovers at different $\phi$ and therefore will 
experience different torques during crossovers. However, this would not change
the qualitative picture, as RATs will drive $J$ to zero
irrespectively of $\phi$. We shall discuss this issue in more details elsewhere.

Let us consider qualitatively
the issue of thermal fluctuations. For the Purcell's torques
those can be ignored during the crossovers for grains with size $a_{eff}\gg a_c$. RATs, however, are different. They tend to decrease $J$ further rather than
spin up the grains after a crossover. The time of relaxation, which is also
the
time over which thermal fluctuations of $\gamma$ take place, increases as
$J$ decreases. Therefore in the absence of external collisions
one can imagine two situations: according
to one, RATs stop the grain completely, so no thermal wobbling of $\gamma$
is possible, another is that $J$ settles at the value that is of the order
of $J_{th}$. The latter seems more probable for irregular grains.

An interesting consequence of the considerations above is that a grain tends
to get to the $J=0$, but the irregularities in the motion
of the grain axes in respect to the
photon flow prevents the angular momentum from reaching zero. Thus, we can
expect that for grains with moments of inertia closer to one of a spheroid the value of
angular momentum at the low-$J$ attractor point is going to be smaller than for
more irregular grains.

To find the actual value of effective $J$, one should recall that grains are subject to substantial random excitations in the typical
astrophysical conditions. Gaseous bombardment, interactions with ions,
absorption of photons (see Draine \& Lazarian 1998 for a quantitative
description of the processes), affects grain rotation at the low-J attractor
point. We expect these
influences to increase the mean value of $J$ above $J_{th}$.

Another effect should also be present in the presence of random bombardment. If
the phase maps have both high and low-$J$ attractor point, the bombardment allow
 more grains to get to a higher-$J$ attractor point. As low-$J$ attractor
points are characterized by higher internal randomization, quite counter-intuitively, the random forcing can increase the degree of alignment.    

All in all, our considerations above justify the simplified model
of crossovers that we adopt in the paper. We shall provide a more detailed
treatment in Hoang \& Lazarian (2007).

\section{Particular Cases}\label{sec9}

Our earlier discussion covered two limiting cases, namely, (a) the
light beam being the axis of alignment, which corresponds to the
precession rate induced by RATs $t_k^{-1}$ much larger than the Larmor
 precession rate $t_B^{-1}$ and (b) when the magnetic field constitutes the
axes of alignment, i.e. $t_B^{-1}\gg t_k^{-1}$. Can this case be ever astrophysically
important? 

\subsection{Criterion for {\bf B} alignment} \label{sec91}

Consider the precession of ${\bf a}_{1}$ or ${\bf J}$ around ${\bf k}$
 driven by the component $Q_{e3}$ which
is perpendicular to the axis ${\bf a}_{1}$ and ${\bf k}$. 
 The timescale for RAT precession is defined by
\bea 
 t_{k}=\frac{2\pi}{|d\phi/dt|}= {\frac{10^{11}}{\hat{Q}_{e3}}}(\hat{\rho}a_{-5}\hat{T})^{1/2}(\frac{1}{\hat{\lambda}\hat{u}_{rad}}) \mbox{s},\label{eq78}
\ena 
 where 
\bea 
 \frac{d\phi}{dt}=\frac{\lambda a_{eff}^{2}\gamma}{I_{1}\omega} \mbox{{\bf
    Q}}_{\Gamma}.\hat{\Phi}=\frac{\lambda a_{eff}^{2}\gamma}{I_{1}\omega} Q_{e3}.\label{eq79}
\ena 
Here $\omega$ is the angular velocity of a grain around  the maximal inertia
 axis ${\bf a}_{1}$. $\hat{Q}_{e3}=Q_{e3}/10^{-2}$,
$\hat{\lambda}=\lambda/1.2 \mu m$, $\hat{u}_{rad}=u_{rad}/8.64\times 10^{-13}
ergs~ cm^{-3}$, and $\hat{\rho}=\rho/3 gcm^{-3}$, $a_{-5}=a_{eff}/10^{-5} cm$, and
$\gamma=0.1$ for anisotropy of ISRF. In
 deriving equation (\ref{eq78}) the assumption $\omega=\omega_{T}$ is used. 
It is easy to see that for axisymmetric grain shapes, 
$Q_{e1}, Q_{e2}$ are
equal to zero, while $Q_{e3}$  is non-zero (see equations \ref{eq3}-\ref{eq5}). Therefore, the third component produces fast precession of grains about ${\bf k}$. 

 The precession rate above should be compared with
the Larmor precession rate. 
 A rotating grain acquires a magnetic moment by the Barnet effect 
which is shown to
be much stronger than that arising from the rotation of its charged body 
(Dolginov \& Mytraphanov 1976). The
interaction of the magnetic moment with the external magnetic field causes the gradual
  precession of the grain around the magnetic field direction. 
The rate of Larmor
precession around magnetic field, $t_{B}$, is given by
  \bea
t_{B}=3\times 10^{7} a_{-5}^{2} \hat{\rho}^{-1/2}\hat{\chi}^{-1}\hat{B}^{-1} ~\mbox{s},\label{eq80}
 \ena
where $\hat{B}=B/5 \mu G$, $\hat{\chi}=\chi/3.3\times 10^{-3}$ are
normalized magnetic field and magnetic susceptibility, respectively. 

The alignment of grains whether with the radiation or magnetic field depends on their 
precession rate around these axes. According to equations (\ref{eq78}) and (\ref{eq80}),
the ratio of precession rate due to radiation 
and magnetic field is given by
\begin{align}  
\frac{t_{k}}{t_{\mbox{B}}}&=\frac{3.33\times 10^{3}}{\hat{Q}_{e3}}\hat{\rho}\hat{\chi}\hat{T}^{0.5}a^{-1.5}_{-5}(\frac{\hat{B}}{\hat{\lambda}\hat{u}_{rad}}).\label{eq81}
\end{align}
It can be easily checked that for typical diffuse ISM,
 $t_{B}/t_{k} \sim 10^{-3}$, i.e. the Larmor precession is much faster than
precession induced by radiation, therefore magnetic field is the
alignment axis.

For AMO, RATs can be
very different as $Q_{e3}$ arises from a spheroidal shape, while
the other two components arise from a mirror; the ratio of
the sizes of the mirror and the spheroidal body may be arbitrary.
For irregular grains, however, we can see all three RAT components to be roughly
comparable (see Figs \ref{f21}-{\ref{f23a}). In this case, $t_{k}$ can be
  used as a proxy for the time of ``fast alignment'' (see \S \ref{fast}).

\subsection{Astrophysical Implications}

It is possible to estimate that for the magnetic field of
5 $\mu$G grains get aligned with the radiation beam when the density of
energy in a beam is 
$\mbox{u}_{\mbox{rad}}>10^{3}\mbox{u}_{\mbox{ISRF}}$. The appropriate
radiation fields is typical near stars and supernovae.

Fig. \ref{f16} shows the variation of the ratio of precession timescales
with distance 
for different stars in which we use $Q_{e3}=Q_{e3}^{max}$ (for shape 1 and ISRF) for equation (\ref{eq81}). It is apparent that near the stars, grains precess 
around ${\bf k}$ much faster than around magnetic field. However, the ratio increases
with distance as the radiation field decreases, so the precession around magnetic field is faster than around light, and grains align
with magnetic field at some distance. 

\begin{figure}
\includegraphics[width=0.49\textwidth]{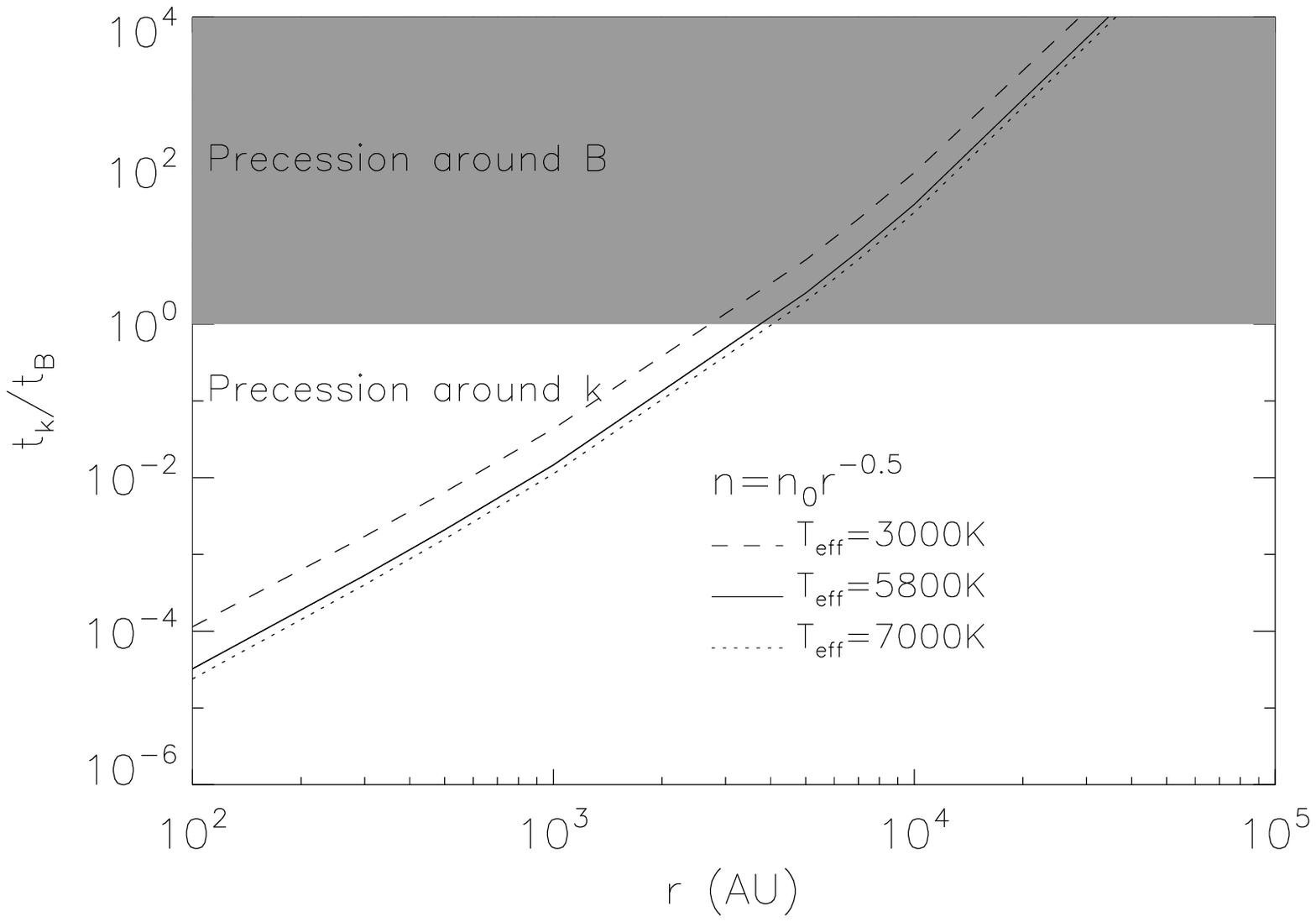} 
\hfill
\includegraphics[width=0.49\textwidth]{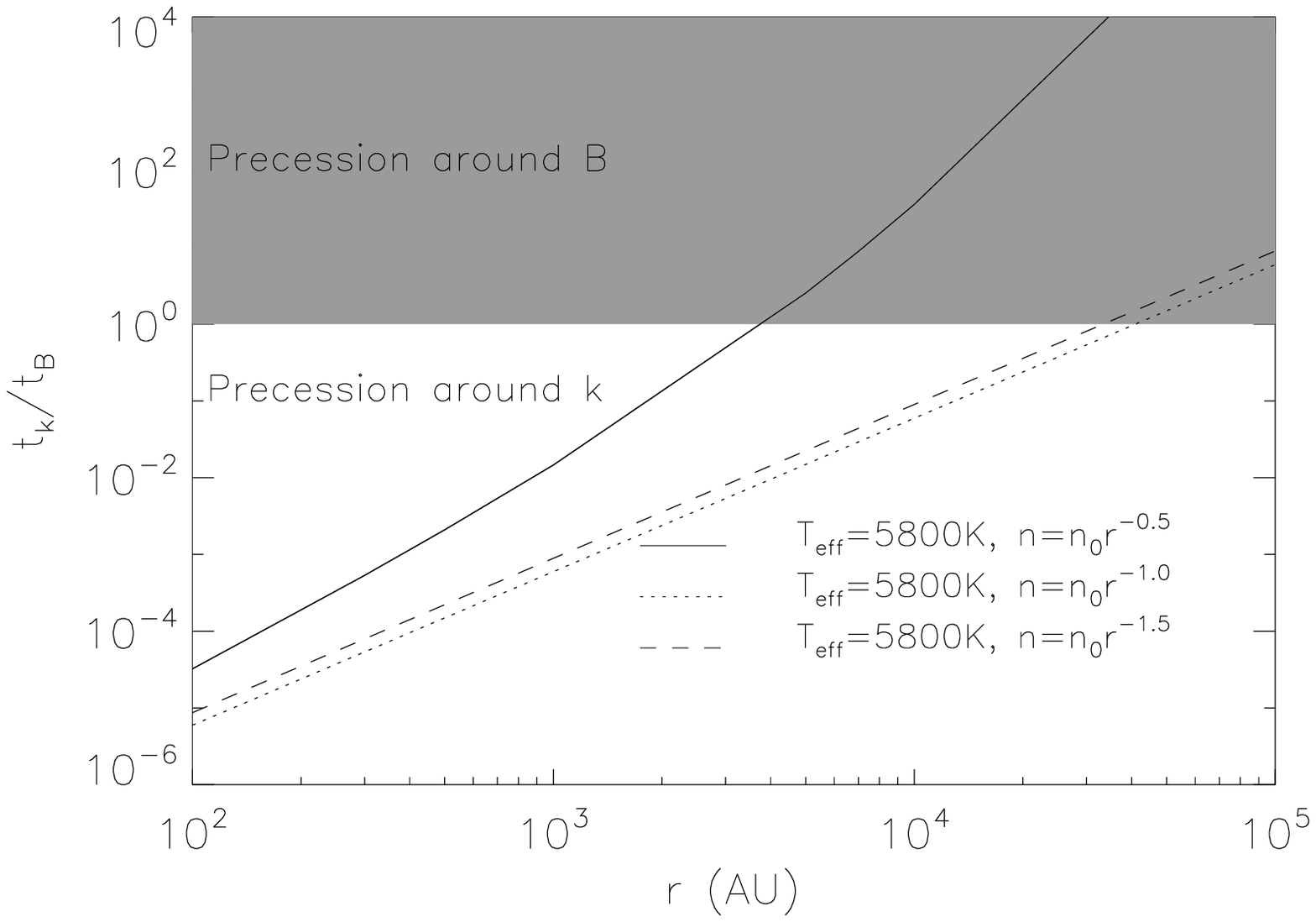} 
\caption{The ratio of timescales for the precession about the light direction and 
about the magnetic field assuming that magnetic field is homogeneous with
  strength $B=0.5 \mu G$ through
  envelopes. {\it Upper Panel}: For three values of star temperature, the density dependence on
the  distance from the star is fixed; {\it Lower Panel}: For a given star with
  different functions of density.} 
\label{f16}
\end{figure} 

Another case when the alignment can happen with respect to the direction of light
is the case of cometary dust. This case may be somewhat more complex, as
electric field can be present near the comet head. This can induce 
precession of grains with dipole electric moment and therefore provide
yet another axis of alignment. We discuss this problem in Hoang \& Lazarian (2007). 

If $t_k^{-1}$ and $t_B^{-1}$ are comparable, the axis of alignment does not 
coincide with either beam or {\bf B}. This case can be relevant to some part
of comet grains, but a discussion of it is beyond the scope of the present
 paper.
 
Note that $t_k^{-1}$ is a function of $\Theta$. Therefore when its amplitude
value is larger than $t_B^{-1}$ this does not guarantee that
the effect of magnetic field is negligible. For AMO the alignment
drives grains into the position corresponding to $\Theta=\pi/2$, for
which the component $Q_{e3}$ gets zero and therefore the magnetic field
again dominates. Thus the alignment is expected with respect to an intermediate
axis. This may be important for explaining circular polarization arising
from comets (Rosenbush et al. 2007). Naturally, not only precession arising
from magnetic field, but also from electric field and mechanical torques (see
\S 11.4) should be taken into account for that case.

\subsection{Fast alignment}\label{fast}
In this section we ignore the damping role of the
 ambient gas, and show that grains can be
aligned by radiation on a timescale shorter than the gas damping time. Such 
alignment can be called ``fast alignment'' in analogy with the ``fast dynamo''
process (see Vishniac, Lazarian \& Cho 2003 and ref. therein) 
that can amplify magnetic field on the timescales shorter than the
magnetic diffusion time. In particular, supernovae flashes can align grains
around them with respect to the direction of light (see \S \ref{kalign}).

\begin{figure}
\includegraphics[width=0.49\textwidth]{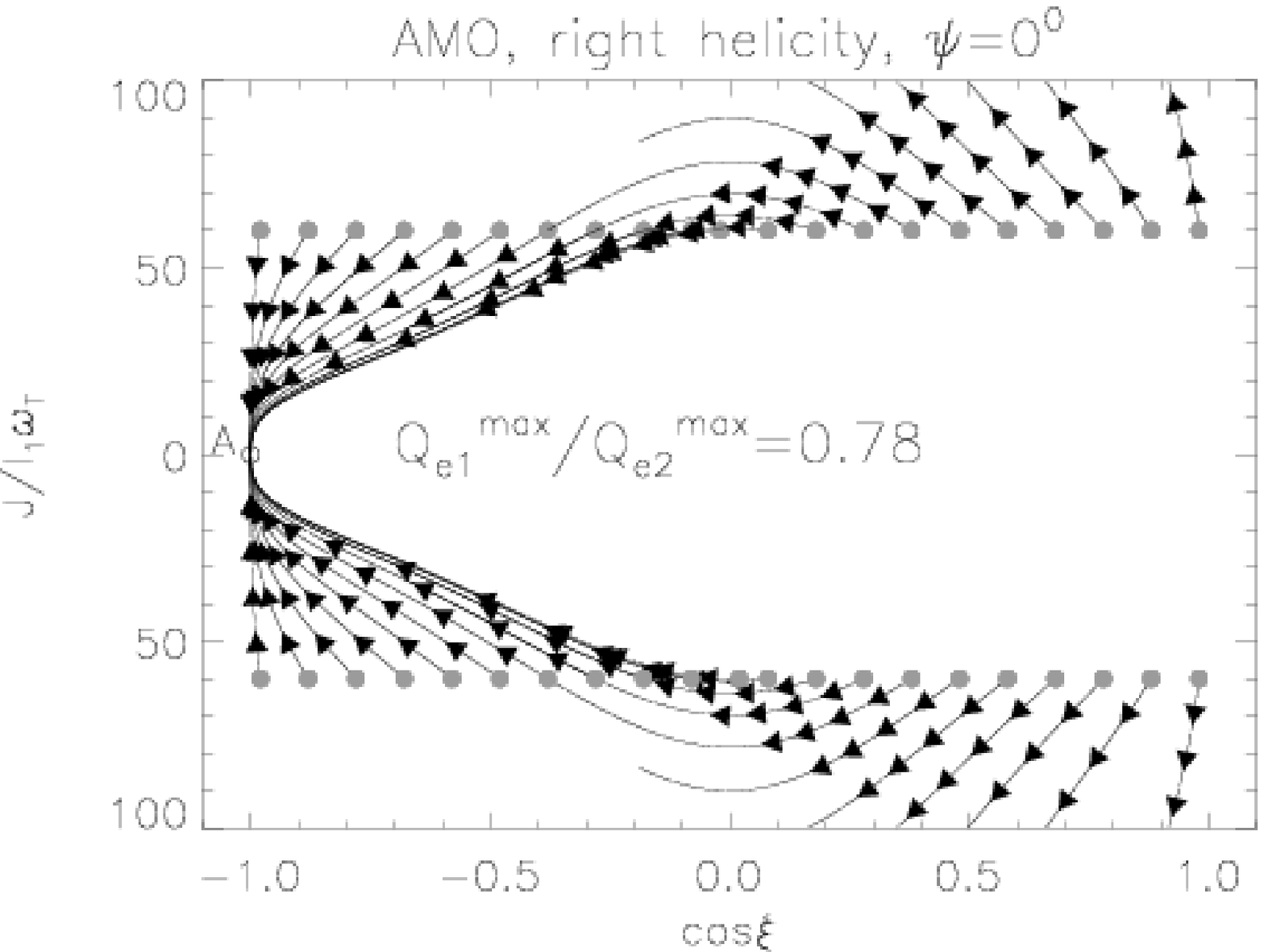}
\includegraphics[width=0.49\textwidth]{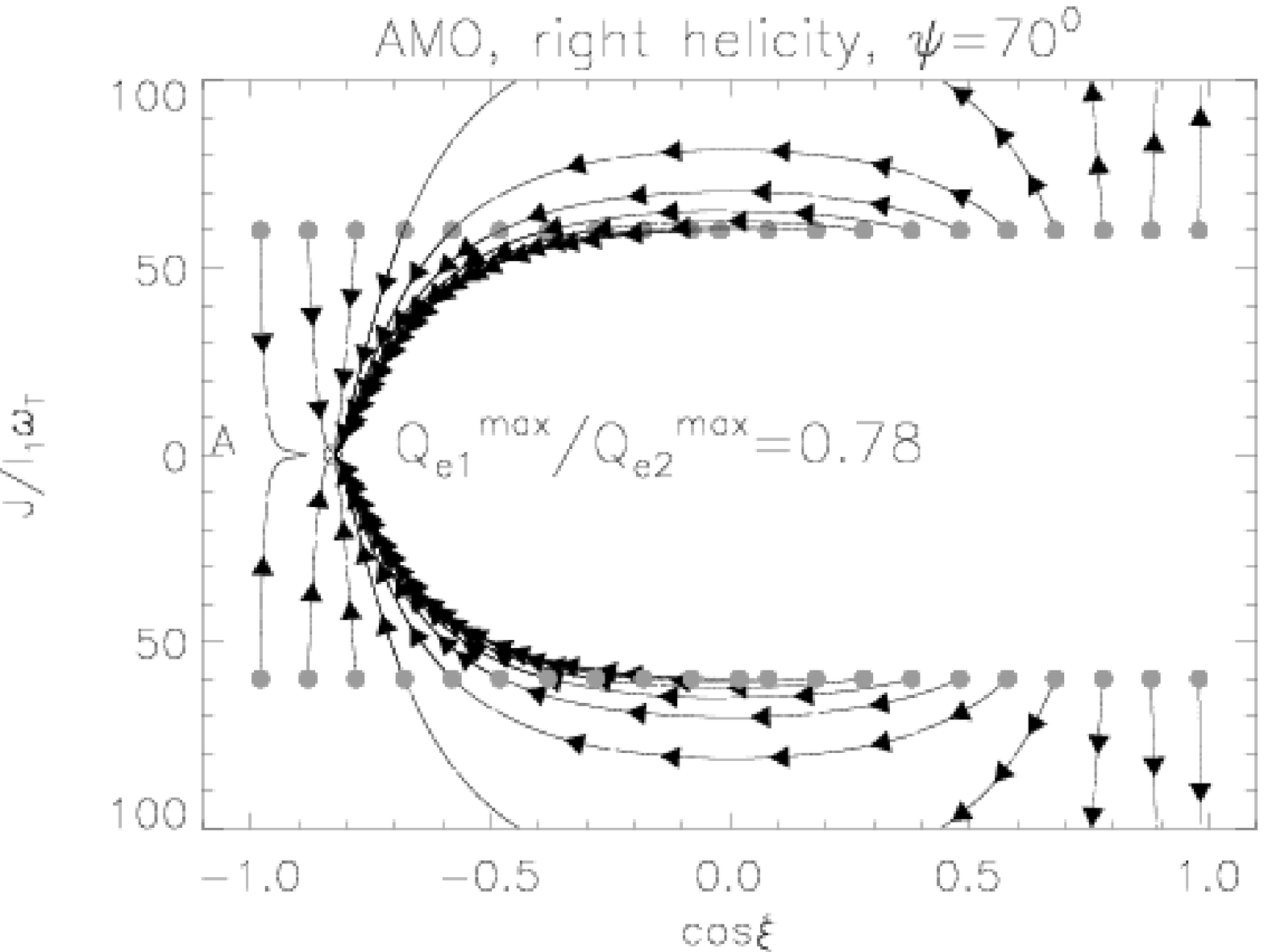}
\caption{Alignment in the absence of gas damping for two directions of
  radiation $\psi=0^{0}$ ({\it Upper panel}) and $ 70^{0}$ ({\it Lower
    panel}). Here, the gray circles represent
    the initial position of grains. In the upper and lower panel, the distance between
    two filled 
arrows corresponds to a time interval $\Delta t= 10 t_{k}$, and $\Delta t=50
    t_{k}$, respectively.
 }
\label{f14}
\end{figure}

\begin{figure}
\includegraphics[width=0.49\textwidth]{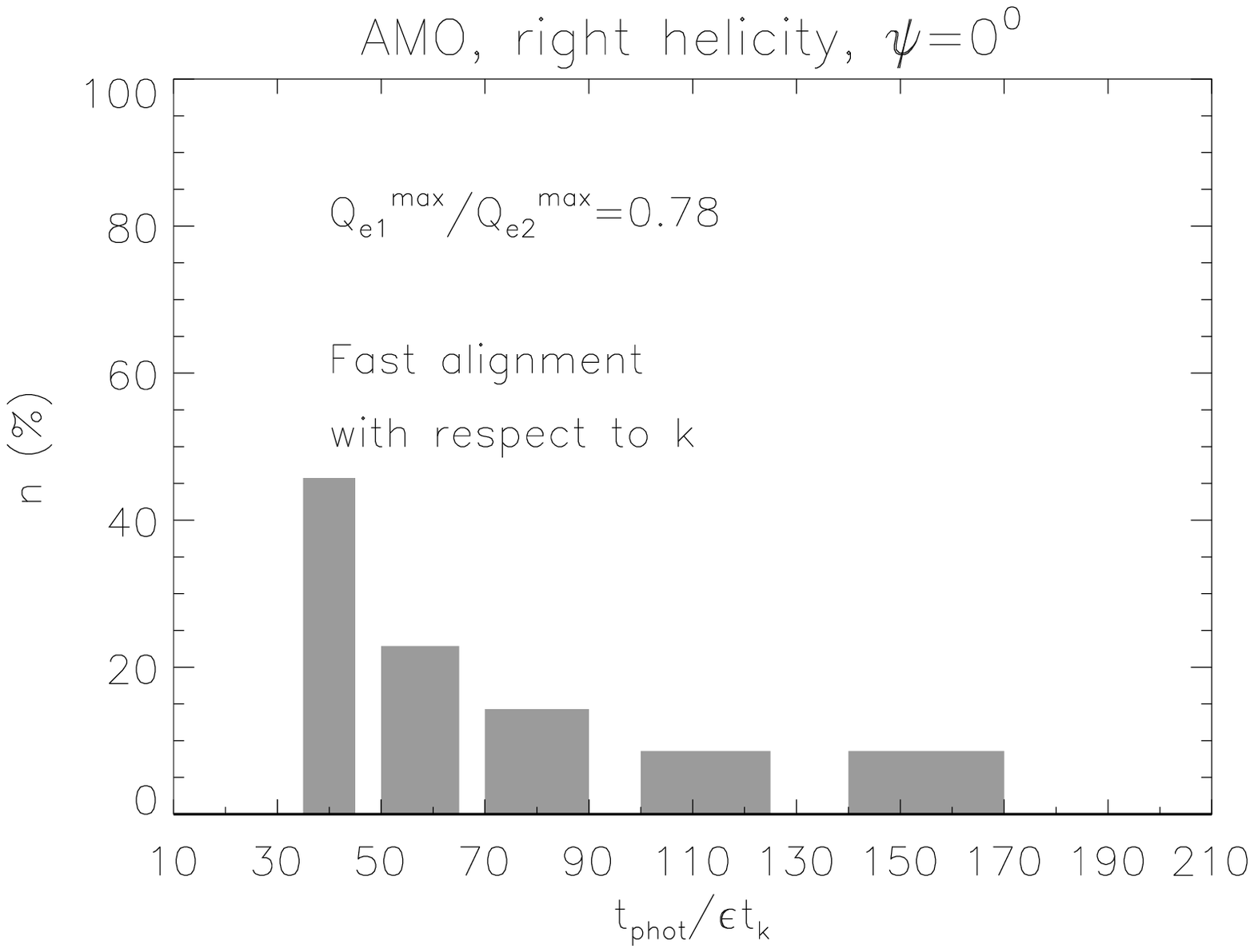}
\includegraphics[width=0.49\textwidth]{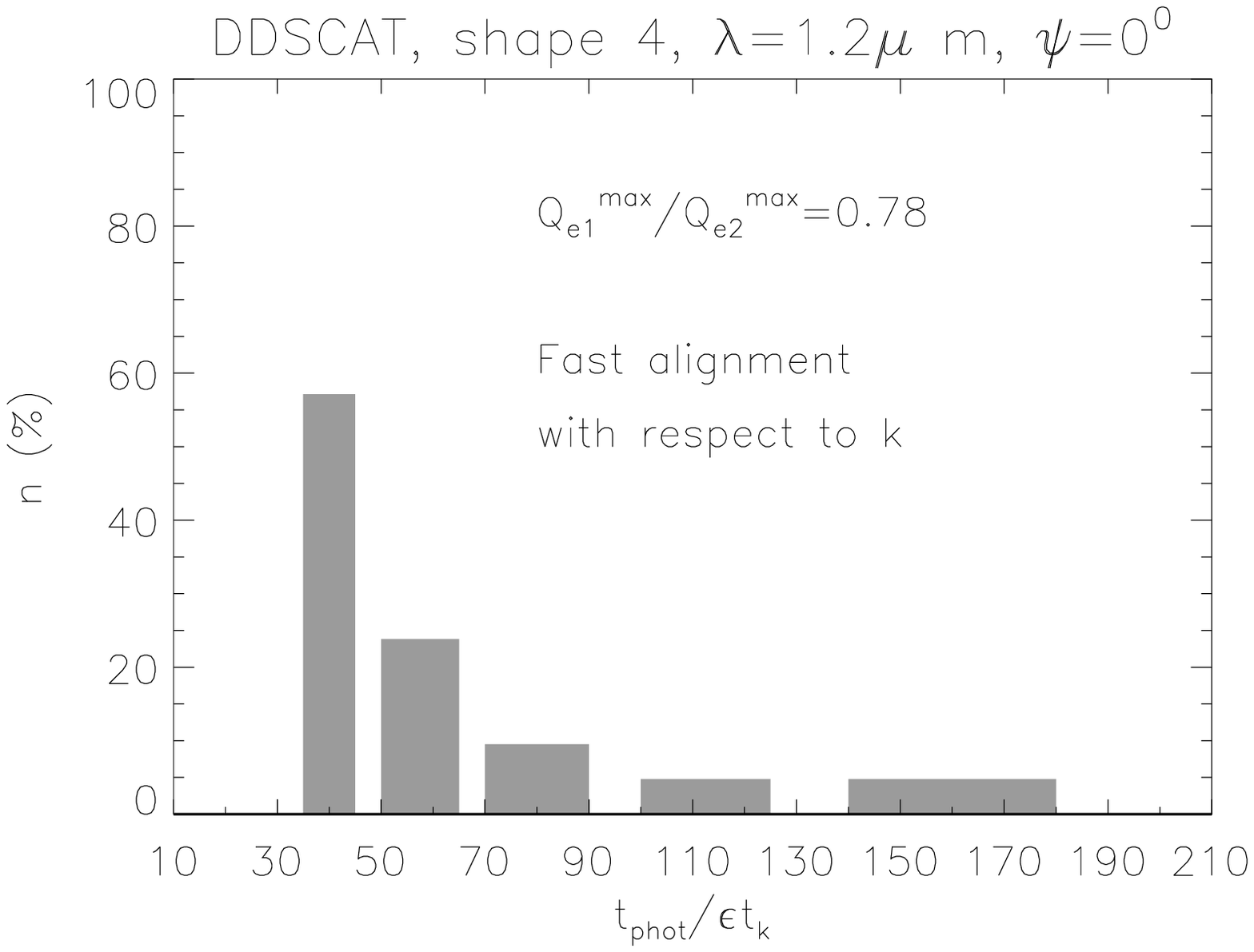}
\caption{Percentage of grains as function of the ratio of the alignment
  timescale due to photon, namely $t_{phot}$ to the precession time about the radiation direction, $t_{k}$, for AMO with
  $Q_{e1}^{max}/Q_{e2}^{max}=0.78$ ({\it upper panel})  and shape 4 ({\it lower panel}). Here $\epsilon=\frac{Q_{e1}^{max}}{Q_{e3}^{max}}$, that is chosen to be equal unity for AMO, and $\epsilon \sim 1$ for shape 4.}
\label{f14*}
\end{figure}

In Fig.~\ref{f14}{\it upper} we show the map for $\psi=0^{0}$ in which the
distance between two arrows represents a time interval equal 10 precession
time $t_{k}$ that is defined in equation
(\ref{eq78}). It can be seen that the grains that have initial orientations close to
the attractor point A require alignment timescales $t_{phot}$\footnote{$t_{phot}$ is defined as the alignment timescale of grains due to photon or radiation} is  approximately  $40 t_{k}$ to get aligned (see Fig. \ref{f14}{\it upper}). In contrast, grains that have
initial angles far away from the attractor point need more time (up to $\sim 160 t_{k}$) to get
there (see Fig. \ref{f14*}). 

Fig. \ref{f14}{\it lower} shows a similar effect for the case $\psi=70^{0}$. However, the interval of two arrows represents a time interval $\Delta t=50t_{k}$.  
The map shows that grains bound to the low-$J$
attractor point A reach it fast, about $70 t_{k}$, i.e., on the time scales much less than the rotational damping time. Some grains that otherwise would go to the 
high-$J$ attractor point stream to infinite $J$ in the absence of damping
({\it lower panel})\footnote{We may observe that phase trajectories in Fig. \ref{f14}{\it lower}
 directed to high attractor points also correspond to the aligned state
  of grains, although the stationary state requires $t_{gas}$ to be achieved. In this sense all grains get aligned fast.}.  In practical terms the latter fact for this regime (i.e., alignment timescale
$t_{phot} < t_{gas}$) is not so important, as most
grains get aligned at the low-$J$ attractor points anyhow. We also observe the shift of low-$J$ attractor point A as in the upper panel of Fig. \ref{f13*}.,

The corresponding time for the fast alignment is proportional to ratio of
grain angular momentum and the component of the torque $\Gamma_{rad}$. In
terms of RATs normalized components $Q_{\Gamma}$ that is related to ${\bf \Gamma}_{rad}$ via equation~ (\ref{eq1}) the relevant combination is given by
equation (\ref{eq78}). The corresponding function $F$ depends only on $Q_{e1}$ and
$Q_{e2}$. The amplitudes of those can be very different from the component
$Q_{e3}$ that causes the grain precession. 
Therefore for such grains, the ratio of
$J/\Gamma_{rad} \propto J/F$ can be measured in units of ``the period of
radiation induced precession'', namely, in $t_{k} \propto J/Q_{e3}$ (see
equation \ref{eq78}). To make AMO more correspond to irregular grains in this respect,
we make the choose the amplitude of the AMO components $Q_{e3}$ to be of the
similar to the amplitude of $Q_{e1}$.

Histograms showing the distribution of grains on the low-$J$ attractor point as a function of alignment timescale corresponding to AMO and an
irregular grain shape 4 are shown in Fig. ~\ref{f14*}. It can be seen that, for AMO, about $45\%$ grains get aligned with respect to
${\bf k}$ over $t_{phot}\sim 35 t_{k}$ to $45 t_{k}$, and about $22\%$ of grains get there over $t_{phot}\sim 50 t_{k}$ to $65 t_{k}$.  A few percent of grains require longer time to get
aligned, up to $170t_{k}$ (see Fig. \ref{f14*}{\it upper}) . This relative inefficiency of alignment is a
consequence of small amplitude of the function $F$ in the vicinity of the
low-$J$ attractor points ($F=0$ at the stationary points). On the other hand, for the shape 4,
Fig. \ref{f14*}{\it lower} shows that about $55\%$ and $21\%$ of grains get aligned with
${\bf k}$ over $35 t_{k}$ and $65 t_{k}$, respectively. Some others can get aligned over $170 t_{k}$. We see that the similar distribution of grains as functions of alignment time between shape 4 and AMO, though the slight difference in percentage of grains corresponding to each $t_{phot}/t_{k}$ present due to the fact that, their functional forms of their torques are not comletely the same. 

Fast alignment happens in {\it respect to magnetic field} provided that
$t_{B}<t_k$ (see equation \ref{eq81}), but $t_{phot}<t_{gas}$. The ratio
$t_{gas}/t_{B}=1.2\times 10^5 \frac{\hat{\rho}^{3/2}\hat{B}\hat{\chi}}{\hat{n}\hat{T}_{g}^{1/2}}a_{-5}^{-1} $  which provides a substantial parameter space if
$t_{phot}$ is much larger
that $t_k$, e.g. $t_{phot}\sim 10^2 t_k$. Comparing $t_{gas}$ in Table~2 and
$t_k$ given by
equation (\ref{eq78}) we may conclude that for typical grains in diffuse
interstellar gas $t_{gas}$ is
marginally smaller than $t_{phot}$ and therefore the grain phase
trajectories are still determined by $t_{gas}$. However, closer to stars
$t_{phot}$ provides the measure of the characteristic time of alignment\footnote{We see that the alignment times are more than $30 t_{k}$. This is due to the fact that the aligning torque gets weaker near the low-$J$ attractor points (see Fig. \ref{f10}{\it upper})}.

 \section{Fitting formulae for RATs}\label{self}

Astrophysically motivated situations require calculations
 of RATs for grains of different sizes and at many wavelengths. This
requires rather intensive numerical computations. Our encouraging
results with AMO motivate us to consider whether we can predict the
scalings of torques.

 Dolginov \& Mytrophanov (1976) associated RATs with the scattering
of right and left handed photons by a grain. For this model one should
 conjecture that
amplitude of RATs decreases rapidly with increasing of $\lambda/a_{eff}$
as the grain-photon interactions get into the random walk regime. In other
words, a sharp peak is expected for the torque efficiency for photons
with $\lambda\sim a_{eff}$. AMO, on the contrary, suggests  of a 
linear increase for $\lambda \ll a_{eff}$. Our computation, however, indicate that
RAT gets constant for $\lambda < a_{eff}$. However, both computational and DDSCAT
intrinsic limitations do not allow us to perform calculations
for $\lambda/a_{eff}<0.1$. More studies with other techniques, e.g., ray-tracing
one, are necessary.

For grains with $\lambda\gg a_{eff}$, Lazarian (1995) suggested that the
scaling of RATs efficiency should be $\sim (\lambda/a_{eff})^{-4}$.
 Such considerations disregard
the variations of the optical constants. Therefore the testing is essential. 
  
We calculated RATs in function of $\lambda/a_{eff}$ for three
 grain shapes. Shape 1 and 2 are shown in Fig. \ref{f18}, and a hollow grain
is produced from the shape 1 by removing the core of grain. The latter is done
to reduce the amount of necessary DDSCAT computations while achieving
smaller $\lambda/a_{eff}$ ratios.

We use both the dielectric function for the smoothed astronomical
 silicate (DW97; Weingartner \& Draine 2001; Cho \& Lazarian 2005) and constant optical constant. Results are shown in
Figs. \ref{f26} and \ref{f27}.  
 \begin{figure}
\includegraphics[width=0.49\textwidth]{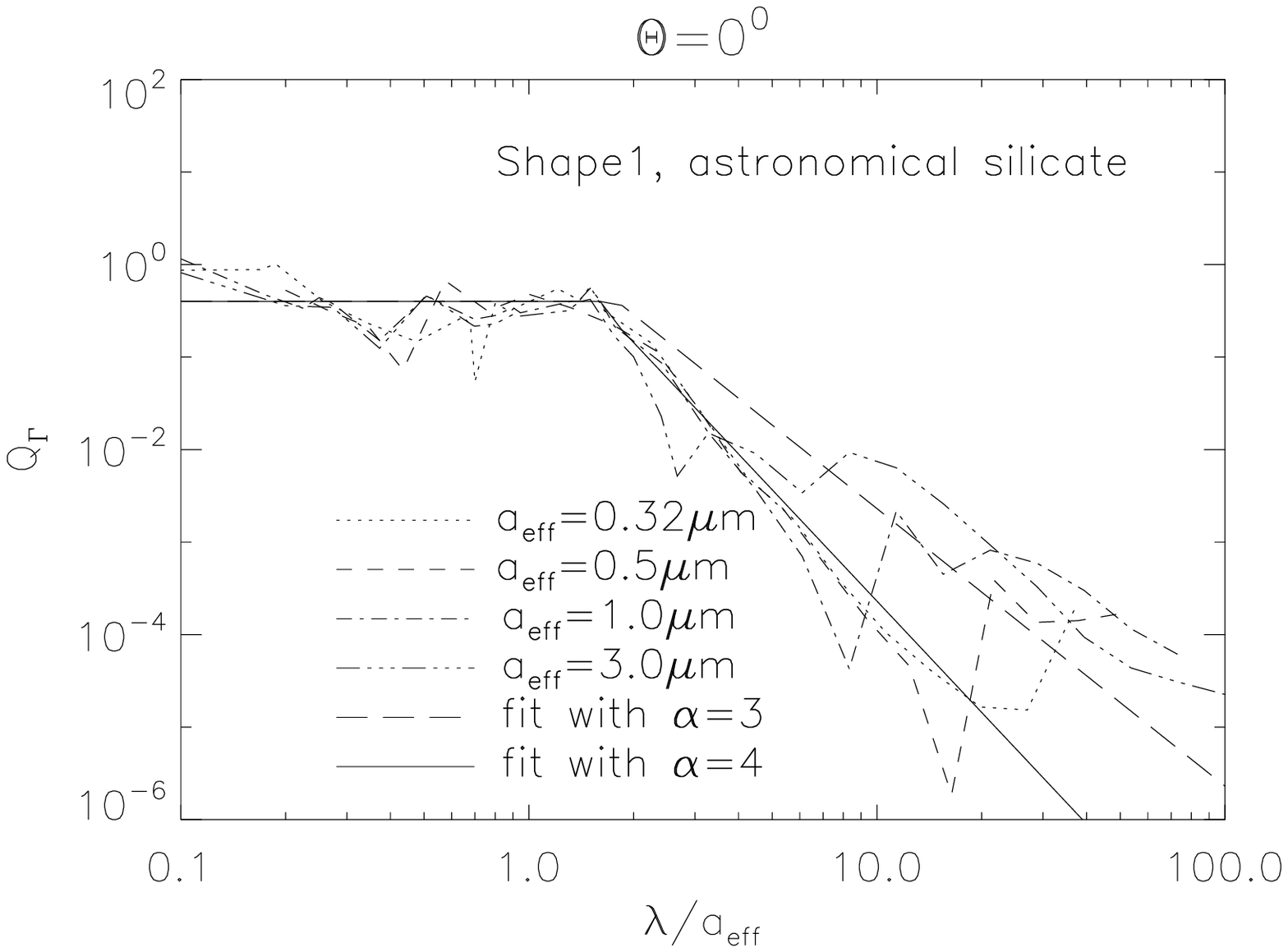}
\includegraphics[width=0.49\textwidth]{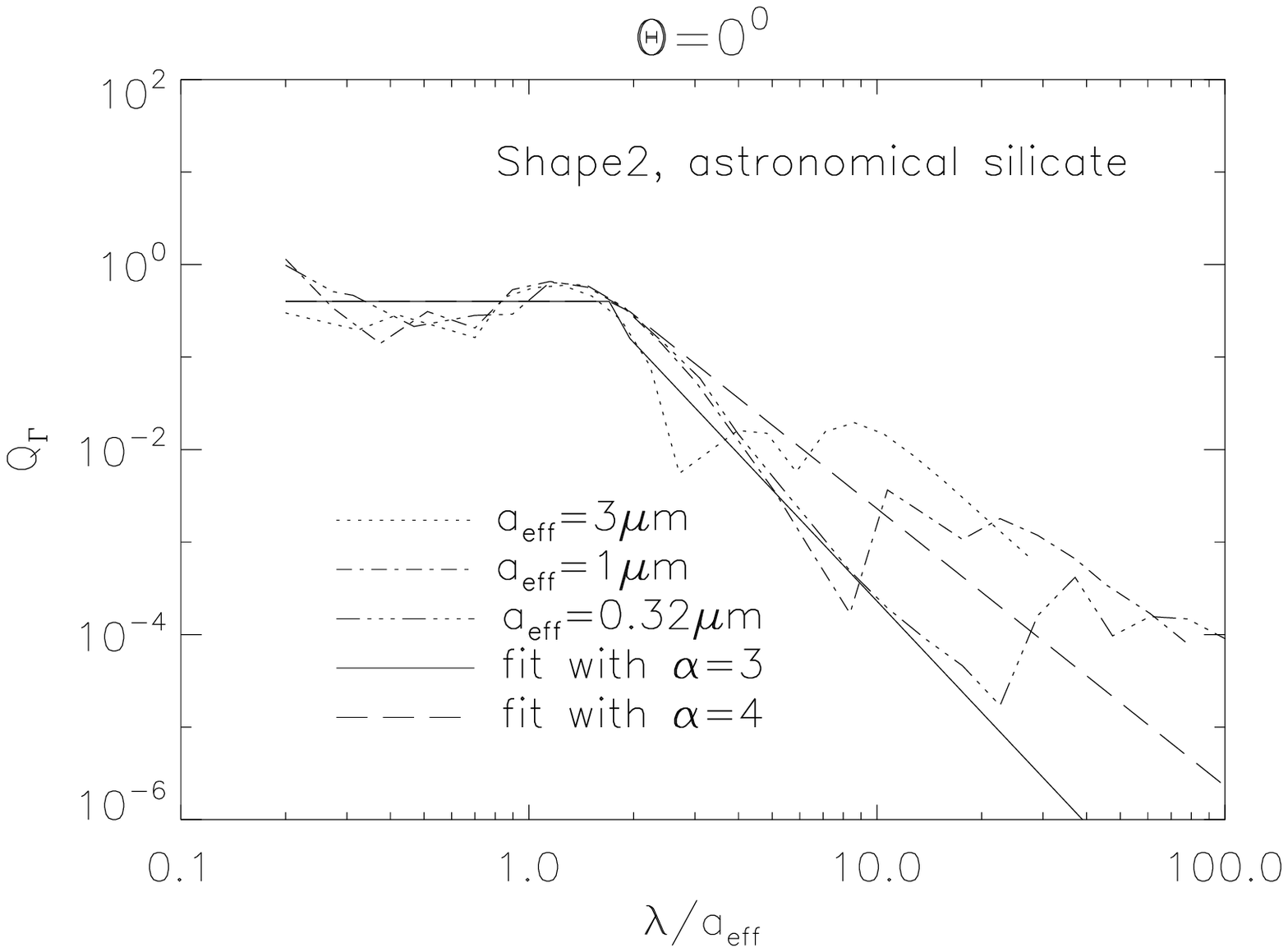}
\caption{{\it Upper panel}: RATs for grain shape 1, as the function of 
wavelength to grain size for the
  direction  $\psi=0^{0}$. We see that RATs decrease steeply as
  $\lambda/a_{eff}$ increases. Around $\lambda/a_{eff}\sim 1$ the dependence
flattens. Several positions at which
  RATs decrease dramatically correspond to the peculiarities of
dielectric susceptibility. If those peculiarities are
ignored, RATs are self-similar. {\it Lower panel}: 
The same as the upper panel but  for
  shape 2. RATs can be fitted by analytical functions with $\eta =3 \mbox{ or }
  4$ (see equation \ref{eq86}).}
 \label{f26}
\end{figure}
\begin{figure}
\includegraphics[width=0.49\textwidth]{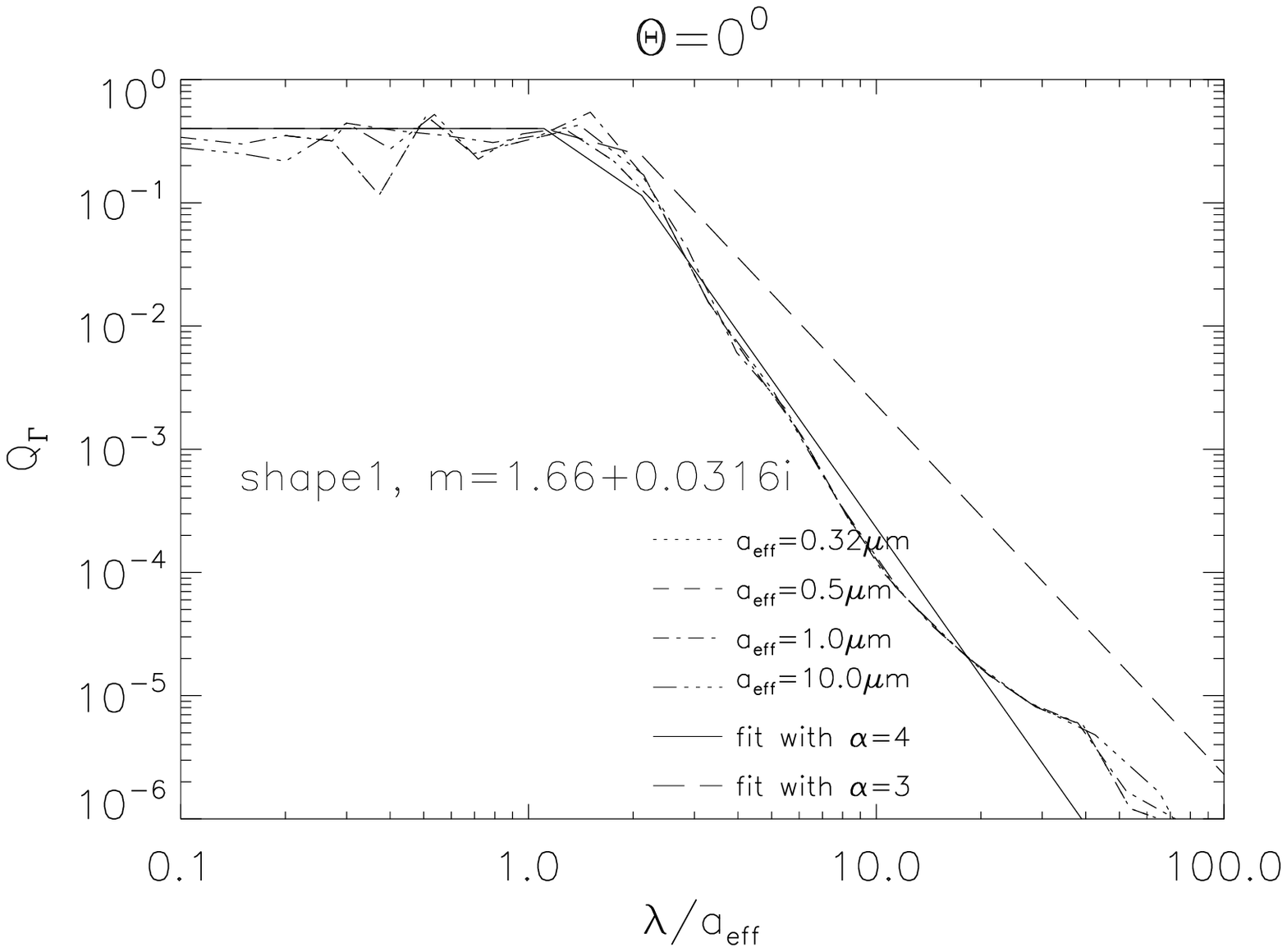}
\includegraphics[width=0.49\textwidth]{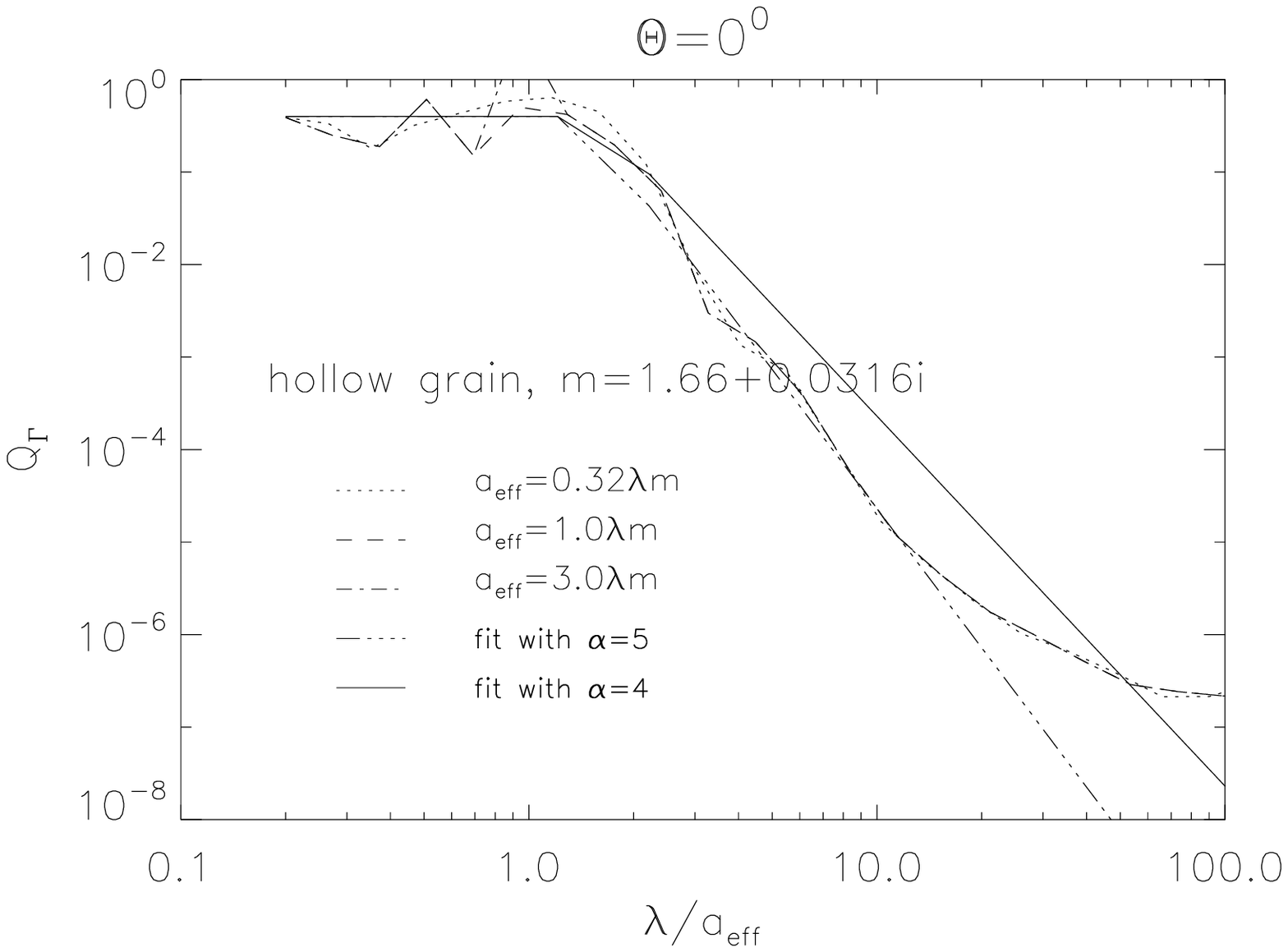}
\caption{RATs for shape 1 and hollow grain with constant refractive index:
   self-similarity exhibits extremely well. They can be fitted by a step function
  and a power function with $\eta=3, 4$ ({\it Upper Panel}) and $\eta=4, 5$
  ({\it Lower Panel}) where $\eta$ is defined in equation (\ref{eq86}).}
 \label{f27}
\end{figure}

 We see that the approximate
 self-similarity (i.e. the dependence on $\lambda/a_{eff}$)
 is an intrinsic property of radiative torques. When optical constant
 changes as a function of wavelength, RAT efficiencies for different grain
sizes mostly
 differ at wavelengths corresponding to resonance absorption features. 

 In addition, RATs have nearly constant magnitude as $\lambda \sim a_{eff}$, and decrease steeply with the ratio of wavelength to grain
size. This is because the scattering of photon by irregular grains are strongest
 as $\lambda \sim a_{eff}$. We can fit our calculations for RATs (see Figs
 \ref{f26}, \ref{f27}) by a simple function
given by
 \begin{align}
 Q_{\Gamma}&=0.4 \mbox{ for } \frac{\lambda}{a_{eff}}<1.8,\nonumber \\
&=0.4(\frac{\lambda}{a_{eff}})^{-\eta} \mbox{ for } \frac{\lambda}{a}>1.8,\label{eq86}
 \end{align}
where $\alpha$ is the spectral index that according to 
 Figs \ref{f26} and \ref{f27} is between 3 and 4. The first case provides
a good fit in the whole range of $\lambda/a_{eff}$. In contrast, the latter case
 gives better fit for the range of $\lambda/a_{eff} <20$. Cho \& Lazarian (2006)
use the former 
 fit formulae to calculate polarization degree for accretion disks because
grains there are widely believed to be very large.

 For the case when ${\bf a}_{1}$ makes an angle $\Theta=45^{0}$ with respect to ${\bf
 k}$, we found that the self similarity is also valid. However, the curve of RATs is
  shallower, and can be fitted by a power index $\eta=-2$.

 To study the efficiency of the self similarity, we calculate rotation velocity of
grains induced by RATs in which RATs are directly computed from DDSCAT and derived
 from the self similarity assumption. We use radiation intensity of a molecular cloud (see Mathis
1983) to calculate RATs for different optical depths $A_v$. Resulting rotation speed obtained with these two methods (see
 Fig. \ref{f28}) shows clearly that the self similarity provides
a fair agreement between exact calculations and those based on
the self-similarity arguments. This
 allows to reduce the DDSCAT computational efforts substantially.

 \begin{figure}
\includegraphics[width=0.49\textwidth]{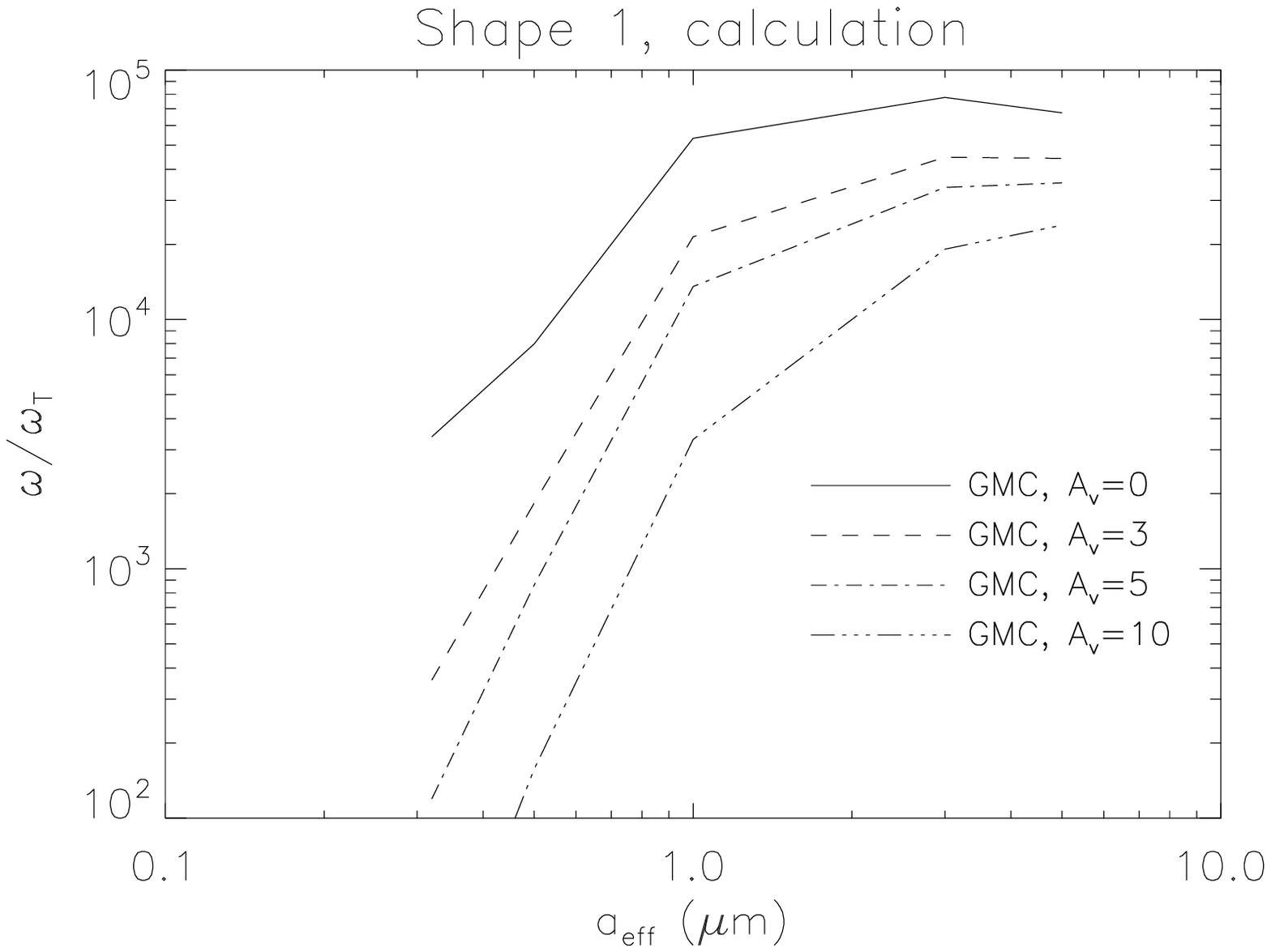}
\includegraphics[width=0.49\textwidth]{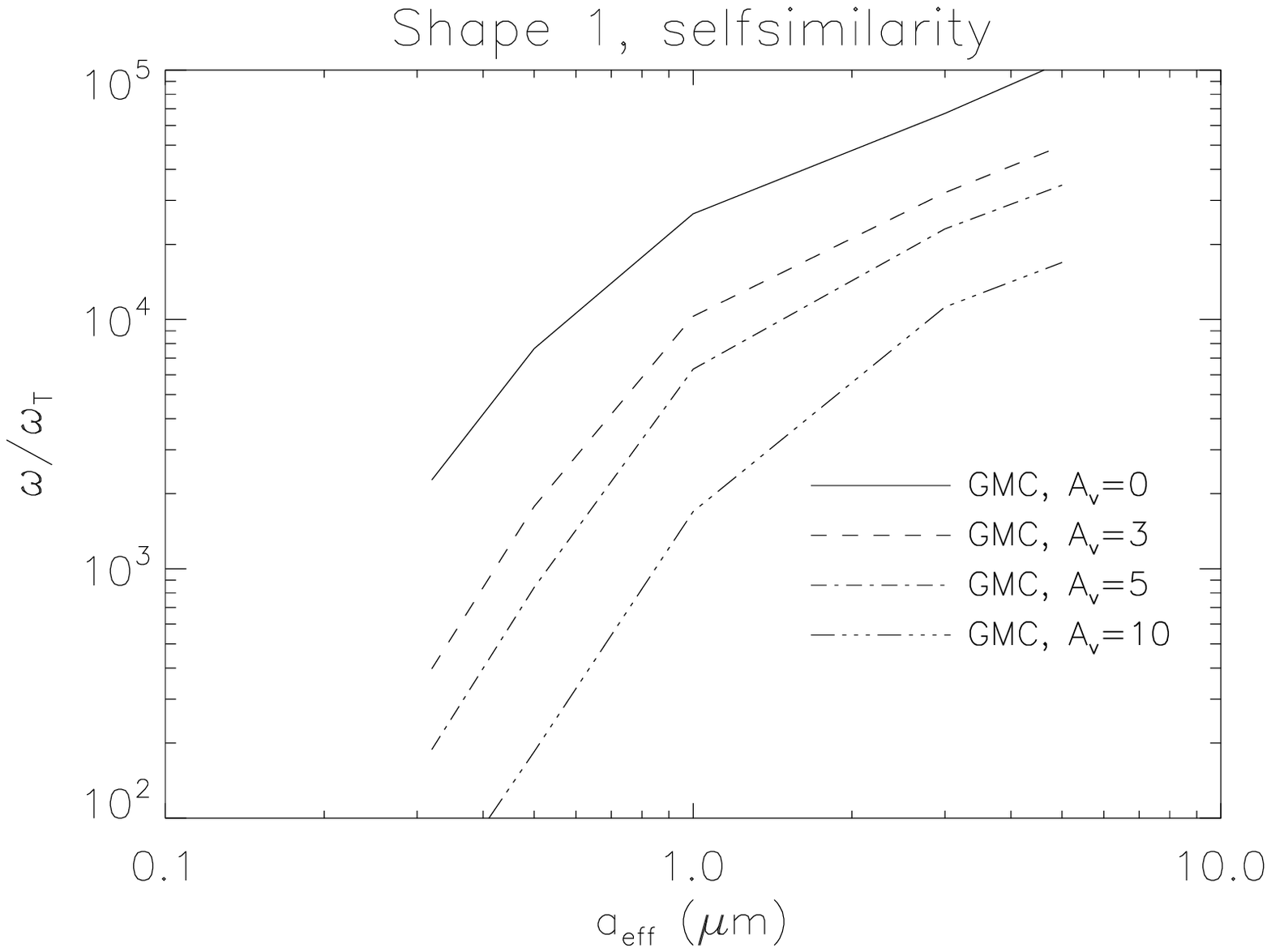}
 \caption{Ratio of rotation speed to thermal rotation speed with respect to
  grain size obtained with exact RATs from DDSCAT ({\it Upper panel}) and using
   the self-similarity ({\it Lower panel}). Here, radiation direction is assumed to be  parallel to the maximal inertia axis
   ${\bf a}_{1}$, that is constrained to be
   frozen to angular momentum ${\bf J}$. The figures show the rapid increase of the rotational
  speed with grain size, and even in the cloud core with high visual
   extinction $A_{V}=10$, large grain still have suprathermal rotation rate,
  ensuring that grain alignment is not susceptible to random collision by atomic gas.}
 \label{f28}
\end{figure}

\section{Discussion}\label{discuss}

Our study above have approached an important problem of the RAT alignment mechanism by studying the fundamental properties of RATs. The goal of such studies
 was to change
the status of the RAT alignment from an empirical fact to a theoretically
understood process. Our work indicates that grain helicity is an essential
property of realistic grains, which suggests that it should be accounted not only for RATs, but
for mechanical alignment as well.

\subsection{Evolution of ideas on RATs}

For the first time grain spin-up due to differential scattering
of left and right handed photons was considered in Dolginov (1972).
The suggestion was limited to quartz grains, however. In Dolginov \&
Mytrophanov (1976) it was realized that grains of made of more accepted 
astrophysical materials can be spun up, provided that they are somewhat
twisted. This ground-breaking
study qualitatively considered both the possibility
of the RAT alignment with respect to magnetic field and the radiation beam. 
Regretfully, it was not appreciated in its time.

A reason of the lukewarm response of the community to this work was
probably related to both the absence of reliable estimates of RATs 
efficiencies and to the fact that some features of grain dynamics were
unknown at that time. For instance, Dolginov \& Mytrophanov (1976) claimed
that prolate grains should align along magnetic field, while oblate grains
should align perpendicular to magnetic field. The former conclusion was based
on the assumption that grain rotation along the axis of minimal inertia is
stable. It is only later that Purcell (1979) 
discovered so-called Barnett relaxation and concluded that only rotation 
about the axis of maximal inertia is stable\footnote{The Barnett magnetization
that induced  Ed Purcell to think about the relaxation
was described in the same Dolginov \& Mytrophanov (1976) paper.}. On the basis of this Lazarian 
(1995) claimed that the alignment of both prolate and oblate helical 
grains should happen with long axes perpendicular to magnetic field. However,
it was not quite clear at that moment what makes the grain helical. In addition, Lazarian (1995) underestimated the importance of the  RAT alignment.

The quantitative stage of RAT studies was initiated by Draine \& Weingartner (DW96 and DW97). 
There RATs were shown to be generic for irregular 
(and therefore 
realistic) grains and the magnitude of RATs was reliably quantified. 
The alignment with respect to magnetic field was
demonstrated in numerical simulations. On the basis of
these simulations DW97 claimed that the RAT alignment
was the dominant process responsible for polarization arising
from dust in the diffuse ISM. However, the RAT alignment was an empirical
fact in DW97 that was based on a limited sample. In this situation
additional important
 complexities arising from including thermal wobbling of grains
(WD03), gaseous bombardment etc. could not help to improving our
physical understanding of the RAT alignment.

What we have done in the present paper is we attempted to clarify the
basics of the RAT alignment by seeking the generic properties of RATs that
induce the alignment. We used analytical modeling which was tested with
numerical DDSCAT simulations. As we discuss further, we hope that this work
contributes to both more intuitive understanding of the alignment and to
further elaborating of the mechanism in order to get precise
 predictions of the alignment degree in different astrophysical situations.
  
It is worth mentioning that even we when know the analytical form of the torques,
the dynamics of the system does not get completely trivial. It definitely
exhibits interesting properties.

\subsection{Our approach}

While RATs were initially treated as a quantum effect arising
from the difference of scattering of left and right handed photons (see
Dolginov \& Mytrophanov 1976), above we presented an entirely
classical model of a grain (see Fig.~\ref{f2}) that reproduces well their
 properties (see Fig~\ref{f61}).
 
The gist of our approach above is to consider the basic generic
properties of RATs and to relate of these properties with
the RAT alignment. AMO plays a central role in our considerations.
Our toy model of a helical grain allowed an analytical description, which
enabled us to treat RATs analytically. 
In our study we concentrated on the properties of RAT components
in the scattering system (see Fig.~\ref{f1}),
 i.e. on the properties of $Q_{e1}$, $Q_{e2}$ and
$Q_{e3}$. These components show a remarkable similarity for grains of very
different shapes (see Fig.~\ref{f23a}) and AMO (see Fig.~\ref{f61}). The chi
squared test we present in \S \ref{sec94} returns the mean value of $\chi^{2}$ for
both $ Q_{e1}$ and $Q_{e2}$ about $0.2$ (see Fig. \ref{chi_test0}). This provided
us with the empirical justification of using AMO for studies of RAT alignment.
AMO provides us with both useful intuitive model
 to think about the alignment and
analytical formulae that allow straightforward quantitative calculations.

We find that the basic properties of RATs obtained with AMO are very similar
to the basic properties of RATs for arbitrary shaped grains. This, for
instance, allows
us to talk about helicity of grains being the most important attribute for the
RAT alignment.

 Our major
goal above was to get a better understanding of the physics of the
RAT alignment. To do this we adopt a model, similarly to one in DW97, that disregards the wobbling
of the grain axes with respect to angular momentum direction
(cf. Lazarian 1994;
Lazarian \& Roberge 1997; LD99ab), but treats 
crossovers differently, i.e. in the spirit of the Spitzer \& McGlynn (1979)
model. This provides a substantial change in the dynamics of grains. For instance,
we do not observe cyclic maps reported in DW97.

While studying the properties of RATs
 we addressed the question of the necessary conditions for
RAT alignment to happen and to fail. 
 We also study the RAT alignment that takes place in the radiative-dominated
environments, where the direction of radiation
defines the axis of alignment (see Dolginov \& Mytrophanov 1979;
Lazarian 2003). Such an alignment is both astrophysically important 
and provides a good insight into the physics of the RAT alignment.

\subsection{Accomplishments and Limitations of the present study}

We feel that our major accomplishment in the paper above was establishment of
the analytical form of RATs and clarification of the role of different RAT
components. We hope that
AMO unveils the mystery that have surrounded the RAT alignment from the time
of the mechanism introduction.

 Another important conclusion that follows from AMO is
that the RAT alignment is not limited to
grains with $\lambda \sim a_{eff}$, as it was believed
before. As the similarities
between the torques that we obtained for AMO in the $\lambda \ll a_{eff}$ limit
and for irregular grains in the $\lambda \ge a_{eff}$ limit are striking,
our work shows that the RAT alignment should take place also for large
grains, that are present in accretion disks and dark cloud cores (see
Cho \& Lazarian 2005, 2007).
                     
Obtaining generic properties of RATs makes the RAT alignment more a
predictable theory and opens avenues for further theoretical advances,
e.g. including thermal fluctuations, random bombardment, H$_2$ torques
etc. Our
establishing of a subdominant nature of one of the RAT components, namely
$Q_{e3}$, simplifies the theoretical treatment of RATs.
Insights into the generic properties of two other components allow
to reduce the amount of numerical computations necessary to determine
the degree of achievable alignment. For instance, in the
current paper, we found that the existence of high-$J$ attractor points
depends on the $Q_{e1}^{max}/Q_{e2}^{max}$ ratio. For practical applications
it is important, that the criterion for this established with AMO works well with irregular grains.
We note parathentically, that while $Q_{e1}$ and $Q_{e2}$ demonstrate universal
behavior, the aligning and spinning torques (see equations (\ref{eq61}) and (\ref{eq62}))
that present their combinations do not demonstrate this.

The present study identifies a parameter space for which the
RAT alignment may be suspected to be ``wrong'', i.e. to happen with long
axes parallel to magnetic field. In addition, it provides simple scalings for
RAT dependences on the ratio of the radiation wavelength to the grain
size. 

Our study reveals new properties of the RAT alignment. First of all, the
alignment may be fast, i.e. happen in a small fraction of gas damping time (see
\S \ref{fast}). This has important consequences for the environments with fast
changing radiation, i.e. circumstellar regions, interstellar medium in the
vicinity of supernovae flashes etc. Moreover, we could see, that the alignment
is different depending whether the initial angular momentum is small or large
(see Figs \ref{f12} and \ref{f12*}).

The approximate self-similarity of RATs (see \S 10) is another
practically useful property of RATs. Combined with the established universality
of the functional form of the components
$Q_{e1}$ and $Q_{e2}$ and the established dependences of
the properties of trajectory maps on the ratio of these components, this allows
radically reduce DDSCAT calculations that may be necessary to find the
expected
degree of alignment for an ensemble of realistic grains subjected to a
realistic
radiation field. In fact, we find that RATs change the ratio of
$Q_{e1}^{max}/Q_{e2}^{max}$ with $\lambda$ and this is the most important
difference that the variation of the radiation wavelength entails. The
functional forms of the torques do not change much and can be well approximated
with those of AMO.

One limitation of AMO is the upper limit of the ratio $Q_{e1}^{max}/Q_{e2}^{max}$, that
makes AMO more appropriate to irregular grains with $\lambda >3 a_{eff}$ and $\lambda<a_{eff}$. It
indicates that, though AMO is established based on the geometric optics, i.e.,
$\lambda \ll a_{eff}$, it is also applicable for the opposite limit.

In more general terms, our study proved that irregular
grains can be characterized by helicity. Grain rotation
provides the averaging that defines the helicity axis, while the
irregularities define whether the helicity is left or right. The phase
trajectories of grains that are the mirror-symmetric images of
each other are mirror-symmetric (see
Fig.~\ref{f8}). As expected, the torque component
$Q_{e3}$, unlike the other two components, coincide
for an irregular grain and its mirror-symmetric image (see the lower panel in Fig.~{\ref{f23a}). Indeed, this component is subdominant for most of the alignment
processes and not related to grain helicity.

In our study we do not directly address the grain alignment efficiency.
Some statements can be made, however. For instance,
when grains are aligned rotating suprathermally the direction of ${\bf J}$
is immune to the randomization arising from the gaseous bombardment. In
addition in this case,
${\bf J}\|{a}_1$ provides a good approximation. We find, however,
that an appreciable subset of grains rotates with thermal velocities. 
For those the internal randomization of grain axes in respect to ${\bf J}$
 may be important. Does this signify a new crisis
of the grain alignment theory? We do not believe so. Even in the absence
of high-$J$ attractor points RATs will drive
${\bf J}$ back to low attractor points, which in most cases, as we discussed
in the paper, correspond to the preferential alignment of grains with long
axes perpendicular to magnetic field. As for the internal alignment,
according to Roberge \& Lazarian (1997) for typical interstellar
grains, this alignment is tangible even for $J\sim J_{th}$.
 A detailed study of the attainable degrees of alignment
is provided elsewhere.

We have not discussed RATs of the strongly absorbing materials, e.g. graphite. We expect the torque components to show more irregularity for such
grains. As the grain alignment theory matures and extends to the
environments different from molecular clouds and diffuse interstellar gas (see
Lazarian 2007),
the importance of the studies of  wider range of materials will get
 more pressing.

\subsection{Rates of Alignment and Rotation}

As we mentioned above the RAT alignment can happen on time scales
much shorter than the gaseous damping time. This finding corresponds to the
notion in Dolginov \& Mytrophanov (1976), that the alignment happens on the
time scale that is required for the radiative torques to deposit a grain
with the angular momentum of the order of its initial angular momentum. Such a
fast alignment makes grains good tracers of magnetic field when radiation
direction changes quickly.

The fast alignment takes place for low-$J$ attractor points of the grain 
phase trajectory map. These are the most probable attractor for the grains
to end up with. Thus, most grains do not rotate suprathermally when subject
to RATs. In this sense the RAT alignment tends to minimize grain angular
momentum.

While we expect that in the presence of thermal wobbling and
gaseous bombardment most grains will rotate thermally,
there is a radical difference between this  effect and the effect
of thermal trapping discussed in LD99ab. The effect
of thermal trapping there is based on the compensation of the Purcell rocket
torques, e.g. those related to H$_2$ formation, by thermal flipping of grains.
The more efficient thermal flipping, the more efficient is the trapping and
the less chance of a grain to get high angular momentum. On
the contrary, we have seen in \S \ref{kalign} and \ref{balign} that without thermal wobbling, the significant fraction of grains
ends up in the state of $J=0$. In other words, thermal fluctuations increase
the value of $J$ to a {\it higher}, i.e. thermal value.

In spite of the fact, that most of the grains tend to rotate with velocities
much less than the maximal velocities, $\omega_{max}$,
 that RATs can spin the grain up, we
believe that the parametrization of the alignment in terms of $\omega_{max}/\omega_T$, where $\omega_T$ is the thermal rotational velocity,
may be a rough practical way of describing alignment. Indeed, the
above ratio
reflects the relative importance of RATs compared with those
related to gas. When RATs force the grain into a low-$J$ attractor
point, their ability to do this would also depend on this ratio. A further
research should reveal more sophisticated and precise parametrization of
the RAT alignment, however. This parametrization is necessary, for instance,
to predict the expected alignment from the numerical simulations of magnetized
molecular clouds (see Cho \& Lazarian (2005); Pelkonen et al. (2007); Bethell et al. (2007)).

We have discussed in \S \ref{sec5} that
for some phase trajectories high attractor points are available.
The suprathermally rotating grains correspond to high-$J$ attractor
points. It takes them about 3 damping times to reach such points (see
also DW97). However, our analysis shows that grains get aligned even before
they reach high-$J$ attractor points. Therefore the RAT alignment can happen over
shorter time scales for all grains provided that the radiation is intensive enough.

The predominance of low-$J$ attractor points has consequences that go beyond the
problems of grain alignment. If grains rotate slowly, then loosely connected
conglomerates constituting fractal grains can exist. Ever since the classical
work by Purcell (1979), the suprathermal rotation had been thought to destroy such
grains. When LD99ab showed that Purcell's torques may not be capable to spin-up
grains less than $10^{-4}$~cm, it was still thought that radiative torques can do
the job. Our work questions this (see also WD03).

\subsection{Direction of Alignment} 

The alignment may happen with respect to radiation
rather than to magnetic field if the precession induced by radiative torques
is faster than the Larmor one (see \S \ref{sec9}). 
We found, that in
 the presence of magnetic field the alignment can still happen in
respect to the direction of the beam or, equivalently, the direction
of the anisotropy of radiation, provided that the rate of
precession arising
from the radiative torques is faster than rate of the Larmor precession.
This is the case of comets sufficiently close to the Sun, ISM in the vicinity of
supernovae and some circumstellar regions. Over vast expanses
of diffuse interstellar medium and molecular clouds, however, the generic RAT
alignment is with respect to magnetic field, thus enabling easy tracing of
magnetic fields via polarimetry.
 Note, that our study shows
that many features characteristic
of the alignment in the absence of magnetic field carry over
to the case when magnetic field is present.

Our important finding is that while the generic alignment is
``right'', i.e. with the long grain axes perpendicular to magnetic
field, for a range of angles $\psi$ between
the magnetic field and the direction of the beam around $\psi=\pi/2$,
the alignment may be ``wrong'', i.e. it happens with long grain axes parallel to magnetic field. However, the
range of the angles is rather narrow.
As a result, we do not expect the effect of ``wrong alignment'' to persist 
when grains undergo thermal wobbling. This wobbling is likely to
 vary the direction of the grain axes with respect to the light direction
beyond the range angles in which the alignment is ``wrong''.

\subsection{Magnetic Field and Gas Streaming}

Unlike DW97, in the paper above we disregarded the effects of paramagnetic
alignment altogether. When we consider dynamically important field, its
only effect is to induce averaging due to Larmor precession.  We believe
that our approach is correct, as
for paramagnetic grains the effects
of paramagnetic relaxation are marginal on times over which the
RAT alignment takes place.

Gas streaming can induce its own alignment direction. 
Dolginov \& Mytrophanov (1976) assumed that
 magnetic field or a gaseous flow defines the axis of
alignment depending on the ratio of Larmor precession time to
that of mechanical 
alignment. On the basis of our study of $Q_{e3}$ with AMO, 
we believe that a more physically 
motivated distinction is related to the rate of Larmor precession
versus the precession arising from the mechanical
analogy of the $Q_{e3}$ torque. Such torque for an spheroidal grain 
can be obtained from formulae in Appendix~B by substituting the value of the
gas atom momentum $mv$ instead of the photon momentum.
The corresponding precession timescales ratio is (see equation \ref{eq79})
\begin{align}
\frac{t_{flow}}{t_{B}}= 3\times 10^{5}\frac{\hat{\rho}\hat{\chi}\hat{T}^{-0.5}\hat{a}_{-5}^{-0.5}}{\hat{\alpha}\hat{v}^{2}_{flow}\hat{n}_{H}}\frac{\hat{B}}{\hat{Q}_{e3, gas}},
\label{t_fb}
\end{align}
where $\hat{Q}_{e3, gas}$ is the third component of torques induced by the
gaseous flow, which is the analog of $Q_{e3}$ for RATs.
In equation (\ref{t_fb}), $\alpha=\hat{\alpha}\times 0.1$ is the probability of elastic collision, $v_{flow}=\hat{v}_{flow}\times v_{thermal}$ is gas flow velocity,
$n_{H}=\hat{n}_{H}\times 30$, $T=\hat{T}\times 100$ is gas density and temperature, respectively.
equation (\ref{t_fb}) indicates that for sufficiently intensive gaseous flows, for
instance, $\hat{v}_{flow}>10^{2} $, the alignment
will indeed happen with respect to the flow direction. Note, that
we predict that gaseous flows may define the direction of alignment for a
wider parameter range compared to that in 
Dolginov \& Mytrophanov (1976). Moreover,
we claim that mechanical flows can define the axis of alignment even for
subsonic flow velocities, i.e. at those velocities for which the Gold alignment
and its modifications (cf. below, however) are marginal.

A conceivable situation is that the gaseous bombardment arising from
grain streaming defines the axis of alignment, while RATs do the alignment
job. This situation takes place when $Q_{e3, RATs}$ is less than $Q_{e3, gas}$,
which defines
\begin{equation}
\frac{t_{flow}}{t_{k}}\sim \frac{u_{rad}}{\alpha m_{H}v_{flow}^{2}n_{H}}\sim 10^{1}\frac{\hat{u}_{rad}}{\hat{\alpha}\hat{v}_{flow}^{2}\hat{T}\hat{n}_{H}},\label{flrad}
\end{equation}  
where $u_{rad}=\hat{u}_{rad}\times u_{ISRF}$.

This can be the case of alignment in a part of comet 
atmosphere\footnote{Another case also relevant to the comet atmosphere 
is  that the grains have electric
dipole moments, while comet atmosphere has electric field. Then the comet
electric field defines the axis of alignment.}. Naturally, combining
equations~(\ref{t_fb}) and (\ref{flrad}) it is possible to establish when streaming
defines the alignment axis in spite of the presence of magnetic field.

\subsection{AMO and Mechanical Alignment of Helical Grains}

Our present paper is devoted to RATs and the alignment that they entail.
However, our consideration of a helical grain is quite general. In fact, the
functional dependence of the torques that we obtain for our model grain 
is valid when atoms rather than photons are reflected from the mirror. Therefore
we may predict that for elastic gas-grain collisions the helical 
grains\footnote{The mechanical alignment of helical grains was briefly
discussed in Lazarian (1995) and
 Lazarian, Goodman \& Myers (1997), but was not elaborated there.} will
align with long grain axes perpendicular to the flow
in the absence of magnetic field and with long axes perpendicular to ${\bf B}$,
when dynamically important magnetic field is present. If atoms attach to the
grain surface and then are thermally ejected from it, this changes the values
of torques by a factor of order unity.  The only way that the uncompensated
torques can vanish for a helical grain is if the correlation is lost between
the place at which the atom strikes the grain surface and leaves it\footnote{Even
in this case the local anisotropies of the surface at the place of atom
impact can result in effective helicity similar to the case of damped 
oscillator in Fig. \ref{f10} (lower panel)}. The latter is rather improbable for sufficiently 
large grains. As the result, one has to conclude that helicity is a generic
property of the interaction of irregular grains and atomic flows.

This conclusion alters substantially the paradigm of grain motions in
diffuse gas. Since the time when Gold (1951) proposed his first simple
model of mechanical alignment, it was considered essential to have
supersonic grain-gas velocities to achieve any tangible alignment.
Indeed, both the Gold original process and the ones proposed for suprathermally
rotating grains, namely cross-sectional and crossover alignments (Lazarian
1995) require the supersonic drift to ensure that the momentum deposited
in a regular way exceeds one deposited due to thermal atomic motions.
This is not a requirement for the alignment of helical grains! For those
the regular momentum is deposited in proportion to the number of collisions,
while the randomization adds up only as a random walk. In fact, the difference
between the mechanical alignment of spheroidal and helical grains is similar
to the difference between the stochastic
Harwit (1970) alignment by stochastic absorption
of photons and the RAT alignment. While the 
Harwit alignment requires very special
conditions dominate, the regular RATs easily beat randomization.

Similarly as in the case of RATs, where it is frequently
possible to disregard the Harwit process, it should be possible to disregard
the Gold-type processes\footnote{
This alignment tends to minimize grain cross section, 
which
means, for instance, that for grains streaming along magnetic fields the non-helical stochastic torques will tend to align grains with longer axes parallel to
magnetic field, while helical torques will tend to align in the perpendicular
direction.}
 for the alignment of helical grains. Note, that because of the property of
helicity not to change sign during grain flipping, we do not expect to
observe the thermal trapping effects described in Lazarian \& Draine (1999a)
to be present for the mechanical spin-up of helical grains. The effects
that decrease the efficiency of the mechanical alignment of helical grains
are discussed elsewhere.  
  
In typical conditions of diffuse interstellar medium,
the mechanical alignment of helical grains tends to act to
align grains in the same direction as the RATs, i.e. with longer axes
perpendicular to magnetic field. The relative role of the two mechanisms
should be revealed by further research. The currently available data
(see Lazarian 2007) agrees with the RAT mechanism being the primary source
of alignment. However, the situations are possible, when mechanical
alignment reveals magnetic field, when RATs fail to do so.

\section{Summary} 
In this paper, we studied the properties of RATs, and how different RAT
components affect the grain alignment. Briefly, our results are as follows:

1. We found that a simple model of a helical grain
which consists of a reflecting spheroidal
grain with an attached 
mirror reproduces well the functional
dependences of RATs obtained for irregular
grains using DDSCAT.

2. From the generic properties of RATs  we predicted
the preferential alignment of grains with long axes 
perpendicular to the direction towards the source of light, provided
that magnetic field effect is subdominant.

3. The magnetic field is important and defines the
axis of alignment when it induces the Larmor precession
that is faster than the precession arising from the $Q_{e3}$-component of RATs,
i.e. the component not related to grain helicity. This component is present for
spheroidal grain, for instance.

4. When magnetic field is important, RATs tend to provide both the 
alignment of long grain axes perpendicular to magnetic field. With or
without magnetic field, most of grains are driven to the low angular momentum
attractor points, as RATs align grains. Grains can be driven to
the low-$J$ attractor points on the time scales much less than the 
gaseous damping time. The very 
existence of the high angular
momentum attractor points, and therefore grains rotating much faster than the
thermal velocity, is not a default and
 depends on the ratio $Q_{e1}$ and $Q_{e2}$ components.

5. RATs can also induce ``wrong alignment'', i.e., the alignment
 with long grain 
axes parallel to magnetic field.
The range of angles 
 for ``wrong 
alignment'' is narrowly centered around the $\pi/2$ angle between 
the direction of light and magnetic field. This
range is expected to be 
diminished when thermal fluctuations are accounted for. Thus the
RAT alignment is capable to account for most of the observed polarization.

6. RATs exhibit approximate
self-similarity that allows one to be expressed them as a function of
the ratio of the wavelength to grain size. The dynamics of the grains 
can be reproduced relatively accurately when the
self-similarity is used. 

7. The RAT alignment is a particular case of the alignment of the
helical grains. Therefore our results can be generalized to describe
the mechanical
alignment of irregular grains. Such an alignment is efficient for both
supersonic and subsonic gaseous flows. The
mechanical alignment may happen either with respect to the magnetic field or
the direction of the flow depending on the rates of precession that are
induce by the flow and magnetic field, respectively.

\section*{Acknowledgments}
We acknowledge the support by the NSF Center for Magnetic Self-Organization in Laboratory and Astrophysical
Plasmas. AL acknowledge a partial support by the NSF grant AST 0507164.
We thank the anonymous referee for his/her comments which improved the paper,
especially for comments on the symmetries of the RAT components.

\appendix

 \section{Reflecting Oblate Spheroid}

The toy model of a grain (see Fig. \ref{f2}) allows us to derive analytical formulae for RATs.
Let us consider RATs for an oblate reflecting spheroid characterized by semi-axes $a, b$
where $s =a/b<1$ (see Fig. \ref{ap1}). ${\bf a}_{1},
{\bf a}_{2}, {\bf a}_{3}$ are principal axes of the spheroid with moments of
inertia $I_{1} > I_{2}=I_{3}$, respectively. Assuming that a photon beam of
wavelength $\lambda$ is shined along
${\bf k} \| \hat{e}_{1}$, the photon reflections happen on the spheroidal surface at the location
determined by a normal unit vector $\mn$ and a radius $\mr$. For the sake of
simplicity, a perfect reflection is always assumed in this study.
 \begin{figure}
\includegraphics[width=0.49\textwidth]{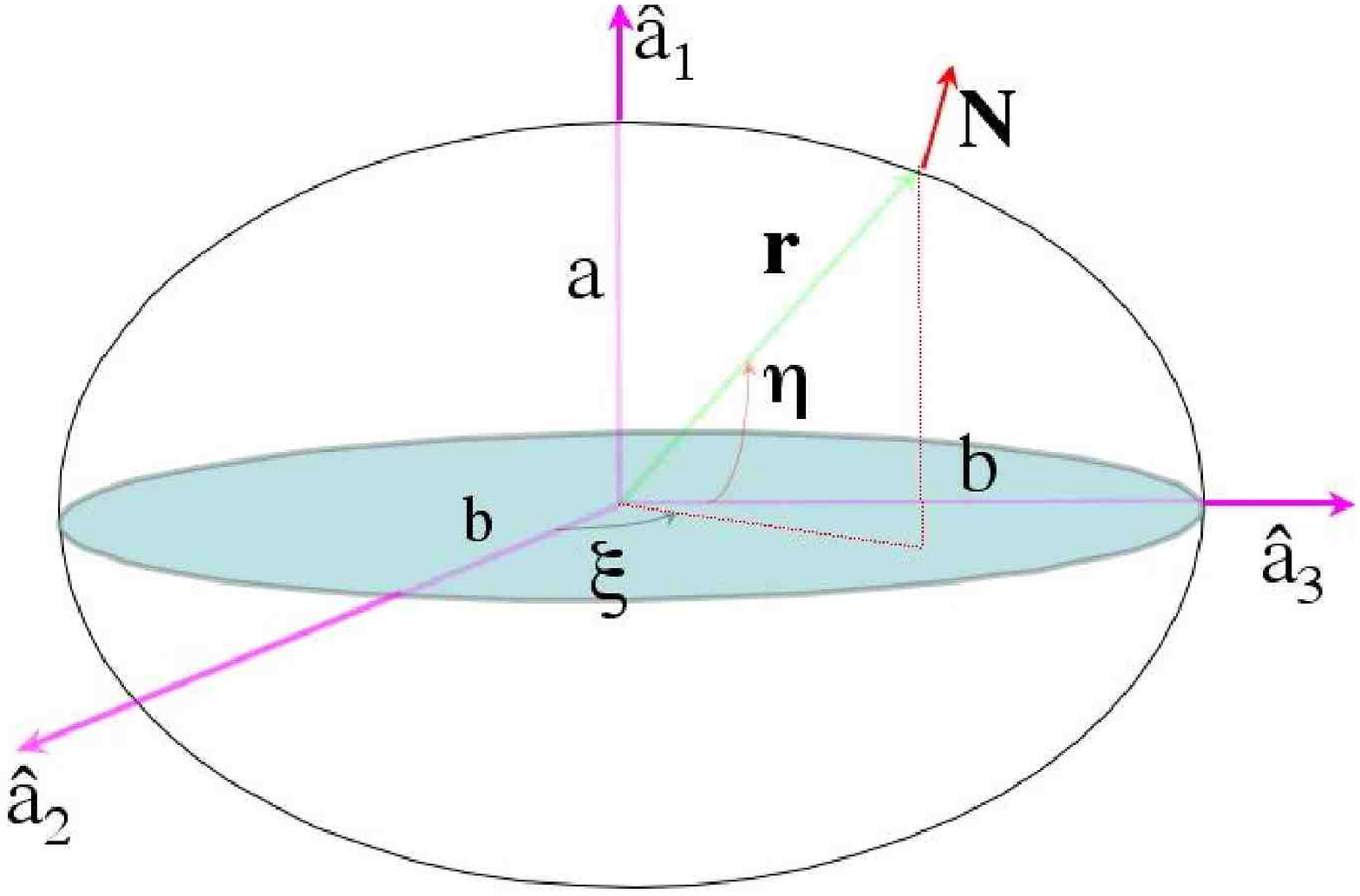}
\includegraphics[width=0.49\textwidth]{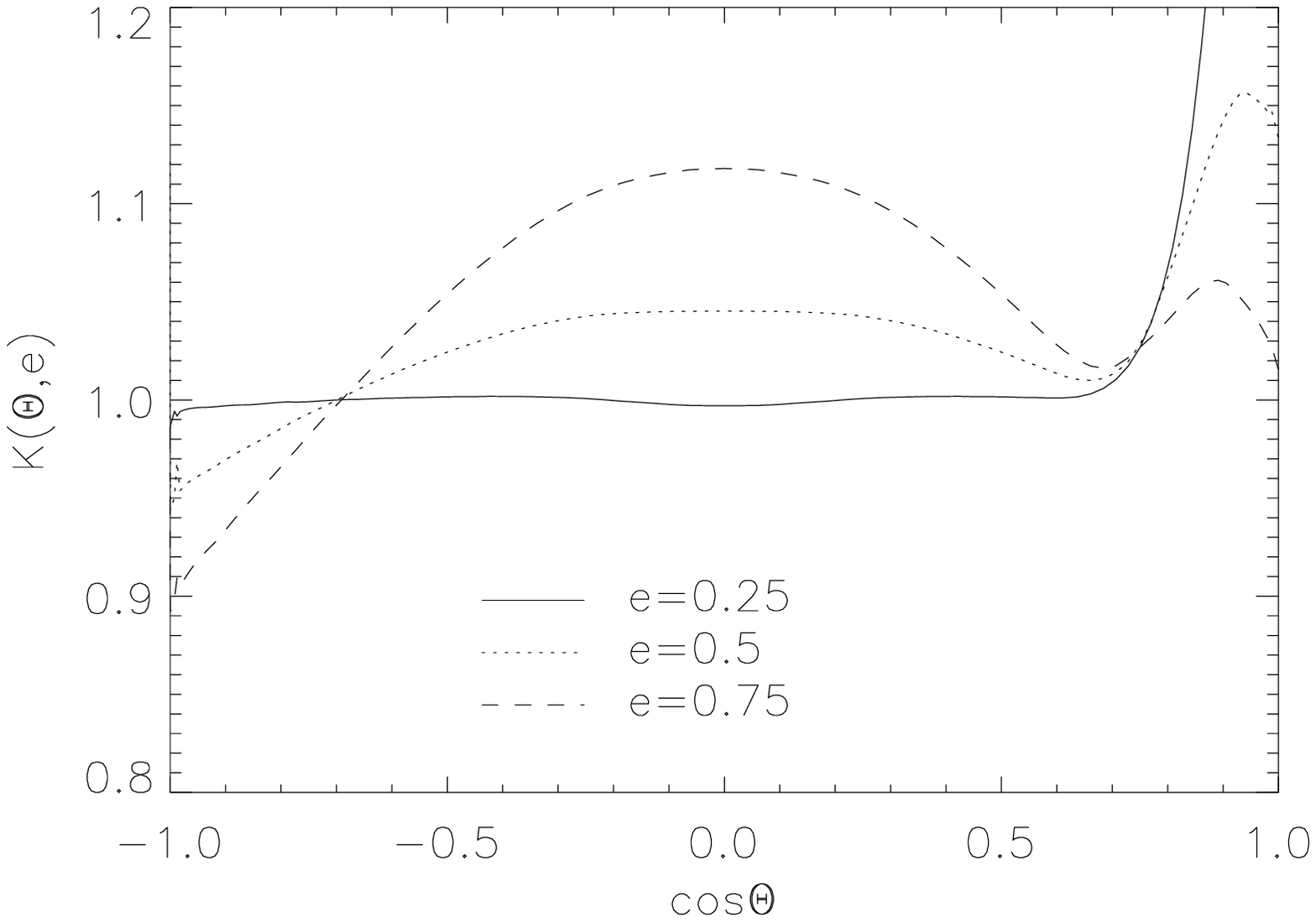}
\caption{{\it Upper panel} represents the coordinates and vectors for the
  spheroid. {\it Lower panel} shows  the fitting function $K(\Theta,e)$ for $Q_{e3}$, depending on the eccentricity of the spheroid $e$ and its angle with light beam $\Theta$.}
\label{ap1}
\end{figure}
The location of impact on the grain surface is specified by the radius
${\mr}$ and the normal vector ${\mn}$, which are, respectively, given by
\begin{align}
{\mr}&=a\ms\eta\ma_{1}+b\mc\eta\mc\xi\ma_{2}+b\mc\eta\ms\xi\ma_{3},\label{a1}\\
{\mn}&=a_{1}\ms\eta\ma_{1}+b_{1} \mc\eta\mc\xi\ma_{2}+b_{1} \mc\eta\ms\xi\ma_{3},\label{a2}
\end{align}
  where $a_{1}=[\ms^{2} \eta +(1-e^{2}\mc^{2}\eta]^{-1/2}, b_{1}=a_{1}(1-e^{2})^{1/2}$,
   and $e$ is the eccentricity of the spheroid, $\xi=[0,\pi], \eta=[-\pi/2,
   \pi/2]$ (see Fig. \ref{ap1}; see also Roberge et al. 1993).

Due to the symmetry around $\ma_{1}$, we only need to find RATs for a single
rotation angle, e.g. for $\beta=0$. Therefore, when the grain axis ${\bf
  a}_{1}$ makes an angle $\Theta$ with
respect to the photon beam, we have
\begin{align}
\ma_{1}&=\mc\Theta\me_{1}+\ms\Theta\me_{2},\label{a3}\\
\ma_{2}&=-\ms\Theta\me_{1}+\mc\Theta\me_{2},\label{a4} \\
\ma_{3}&=\me_{3}.\label{a5}
\end{align}
Substituting equations (\ref{a3})-(\ref{a5}) into (\ref{a1}) and (\ref{a2}) we
obtain
\begin{align}
{\mr}&=(a\mc\Theta\ms\eta-b\ms\Theta\mc\eta\mc\xi)\me_{1}\nonumber\\
&+(a\ms\Theta\ms\eta+b\mc\Theta\mc\eta\mc\xi)\me_{2}\nonumber\\
&+b\mc\eta\ms\xi\me_{3},\label{a6}\\
{\mn}&=(a_{1}\mc\Theta\ms\eta-b_{1}\ms\Theta\mc\eta\mc\xi)\me_{1}\nonumber\\
&+ (a_{1}\ms\Theta\ms\eta+b_{1}\mc\Theta\mc\eta\mc\xi)\me_{2}\nonumber\\
&+b_{1} \mc\eta\ms\xi\me_{3}.\label{a7}
\end{align}   
Hence, RAT produced by the photon beam is defined by
\bea
d{\bf \Gamma}_{rad}=\gamma \mr \times \Delta {\bf P}={-2p} \gamma F dA ({\bf k}.\mn)[\mr
\times \mn], \label{a8}
\ena
where $\gamma$ is the anisotropy degree of radiation field, $p_{ph}$ is the
momentum of each photon, $F$ is the flux of the incident light beam, and $dA$ is  an area
element on the grain surface given by
\begin{align}
dA=eb^{2} f(\eta)\mc\eta d\eta d\xi,\label{a9}
\end{align}
where $f(\eta)=\sqrt{\frac{1-e^{2}}{e^{2}}+\mss\eta}$.

Therefore
 \begin{align}
d{\bf \Gamma}_{rad}&=-2p_{ph} \gamma F (\me_{1}.\mn) dA
 [(r_{2}N_{3}-r_{3}N_{2})\me_{1}+(r_{3}N_{1}-r_{1}N_{3})\me_{2}\nonumber\\
&+(r_{1}N_{2}-r_{2}N_{1})\me_{3}] .\label{a10}
 \end{align}
Substituting $p_{ph}=\frac{h}{\lambda}, F=n_{ph}
c=\frac{u_{rad}}{hc/\lambda} c=\frac{u_{rad}\lambda}{h}$, and
$dA$ from equation (\ref{a9}) into the above equation, we obtain
\begin{align}
d{\bf \Gamma}_{rad}&=\frac{\gamma u_{rad}\lambda b^{2}}{2} [-\frac{4e}{\lambda}(\me_{1}.\mn)
 [(r_{2}N_{3}-r_{3}N_{2})\me_{1}\nonumber\\
&+(r_{3}N_{1}-r_{1}N_{3})\me_{2}+(r_{1}N_{2}-r_{2}N_{1})\me_{3}]f(\eta)\mc\eta d\eta d\xi .\label{a11}
\end{align}
From equations (\ref{a7}) and (\ref{a8}), we get
\begin{align}
r_{2}N_{3}-r_{3}N_{2}&=\frac{ab_{1}-ba_{1}}{2}\ms2\eta\ms\xi\ms\Theta,\label{a12}\\
r_{3}N_{1}-r_{1}N_{3}&=\frac{ab_{1}-ba_{1}}{2}\ms2\eta\ms\xi\mc\Theta,\label{a13}\\
r_{1}N_{2}-r_{2}N_{1}&=\frac{ab_{1}-ba_{1}}{2}\ms2\eta\mc\xi.\label{a14}
\end{align}
Plugging equations (\ref{a12})-(\ref{a14}) into equation (\ref{a11}), and
integrating over the surface, we obtain
\begin{align}
{\bf \Gamma}_{rad}&=\frac{\gamma u_{rad}\lambda b^{2}}{2}(Q_{e1}\me_{1}+Q_{e2}\me_{2}+Q_{e3}\me_{3}),\label{a15}
\end{align}
where the RAT components are given by
\begin{align}
Q_{e1}&=\int_{-\pi/2}^{\pi/2}\int_{0}^{\pi}\frac{-4e}{\lambda}\frac{ab_{1}-ba_{1}}{2}|a_{1}\ms\eta \mc\Theta\nonumber\\
&-b_{1}\mc\eta\mc\xi\ms\Theta|
\ms 2\eta \ms\xi \ms\Theta f(\eta) \mc\eta d\xi d\eta,\label{a16}\\
Q_{e2}&=\int_{-\pi/2}^{\pi/2}\int_{0}^{\pi}\frac{-4e}{\lambda}\frac{ab_{1}-ba_{1}}{2}|a_{1}\ms\eta \mc\Theta\nonumber\\
&-b_{1}\mc\eta\mc\xi\ms\Theta|
\ms 2\eta \ms\xi\mc\Theta f(\eta) \mc\eta d\xi d\eta,\label{a17}\\
Q_{e3}&=\int_{-\pi/2}^{\pi/2}\int_{0}^{\pi}\frac{-4e}{\lambda}\frac{ab_{1}-ba_{1}}{2}|a_{1}\ms\eta \mc\Theta\nonumber\\
&-b_{1}\mc\eta\mc\xi\ms\Theta|
\ms 2\eta \mc\xi  f(\eta) \mc\eta d\xi d\eta.\label{a18}
\end{align}
Note, that we use the absolute value for ${\me}_{1}.{\mn}$ because
in the integral over $\eta$, we always use the range $[-\pi/2, \pi/2]$.

Due to the presence of the term $\ms\xi$ in equations (\ref{a16}) and
(\ref{a17}), their integrals over the range $\xi=[0, \pi]$ vanish. The
integral given by equation (\ref{a18}) provides us a function of $\Theta$ which can be fitted
 by a function $\ms 2\Theta$ and a fitting factor
\bea
 Q_{e3}(\Theta)= \frac{2e  a}{\lambda}(s^{2}-1) K(\Theta, e)\ms2\Theta,\label{ap19}
\ena
where $ K(\Theta, e)$ is a function of $e$ and $\Theta$ (see the lower panel
in Fig.~\ref{ap1}).

Thus, unpolarized
radiation produces only the third component of RATs, i.e. $Q_{e3}$ for the
spheroid, while two first components $Q_{e1}, Q_{e2}$ vanish.  

\section{Mirror on a pole model}
\subsection{RATs calculations}\label{mirror}
Consider now a grain consisting of a square reflective mirror of side $l_{2}$,
 attached to the spheroid by a pole of the length $l_{1}$ (see
 Fig. \ref{f2}). Here we calculate RATs acting on the mirror. Its orientation in the grain coordinate system, ${\ma}_{1},
{\ma}_{2}, {\ma}_{3}$, is characterized by a normal unit vector $\mn$, given by
\bea
 \mn=n_{1}\ma_{1}+n_{2}\ma_{2},\label{b1}
\ena
where  $n_{1}=\ms \alpha, n_{2}=\mc\alpha$ with $\alpha$ is the
angle between $\mn$ and $\ma_{2}$. (see Fig. \ref{f2}). The use of the grain
 coordinate system is appropriate as the spheroidal body determines the grain
 inertia.

Since $l_{1} \gg l_{2}$, the radius vector determining the position of
each reflecting event on the mirror surface $\mr$ is nearly parallel to ${\bf
 \ma}_{3}$, thus
\bea
\mr =l_{1}\ma_{3}.
\label{b2}
\ena
The orientation of the grain in the lab coordinate system is determined by
$\Theta$ and $\beta$ as follows, 
\begin{align}
\ma_{1}&=\mc \Theta \me_{1}+\ms \Theta \me_{2},\label{b3}\\
\ma_{2}&=\mc\beta [-\ms \Theta \me_{1}+\mc \Theta \me_{2}]+\ms\beta \me_{3},\label{b4}\\
\ma_{3}&=-\ms\beta [-\ms \Theta \me_{1}+\mc \Theta \me_{2}]+\mc\beta \me_{3}.\label{b5}
 \end{align}
Plugging equations (\ref{b3}-\ref{b5}) into equations (\ref{b1}) and (\ref{b2}) we get
 \begin{align}
\mn&=\me_{1}(n_{1}\mc\Theta-n_{2}\mc\beta\ms\Theta)\nonumber\\
&+\me_{2}(n_{1}\ms\Theta+n_{2}\mc\beta\mc\Theta)\nonumber\\
&+\me_{3}n_{2}\ms\beta,\label{b6}\\
\mr&=l_{1}(\me_{1}\ms\beta\ms\Theta-\me_{2}\ms\beta\mc\Theta+\me_{3}\mc\beta).\label{b7}
\end{align}
Hence, RAT produced by the photon beam acting on the mirror is 
\bea
d{\bf \Gamma}_{rad}=\gamma \mr \times \Delta {\bf P}=(-2p_{ph}) \gamma F dA ({\bf k}.\mn)[\mr\times \mn]. \label{b8}
\ena
Integrating over the full mirror, RAT becomes 
\bea
{\bf \Gamma}_{rad}={-2p} \gamma F A_{\perp} ({\bf k}.\mn)[\mr
\times \mn],\label{b9}
\ena
where $A_{\perp}=A|\me_{1}.\mn|$ is the cross section of the mirror with
respect to the photon flux.  

Substituting $p_{ph}$, $F$, ${\bf k}\equiv\me_{1}$, and
 $A_{\perp}=l_{2}^{2}|\me_{1}.\mn|$ into equation (\ref{b9}), we get
\begin{align}
{\bf \Gamma}_{rad}&=\frac{\gamma u_{rad}\lambda l_{2}^{2}}{2} (-\frac{4}{\lambda})|\me_{1}.\mn|(\me_{1}.\mn)
 [(r_{2}N_{3}-r_{3}N_{2})\me_{1}\nonumber\\
&+(r_{3}N_{1}-r_{1}N_{3})\me_{2}+(r_{1}N_{2}-r_{2}N_{1})\me_{3}].\label{b10}
\end{align}
From equations (\ref{b6}) and (\ref{b7}), we obtain
\begin{align}
\frac{r_{2}N_{3}-r_{3}N_{2}}{l_{1}}&=-n_{2}\ms\beta \mc\Theta \ms\beta\nonumber\\
&-\mc\beta(n_{1}\ms\Theta+n_{2}\mc\beta \mc\Theta ) \nonumber\\
&=(-n_{1} \mc \beta\ms\Theta -n_{2}\mc\Theta),\label{b11}\\
\frac{r_{3}N_{1}-r_{1}N_{3}}{l_{1}}&=
\mc\beta(n_{1}\mc\Theta-n_{2}\mc\beta\ms\Theta)\nonumber\\
&-\ms\beta \ms\Theta n_{2}\ms\beta \nonumber\\
&=n_{1}\mc\beta\mc\Theta-n_{2}\ms\Theta\label{b12},\\
\frac{r_{1}N_{2}-r_{2}N_{1}}{l_{1}}&=n_{1}\ms\beta.
\end{align}
Therefore, we can write
\begin{align}
{\bf \Gamma}_{rad}&=\frac{\gamma u_{rad}\lambda l_{2}^{2}}{2}(Q_{e1}\me_{1}+Q_{e2}\me_{2}+Q_{e3}\me_{3}),\label{b13}
\end{align}
where the RAT components are given by
\begin{align}
Q_{e1}&=-\frac{4l_{1}}{\lambda}|\me_{1}.\mn|(n_{1}\mc\beta\ms\Theta+n_{2}\mc\Theta)[-n_{1}\mc\Theta\nonumber\\
&+(n_{2}\mc\beta)\ms\Theta],\label{b14}\\
Q_{e2}&=-\frac{4l_{1}}{\lambda}|\me_{1}.\mn|(n_{1}\mc\beta\mc\Theta-n_{2}\ms\Theta)[n_{1}\mc\Theta\nonumber\\
&-n_{2}\mc\beta \ms\Theta],\label{b15}\\
Q_{e3}&=-\frac{4l_{1}}{\lambda}|\me_{1}.\mn|n_{1}\ms\beta[n_{1}\mc\Theta-n_{2}\mc\beta\ms\Theta].\label{b16}
\end{align}
Simplifying the above equations, we get
\begin{align}
Q_{e1}&=\frac{4l_{1}}{\lambda}|n_{1} \mc\Theta-n_{2}
\ms\Theta\mc\beta|[n_{1}n_{2}\mcs\Theta\nonumber\\
&+\frac{n_{1}^{2}}{2}\mc\beta\ms 2\Theta-\frac{n_{2}^{2}}{2}\mc\beta\ms 2\Theta-n_{1}n_{2}\mss\Theta \mcs\beta],\label{b17}\\
Q_{e2}&=-\frac{4l_{1}}{\lambda}|n_{1}
\mc\Theta-n_{2}\ms\Theta\mc\beta|[n_{1}^{2}\mc\beta\mcs\Theta\nonumber\\
&-\frac{n_{1}n_{2}}{2}\mcs\beta\ms 2\Theta-\frac{n_{1}n_{2}}{2}\ms 2\Theta+n_{2}^{2}\mc\beta\mss\Theta],\label{b18}\\
Q_{e3}&=-\frac{4l_{1}}{\lambda}|n_{1} \mc\Theta-n_{2} \ms\Theta \mc\beta| n_{1}\ms\beta[n_{1}\mc\Theta\nonumber\\
&-n_{2}\mc\beta\ms\Theta]. \label{b19}
\end{align}

  Averaging over the rotational angle $\beta$ in the range $[0, 2\pi]$, we get 
\begin{align}
Q_{e1}&=\frac{4\pi l_{1}n_{1}n_{2}}{\lambda} (3 \mbox{cos}^{2} \Theta - 1) f(\Theta, \alpha),\label{b20}\\ 
Q_{e2}&= \frac{4\pi l_{1}n_{1}n_{2}}{\lambda} \ms 2\Theta g(\Theta, \alpha),\label{b21} \\
Q_{e3}&=0,\label{b22}
\end{align}
where $f(\Theta, \alpha), g(\Theta, \alpha) $ are fitting functions  depending
on $\alpha$ and $\Theta$. The dependence on $\Theta$ characterizes the
influence of variation of the mirror cross section on RATs. We will find these fitting functions in the following section.

\subsection{Fitting functions}
 As we have seen above, RATs can be decomposed into analytical terms and fitting functions which are 
functions of both $\Theta$ and $\alpha$ (see equations \ref{b17} and
\ref{b18}). Below we discuss an analytical approximation to
the fitting functions $f(\Theta, \alpha), g(\Theta, \alpha)$.

In the vicinity of $\Theta=0, \pi$, $|n_{1}\mc\Theta-n_{2}\ms\Theta\mc\beta| \sim n_{1}$, so
this factor does not make  $\beta$ -averaging more involved, however.

As $\Theta \sim \pi/2$, equations (\ref{b14}) and (\ref{b15}) can be written
as
\begin{align}
Q_{e1}&=\frac{4l_{1}}{\lambda}|-n_{2}\mc\beta|[n_{1}n_{2}\mcs\Theta+\frac{n_{1}^{2}}{2}\mc\beta\ms2\Theta\nonumber\\
&-\frac{n_{2}^{2}}{2}\mc\beta\ms2\Theta-n_{1}n_{2}\mss\Theta\mcs\beta],\label{eq8a}\\
Q_{e2}&=-\frac{4l_{1}}{\lambda}|-n_{2}\mc\beta|[n_{1}^{2}\mc\beta\mcs\Theta
-\frac{n_{1}n_{2}}{2}\mcs\beta\ms2\Theta\nonumber\\
&-\frac{n_{1}n_{2}}{2}\ms2\Theta+n_{2}^{2}\mc\beta\mss\Theta],\label{eq9a}
\end{align}
Integrating equations (\ref{eq8a}) and (\ref{eq9a}) over $\beta $ in a range
$[0, 2\pi]$, we get
\begin{align}
Q_{e1}&=\frac{4l_{1}}{\lambda}\frac{4n_{1}n_{2}|n_{2}|}{3}(5\mcs\Theta-2),\label{eq8b}\\
Q_{e2}&=\frac{4l_{1}}{\lambda}\frac{10 n_{1}n_{2}|n_{2}|}{3}\ms2\Theta.\label{eq9b}
\end{align}
Comparing equations (\ref{b17}) and (\ref{b18}) with (\ref{eq8b}) and
(\ref{eq9b}), we have 
\begin{align}
f_{\pi/2}(\Theta,\alpha)&=\frac{|\mbox{cos}\alpha| (5 \mcs\Theta-2)}{3\pi(3\mcs\Theta-1)},\label{eq16}\\
g_{\pi/2}(\Theta,\alpha)&=\frac{10|\mbox{cos}\alpha|}{3\pi}.\label{eq17}
\end{align}
  
The fitting functions $f_{\pi/2}(\Theta, \alpha)$ and the tabulated function $f(\Theta, \alpha)$ for $\alpha=45^{0}$ are shown in Fig. (\ref{f4}).
\begin{figure}
\includegraphics[width=0.49\textwidth]{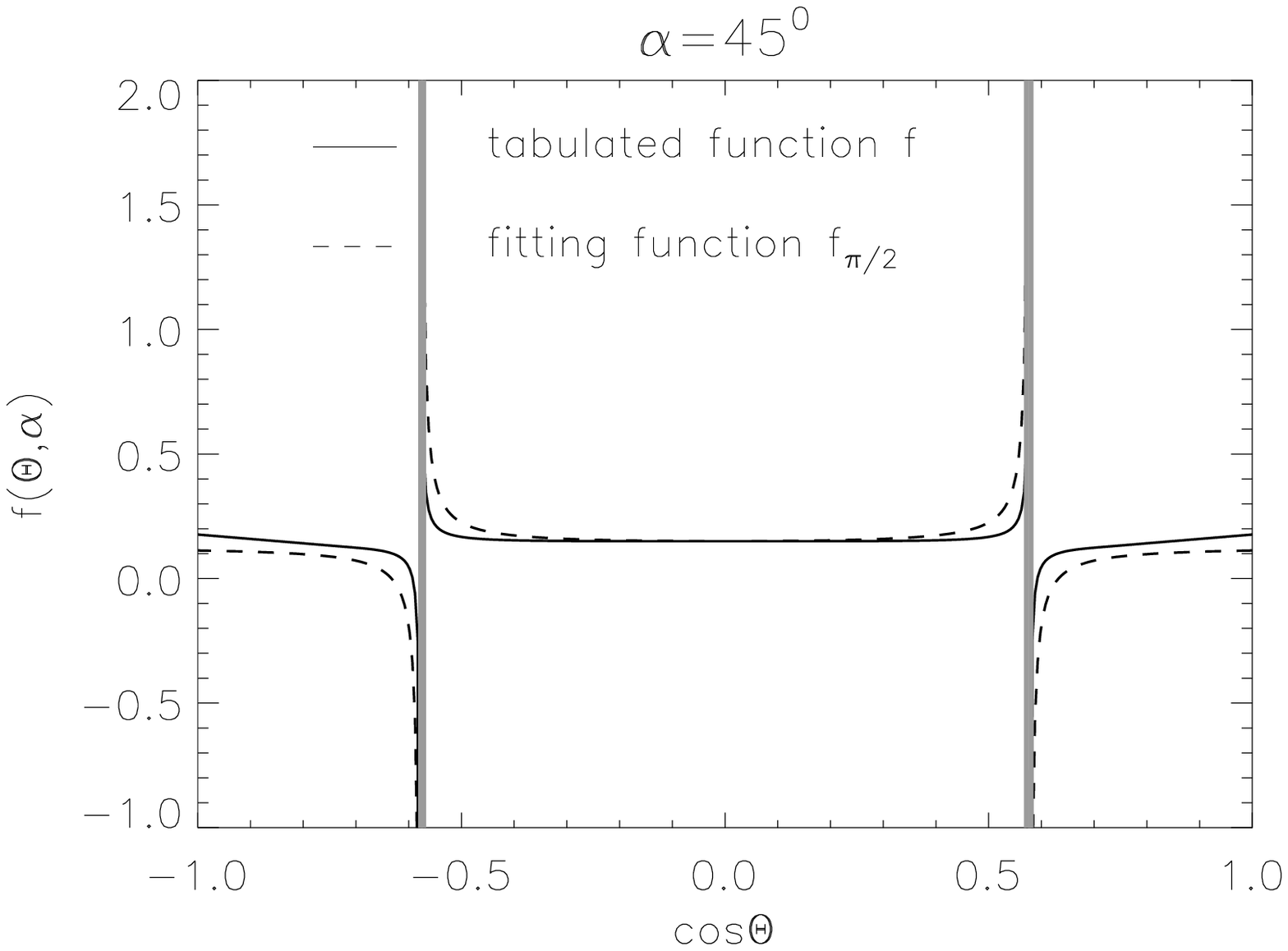}
\caption{Tabulated and fitting functions $f$ for $\alpha = 45^{0}$. Shaded lines show the vicinity of the singular line $\mc\Theta=1/\sqrt{3}$} 
 \label{f4} 
\end{figure}

\begin{figure}
\includegraphics[width=0.49\textwidth]{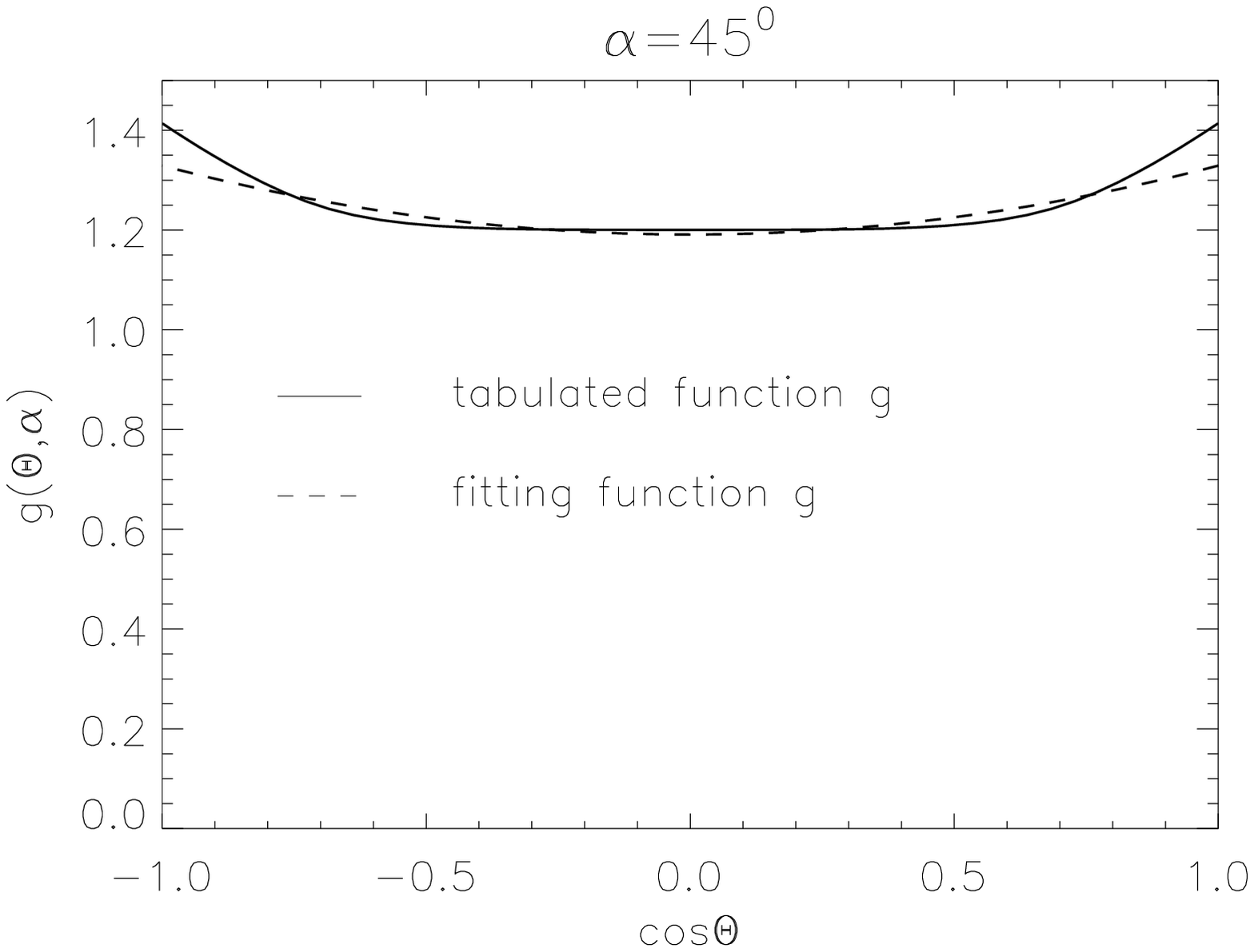}
\caption{Tabulated and fitting functions $g$ for $\alpha =45^{0}$.}
 \label{f4*} 
\end{figure}

\begin{figure}
\includegraphics[width=0.49\textwidth]{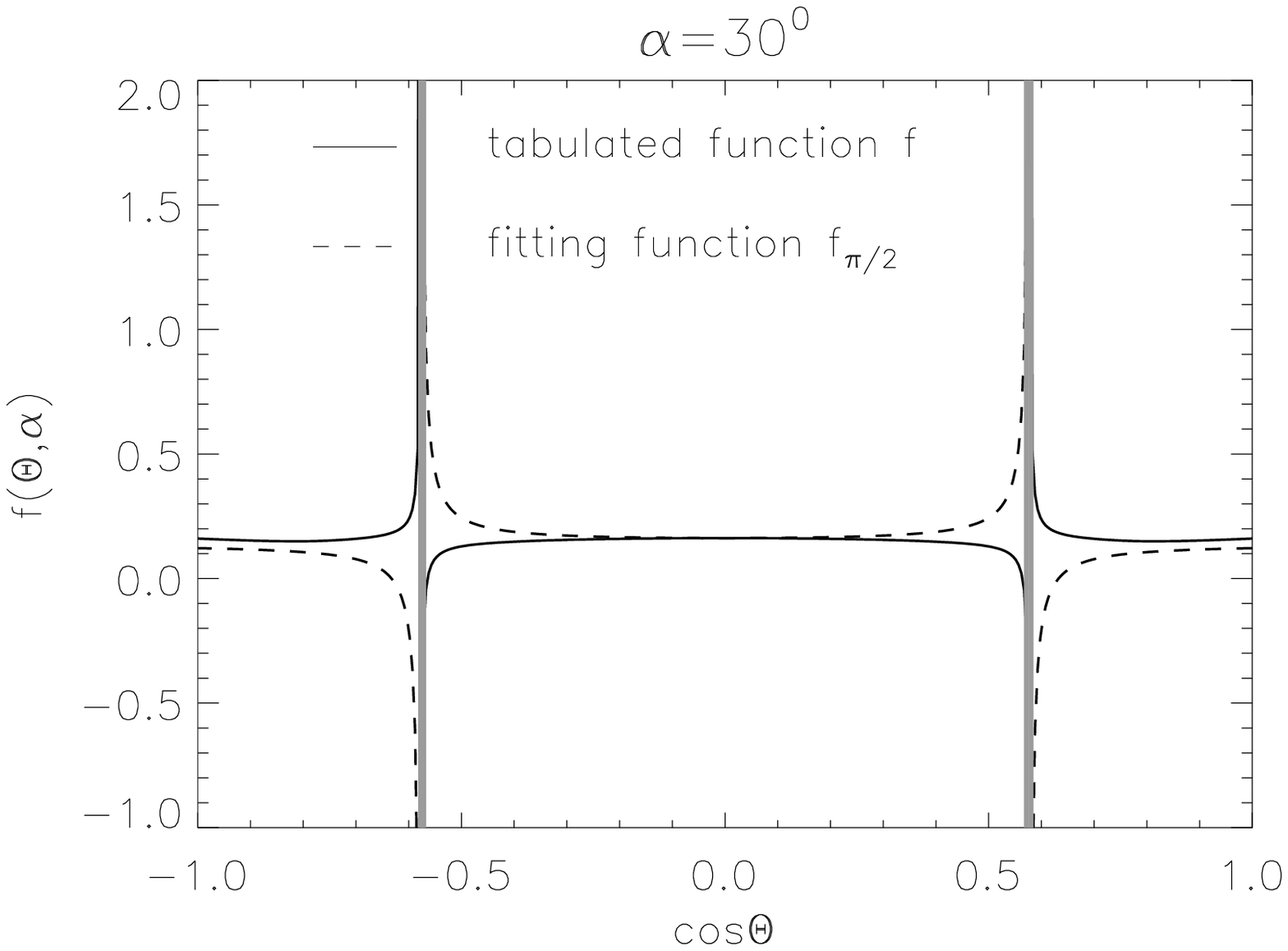}
\includegraphics[width=0.49\textwidth]{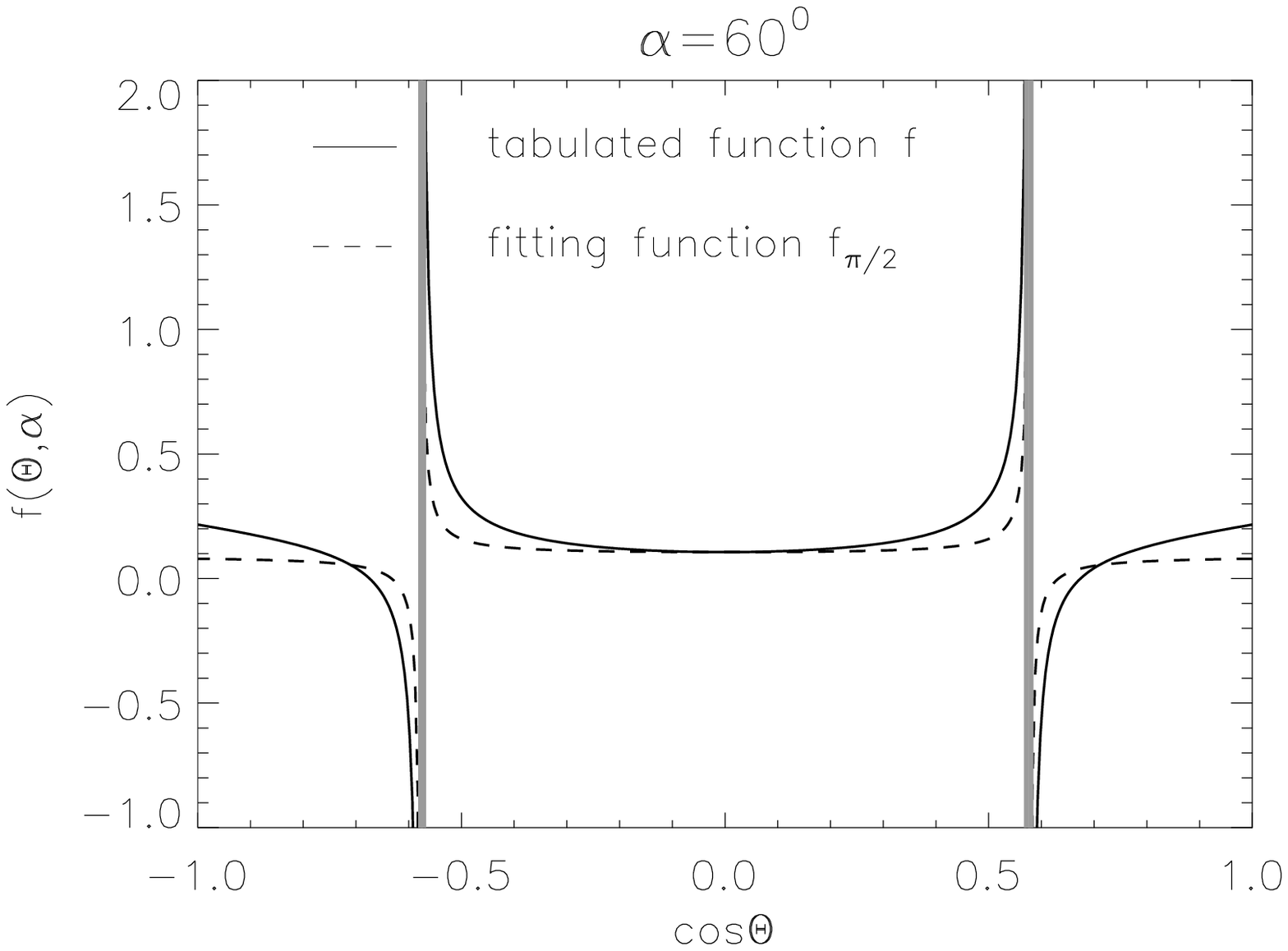} 
\caption{Fitting function $f$ for two tilted angles of the mirror
 in the body system: upper and lower panel correspond to $\alpha =30^{0}, 60^{0}$. Shaded lines show the vicinity of the singular line $|\mc\Theta|=1/\sqrt{3}$} 
\label{fap3} 
\end{figure}

\begin{figure}
\includegraphics[width=0.49\textwidth]{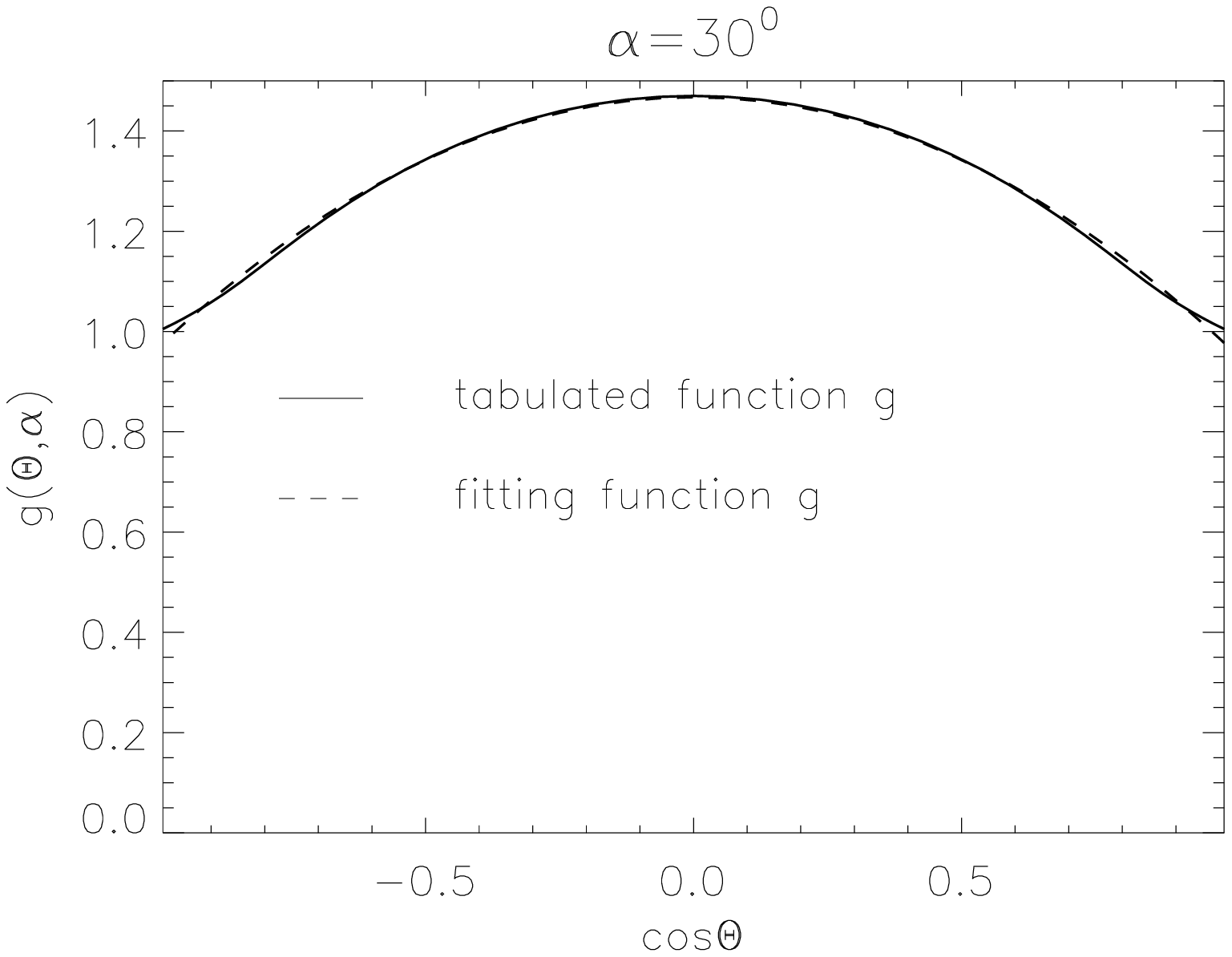}
\includegraphics[width=0.49\textwidth]{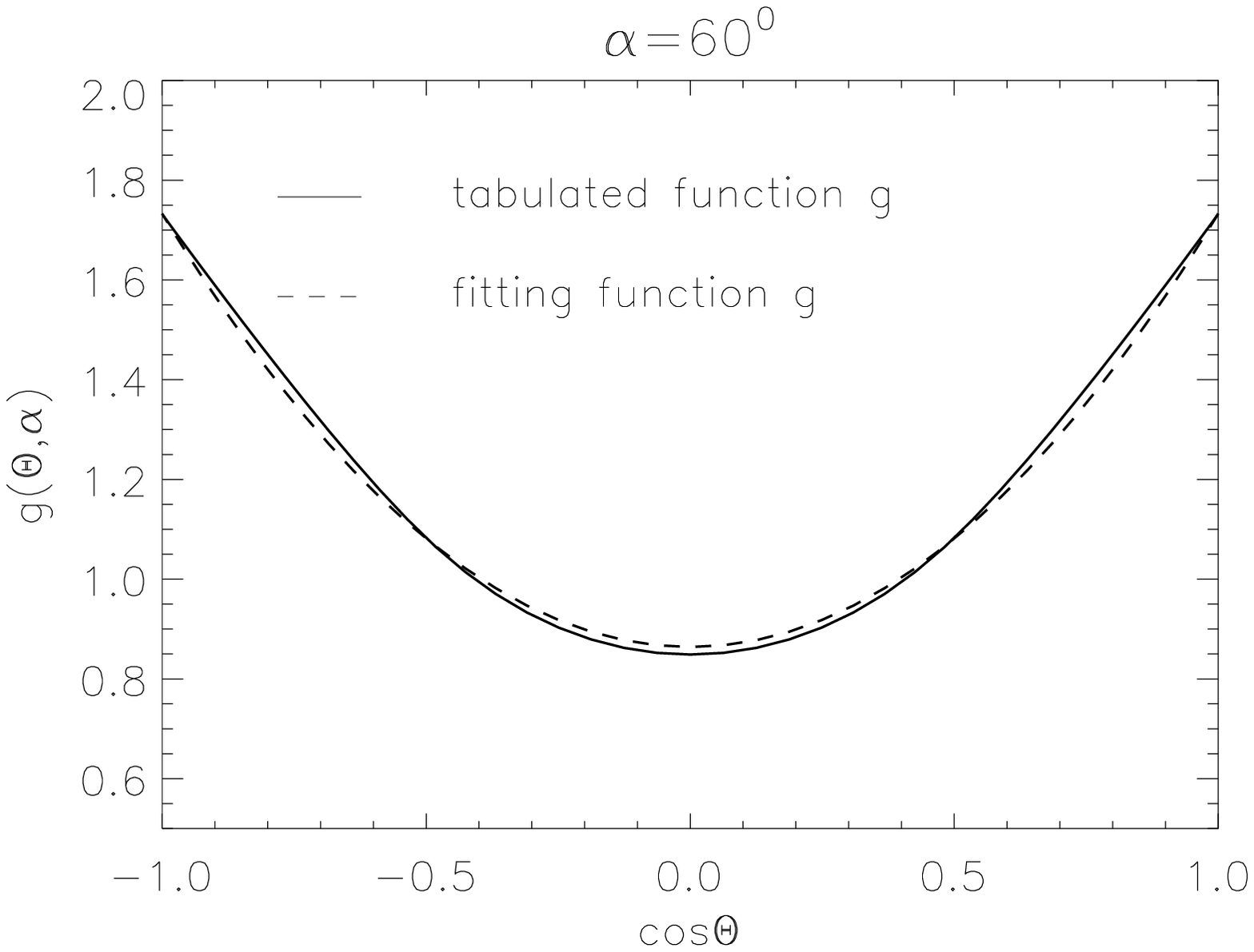}
\caption{Fitting and tabulated functions $g$ for two tilted angles of the mirror
 in the body system: upper and lower panel correspond to $\alpha =30^{0}, 60^{0}$.}
 \label{fap4} 
\end{figure}

Interestingly enough, the approximation of the fitting
functions obtained in the vicinity of $\Theta \sim  \pi/2$ presents
a reasonable approximation for the fitting function $f$ over the
entire range of $\Theta$.
Indeed, Fig.~\ref{f4} shows the tabulated function $f(\Theta, \alpha)$
obtained by substituting $Q_{e1}, Q_{e2}$ resulted from numerically averaging
equations (\ref{b14}) and (\ref{b15}) into equations (\ref{b17}) and (\ref{b18})(solid lines) and the fitting function $ f_{\pi/2}(\Theta,\alpha)$ given by equation (\ref{eq16}) (dashed line) for a particular case of $\alpha=45^{0}$. It is shown that the fitting function $f_{\pi/2}(\Theta,\alpha)$ has a good agreement with $f(\Theta, \alpha)$. Similar results are also found for other $\alpha$ angles (see Fig.~\ref{fap3}). 

However, the fitting function $g_{\pi/2}$ given by equation~(\ref{eq17}) is independent
of $\Theta$ and therefore does not provide a
good fit with the tabulated function $g$. In this case, we perform a quadratic
fitting for tabulated functions for different $\alpha$ angles. The results for
$\alpha=30^{0}$, $45^{0}$ and $60^{0}$ are given by
\begin{align}
g(\Theta, \alpha=30^{0})&=1.4675-0.5\mcs\Theta,\label{eq17a}\\
g(\Theta, \alpha=45^{0})&=1.191+0.1382\mcs\Theta,\label{eq17b}\\
g(\Theta, \alpha=60^{0})&=0.864+0.869\mcs\Theta.\label{eq17c}
\end{align}
Tabulated and fitting functions g are shown in Fig.\ref{f4*} for $\alpha=45^{0}$ (see Fig.~\ref{fap4} for $g$ corresponding to other angles $\alpha$).  

Figs (\ref{fap3}) and (\ref{fap4}) show $f, g$ for $\alpha=30^{0}$ and $60^{0}$.
Fig. \ref{fap3} shows that there is a change in sign of the tabulated function $f$ as $|\mc\Theta| \rightarrow 1/\sqrt{3}$ between $\alpha=30^{0}$ and $60^{0}$. This stems from the fact that, for $\alpha <45^{0}$, the influence of $A_{\perp}$ shifts the zero of $Q_{e1}$ from $|\mc\Theta|=1/\sqrt{3}$ to $|\mc\Theta|<1/\sqrt{3}$. As a result, when the term $3 \mbox{cos}^{2} \Theta - 1$ approaches the zero from the left of $\mc\Theta=-1/\sqrt{3}$, for instance, then $Q_{e1}$ still positive, which causes $f$ rises sharply to the positive direction as in Fig. \ref{fap3}{\it upper}.

As an example, in the paper we mostly discuss the case in which the mirror is tilted by an angle
of $\alpha=\pi/4$ for which there is the best fit between the fitting function
$f_{\pi/2}$ and the tabulated function $f$ except a narrow range in the
vicinity of $|\mc \Theta|=1/\sqrt{3}$ (see Fig. \ref{f4}). Similarly,
$g$ given by equation (\ref{eq17b}) also provides a good fit for $Q_{e2}$
(see Fig. \ref{f4*}).
Therefore, RATs acting upon the mirror can be roughly approximated by equations (\ref{b17})-(\ref{b18}) with
$f, g$ are given in equations (\ref{eq16}) and (\ref{eq17b}).

Needless to say, that such an approximation entails errors in calculating
the torques $Q_{e1}$ and $Q_{e2}$. Fig.~\ref{f6*} shows the torques with $f,
g$ being the tabulated and fitting functions. It can be seen that there is a very good fit for the component $Q_{e2}$, but there exhibit some deviation for the component $Q_{e1}$ toward the range $\mc\Theta \sim \pm 1$. These errors could be
decreased by using more sophisticated analytical fits, e.g. piecewise
analytical fit. However, this would decrease the heuristic value of the
formulae. Therefore while using the approximate fits (\ref{eq16}) and
(\ref{eq17b})
for some analytical calculations below, for most of the quantitative estimates,
including those related to evolving the phase trajectories, we tabulate
the fitting functions $f$ and $g$. Naturally, the latter
 is equivalent to the direct use of averaged
equations (\ref{b14})-(\ref{b15}) for RAT components.
\begin{figure}
\includegraphics[width=0.49\textwidth]{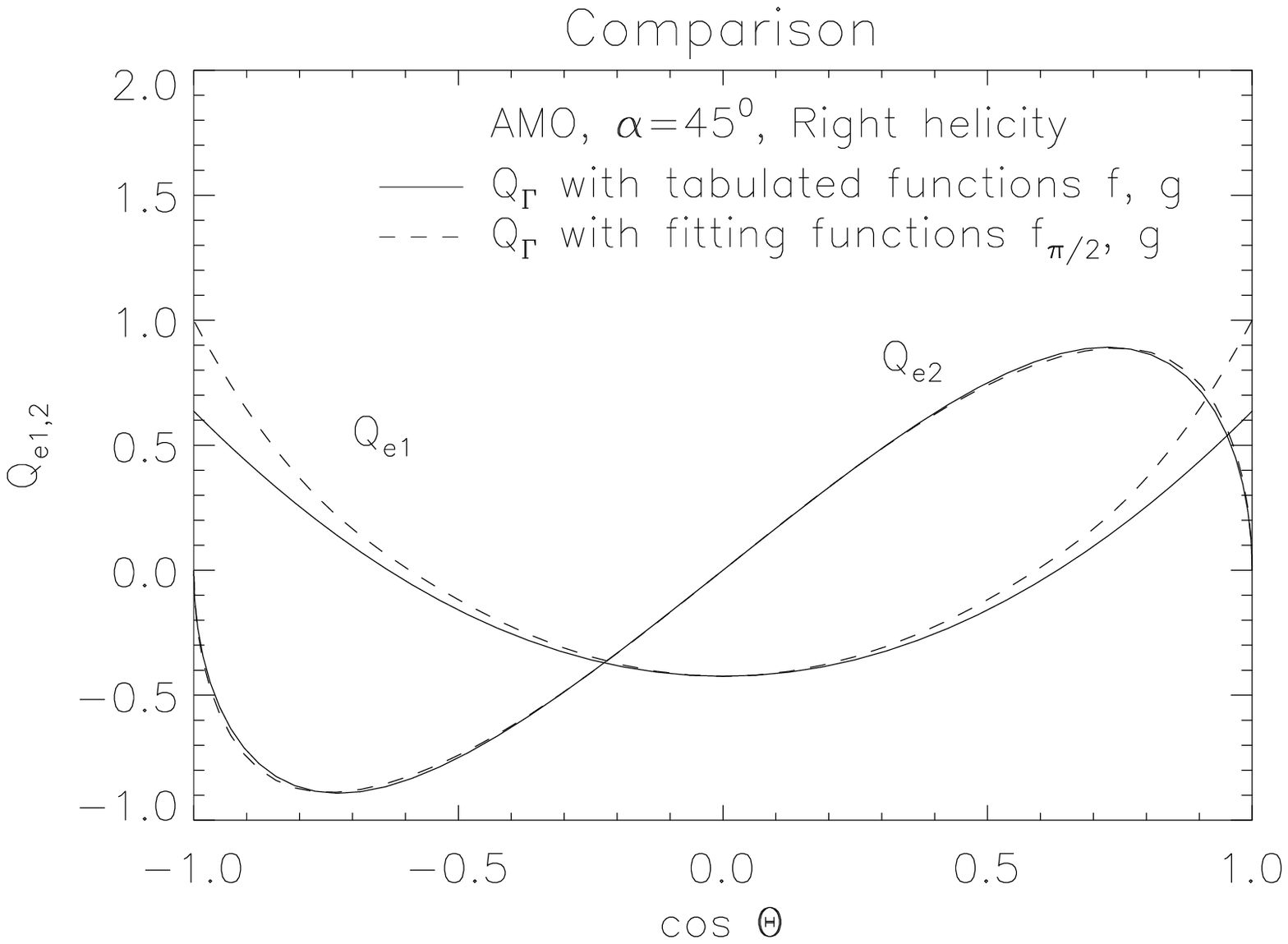}
\caption{Fig shows the comparison of RAT components. Dashed lines show  RATs obtained from the analytical approximation given by equations
 (\ref{b17})-(\ref{b18}) and solid
lines show RATs obtained by numerically averaging equations (\ref{b14})-(\ref{b15}) over $\beta$. }
\label{f6*}
\end{figure}

What is the purpose of having the analytical form of the fitting functions, if
they do not provide exact fits for the torque components (see Fig. \ref{f6*})? We
may see the actual RATs acting on irregular grains (e.g. Fig. \ref{f61}) also differ
somewhat from the corresponding AMO RATs. These differences, however, do not
change substantially, as we show in the rest of the paper, the alignment of
irregular grains in comparison with the alignment predicted by AMO. Thus we
believe that the slightly distorted AMO with the appropriate analytically
fitting functions should also reflect correctly the generic properties of the
RAT alignment.
\subsection{Dependences on $\alpha$}
\begin{figure}
\includegraphics[width=0.49\textwidth]{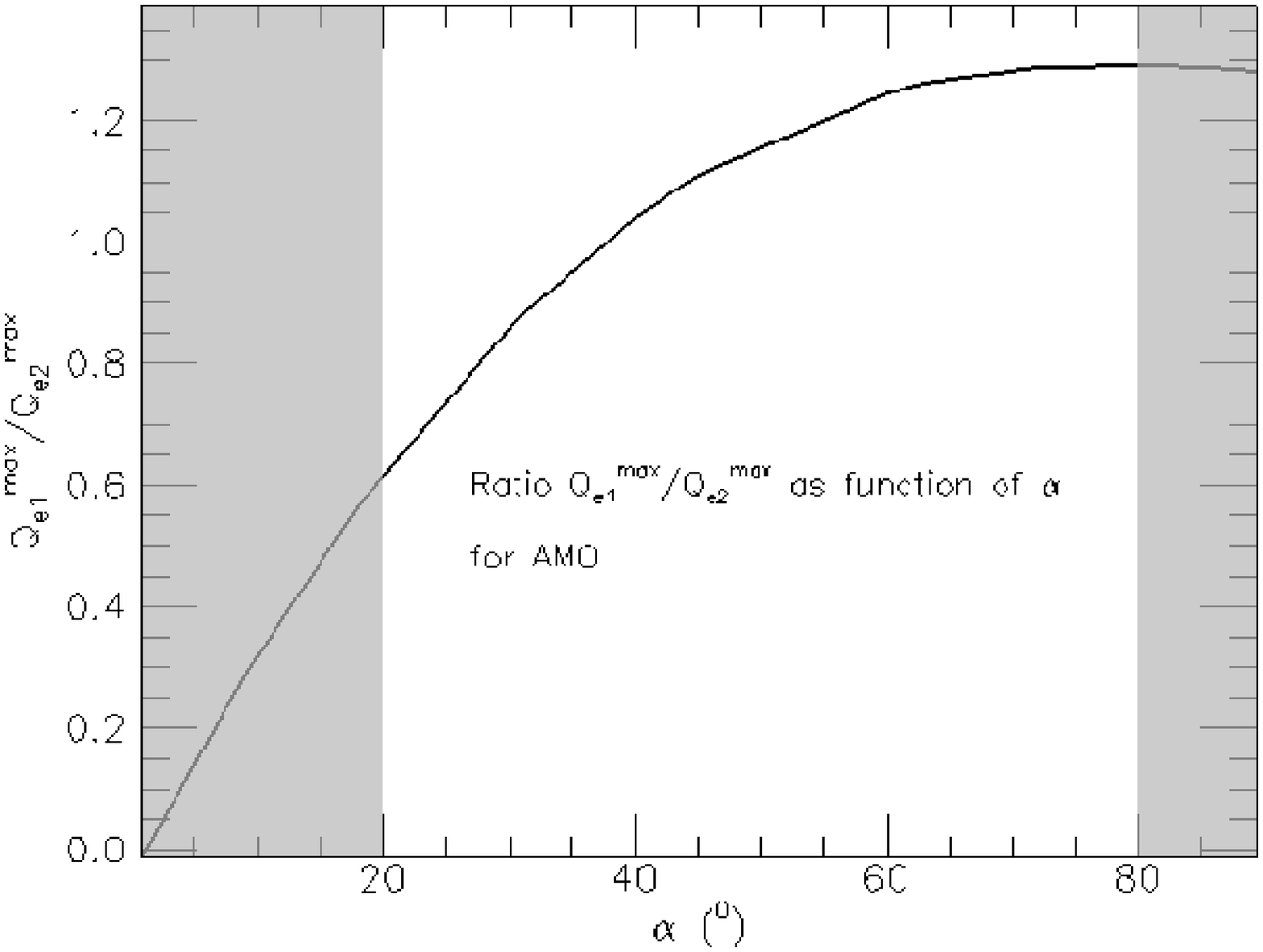}
\caption{Variation of $Q_{e1}^{max}/Q_{e2}^{max}$ as a function of $\alpha$.
This ratio determines the existence of high attractor points and the shift
of a crossover point as we describe in \S \ref{qratio}. Shaded areas show the range of $\alpha$ in which the symmetry of RATs are significantly affected by the cross section.}
\label{f5}
\end{figure}
Fig. \ref{f5} shows that the variation of the mirror orientation can change the ratio of their maximum $Q_{e1}^{max}/Q_{e2}^{max}$. Here we denote $Q_{e1}^{max}$ the maximum of
$|Q_{e1}|$ which is exactly its amplitude at $\Theta=0$ or $\pi$, and
$Q_{e2}^{max}$ the maximum of $Q_{e2}$. In Fig. \ref{f5} it is obviously shown that
$Q_{e1}^{max}/Q_{e2}^{max}$ increases with $\alpha$ increasing due to the increase of the cross-section $A_{\perp}$.
However, the ratio $Q_{e1}^{max}/Q_{e2}$ is limited to the upper limit of $1.3$. We will see later
that the ratio plays an important role on the alignment that is discussed in \S~ \ref{sec5}. 

Moreover, the variation of $\alpha$ does not only give rise to the change of the ratio $Q_{e1}^{max}/Q_{e2}^{max}$, but it can also affect the symmetric functional form of RATs. Our calculations show that for $\alpha$ smaller than $20^{0}$ or larger than $80^{0}$, the RATs have some changes in the
functional form of the $Q_{e1}$ and $Q_{e2}$ compared to what is shown in Fig. \ref{f6}. 

Let us explain why for $\alpha$ small, the 
symmetry of RATs for AMO is affected. Consider equation (\ref{b17}) with $n_{1}=\ms\alpha, n_{2}=\mc\alpha$
\begin{align}
Q_{e1}&=\frac{4l_{1}}{\lambda}|\ms\alpha \mc\Theta-\mc\alpha
\ms\Theta\mc\beta|[\ms\alpha \mc\alpha\mcs\Theta\nonumber\\
&+\frac{\mss\alpha}{2}\mc\beta\ms 2\Theta-\frac{\mcs\alpha}{2}\mc\beta\ms 2\Theta\nonumber\\
&-\ms\alpha\mc\alpha\mss\Theta \mcs\beta].\label{apc1}
\end{align}
The first term is the cross-section $A_\perp$, and the second term is $Q_{e1}$ without 
cross-section. 
The second term when averaged gives rise to $Q_{e1} \sim 5\mcs\Theta-2$. It indicates that if the cross-section is constant, $Q_{e1}$ is fully symmetric. Consider $\alpha$ small, $cos\alpha$ is significant, and equation (\ref{apc1}) reduces to
\begin{align}
Qe1&=\frac{4l_{1}}{\lambda}|-\mc\alpha \ms\Theta \mc\beta|[\ms\alpha \mc\alpha\mcs\Theta\nonumber\\
&+\frac{\mss\alpha}{2}\mc\beta\ms 2\Theta-\frac{\mcs\alpha}{2}\mc\beta\ms 2\Theta\nonumber\\
&-\ms\alpha\mc\alpha\mss\Theta \mcs\beta].\label{apc2}
\end{align}
Obviously, when averaging over $\beta$, the absolute term containing $\ms\Theta$ contributes to modify substantially the symmetry of the resulting torques. The same problem also occurs for $Q_{e2}$. We found that for both $\alpha \le 20^{0}$ or $\alpha \ge 80^{0}$, the functional forms of RATs are very influenced. However, for $\alpha$ within $[20^{0},80^{0}]$, their functional forms are not much different with what shown in Fig. \ref{f6} for $\alpha=45^{0}$.

\section{RATs with DDSCAT}
In the discrete electric dipole approximation (Draine \& Flateau 1994), a grain is presented as an
ensemble of electric dipoles. The interaction between the electric field of
incident light and the
dipoles produces radiative forces and torques.
RATs produced by radiation on a grain consisting of $N$ electric
dipoles are 
\bea
{\bf \Gamma}_{rad}=\sum_{j=1}^{N}{\bf r}_{j}\times {\bf F}_{j}+\sum_{j=1}^{N} {\bf p}_{j}\times {\bf E}_{j},\label{rad0}
\ena
where ${\bf r}{j}$ is radius of $j^{th}$ dipole, $ {\bf E}_{j}, {\bf F}_{j}$
are electric field at the location of, and radiative force
which acts on the $j^{th}$ dipole. Radiative force ${\bf F}$ present in equation (\ref{rad0}) is produced by
the gradient of electric field in the grain and the Lorentz force due to vibration of electric
dipole in magnetic field.

 Radiation field inside the
grain consists of that of incident and scattered light. Each
dipole receives the incident light which induces its vibration and scattered
light produced by all electric dipoles except the dipole under study. Hence,
the total RAT can be written in a different form 
\bea
{\bf \Gamma}_{rad}={\bf \Gamma}_{inc}+{\bf \Gamma}_{sca}.\label{rad1}
\ena
RAT efficiency ${\bf Q}_{\Gamma}$ is defined as followings,
\bea
{\bf \Gamma}_{rad}=\frac{u_{\lambda}a_{eff}^{2}\lambda}{2}\gamma{\bf Q}_{\Gamma},\label{rad11}
\ena
where  $\gamma$ is the anisotropy, and  $u_{\lambda}$ is the energy density of radiation field of wavelength
$\lambda$.
Hence, equation (\ref{rad1}) can be rewritten
\bea
{\bf Q}_{\Gamma}={\bf Q}_{inc}+{\bf Q}_{sca}.\label{rad2}
\ena
Here ${\bf Q}_{inc} \equiv {\bf Q}_{abs} $ and ${\bf Q}_{sca}$ are given by (DW96)
\bea
{\bf Q}_{inc}=\frac{4k}{a_{eff}^{2}|E_{inc,0}|^{2}} {\bf Re}\sum_{j=1}^{N}{\bf p}_{j}(0)\times {\bf E}_{inc,0}e^{i{\bf k}.{\bf r}_{j}} \nonumber \\
-i{\bf k}\times \sum_{j=1}^{N}{\bf r}_{j}[{\bf p}_{j}.{\bf E}_{inc,0}]e^{i{\bf k}.{\bf r}_{j}},\label{rad3}\\
{\bf Q}_{sca}=\frac{-k^{5}}{\pi a_{eff}^{2}|{\bf E}_{inc,0}|^{2}}\int d\Omega {\bf Re}(S_{E}^{*}V_{B}+S_{B}^{*}V_{E}).\label{rad4}
\ena
In above equations, ${\bf E}_{inc}$ denotes electric field of incident light,
$S_{E}, S_{B}$, $V_{E}, V_{B}$ are given by
\begin{align}
S_{E}&=\sum_{j=1}^{N}[\mr_{j}-(\mm.\mr_{j})\mm-\frac{2i}{k}\mm].\mp_{j}(0)exp(-ik\mm.\mr_{j}),\\
S_{B}&=\mr.\sum_{j=1}^{N}\mp_{j}(0)\times \mr_{j}exp(-ik\mm.\mr_{j}),\\
V_{E}&=\sum_{j=1}^{N}{\mp_{j}(0)-\mm[\mm.\mp_{j}(0)}exp(-ik\mm.\mr_{j})\\
V_{B}&=-\mn\times V_{E},
\end{align}
where $\mp_{j}$ is $j^{th}$ electric dipole moment, k is wave number, $\mr$ is the radius vector and $\mm$ is normal unit vector.

We use DDSCAT code to compute separately the components arising from
 absorption, scattering
and total RAT for a grain in which its direction with respect to ${\bf k}$ is
determined by the angles $\Theta,\beta,\Phi$ (see Fig. \ref{f1}). Here $\Theta$ is the angle between ${\bf a}_{1}$
and ${\bf k}$; $\beta$ is the rotational angle of the grain around ${\bf a}_{1}$; and
$\Phi$ is the precession angle of ${\bf a}_{1}$ about ${\bf k}$. 
For our study (see Table \ref{tab2}), we compute RATs for the 
spectrum of the interstellar radiation field, over 21 directions of $\Theta$ from $0$ to $\pi$ and 21 values of $\beta$ from
$0$ to $2\pi$, at $\Phi=0$.  

RAT can be decomposed into components in the scattering system via
\bea
{\bf Q}_{\Gamma}=Q_{e1}\hat{e}_{1}+Q_{e2}\hat{e}_{2}+Q_{e3}\hat{e}_{3}, \label{q4}
\ena
where $\hat{e}_{1}$, $\hat{e}_{2}$, $\hat{e}_{3}$ are shown in Fig.
\ref{f1}.

Mean radiative torque over wavelengths, $\overline{\bQ(\Theta,\beta,\Phi)}$ is defined by
\bea
\overline{\bQ}=\frac{\int \bQ_{\lambda}u_{\lambda}d\lambda}{\int
  u_{\lambda}d\lambda} \label{ra1}.
\ena

Since $\beta$ varies very fast due to the swift rotation of the
grain around the axis
of major inertia ${\bf a}_{1}$, we can average RATs over $\beta$ from $0$ to
$2\pi$. 

\section{Attractor and Repellor points}
Here we derive the condition for which a stationary point becomes an attractor 
and a repellor point following the approach in DW97.
Assuming that the stationary point has the angle $\xi_{s}$ with respect to
magnetic field and angular momentum $J_{s}$, one can expand the right hand sides 
of equations of motion around this point. As a result, equations (\ref{eq59})
and (\ref{eq60}) give
\bea
\frac{d\xi}{dt}=\frac{\langle
  F(\xi_{s})\rangle_{\phi}}{J_{s}}+\frac{d\langle
  F\rangle_{\phi}}{J d\xi}(\xi_{s},J_{s})-\frac{\langle F(\xi_{s})\rangle_{\phi}}{J^{2}_{s}}(J-J_{s}),\label{ap16}\\
\frac{dJ}{dt}=\langle
H(\xi_{s})\rangle_{\phi}-J_{s}+\frac{d\langle H\rangle_{\phi}}{d\xi}(\xi-\xi_{s})-(J-J_{s})
,\label{ap17}
\ena
where $\langle H(\xi)\rangle_{\phi},\langle F(\xi)\rangle_{\phi}$ are spinning
and aligning torques already averaged over the precession angle $\phi$. 
Since for stationary points $\xi_{s}, J_{s}$, we have $\langle
F(\xi_{s})\rangle_{\phi}=0, \langle H(\xi_{s})\rangle_{\phi}-J_{s}=0$, equations
(\ref{ap16}) and (\ref{ap17}) become

\bea
\frac{d\xi}{dt}=A(\xi-\xi_{s})+B(J-J_{s}),\label{ap18}\\
\frac{d\omega}{dt}=C(\xi-\xi_{s})+D(J-J_{s}),\label{ap19}
\ena
where
\bea
A=\frac{d\langle F\rangle_{\phi}}{Jd\xi}(\xi_{s}, J_{s}),\\
B=-\frac{\langle F(\xi_{s})\rangle_{\phi}}{J^{2}_{s}},\\
C=\frac{d\langle H\rangle_{\phi}}{d\xi}(\xi_{s}, J_{s}),\\
D=-1.
\ena
To have an attractor point, one requires
\bea
A+D<0,\label{ap20}\\
BC-AD<0.\label{ap21}
\ena
In other words,
\bea
\frac{d\langle F\rangle_{\phi}}{\langle H\rangle_{\phi}d\xi}(\xi_{s},J_{s})<1,\label{ap22}\\
H\frac{d\langle F\rangle_{\phi}}{d\xi}(\xi_{s},J_{s})<\langle
F\rangle_{\phi}\frac{d\langle H\rangle_{\phi}}{d\xi}(\xi_{s},J_{s}),\label{ap23}
\ena
where we have substituted $J_{s}=H(\xi_{s})$.

\section{Effective grain size for AMO}

For phase maps, we numerically average equations (\ref{eq8})-(\ref{eq10}) to obtain
exact RATs for AMO, rather than using approximate formulae as in the analysis. However, for this case, the absolute magnitude of torques
matters. Therefore, we normalize AMO
in the following way. Assuming that the size $l_{2}$ of the mirror and $l_{1}$ are chosen so that the RAT for AMO has the magnitude equal to that of irregular grain of an effective size $a_{eff}$. Thus, following equations (\ref{eq1}) and (\ref{eq7b}) we have
\begin{align}
\frac{\lambda u_{rad}l_{2}^{2}}{2}Q_{AMO}&=\frac{\lambda u_{rad}a_{eff}^{2}}{2}Q_{DDSCAT},\label{eq66*}
\end{align}
where $Q_{AMO}$ and $Q_{DDSCAT}$ are the magnitudes of RATs for AMO and an irregular grain, respectively.
We can simplify further by normalizing RAT components over the maximum of $Q_{e1}$, and let $Q_{AMO}, Q_{DDSCAT}$ equal the maximum of $Q_{e1}$ for AMO and irregular grain. As a result, for AMO, we have  $Q_{AMO}=|Q_{e1}|^{max}=\frac{16\pi n_{1}n_{2}^{2}l_{1}}{\lambda}$, and substituting into equation (\ref{eq66*}), we get
\begin{align}
l_{2}^{2}l_{1}&=\frac{\lambda}{16\pi n_{1}n_{2}^{2}}a_{eff}^{2} Q_{DDSCAT}.\label{eq67*}
\end{align}
We can define the effective size of AMO subject to the mirror size and the rod length as
\begin{align}
a_{AMO}^{2}&=l_{2}^{2}(\frac{l_{1}}{\lambda}).\label{eq67_1}
\end{align}
Equations (\ref{eq67*}) and (\ref{eq67_1})  enables us to find the effective size of AMO that produce the same the RAT magnitude with the irregular grain of size $a_{eff}$, given by
\begin{align}
a_{AMO}^{2}&=\frac{1}{16\pi n_{1}n_{2}^{2}}a_{eff}^{2} Q_{DDSCAT}.\label{eq68*}
\end{align}

As we discuss in \S \ref{self} that the magnitude of RATs for irregular grains can be crudely approximated as 
\begin{align}
Q_{DDSCAT}&=0.4(\frac{\lambda}{a_{eff}})^{\eta},\label{eq69*}
\end{align}
where $\eta=0$ for $\lambda< 1.8~a_{eff}$ and $\eta=3$ or $4$ for $\lambda>1.8~ a_{eff}$ (see \S \ref{self}).

Equations (\ref{eq69*}) and  (\ref{eq68*}) allow us to roughly estimate the the effective size of AMO, $a_{AMO}$ as a function of the effective size of irregular grain, $a_{eff}$, provided that the wavelength of radiation field is known.

 \newpage
\clearpage


\begin{thebibliography}{8.}
 \bibitem{} Abbas M.M., Craven P. D., Spann J. F. et al. 2004, ApJ, 614, 781
\bibitem{} Bastien P., Jenness T., Molnar J. 2005, ASPC, 343, 69
\bibitem{} Bethell T., Cherpunov A., Lazarian A., Kim J. 2006, ApJ, in press
\bibitem{} Bohren, C Craig 1974, Chemical Physics Letters, 29, 458
\bibitem{} Cho J., Lazarian A. 2005, ApJ, 631, 361
\bibitem{} Cho J., Lazarian A. 2007, ApJ, submitted
 
 \bibitem{} Davis L., Greenstein J.L. 1951, ApJ, 114, 206
 \bibitem{} Dolginov A.Z. 1972 Ap\&SS, 16, 337
 \bibitem{} Dolginov A.Z. Mytrophanov I.G. 1976, Ap\&SS, 43, 291
 \bibitem{} Dolginov A.Z. Silantev N.A. 1976, Ap\&SS, 43, 337
  \bibitem{} Draine B., Flatau P. 1994, J. Opt. Soc. Am. A., 11, 1491
 \bibitem{} Draine B., Lee H. 1984, ApJ, 285, 89                   
  \bibitem{} Draine B. 1985, ApJS, 57, 587
 \bibitem{} Draine B., Lazarian, A. 1998, ApJ, 508, 157
  \bibitem{} Draine B., Weingartner J. 1996, ApJ, 470, 551 (DW96)
  \bibitem{} Draine B., Weingartner J. 1997, ApJ, 480, 633 (DW97)
 \bibitem{} Gold T. 1951,  Nature, 169, 322  

\bibitem{} Goodman A., Jones T., Lada E., Myers P. 1995, ApJ, 448, 748

 \bibitem{} Hall J. 1949, Science, 109, 166


\bibitem{} Harwit M. 1970, Nature, 226, 61-63

 \bibitem{} Hildebrand R., Davidson J. A., Dotson J.L, Wovell C.D., \\
  Novak G., Vaillancourt J.E. 2000, PASP, 112, 1215
 \bibitem{} Hildebrand R. 2002, in {\it Astrophysical Spectropolarimetry},\\
                                 ed. by J. Trujillo-Bueno,
        F. Moreno-Insertis, \\\& F. Sanchez (Cambridge, UK: Cambridge\\
        Univ. Press), p. 265
 \bibitem{} Hiltner W. 1949, Science, 109, 165
\bibitem{} Hoang T., Lazarian A., 2007, MNRAS, submitted
\bibitem{} Hoang T., Lazarian A., in preparation

\bibitem{} Jones M.,  Spitzer L., 1967, ApJ, 147, 943
 \bibitem{} Lazarian A. 1994, MNRAS, 268, 713
\bibitem{} Lazarian A. 1995, ApJ, 453, 229
  \bibitem{} Lazarian A. 1997a, MNRAS, 288, 609
\bibitem{} Lazarian A. 1997b, ApJ, 483, 296 
  \bibitem{} Lazarian A. 2003, Journal of \\
Quantitative Spectroscopy and Radiative Transfer, 79, 881
 \bibitem{} Lazarian A. 2007, Journal of \\
Quantitative Spectroscopy and Radiative Transfer, 106,225
\bibitem{} Lazarian A., Goodman A.A., Myers P.C. 1997, ApJ, 490, 273
\bibitem{} Lazarian A., Roberge W. 1997, ApJ, 484, 230
\bibitem{} Lazarian A., Draine B. 1999a, ApJ, 516, L37 (LD99a)
\bibitem{} Lazarian A., Draine B. 1999b, ApJ, 520, L67 (LD99b)
\bibitem{} Lebedev P. 1901  Ann. der Physik, 6, 433
\bibitem{} Lee H., Draine B. 1985, ApJ, 290, 211
\bibitem{} Mathis J., 1986, ApJ, 308, 281 
\bibitem{} Mathis J., Mezger P., Panagia N. 1983, A\&A, 128, 212
           1977, ApJ, 217, 425 (MRN)
\bibitem{} Novak G., et al. Millimeter and \\
           Submillimeter Detectors for Astronomy II. In Antebi J, D. Lemke, editors.\\
 Proceedings of the SPIE, vol. 5498; 2004 p. 278

\bibitem{} Pelkonen V., Juvela M., Padoan P. 2007, A\&A, 461, 551
 \bibitem{} Purcell E. 1979, ApJ, 231, 404
 \bibitem{} Purcell E., Spitzer L. 1971, ApJ, 167, 31
 \bibitem{} Roberge W., Hanany S. 1990, B.A.A.S., 22, 862
 \bibitem{} Roberge W., Hanany S., Messinger D. 1995, 453, 238
 \bibitem{} Roberge W.G., Lazarian A. 1999, MNRAS, 305, 615
\bibitem{} Rosenbush V., Kolokolova L.,Lazarian A., Shakhovskoy N., \&
  Kiselev N. 2007, Icarus, 186, 317 
\bibitem{} Serkowski K, Mathewson DS, Ford VL 1975, ApJ, 196,261 ≈         
\bibitem{} Spitzer L., McGlynn TA. 1979, ApJ, 231, 417
\bibitem{} Vishniac E., Lazarian A., \& Cho, J. 2003, in Turbulence and
Magnetic Fields in Astrophysics, Eds. E. Falgarone, and T. Passot, LNP, 614,
376
 \bibitem{} Weingartner J., Draine B. 2001, ApJ, 548, 296
\bibitem{} Weingartner J., Draine B. 2003, ApJ, 589, 289 (WD03)
\bibitem{} Ward-Thompson D., Kirk J.M., Crutcher R.M., Greaves J.S., Holland W.S., Andre P. 2000, ApJ, 537, L135
\end{thebibliography}
\end{document}